\documentclass[structabstract]{aa} % for double columns
\usepackage{amsmath}
\usepackage{rotating}
 \usepackage{color}
\usepackage{graphicx} 
\usepackage{natbib}

\usepackage{overpic}
\usepackage{mathptmx}
\usepackage{anyfontsize}

\usepackage{appendix}

\usepackage{lscape} \usepackage{longtable, needspace} \usepackage{ulem}
\def\new#1 {{\bf #1} } \def\cut#1 {\sout{#1}}

\usepackage{subfigure}
\usepackage{txfonts}
\usepackage{tabu}
\usepackage{multirow}
\usepackage{url}
\usepackage{xcolor}
\usepackage[labelfont=bf,singlelinecheck=false]{caption}% Allows to modify the captions

\usepackage{lipsum}

\begin{document}
\title{Resolving the chemical substructure of Orion-KL
\protect\footnote{ Reduced maps as FITS files are only available at the CDS via
anonymous ftp to cdsarc.u-strasbg.fr (130.79.128.5) or via
http://cdsarc.u-strasbg.fr/viz-bin/qcat?J/A+A/
tow}
} \author{S. Feng\inst{1}, 
H.  Beuther\inst{1},   Th.  Henning\inst{1},  D.  Semenov\inst{1}, Aina. Palau\inst{2}, E. A. C. Mills\inst{3}}
\institute{1. Max-Planck-Institut f\"ur Astronomie, 
K\"onigstuhl 17,  D-69117,  Heidelberg, Germany. syfeng@mpia.de \\
2. Centro de Radioastronomia y Astrofisica, Universidad Nacional Autonoma de Mexico, Morelia Michoacan, Mexico\\
3. National Radio Astronomy Observatory, 1003 Lopezville Road, Socorro, NM 87801\\}

\offprints{syfeng@mpia. de} 
\date{\today} 

 \abstract {The Kleinmann-Low nebula in Orion (Orion-KL) is the nearest example of a high-mass star-forming environment. Studying the resolved chemical substructures of this complex region provides important insight into the chemistry of high-mass star-forming regions (HMSFRs),  as it relates to their evolutionary states.  }
{The goal of this work is to resolve the molecular line emission from individual substructures of Orion-KL at high spectral and spatial resolution and to infer the chemical properties of the associated gas.}
{We present a line survey of Orion-KL obtained from  combined Submillimeter Array (SMA) interferometric and IRAM 30\,m single-dish  observations. Covering a 4 GHz bandwidth in total, this survey contains over 160 emission lines from 20 species (25 isotopologues), including 11 complex organic molecules (COMs).
Spectra are extracted from individual substructures and the intensity-integrated distribution map for each species is provided. We then estimate the rotation temperature for each substructure, along with their molecular column densities and abundances.} 
{  For the first time, we complement 1.3\,mm interferometric data with single-dish observations of the Orion-KL region and study small-scale chemical variations in  this region.  
(1) We resolve continuum substructures on $\sim$3\arcsec angular scale. 
(2) We identify lines from the low-abundance COMs $\rm CH_3COCH_3$ and $\rm CH_3CH_2OH$, as well as tentatively detect $\rm CH_3CHO$ and  long carbon-chain molecules $\rm C_6H$ and $\rm  HC_7N$. 
(3) We find that while most COMs are segregated by type, peaking either towards the hot core (e.g., nitrogen-bearing species) or the compact ridge (e.g., oxygen-bearing species like $\rm HCOOCH_3$ and $\rm CH_3OCH_3$),  the distributions of others do not follow this segregated structure (e.g., $\rm CH_3CH_2OH$, $\rm CH_3OH$, $\rm CH_3COCH_3$).  
(4) We find a second velocity component of HNCO, $\rm SO_2$, $\rm ^{34}SO_2$, and SO lines, which may be associated with a strong shock event in the low-velocity outflow.
(5) Temperatures and molecular abundances show large gradients between central condensations and the outflow regions, illustrating a transition between hot molecular core and shock-chemistry dominated regimes.
}
{Our observations of spatially resolved abundance variations in Orion-KL provide the nearest reference source for hot molecular core and outflow chemistry,  which will be an important example for interpreting the chemistry of more distant HMSFRs. }

\keywords{Stars: formation; Stars: massive; ISM: lines and bands; ISM: molecules; ISM: abundance; Submillimeter: ISM} \titlerunning{Resolving the chemical substructure of Orion-KL} \authorrunning{Feng et al.} \maketitle

%______________________________________________________________
\section{Introduction}

The birth sites of high-mass star-forming regions (HMSFRs) are highly complex structures,  consisting of different gas and dust cores, embedded in a less dense gas envelope. While the complexity of dynamic processes within these regions has been characterized well (e.g.,  \citealt{zinnecker07, beuther07,tan14}), our knowledge of the chemical properties of these regions is still poor. 
Since specific molecules can act as tracers of physical processes (e.g., \citealt{schilke97, charnley97,  bachiller01,  viti04,  garrod13}), characterizing the chemistry of gas associated with  the early stages of high-mass stars is crucial for understanding how these stars might form. 
 \\

Line surveys of HMSFRs in mm/sub-mm atmospheric windows  exhibit rich molecular emission spectra (see  \citealt{herbst09} for a comprehensive review). The multitude of detected lines allow for an unbiased look at the chemical composition of a particular region, from which its physical properties  can be inferred, including temperature and density. However, HMSFRs are typically distant (several kpc), so that spatially disentangling the physical and chemical substructures is challenging with the resolution of most facilities.
This makes  the Becklin-Neugebauer and Kleinmann-Low Nebula in Orion (Orion BN/KL), at a distance of $\rm 414\pm7$ pc \citep{sandstrom07, menten07}, ideal for studies of the physical and chemical properties of HMSFRs.  As the nearest HMSFR, it has been extensively studied at both far-infrared to (sub) millimetre wavelengths accessible from space (e.g.,   \citealt{olofsson07, bergin10, crockett14}) and at radio, millimetre, and submillimetre wavelengths accessible from the ground. 
A broad range of  studies on this source  have been conducted either with  single-dish telescopes 
(e.g., \citealt{johansson84, sutton85, blake87, turner89, greaves91, schilke97, schilke01, lee02, comito05, remijan08, tercero10}) or interferometers 
(e.g., \citealt{wright96, blake96, beuther05, beuther06, friedel08, guelin08}).  Strong emission lines in these observations have revealed complex dynamics (e.g., \citealt{sutton85, blake87, turner89, beuther08, beuther10}) and chemistry (e.g., \citealt{wright96, blake96, beuther05, beuther06, friedel08, guelin08, widicus12, friedel12}). \\

Both single-dish-only and interferometer-only observations, however, have limitations.  Single-dish observations have low spatial resolution, making it difficult to detect spatial variations of chemistry on small scales. In contrast, interferometric observations can probe spatial variations on small scales, but suffer from  filtering-out to large scale structures (the ``missing short spacing'' problem; \citealt{bajaja79}), making measurements of the chemistry of more extended features unreliable. 
In this paper, we aim to study the chemistry of Orion-KL by obtaining a combined dataset from an interferometer (Submillimetre Array, SMA)  and a single-dish (IRAM 30\,m) telescope.  These combined data, which resolve small-scale substructures and are sensitive to more extended  components, allow us to accurately map the distribution of molecular species, determine their abundances, and ultimately analyse spatial variations in the chemistry. \\

One unique feature of HMSFRs is  hot molecular cores,  which are compact objects  ($\le 0. 05$ pc)  that exhibit  high temperatures ($\gtrsim100$ K) and densities ($\gtrsim10^6 ~\rm cm^{-3}$; \citealt{palau11}).  Numerous spectral surveys (see, e.g., \citealt{turner89,schilke01,belloche09,herbst09,tercero10}) have revealed that hot molecular cores  are characterized by a large abundance of simple organic molecules,  including  $\rm H_2CO$ (formaldehyde),  HCOOH (formic acid), and complex organic molecules (COMs, organics containing $\ge 6$ atoms), such as $\rm CH_3OH$ (methanol), $\rm HCOOCH_3$ (methyl formate), and $\rm CH_3OCH_3$ (dimethyl ether).   Although some larger COMs have low abundances in hot molecular cores, e.g., $\rm CH_3CH_2OH$ (ethanol; \citealt{rizzo01}),  $\rm CH_3COCH_3$ (acetone;  \citealt{snyder02}),  $\rm CH_3CHO$ (acetaldehyde; \citealt{nummelin00}), $\rm (CH_2OH)_2$ (ethylene glycol;  \citealt{hollis02}),  and $\rm CH_2OHCHO$ (glycolaldehyde;  \citealt{hollis00, hollis04}), they have  been previously detected towards the Galactic-centre hot molecular core source SgrB2 (N-LMH).
Because of their high degree of chemical and structural complexity, the precise origins and formation mechanisms of these COMs is still a subject of debate.  Nevertheless,  understanding these topics is particularly important for understanding the chemical evolution of the interstellar medium (ISM;  e.g.,  \citealt{herbst09}).  
Recently, a number of interferometric observations of COMs in Orion-KL  (e.g., \citealt{friedel05, friedel08, guelin08, favre11a, favre11b, widicus12}) have shown spatial-related chemical differentiation. However,  the relatively low sensitivity of these observations means that the lowest-abundance COMs (e.g.,   $\rm CH_2OHCHO$,  $\rm CH_3CH_2OH$,   and $\rm CH_3CHO$) are below the $\rm 3\sigma$ detection limit or are blended by strong neighbouring lines.  \\

In addition to the hot molecular core chemistry, the shock chemistry driven by outflows  is another important aspect of  chemistry in HMSFRs.  The region in Orion-KL that is most affected by  outflow and shock events is called the ``plateau" \citep{blake87,wright96,lerate08,esplugues13} and is characterized by spectral lines with broad line wings.  Two roughly perpendicular outflows are seen in Orion-KL, traced by the proper motion of $\rm H_2O$ masers at 22 GHz \citep{genzel81}. One is a high-velocity outflow ($\rm 30\text{--}100~km\,s^{-1}$, along the  north-west--south-east (NW--SE) axis); this outflow is also detected in OH (hydroxyl) masers \citep{norris84, cohen06}, broad bipolar CO line wings \citep{zuckerman76,  kwan76, erickson82, masson87, chernin96, beuther08,zapata09}, and lobes of poorly collimated shock-excited $\rm H_2$ \citep{beckwith78,scoville82,taylor84, allen93,sugai94}. 
Although the source responsible for driving the outflow is still uncertain, it has been suggested that the outflow is only about 500--700 years old \citep{lee00, doi02, gomez05, nissen07, gomez08, beuther08, bally11, goddi11a, nissen12}.   
The other is a  low-velocity outflow ($\rm \sim 18\, km\,s^{-1}$, $\rm \triangle V\sim 35~ km\,s^{-1}$, along the north-east--south-west (NE--SW) axis; \citealp{genzel81, gaume98}), which is also detected in SO \citep{plambeck82}, SiO masers \citep{wright83,plambeck09,greenhill13} and $\rm H_2O$ maser \citep{greenhill13}.  This outflow is believed to be closely associated with an expanding gas shell or torus around Source I, but whether its age is  $>3\,000$ years  \citep{menten95, greenhill98,  doeleman99, reid07} or much younger \citep{plambeck09,greenhill13} is still not clear.  
Another goal of this study is to look for additional chemical signposts of these outflows, apart from the known masers, in order to better understand their physical and chemical properties.\\

In this paper,  we present the first combined single-dish and interferometric  line survey at 1.3\,mm  towards Orion-KL and  study the chemical properties of this region at a linear resolution of $\sim$1\,200\,AU (at a distance of $\rm \sim 414$ pc).  In Section \ref{observations},  we describe the observations and their combination.    In Section \ref{obs-res}, we present a continuum map and maps of the distribution  of different molecules, including simple molecules and COMs, where the presences of COMs are identified by synthetic spectrum fittings. The spatial distributions, column densities, and abundances for all the detected species are derived in Section \ref{calcu}. A discussion of the chemistry in the hot molecular core and outflows of Orion-KL can be found in Section \ref{dis}.

\section{Observations and data reduction}\label{observations}
\subsection{Submillimetre Array (SMA)}

We carried out observations of  Orion-KL with the SMA on Feb. 15,  2005 and Feb. 19,  2005.  At
1.3\,mm (220/230\,GHz) in the  compact 
 configuration with seven 
antennas,  the baselines ranged between 16 and 69 m.  
The short baseline cutoff causes source structures $\ge 20$\arcsec to be filtered out.  
The primary beam is 52\arcsec, and  the phase centre of the observations is 
$\alpha_{2000}=\rm 05^h35^m14.5^s$ and
$\delta_{2000}$=$-$5$^{\circ}$22$'$30.45$''$ with 
$\rm V_{lsr}\sim7~km\,s^{-1}$.  Bandpass calibration was done with
{\color{black} Callisto},  and  the  flux  calibration is estimated to be accurate within
15\%,  based on the SMA monitoring of quasars.   Phase and amplitude calibration was performed via frequent
observations of the quasar 0607-157.  The zenith opacities,  measured with the NRAO tipping
radiometre located at the Caltech Submillimetre Observatory,  were  excellent
during both 
tracks with $\tau (\rm 1.3 ~mm)\sim0.03-0.04$,  allowing us to observe at both 1.3\,mm and $\rm 440~\mu$m simultaneously \citep{beuther06}. 
More technical descriptions of the SMA and its calibration schemes can be found in \citet{ho04}. 
The receiver operated in a double-side band mode
with an intermediate frequency of 4--6\,GHz so that the upper and lower
side bands were separated by 10\,GHz.  The correlator had a bandwidth of
2\,GHz, and the channel separation was 0.8125\,MHz,  corresponding to a velocity resolution of $\rm \sim1.2~km\,s^{-1}$.  
Double-sided band system temperatures corrected for 
the atmosphere were between 150 and 300\, K (depending on the elevation of the source).   \\

The initial flagging and calibration was done with the IDL superset
MIR, which is originally developed for the Owens Valley Radio Observatory
and adapted for the SMA\footnote{
http://cfa-www.harvard.edu/$\sim$cqi/mircook.html. }.  The imaging and
data analysis was conducted in MIRIAD
\citep{sault95} and CASA\footnote{http://casa.nrao.edu}.  Between the observations on two nights,  the frequency setups were shifted by 0.4 GHz, yielding a slightly larger total bandwidth for the combined dataset ($\rm 218.870-221.234~ GHz$ for the lower sideband,  and $\rm 228.870-231.234~ GHz$ for the upper sideband).   The synthesised beams of the combined data are $3.49'' \times2.57'' $ for the lower sideband and $3.68'' \times2.69'' $  for the upper sideband. 

\subsection{Single-dish observations with the IRAM 30~m telescope}
In addition to the high spatial  resolution data from the SMA, we mapped an area of
$1'\times1'$, with the IRAM 30~m telescope,  to complement the short-spacing information missing from the interferometric observations. Observations were performed in the on-the-fly mode on  Nov. 30, 2012, covering
a broad bandpass (8 GHz  bandwidth for each sideband) of EMIR at 220/230 GHz.  
 Observations were conducted under good weather conditions ($\rm T_{sys}=266$ K,  $\rm \tau \sim 0.152$ at 220 GHz).  
The focus was on Uranus, and pointing was checked on Uranus and 0420-014.    
Using a forward efficiency of $92\%$ and a main beam efficiency of $58\%$,  we converted the data from
antenna temperature (K)  to flux (Jy) by using an efficiency\footnote{
http://www.iram.es/IRAMES/mainWiki/Iram30mEfficiencies}  of  7.8.   The
beam of the 30~m telescope at 1.3\,mm is
$\sim11.3''$ in the upper sideband and $\sim11.8''$ in the lower sideband. The observations cover frequencies $\rm 213.370\text{--}221.150$ GHz and $\rm 229.051\text{--}236.831$, with a velocity resolution of 0.255$\rm ~km\,s^{-1}$.
We used the Gildas\footnote{http://www.iram.fr/IRAMFR/GILDAS} software for data reduction,  and then converted the 30\,m data to the MIRIAD data format.\\

\subsection{Data merging}
The 30\,m single-dish spectral-line data were first converted to visibilities, and then combined with the SMA data using the MIRIAD package task UVMODEL\footnote{https://www.cfa.harvard.edu/sma/miriad/manuals/SMAuguide/ smauserhtml/uvmodel.html}.   To recover large-scale emission,  while simultaneously maintaing high spatial resolutions for resolving small scale structure,  we chose an intermediate weighting between natural and uniform (``robust"$=-0.5$ value in MIRIAD).  However, as continuum information is not available from the 30\,m observations, only the spectral-line data can be corrected in this way.\\

\begin{figure}
\centering
\includegraphics[width=8cm]{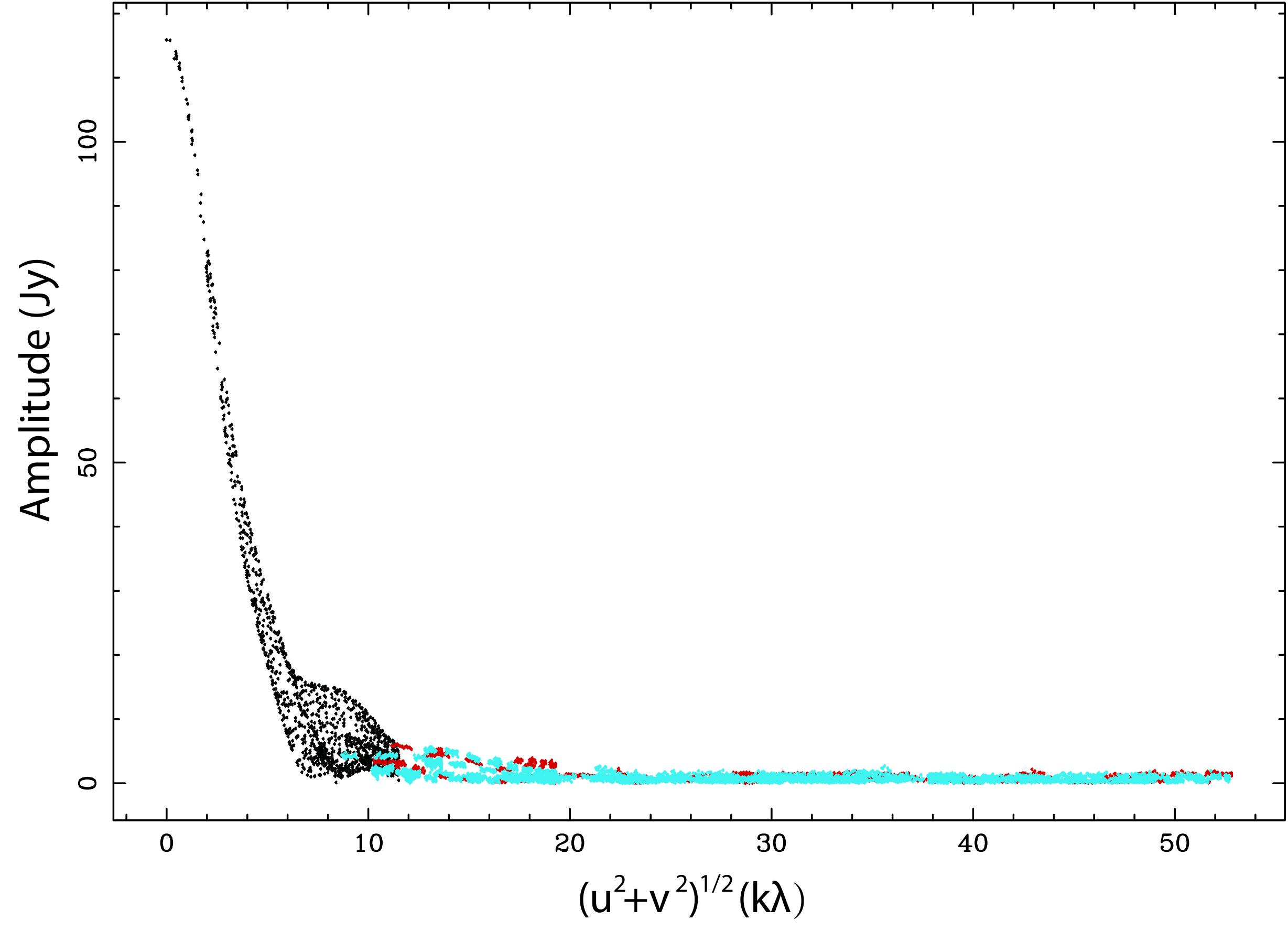}
\includegraphics[width=8cm]{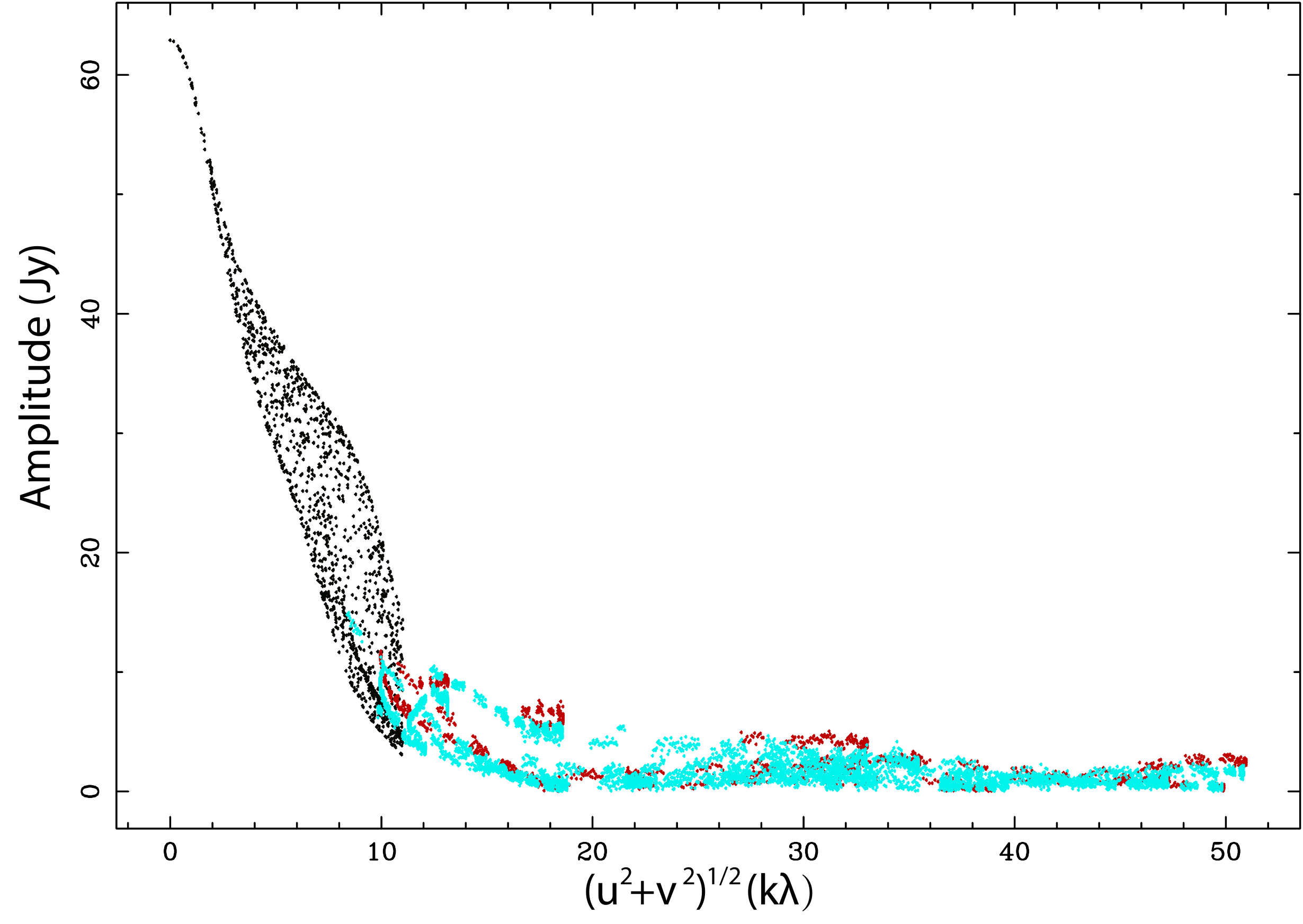}
\caption{ Amplitude in Jy versus the projected baseline  showing the combination of the interferometer and single-dish data,  from the upper sideband (top) to lower sideband (bottom).  Red and blue dots present data from the SMA on Feb. 15,  2005 and Feb. 19,  2005,  respectively,  showing the lack of baselines shorter than 10$\rm \,k \lambda$.  Black dots are data converted from 30\,m single-dish observations,  which fill in  the missing short spacing information. 
\label{uvplt}}
\end{figure}

The combined data have a field of view of 52\arcsec and a beam of $5.55'' \times4.35'' $ with position angle $\rm P.A.=+8\degr$ in the lower sideband and $5.44'' \times4.43'' $ with $\rm P.A.=+2\degr$ in the upper sideband. The frequency overlap between the SMA and 30\,m data results in the combined data having a frequency range
$218.870\text{--}221.150\rm\, GHz$  and $229.051\text{--}231.234\,\rm GHz$.   Figure~\ref{uvplt} presents an amplitude-versus-projected uv distance plot, showing the good combination of the single-dish and interferometer data.
 Missing projected baselines shortwards of 10$\rm~k \lambda$ from the SMA have been complemented  by using the 30\,m data. After combination, the $1\sigma$ rms per $\rm 1.2~km\,s^{-1}$ channel is measured to be $\rm 80~mJy~beam^{-1}$ for the lower sideband  and $\rm 70~mJy~beam^{-1}$ for the upper sideband.\\

\section{Observational results}\label{obs-res}

\subsection{Continuum emission}\label{continuum}
\begin{figure*}[!ht]
%\centering
%\begin{overpic}[scale=0.8, grid, tics=10]{orion_conti_jet}
\begin{overpic}[scale=0.75]{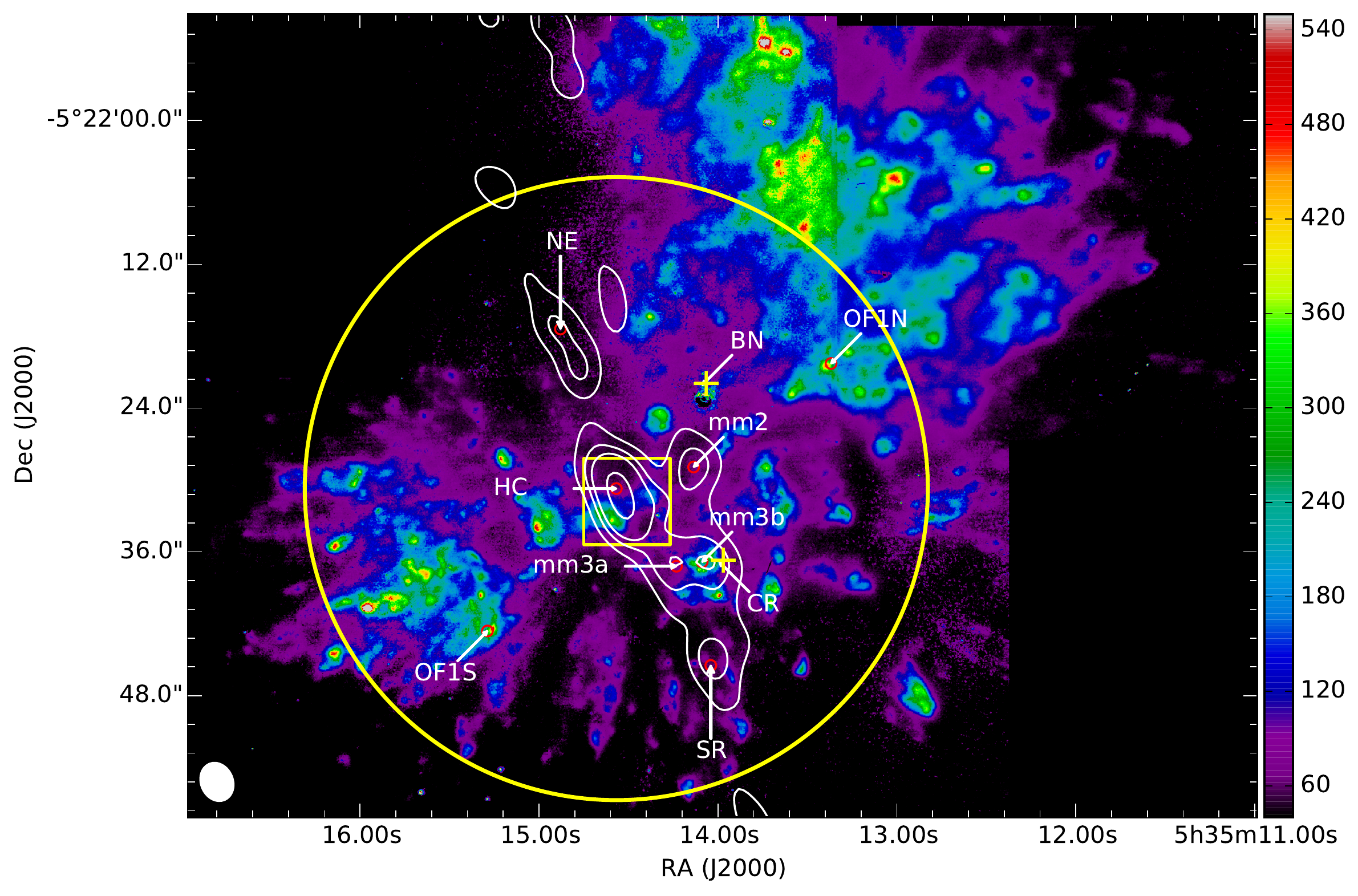}
  \put(63.5, 7){%
    \includegraphics[scale=.23]%
                    {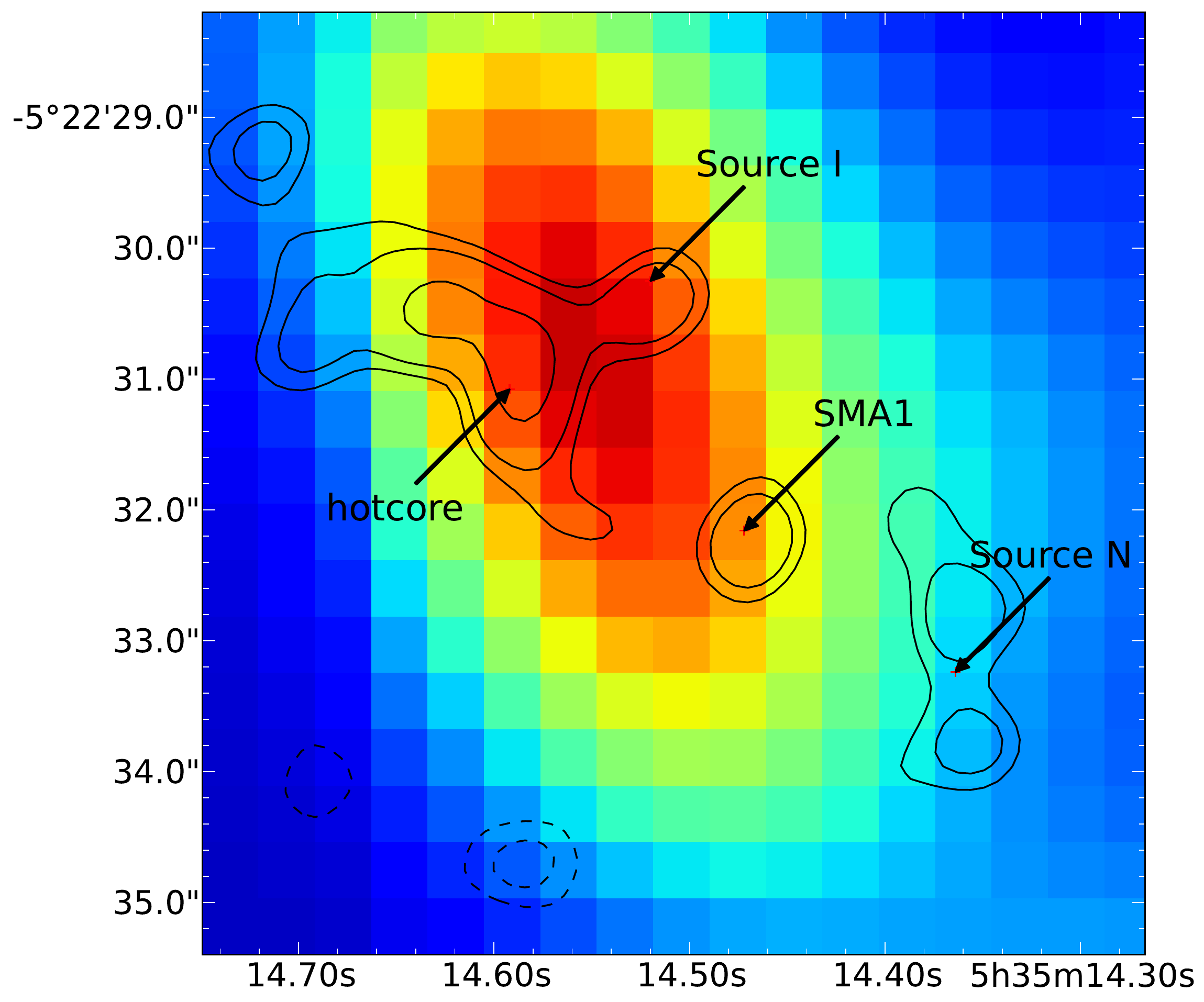}}

\end{overpic}
\caption{Substructures resolved by SMA-only observations at a spatial resolution of $\rm \sim1200\,AU$. White contours show the 1.3\,mm continuum emission at -5$\sigma$,  5$\sigma$,  15$\sigma$, 25$\sigma$,  and 60$\sigma$ rms  levels 
with the $\rm \sigma=0.04~Jy~beam^{-1}$.  The yellow circle denotes the primary beam of SMA at 1.3\,mm.
Red circles denote the peaks of hot core (HC),  mm2,  mm3a, mm3b,  the southern region (SR),  north-east clump (NE),  the NW and SE parts of the high-velocity outflow (OF1N, OF1S). Yellow crosses denote the BN object and the compact ridge (CR). The beam in the bottom left corner is from the SMA-only data.  Colourmap shows the  brightness (in counts per 400 s) of shocked $\rm H_2$ emission \citep{nissen07}.  Corresponding to the yellow square,  the insert panel shows the HC continuum from the SMA at 1.3\,mm   in detail as colourmap,  with black contours of the 865 $\mu m$ continuum emission at a spatial resolution of $\rm \sim300\,AU$ from \cite{beuther04} overlaid. 
}\label{conti}
\end{figure*}

Orion-KL has a rich  spectrum of molecular emission lines,  so extracting the continuum requires some care.  To construct our continuum map, we averaged  the  line-free parts of the upper and lower sidebands from the SMA-only data. In Figure~\ref{conti}, we present  the resulting 1.3\,mm  continuum image as contours overlaid on a map of the outflow traced by shock-excited  $\rm H_2$ \citep{nissen07}.  
 We detected the previously known compact continuum substructures of this region and moreover, resolved some of these substructures into multiple distinct condensations. 
We list the following resolved condensations in Table \ref{source}: 
\begin{itemize}
\item The brightest hot core (HC):  Located close to a well-studied infrared source IRc2 \citep{downes81,wynn84}, the brightest HC  
is the first clump in Orion-KL detected by the high spatial resolution mm observations \citep {masson85, mundy86, woody89, wright92}, and it exhibits the highest  peak specific intensity ($\rm Jy~beam^{-1}$) in the 1.3\,mm SMA continuum map. With the temperature estimated to be 100--300 K (e.g.,  \citealt{wilner94, wilson00, beuther05}),  this source has been found to be abundant in hydrogenated species and nitrogen- (N-) bearing molecules \citep{sutton95, blake96,wright96, wilson00}. 
With even higher resolution ($\rm 0.7''$, $\sim$300\,AU), observations from SMA at $\rm 865~\mu m/348~ GHz$ expose additional substructures within  the HC \citep{beuther04}. Presented as black contours overlaid on the 1.3\,mm colourmap in the inner panel of  Figure~\ref{conti}, these 865 $\mu$m structures are: 
SMA1 (a high-mass protostellar source, suggested to be the driving source of the high-velocity outflow; \citealt{beuther08}),  Source I (located at the centre of the SiO masers  
and believed to drive the bipolar low-velocity outflow;  \citealt{menten95,matthews10}), Source N (a Herbig Ae/Be or mid-B star with a luminosity of $\sim 2\,000\,L_{\odot}$; \citealt{menten95, greenhill04, gomez05, rodrquez05, goddi11a, nissen12}), and hotcore. Hereafter, we use ``HC" to denote the whole region resolved by 1.3\,mm continuum at a spatial resolution of $\sim$1\,200 AU, the ``hotcore" to denote the position resolved by $\rm 865~\mu m$ continuum at a spatial resolution of  $\sim$300 AU, and the ``hot molecular core" to describe an evolutionary status.\\

 \item mm2: Located $\rm \sim7''$ west of the HC, and $\rm \sim5''$ south of the BN object.
 Coincident with infrared sources IRc3, IRc6, and IRc20 \citep{dougados93}, this condensation is also known as the western clump (WC;  \citealt{wright92}), or the north-west clump (NWC; \citealt{blake96,tang10,favre11a}). The mass (2--3\,$\rm M_{\sun})$, $\rm H_2$ column
density, and  high temperature (a few hundred K) in this source suggest that it may be the site of ongoing low-mass star formation \citep{schreyer99}.\\
\item mm3a and mm3b: These two condensations are located south-west of the HC. Coincident with infrared IRc5  \citep{dougados93},  they were previously considered to be a single clump called the south-west  clump (SWC; \citealt{tang10}), or CntC \citep{murata92},  situated at the compact ridge (CR; \citealt{blake96,wright96,tang10}). Similar dual continuum peaks have been detected in other  high spatial resolution observations of HMSFRs \citep{tang10,favre11a,friedel11}, so we consider them to be separated. The CR  is known to have different kinematics and chemistry than in  the HC: it is a vast region ($\sim 15\arcsec\text{--}25\arcsec$; \citealt{plume12,esplugues13}) with  lower temperature \citep{genzel89}, is dominated by oxygen- (O-) bearing molecules \citep{blake87, wright96, liu02}, and is believed to be the region where two outflows  interact with the dense ambient material \citep{blake87, liu02}. Therefore, to ensure that we fully characterise its chemical properties, the following study focuses separately on mm3a and mm3b, instead of considering CR as a whole. \\
\item The southern ridge (SR):  the southernmost condensation detected in the 1.3\,mm continuum map. Although it has been detected in other observations \citep{eisner08,favre11a}, it has not been characterized in detail.\\
\item The north-east clump (NE): a strong continuum emission peak which is adjacent the HC and the north extended ridge (ER; \citealt{sutton85,blake87}). This clump is usually filtered out  in high spatial resolution observations, but can be clearly detected in interferometric observations made with large scale configurations \citep{murata92,tang10}. Comparison with the existing single-dish maps suggests it  is part of a large-scale dusty filament along the NE--SW direction (e.g.,  detected by JCMT at $450~\mu$m, \citealp{wright92}; SCUBA at $450~\mu$m and $850~\mu$m, \citealp{di08}). In addition, this clump is in the plateau of the low-velocity outflow, so its chemistry may be affected by shocks (Section~\ref{outflow}). \\
\end{itemize}
Because of the limited UV coverage and  field of view, the high-velocity outflow and the BN object are not detected above $\rm 5\sigma$ in the SMA-only continuum map.  
To study the chemistry in these regions from the combined data, we take the positions of the high-velocity plateau  ($\rm  V_{lsr}\sim 11~km\,s^{-1}$) and BN object from  \citet{beckwith78} and denote them as OF1N, OF1S, and BN in Figure~\ref{conti}.\\

Table~\ref{source} lists the nominal absolute positions of these substructures, as well as the peak specific intensity ($\rm Jy ~beam^{-1}$) for each continuum  substructure, obtained from the SMA-only 1.3\,mm continuum map (Figure~\ref{conti}, for more detail about this table, see Section \ref{continu}). The HC is the brightest condensation; mm2, mm3a, and mm3b  are fainter; while NE, SR, OF1N, and OF1S are even  weaker. \\

\begin{table*}[!ht]

\centering
\caption{Properties of substructures, which are denoted on the continuum image obtained from SMA observations at 1.3\,mm (Figure~\ref{conti}).   
\label{source}}
\scalebox{0.95}{
\begin{tabular}{lc|cccccc|p{4cm}}
\hline\hline
Source & &R. A. ($\alpha$) &Dec. ($\delta$)   &$\rm I_\nu$$^a$ &$\rm T_{rot}$     &$\rm N_{H_2,1}^{\it d}$    &$\rm N_{H_2,2}^ {\it e}$  &Alternative Designations \\
&&[J2000] &[J2000]    &(Jy/beam) &(K)                  &($\rm 10^{23}cm^{-2}$)    &($\rm 10^{23}cm^{-2}$)  \\
&&$\rm 5h35m\sim$   &$\rm -5^{o}22^{'}\sim$ &&&$\rm 3.49\arcsec \times 2.57\arcsec$ &$\rm 5.55\arcsec \times 4.35\arcsec$ &\\
\hline
                            & &      &        &                                                 &$155\pm16^{b_1}$  &$23.7\pm4.6$        &                              &\\
                           & &      &        &                                                   &$121\pm16^{b_2}$   &$31.0\pm6.7$        &                          &\\
 HC                     & &14.6s      &$31.0^{''}$        &$2.75\pm 0.3$   &$126\pm13^{b_3}$  &$29.6\pm5.8$        &$11.9\pm0.7$     &Ca$^{j}$, SMM3$^{l}$, CntB$^{k}$\\
                             & &      &        &                                                 &$225\pm19^{b_4}$  &$16.3\pm2.9$        &                              &\\
                            & &      &        &                                                 &$199\pm45^{b_5}$  &$18.4\pm5.1$        &                              &\\
\cline{2-8}
&hotcore$^f$ &14.58s   &$30.97^{''}$  &&&& &\\
&SMA1$^f$ &14.46s   &$32.36^{''}$ &&&&&\\
&Source I$^g$   &14.51s   &$30.53^{''}$ &&&&&\\
&Source N$^g$   &14.35s  &$33.49^{''}$ &&&&&\\

\hline
         & &      &        &                                                     &$108\pm4^{b_1}$  &$11.0\pm1.7$        &                          &\\
mm2 &  &14.1s      &$27.1^{''}$        &$0.87\pm 0.10$   &$112\pm7^{b_3}$  &$10.5\pm1.8$           &$9.3\pm0.1$       &NWC$^{h,i}$, Cc$^{j}$, SMM2$^{l}$, \\
         & &      &        &                                                     &$174\pm14^{b_4}$  &$6.7\pm1.2$        &                          &WC$^{n}$, CntD$^{k}$\\
         & &      &        &                                                     &$173\pm40^{b_5}$  &$6.7\pm1.9$        &                          &\\
\hline

CR$^l$  & &14.0s      &$36.9^{''}$        &  &  &           &    &SWC$^{i}$, SMM1$^{l}$\\
\cline{2-9}
&             &                &                        &                           &$103\pm3^{b_1}$  &$12.8\pm1.6$        &                          &\\
&mm3a   &14.3s      &$37.9^{''}$        &$0.96\pm 0.09$  &$101\pm5^{b_3}$  &$13.1\pm1.8$           &$6.2\pm0.07$    & Cb1$^{j}$, CntC$^{k}$\\
&             &                &                        &                           &$159\pm12^{b_4}$  &$8.1\pm1.3$        &                          &\\
&             &                &                        &                           &$101\pm26^{b_5}$  &$13.1\pm3.7$        &                          &\\

\cline{2-9}
&             &                &                        &                           &$88\pm6^{b_1}$  &$15.9\pm2.7$        &                          &\\
&mm3b   &14.1s      &$36.4^{''}$        &$1.01\pm 0.11$   &$102\pm5^{b_3}$  &$13.5\pm2.1$           &$6.7\pm0.05$         & Cb2$^{j}$, CntC$^{k}$\\
&             &                &                        &                           &$160\pm17^{b_4}$  &$8.5\pm1.6$        &                          &\\
&             &                &                        &                           &$89\pm24^{b_5}$  &$15.6\pm4.8$        &                          &\\

\hline
NE & &14.8s      &$19.1^{''}$        &$0.38\pm0.09$   &$43\pm3^{b_1}$  &$20.1\pm4.2$         &$3.3\pm0.06$      &CntH$^{k}$\\

\hline 
SR & &14.0s               &$45.4^{''}$        &$0.41\pm 0.06$     &$43\pm3^{c}$  &$14.2\pm2.9$        &$1.0\pm0.02$      &Cd$^{j}$, MM23$^{m}$\\
\hline
OF1N  & &13.4s      &$20.5^{''}$        &$\rm \sigma=0.04$  &$43\pm3^c$   &$<4.2$             &$\sim1.1$                 &PK1$^{o}$\\
\hline
OF1S  & &15.3s      &$42.5^{''}$        &$\rm \sigma=0.03$   &$43\pm3^c$ &$<3.1$            &$\sim0.6$                   &PK2$^{o}$\\
\hline
BN$^o$ & &14.11s &$22.7^{''}$ &&&&&\\
\hline
\hline
\end{tabular}
}

\begin{tabular}{l}
\scriptsize{Notes. Positions with upper case are from the references listed in the footnotes,  without are from 1.3\,mm SMA continuum observations.}\\
\scriptsize{$a$. Peak specific intensity $\rm I_\nu$ is obtained from each substructure of SMA-only continuum.} \\
\scriptsize{$b_1$. Excitation temperature is from $\rm CH_3CN$ rotation map fitting in Figure~\ref{rotation}.} \\
\scriptsize{$b_2$. Excitation temperature is from $\rm CH_3^{13}CN$ rotation map fitting in Figure~\ref{rotation}.} \\
\scriptsize{$b_3$. Excitation temperature is from $\rm HCOOCH_3$ rotation map fitting in Figure~\ref{fig:trot_other}.} \\
\scriptsize{$b_4$. Excitation temperature is from $\rm CH_3OH$ rotation map fitting in Figure~\ref{fig:trot_other}.} \\
\scriptsize{$b_5$. Excitation temperature is from $\rm ^{34}SO_2$ rotation map fitting in Figure~\ref{fig:trot_other}.} \\
\scriptsize{$c$. Excitation temperature is assumed the same as that in NE from $\rm CH_3CN$.}\\ 
\scriptsize{$d$. $\rm H_2$ column density is calculated from the SMA-only continuum at substructure peak using Eq.~\ref{gas}, performed in Section \ref{continu}. }\\ 
\scriptsize{$e$. $\rm H_2$  column density is calculated from the conversion of combined SMA-30\,m $\rm C^{18}O$ at temperatures derived from $\rm HCOOCH_3$, performed in Section  \ref{continu}.}\\ 

\scriptsize{{\bf References. }$f$. \citet{beuther04}; }\\
\scriptsize{$g$. \citet{menten95}; }\\
\scriptsize{$h$. \citet{blake96};} \\
\scriptsize{$i$. \citet{tang10};}\\
\scriptsize{$j$. \citet{favre11a};}\\ 
\scriptsize{$k$. \cite{murata92};} \\
\scriptsize{$l$. \citet{zapata11}; }\\
\scriptsize{$m$. \citet{eisner08}};\\
\scriptsize{$n$. \citet{wright92};}\\
\scriptsize{$o$. \citet{beckwith78}} \\
\end{tabular}

\end{table*}

\subsection{Spectral line emission}\label{lineemi}

\subsubsection{Line identification}\label{identify}
\begin{figure*}[!ht]
\centering
\includegraphics[width=17.5cm]{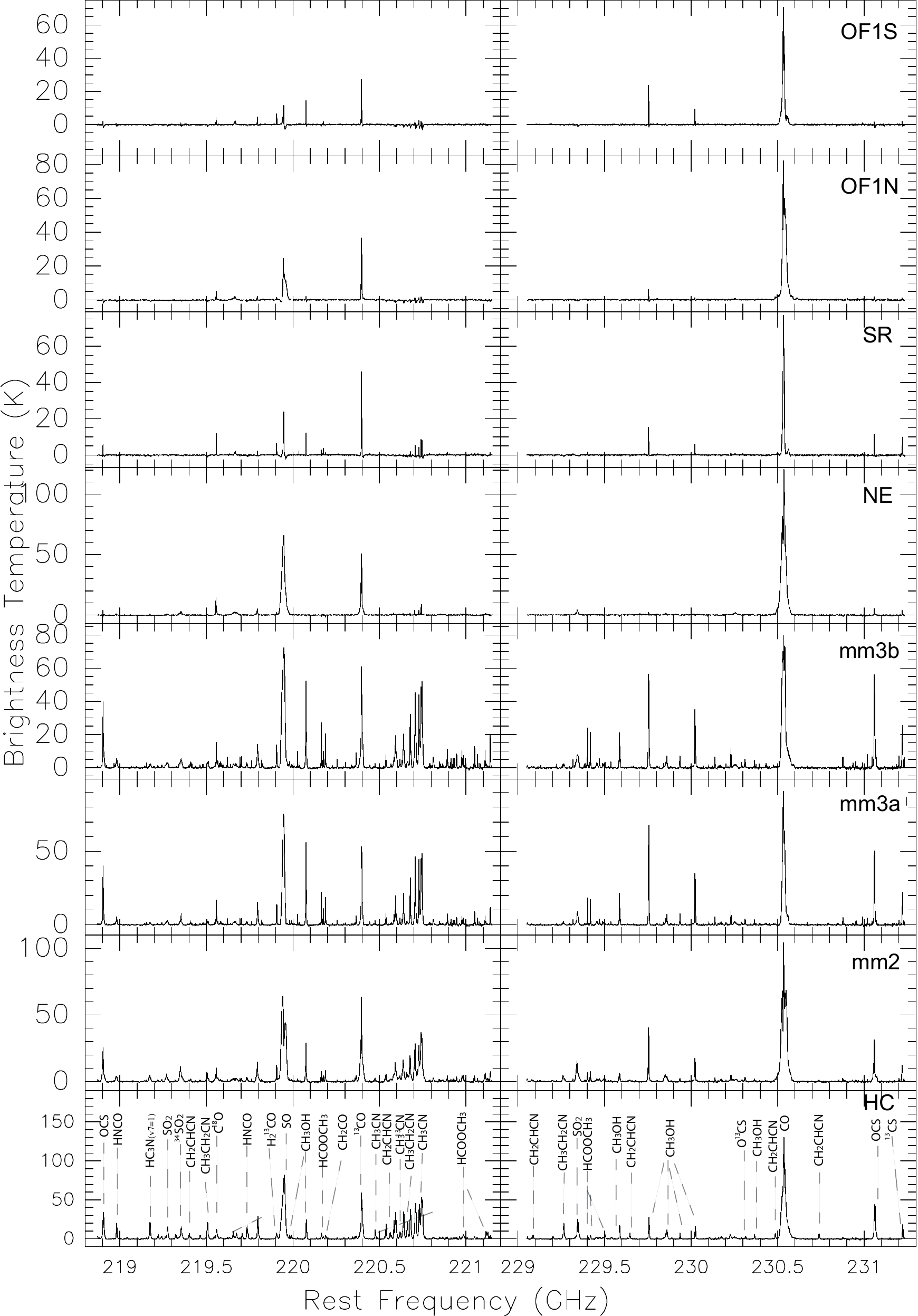}
\caption{Spectra taken towards selected positions (denoted by red circles in Figure~\ref{conti}) after imaging the whole data cube. The spectral resolution is smoothed to $\rm 1.2~km\,s^{-1}$ after the combination of SMA and 30\,m data, and strong emission lines are labelled. A blow-up figure of the spectrum in the HC is Figure~A1.
 \label{co_spec} }
\end{figure*}

After imaging the whole SMA-30\,m combined data cube,   we extracted the double sideband spectrum from each substructure denoted  in Figure~\ref{conti} and show them in Figure~\ref{co_spec}.  
The majority of lines are strongest towards  the HC,  so we use the spectrum from the HC to identify lines.  
 However, line identification may be ambiguous in a region such as the Orion-KL HC,  because of a high density of lines (that can have different line widths according the species),  also because the  $\rm \sim 1.2~km\,s^{-1}$ spectral resolution of the observations does not allow us to separate all the blending.
For strong lines,   
we first compare their rest frequencies with an existing single-dish survey catalogue of Orion-KL by \cite{sutton85}. Next, we cross-check their rest frequencies with the molecular database at ``Splatalogue"\footnote{www.splatalogue.net} (a compilation of the Jet Propulsion Laboratory (JPL\footnote{http://spec.jpl.nasa.gov},  \citealp{pickett98}),   Cologne Database for Molecular Spectroscopy catalogues (CDMS\footnote{http://www.astro.uni-koeln.de/cdms/catalog},  \citealp{muller05}),  and Lovas/NIST catalogues, \citealp{lovas04}), and confirm the identification  if the CDMS/JPL intensity is $\rm >10^{-6} ~nm^{2}MHz$ \footnote{ This is a coarse cutoff for identifying the strong emission lines, corresponding to spontaneous emission rate $\rm A_{ul}\sim10^{-5.5\sim-6}$. Taking  the frequency resolution of our observations  into consideration ($\sim0.8$ MHz), we found that 50\% number possibility at certain rest frequency of a particular line has CDMS/JPL intensity $>10^{-6}$.}. \\

\onecolumn
\twocolumn

Using the above method, the detection of strong lines are shown in Figure~\ref{blow}. In addition,  we can also detect multiple transitions of low-abundance COMs in Orion-KL, owing to the high sensitivity and  recovering the full intensity for spatially extended molecular emission from the combined dataset.  We use a synthetic spectrum fitting program \citep{sanchez11, palau11} in conjunction with molecular data from the JPL/CDMS catalogues, to simultaneously fit multiple COM lines under the assumption that all transitions are optically thin, are in local thermodynamic  equilibrium (LTE), and have the same line width\footnote{The synthetic fitting program we use here is for line detection, but not for finding the best fit of column density or temperature to the observations.  Since the synthetic fitting is performed with optically thin assumption, minor lines  of a particular species which cannot fit as well as the majority may be optically thick.}. Given these assumptions, the fitted flux densities are not always a perfect match to the observed spectra.  However, in many cases we are able to robustly identify multiple transitions of a COM (Figure~\ref{COMspec}). We discuss the properties of the identified COM species further in Section \ref{image}.  \\

With these two approaches,  
we identified more than  160 lines that we  assigned to 20 molecules (including 25 isotopologues) over the entire band; we labelled the unblended strong emissions in Figure~\ref{co_spec},  and list all the lines in Table~\ref{tab:line}.   This table includes the rest frequency,  quantum number,  symmetry label,  and upper state energy for each transition. Lines marked with ``$\dag$" are from molecules containing less than six  atoms as shown in  Figure~\ref{into},  and those marked with  ``$\ddag$" are from molecules we designated as  COMs and carbon chains, which are shown in  Figure~\ref{COMdis}. 
 Our observations have a frequency resolution of $\rm 0.8125~ MHz$ ($\rm \sim1.2~ km\,s^{-1}$ at 1.3\,mm). Thus, some observed lines with broad linewidths can be attributed to multiple transitions, when   
compared to the laboratory-measured rest frequencies recorded in the ``Splatalogue" database. If  potentially blended transitions for one line  are from different isotopologues but have similarly strong CDMS/JPL intensities, we compared the spatial distribution of this line to those of the known isotopologues. Then, we either confirm  a match or indicate all possibilities  with ``*" in Table~\ref{tab:line}. If there are multiple possible transitions from the same isotopologue at this rest frequency, they are also marked  with ``*".

%%%%%%%%%%%%%%%%%%%%%%%%%%%%%%%%%%%%%%%
%%%%%%%%%%%%%%%%%%%%%%%%%%%%%%%%%%%%%%%%
%%%%%%%%%%%%%%%%%%%%%%%%%%%%%%%%%%%%%%%%

\subsubsection{Line imaging}\label{image}

In general, all the identified species show different spatial distributions, velocity widths (FWHM  $\rm \Delta V$), and velocity at line centre ($\rm V_{peak}$) in each substructure (see Figure~\ref{velpro}, on average, $\rm V_{peak}\sim 6\text{--}8~km\,s^{-1}$ in the HC; $\rm V_{peak}\sim 8\text{--}10~km\,s^{-1}$ in mm2, mm3a, and mm3b;  $\rm V_{peak}\sim 10\text{--}12~km\,s^{-1}$ in NE).  
However,  transitions from a single isotopologue have the same spatial distribution (except for $\rm CH_3OH$). In this paper, we have
chosen one of the strongest unblended transitions from each isotopologue  and produced their intensity distribution maps from the SMA-30\,m combined data. In avoiding contamination from other strong  lines, we integrate the intensity of each line through 40\%-60\% of their profile range in Figure~\ref{velpro}\footnote{The chosen velocity ranges are slightly larger than the FWHM of the lines, when  accounting for the small variation in the peak velocity between different substructures.}.\\

\onecolumn
\begin{figure*} [!ht]
\small
\begin{center}

%\begin{tabular}{lll}
%\begin{sideways}
%\Large$\triangle\delta ['']$
%\end{sideways}
%&\includegraphics[scale=.52] {S_bearing}
 %&\includegraphics[scale=.52]{CO_bearing}\\
 %&\includegraphics[scale=.52] {N_bearing}
%&\\
 %&&\Large$\triangle\alpha ['']$
%\end{tabular}

 \begin{tabular}{l}
 \includegraphics[width=18cm]{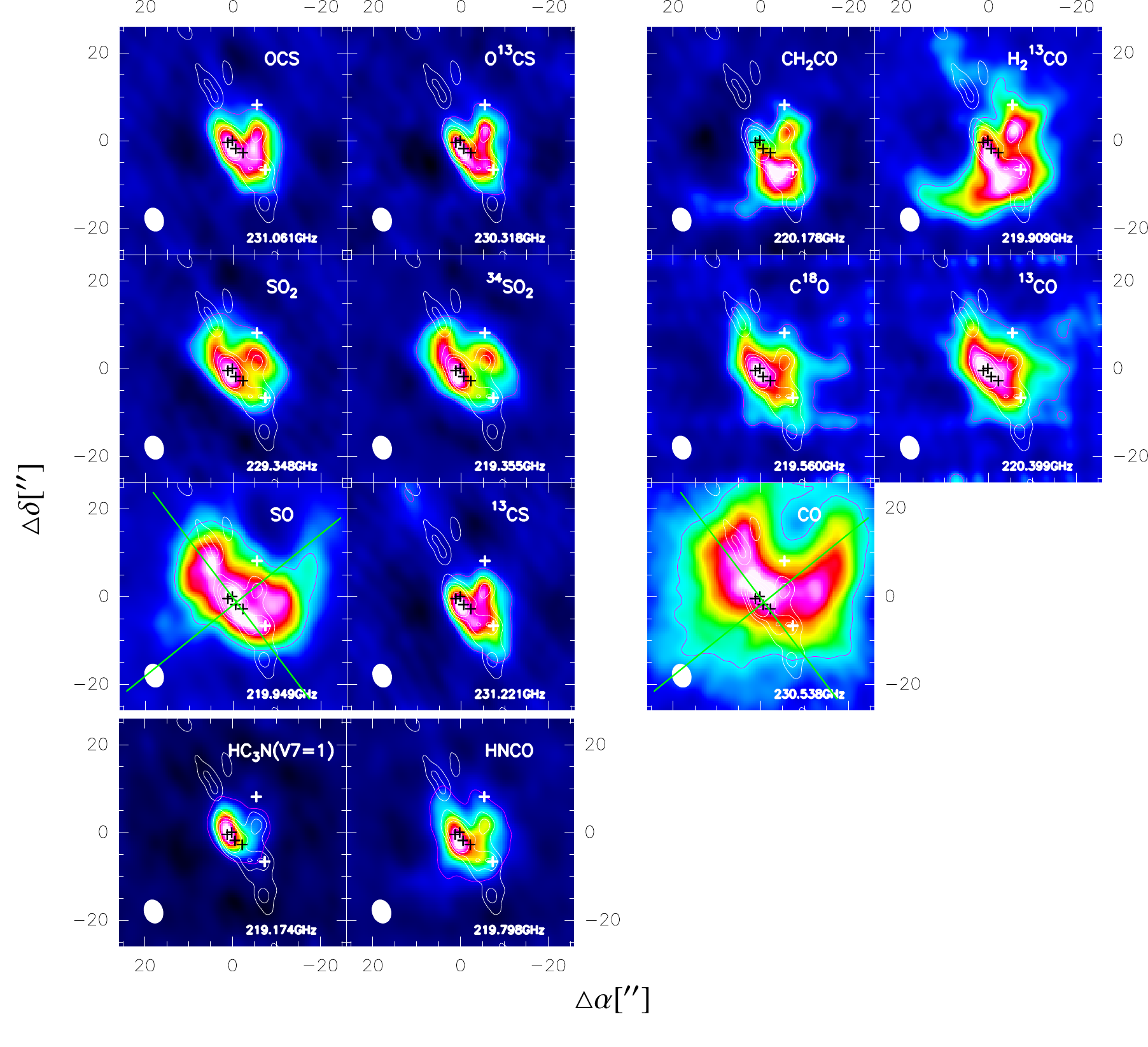}
\end{tabular}

\caption{ Intensity distribution maps of  transitions from molecules containing $<$6 atoms, which are detected in the combined SMA-30\,m data at 1.3\,mm  (corresponding quantum numbers are listed in Table~\ref{tab:line}, with ``$\dag$" marks). The intensities have been derived by integrating the line emission over the velocity range shown in Figure~\ref{velpro}.
The filled beam in the bottom left corner is from  SMA-30\,m data.   White contours show the  continuum from the SMA-only data  (at -5$\sigma$,  5$\sigma$,   15$\sigma$, 25$\sigma$,  and 60$\sigma$ levels).  Purple contours show  $\rm \pm 1\sigma$ levels of the molecular emission (solid contours indicate positive flux, and dashed contours indicate negative flux).  Green lines in the SO and CO maps sketch  the outflow directions.  The white crosses denote the BN object and the CR, and the black crosses denote the positions of the hotcore,  Source I,  Source N, and SMA1,  as in Figure~\ref{conti}. {\color{black}All images have different colour scales  (in $\rm Jy\, beam^{-1}\, km s^{-1} $), increasing from black to white, which are optimized to emphasize the features in the distribution of each molecules.}
}\label{into}
%\end{figure}
\end{center}
\end{figure*}

\begin{figure*}[!ht]
\small
\begin{center}
%\begin{tabular}{lll}
 %\begin{sideways}
%\Large$\triangle\delta ['']$
 %\end{sideways}
%&\includegraphics[width=8cm]{CHO-COM_bearing-eps-converted-to.pdf}
%&\includegraphics[width=8cm]{CHN-COM_bearing-eps-converted-to.pdf}\\
%&\includegraphics[width=4.35cm]{CH3CHO-COM_bearing-eps-converted-to.pdf}\\
 %&&\Large$\triangle\alpha ['']$\\
 % \multicolumn{1}{l}{* Channel maps of more $\rm CH_3OH$ lines are shown in Figure~\ref{ch3oh_cha}.}\\
% \end{tabular}
 
 \begin{tabular}{l}
 \includegraphics[width=18cm]{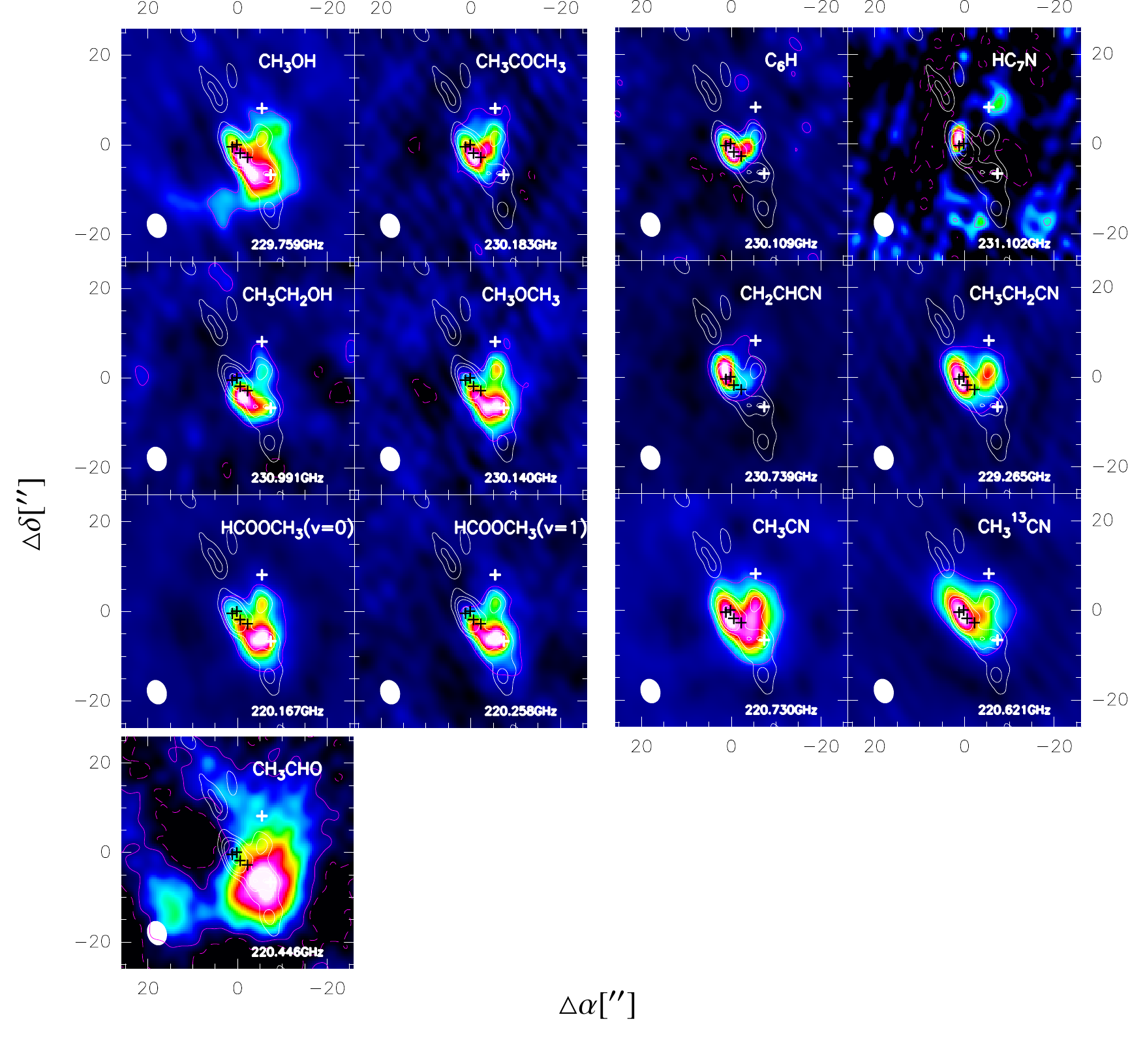}\\
* Channel maps of more $\rm CH_3OH$ lines are shown in Figure~\ref{ch3oh_cha}.\\
\end{tabular}

\caption{ Intensity distribution maps of  transitions from COMs and carbon chains detected in the combined SMA-30\,m data at 1.3\,mm  (corresponding quantum numbers are list in Table~\ref{tab:line}, with ``$\ddag$" marks). 
The intensities have been derived by integrating the line emission over the velocity range shown in Figure~\ref{velpro}.
The filled beam in the bottom left corner is from  SMA-30\,m data.   White contours show the  continuum from the SMA-only data  (at -5$\sigma$,  5$\sigma$,   15$\sigma$, 25$\sigma$,  and 60$\sigma$ levels).  Purple contours show  $\rm \pm 1\sigma$ levels of the molecular emission (solid contours indicate positive flux, and dashed contours indicate negative flux).   The white crosses denote the BN object and the CR, and the black crosses denote the positions of the hotcore,  Source I,  Source N, and SMA1,  as in Figure~\ref{conti}. {\color{black}All images have different colour scales  (in $\rm Jy\, beam^{-1}\, km s^{-1} $), increasing from black to white, which are optimized to emphasize the features in the distribution of each molecules.}
}\label{COMdis}
\end{center}
\end{figure*} 
\twocolumn

The combined data at 220/230 GHz covers a  field of view of 52\arcsec,  which is greater than the extent of the source (not including the outflows),  so our maps exhibit negligible artefacts caused by residual missing flux. The high quality of our data allows us to map molecules, such as  $\rm ^{34}SO_2$, $\rm O^{13}CS$,  HNCO, $\rm H_2^{13}CO$, $\rm ^{13}CO$, and $\rm CH_2CO$ for the first time with both high resolution and sensitivity to spatial information on all scales. After sorting by different groups, we present the intensity integration map for each identified molecule in Figures~\ref{into} and~\ref{COMdis}. Compared to previous sub-mm and mm (especially interferometric) observations in Orion-KL (listed in the following paragraphs), the main features of our molecular distribution maps are:

\begin{enumerate}
\item All the identified sulphur- (S-) bearing species in our dataset are diatomic/triatomic containing no hydrogen.  Most of them have extended distributions (except for $\rm SO_2$ and $\rm ^{34}SO_2$),
exhibiting a V-shaped morphology that extends from the emission peak at HC,  through  mm3a, mm3b to mm2.  Their velocities at line centre also vary among  substructures from $\rm \sim 6~ km\,s^{-1}$ (HC) through 8--10  $\rm ~km\,s^{-1}$  (mm2, mm3a, mm3b, and SR)  to 11--12  $\rm ~km\,s^{-1}$ (NE). \\

\begin{itemize}
\item OCS and $\rm O^{13}CS$ (carbonyl sulphide) peak at SMA1 and have emission that extend to the CR. In addition, OCS is also detected in NE and SR. Although only one line of $\rm O^{13}CS$ is detected in our data ($\rm 19\rightarrow18$, Fig,~\ref{1line}),  other lines of this isotopologue have been previously confirmed and imaged by single-dish observations (e.g., \citealt{sutton85, ziurys93, tercero10}). In addition to the  detections by single-dish surveys (e.g., \citealt{sutton85, turner91, schilke97, tercero10}), previous interferometric surveys of OCS also reproduce the same V-shaped widespread intensity distribution of this molecule, peaking at the HC (e.g.,   \citealt{wright96,friedel08}).\\

 \begin{figure}[!t]
\centering
\includegraphics[width=9cm]{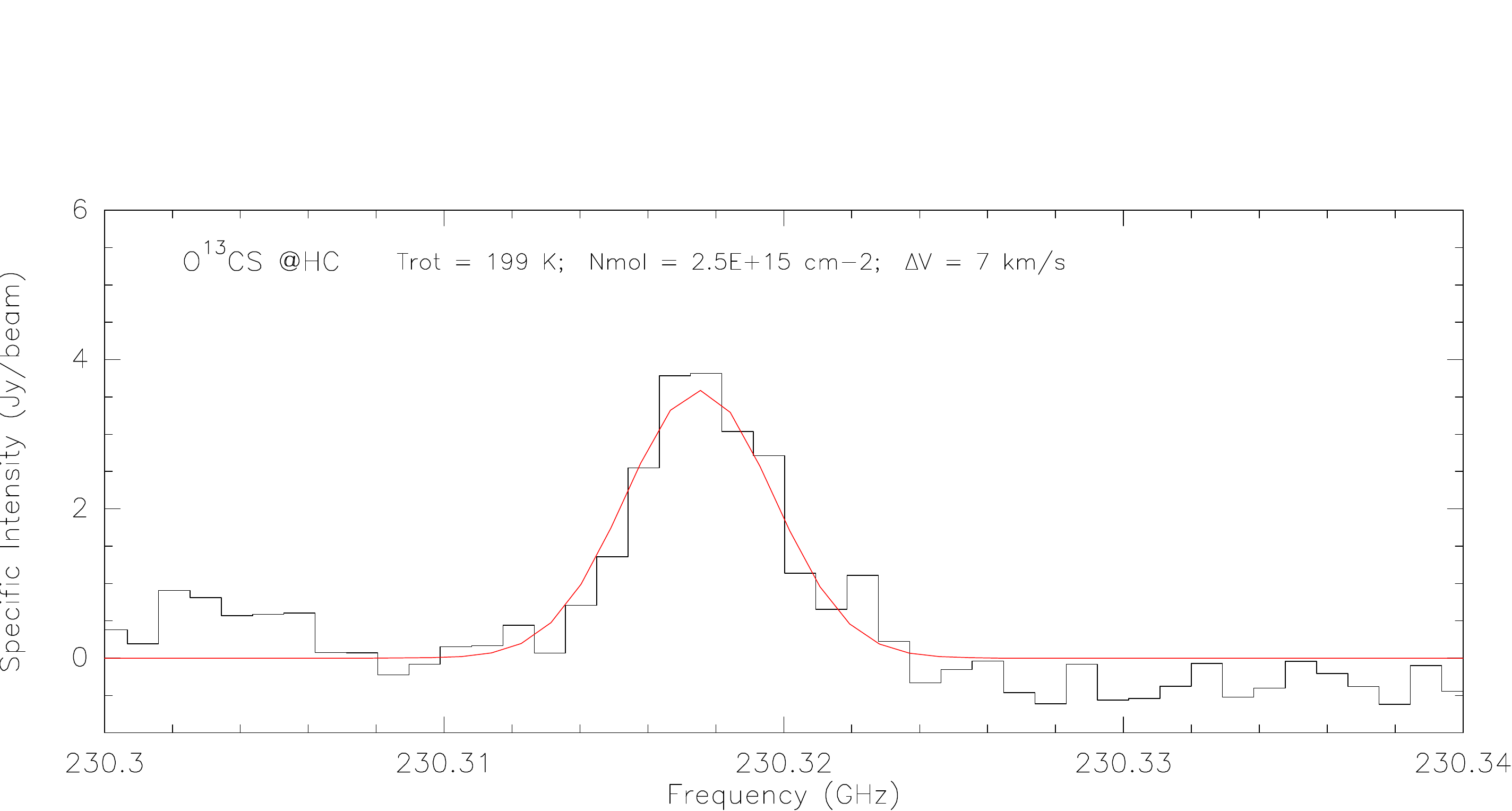}\\
\includegraphics[width=9cm]{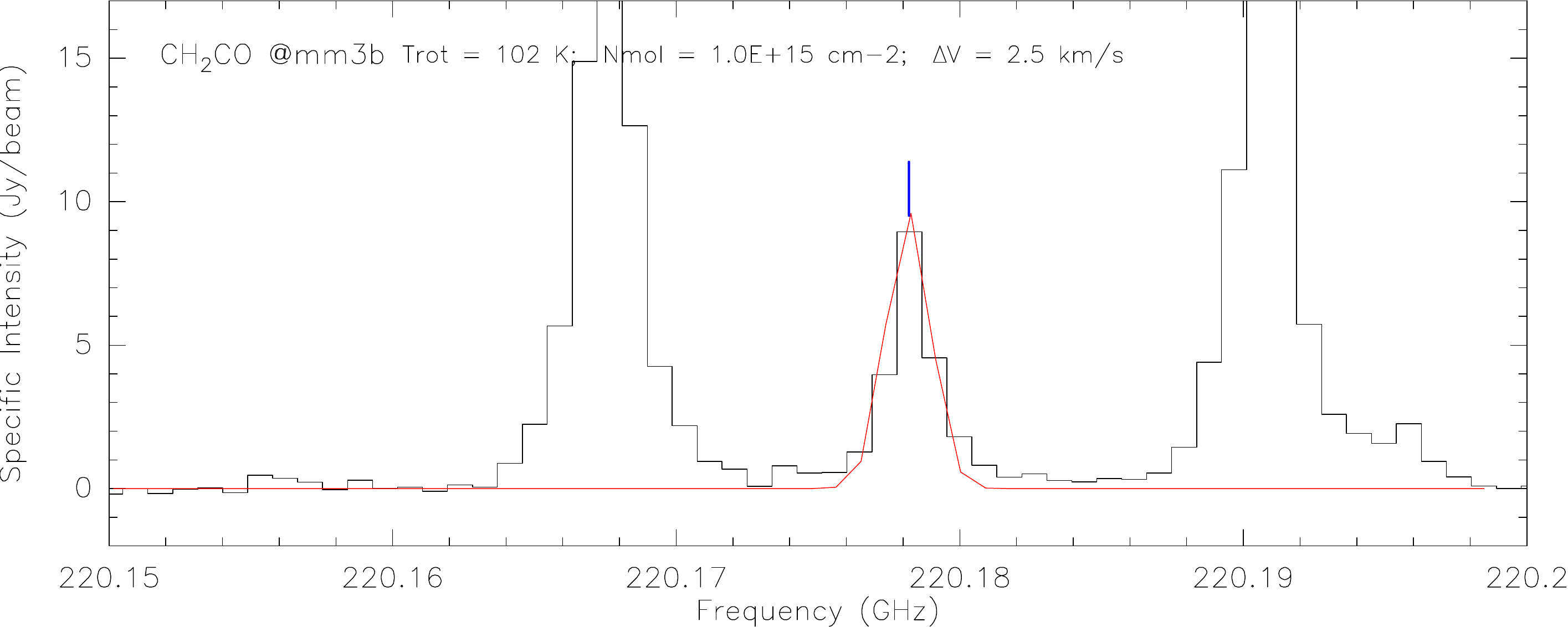}
\caption{Lines that only have one transition with low specific intensity ($\rm \le10~ Jy~beam^{-1}$) in our data, fitted with synthetic spectrum from their emission peaks.
 \label{1line} }
\end{figure}

\item $\rm SO_2$ and $\rm ^{34}SO_2$ (sulphur dioxide) peak at the hotcore and SMA1 and exhibit extended emission in the NE direction, with $\rm V_{peak}\sim 9\text{--}11~km\,s^{-1}$. Unlike the other S-bearing species, no emission has been detected from these species in  SR, and they do not display a  clear V-shaped feature.   
However, there is a clear second velocity component in the line profile of $\rm ^{34}SO_2$ (at $\rm \sim22.7~ 
km\,s^{-1}$, Figure~\ref{wing}) and $\rm SO_2$ (at $\rm 22.1~ 
km\,s^{-1}$, Figure~\ref{wing}). In addition, red- and blue-shifted   $\rm ^{34}SO_2$ trace outflows in the NW--SE direction  (Figure~\ref{shift}). Single-dish studies  (e.g., $\rm SO_2$; \citealt{blake87, persson07,esplugues13}, $\rm ^{34}SO_2$; \citealt{esplugues13})  conclude that this species is a good tracer not only for the warm dense gas, but also for  regions affected by shocks.   While based on the velocity around $\rm 22~ km\,s^{-1}$,  \citet{esplugues13}\footnote{$\rm V_{peak}=20.5~km\,s^{-1}$ in \citet{esplugues13}}  suggests that this velocity component may be associated with shocks caused by the explosive (NW--SE) outflow and/or the BN object, our data clearly show that this velocity component of $\rm SO_2$ is spatially offset from BN and seems to be associated much more  with the main NE--SW outflow (Figure~\ref{so2_cha}).
On small scales, interferometric observations of Orion-KL also  detect a compact $\rm SO_2$ emission peak \citep{wright96, zapata11} and strong extended emission of  $\rm ^{34}SO_2$ to mm2 \citep{beuther05}. \\

\item SO (sulphur monoxide) has  multiple strong emission peaks towards HC, NE, mm3a, and mm3b and exhibits strong self-absorption in mm2 (Figure~\ref{velpro}). There is also significant emission extending along both the plateau of high-  ($\rm  V_{peak}\sim 8\text{--}10~km\,s^{-1}$) and low-velocity  ($\rm  V_{peak}\sim  7\text{--}8~km\,s^{-1}$) outflows.  Similar to  $\rm ^{34}SO_2$, its line wings are broad ($\rm \triangle V\sim23~km\,s^{-1}$), and a second velocity component  is clear in the line profile at $\rm \sim21~km\,s^{-1}$ (see Figure~\ref{velpro} mm3b), which may be associated with  shock events\footnote{ The line wing of SO  at $\rm \sim 7~km~s^{-1}$ is broad, and its second velocity component is blended, so we cannot  tell the origin of the shock precisely from our observations, Figure~\ref{so2_cha}.}. These morphological features (elongated emission peak(s) and broad line wings of SO) have also been detected by both single-dish \citep{persson07, esplugues13} and interferometric \citep{wright96} line studies.\\

\item $\rm ^{13}CS$ (carbon monosulphide) has only one transition in our data set, but it is very strong, peaking at SMA1, with a similar distribution to  $\rm O^{13}CS$ --extended emission from NE through HC to the CR. Owing to its high dipole moment, this molecule is considered to be  a dense gas tracer (e.g., \citealt{bronfman96}), 
and its detection in both CR and NE is confirmed by \citet{persson07} and \citet{tercero10}.\\ 
\end{itemize}

\item The distribution of simple $\rm C_xH_yO_z$ (x,z=1,2...; y=0,1,2...; $x+y+z\le5$) molecules are co-spatially extended. 
\begin{itemize}
\item CO, $\rm C^{18}O$, and $\rm ^{13}CO$ (carbon monoxide) have a common morphological peak at the HC (though the integrated intensity of CO peaks north of the HC due to its strong self-absorption in this source, see Figure~\ref{velpro}).  Velocity at line centre $\rm V_{peak}$ ranges from $\rm 6~km\,s^{-1}$ (HC) through $\rm 8~km\,s^{-1}$ (mm2, mm3a, mm3b, and SR) to $\rm 10~km\,s^{-1}$ (NE, OF1N, and OF1S), with broad red- and blue-shifted line wings trace both high-velocity (CO, $\rm ^{13}CO$) and low-velocity ($\rm C^{18}O$) outflows (Figure~\ref{shift}). As the second most abundant species in the ISM, this species has been detected in numerous other observations of Orion-KL ($\rm C^{18}O$; \citealt{beuther08,plume12}, CO; \citealt{zuckerman76,  kwan76, erickson82, blake87} (single-dish), \citealt{masson87,chernin96,zapata09,peng12b} (interferometry), $\rm ^{13}CO$; \citealt{wilson11} (single-dish)), showing the same morphology as we see:  extended structure, with bipolar line wings covering from the NE through HC to the CR and tracing outflows.\\

\item $\rm CH_2CO$ (ketene) has a single transition ($\rm 11_{1,11}\rightarrow10_{1,10}$) detected (Figure~\ref{1line}), peaking at mm3a and mm3b, with strong emission extending to the  OF1S and the NE filament.  With $\rm V_{peak}\sim8~km\,s^{-1}$ and $\rm \triangle V\sim2.5~km\,s^{-1}$ in mm2, mm3a, and mm3b, this line has also been confirmed by \citet{sutton85}. Some other lines of  this species are also detected in Orion-KL by \citet{lee02}.\\

\item $\rm H_2^{13}CO$ (formaldehyde) exhibits an elongated peak stretching from SMA1 towards mm3a. Similar to the simple organic $\rm CH_2CO$, it also shows extended structure along the SE lobe of the high-velocity outflow and NE. Single-dish detections (e.g., \citealt{persson07}) confirm that this molecule is  present in both the CR and the outflows.\\

\end{itemize}

 \begin{figure}
 \centering
\includegraphics[width=7cm, angle=0]{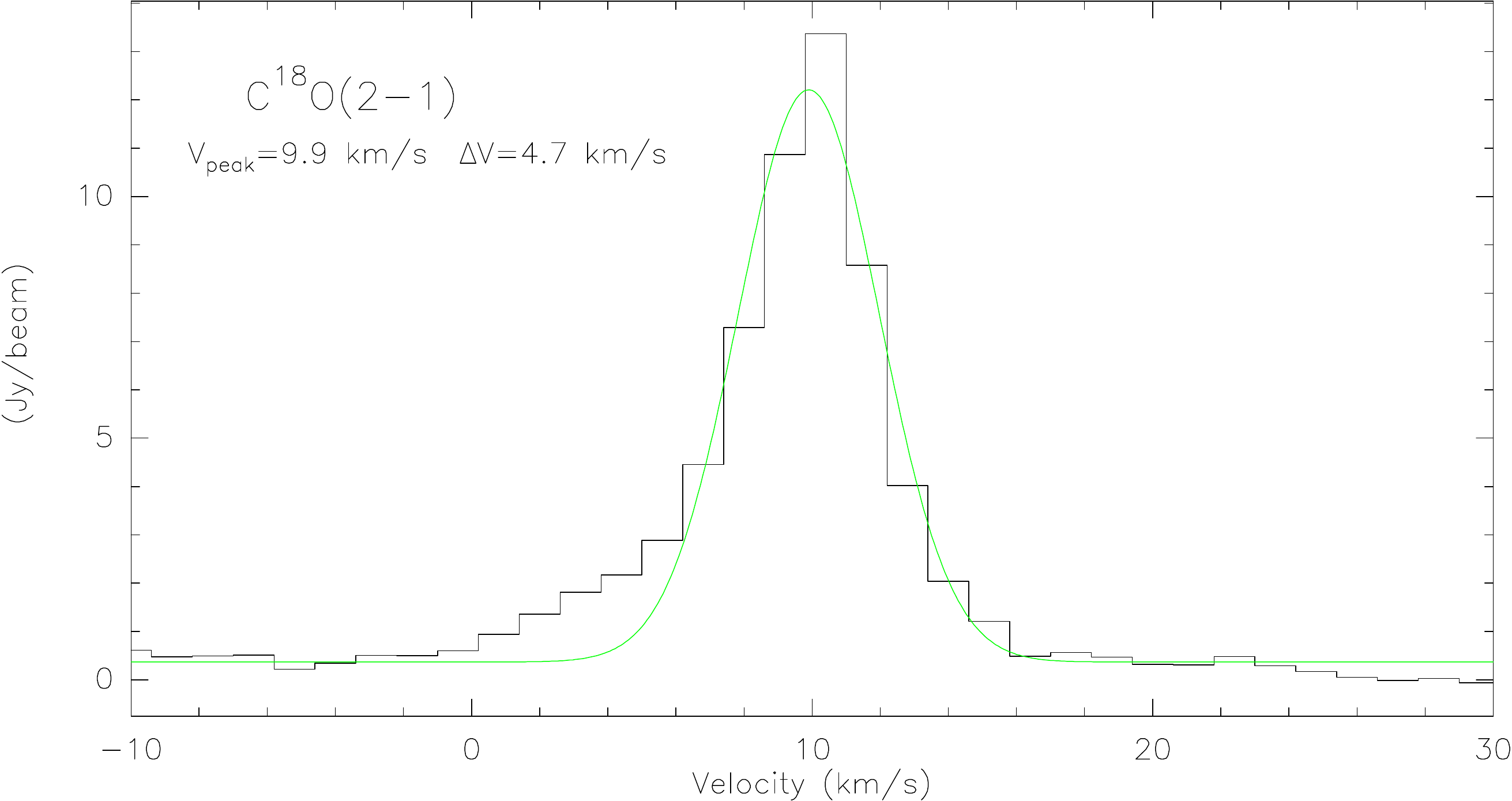}\\
\includegraphics[width=7cm, angle=0]{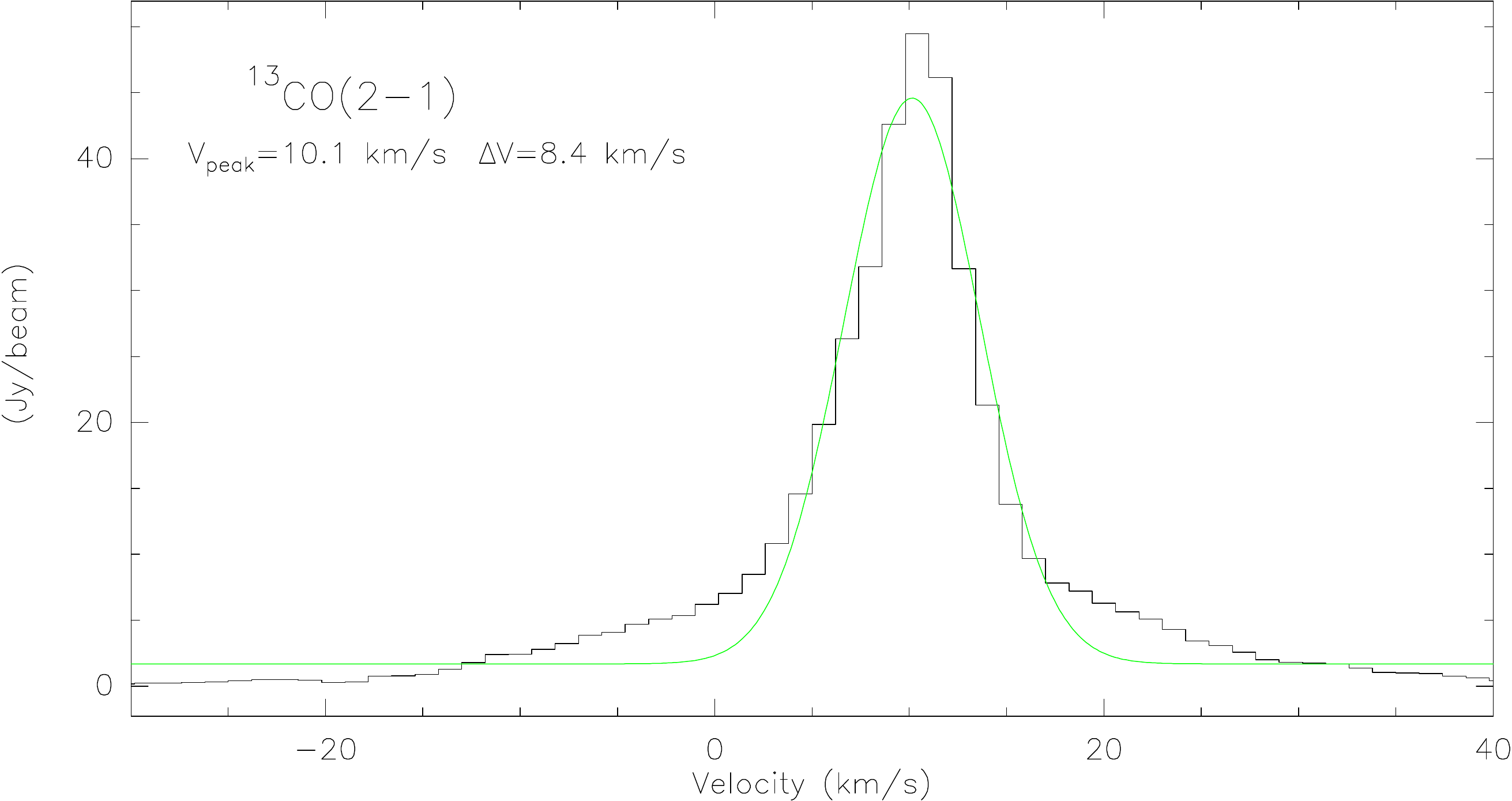}\\
\includegraphics[width=7cm, angle=0]{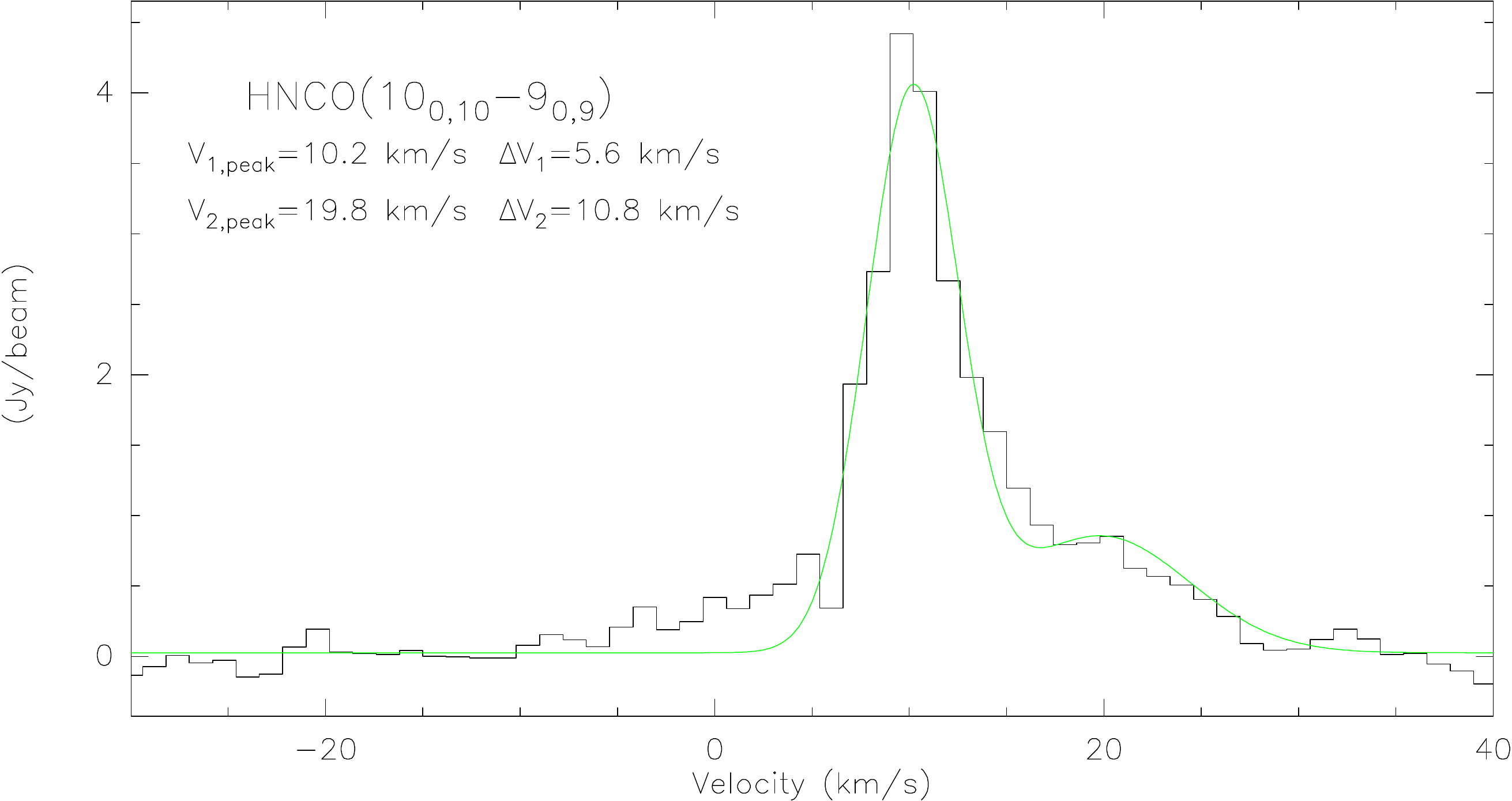}\\
\includegraphics[width=7cm, angle=0]{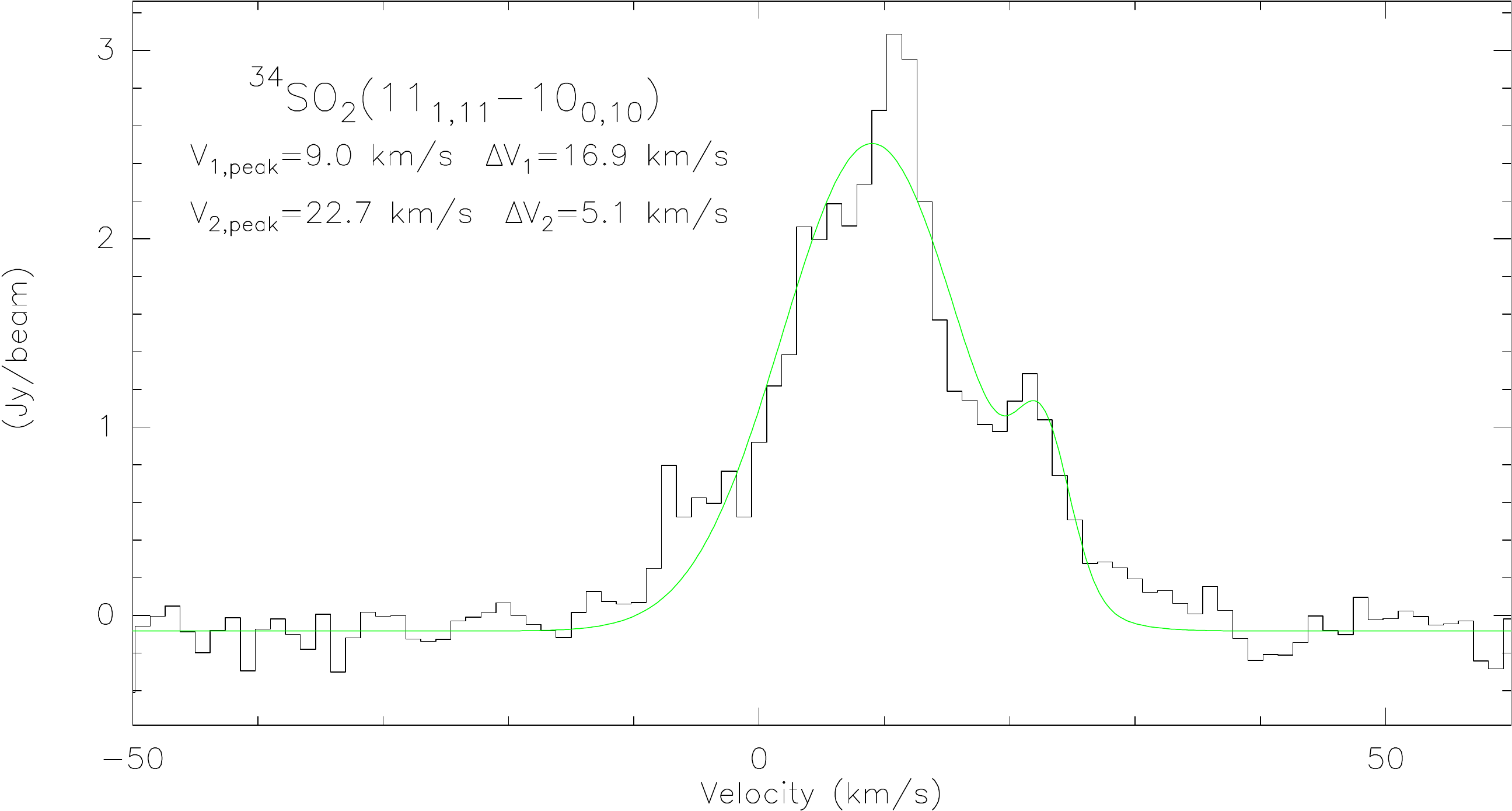}\\
\includegraphics[width=7cm, angle=0]{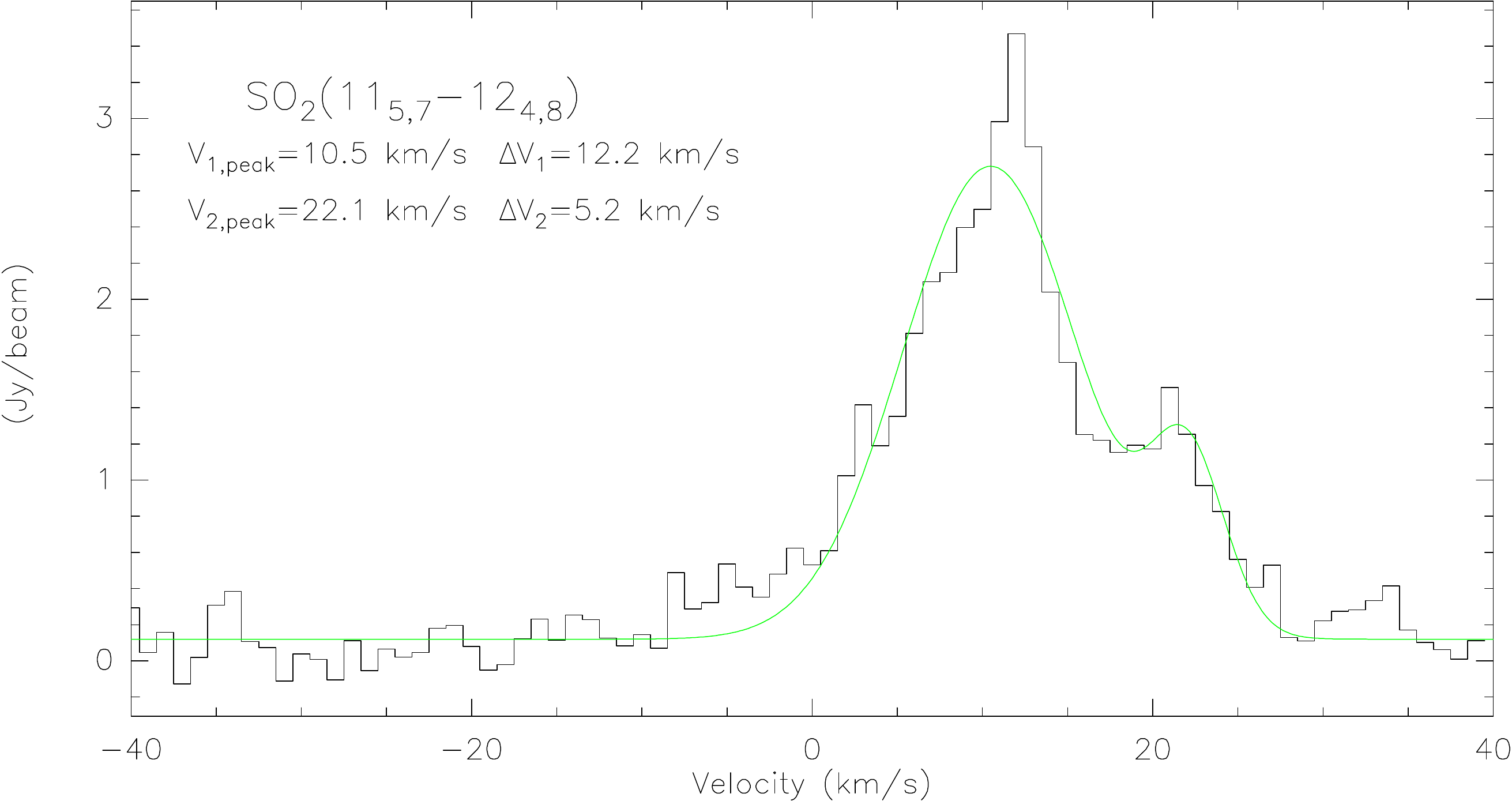}\\
\caption{Gaussian fitting to the line profile of $\rm ^{13}CO$, $\rm C^{18}O$, HNCO, $\rm SO_2$, and $\rm ^{34}SO_2$ extracted from NE. Outflow signatures are seen in the broad line wings and the  second velocity component in the HNCO, $\rm SO_2$, and $\rm ^{34}SO_2$ line profile. }\label{wing}
\end{figure}

\item $\rm HNCO$ (isocyanic acid) is the simplest organic molecule containing C, H, O, and N elements. Peaking at SMA1, it shows extended emission towards all the substructures, with $\rm V_{peak}$ ranging from $\rm 6~km\,s^{-1}$ (HC) through 7--8 $\rm km\,s^{-1}$ (mm3a, mm3b, SR, and OF1S) to 9--10  $\rm ~km\,s^{-1}$ (mm2, NE, OF1N), and  a second velocity component  at $\rm \sim19.8~km\,s^{-1}$ (Figure~\ref{wing}). Although the formation pathway of HNCO  is still not clear \citep{hasegawa93,turner99,garrod08,tideswell10}, its distribution is observed to be correlated with shock tracing molecules SiO (e.g.,  \citealt{zinchenko00}) and $\rm CH_3OH$ (e.g., \citealt{meier05}), suggesting that it is also produced in shocks. \\

\item  In contrast,  
highly saturated COMs\footnote{A COM that has a chain of carbon atoms linked together by single bonds and has rich hydrogen atoms filling  the other bonding orbitals of the carbon atoms.}  
 are the typical COMs 
 detected in HMSFRs, and they exhibit a variety of morphological structures in Orion-KL.
 In general, N-bearing COMs  have a similar peak to HNCO, at or near HC; while most of the O-bearing  COMs are co-spatial with  $\rm CH_2CO$--peaking at mm3a and mm3b, and extending even to southern CR (e.g., $\rm HCOOCH_3$, $\rm CH_3OCH_3$, and $\rm CH_3CHO$). This spatial segregation of the N- and O-bearing COM peaks
 has also been reported by other single-dish \citep{blake87,persson07} and interferometric studies with high spatial resolution (e.g., \citealt{wright96, beuther05, favre11a, friedel12}). In addition, there are several other O-bearing COMs  identified in this study, which have unique morphologies. To understand the spatial differentiation of COMs on small scales (Figure~\ref{COMdis}), we compared the distributions of  these molecules between our observations to other recent interferometric studies (Table~\ref{COMcol}).\\

\begin{itemize}
\item $\rm CH_3CN$ and $\rm CH_3^{13}CN$ (methyl cyanide) are two abundant N-bearing COMs. In our dataset, we detect the $\rm J=12\rightarrow11$ ladder of both isotopologues around 220 GHz. Most of the $\rm CH_3^{13}CN$ lines are blended, but lines with $\rm K=2, 4, 5, 6$ are clearly detected in the HC (Figure~\ref{rotation}). In particular, $\rm K=2$ line of $\rm CH_3^{13}CN$ and   $\rm CH_3CN$  are clearly co-spatial (Figure~\ref{velpro}), peaking at SMA1 ($\rm V_{peak}\sim 5\text{--}7~km\,s^{-1}$) and extending to mm2, mm3a, mm3b ($\rm V_{peak}\sim 8\text{--}10~km\,s^{-1}$), with $>3\sigma$ detection in NE ($\rm V_{peak}\sim 10\text{--}13~km\,s^{-1}$), exhibiting a V-shaped morphology. $\rm CH_3CN$ is detected in SR and  OF1N (S), but not for $\rm CH_3^{13}CN$. This V-shaped morphology of $\rm CH_3CN$ has also been detected in e.g., \citet{beuther05,zapata11,widicus12,bell14}, as well as the NE extension in \citet{wright96}. This species is considered to be a warm and dense gas tracer (e.g., \citealt{kalenskii00,araya05}), with gas phase and grain surface chemical models predicting that it is only detected in an environment with elevated temperatures (e.g., \citealt{purcell06}). We use the $\rm CH_3CN$ ladder and the $\rm CH_3^{13}CN (12_2\rightarrow11_2)$ line to estimate the temperature of the substructures where these lines are detected (see Section~\ref{tem:ch3cn}). \\

\item $\rm CH_2CHCN$ (vinyl cyanide) exhibits strong compact emission  north of the hotcore, with $\rm V_{peak}\sim5~km\,s^{-1}$, and a slight extension towards mm2. 
The northern peak is also coincident with the one in higher resolution maps of this species from SMA at $\rm 865\,\mu m$ \citep{beuther05}, CARMA at 1\,mm \citep{friedel08}, and PdBI  at 3\,mm \citep{guelin08}.\\

\item $\rm CH_3CH_2CN$ (ethyl cyanide) emission also makes a V-shaped structure. Unlike the single-peaked structure of $\rm CH_2CHCN$, $\rm CH_3CH_2CN$ exhibits a dual-peaked structure,  with a first emission peak at the hotcore ($\rm \sim6~km\,s^{-1}$), and a second peak at mm2 ($\rm \sim9~km\,s^{-1}$). Although there are two velocity components in both mm3a and mm3b ($\rm 6~km\,s^{-1}$, $\rm 10~km\,s^{-1}$), there is no detection in southern CR, which is consistent with \citet{blake87}. This observed  morphology  is consistent with  other interferometric observations at high spatial resolution \citep{wright96,beuther05,friedel08,guelin08,widicus12,peng13}.\\

\item $\rm CH_3OH$ (methanol) has multiple transitions in our data (Figure~\ref{COMspec}).  We obtained the channel maps of all the unblended strong $\rm CH_3OH$ transitions in our dataset and found two types of spatial distributions (Figure~\ref{ch3oh_cha}). One type of line clearly shows  extended emission  from mm2 and SR (7.2--8.4$\rm ~km\,s^{-1}$, e.g.,  lines of $\rm 8_{0,8}\rightarrow7_{1,6} E$, $\rm 8_{-1,8} \rightarrow7_{0,7} E$, and $\rm 3_{-2,2}\rightarrow4_{-1,4} E$) towards the high-velocity outflow OF1N(S), exhibiting a similar SE tail to $\rm H_2^{13}CO$ at 7--10$\rm ~km\,s^{-1}$, which has not been detected before; whereas the other type of line (e.g.,  lines of $\rm 15_{4,11}\rightarrow16_{3,13} E$, $\rm 22_{4,18} \rightarrow21_{5,17} E$, and $\rm 19_{5,14}\rightarrow20_{4,17} A--$) does not show such a SE tail. However, both of them show multiple velocity-dependent emission peaks: towards SMA1 at  5$\rm ~km\,s^{-1}$, mm3a at 8$\rm ~km\,s^{-1}$, and mm3b at 10 $\rm ~km\,s^{-1}$. The multiple-peak structure is different from the traditional segregated dichotomous distribution of COMs in Orion-KL (i.e., the main emission peak is either at HC or mm3a(b)), though it has also been observed in other interferometric surveys \citep{wright96,beuther05,friedel08,guelin08,friedel12,peng12a}.
Methanol has previously been detected in the Orion outflow  \citep{wangs11}, and furthermore,  line  $\rm 8_{-1,8} \rightarrow7_{0,7} $ is known to be a potential Class I maser that traces shocks \citep{fish11}. As a result, the SE tail  can be reasonably explained as a feature of the outflow.  In addition, it is also suggested that the SMA1-Source N area is the launching point of the Orion-KL outflows (e.g.,  \citealt{menten95, beuther08}), which may explain $\rm CH_3OH$ and $\rm H_2^{13}CO$ peaking in this area.\\

\item $\rm HCOOCH_3$ (methyl formate) has the largest number of detected lines in our data (Figure~\ref{COMspec}). There are two types of $\rm HCOOCH_3$ in the whole observational band, torsionally excited lines  ($\rm \nu=1$) and lines in ground state ($\rm \nu=0$). Both of them have identical distributions (Figure~\ref{COMdis}) that peak at both mm3a and mm3b, with a clear detection in the southern CR, which is  similar to $\rm CH_2CO$. This morphology has also been reported by \citet{wright96, beuther05, guelin08, friedel08,favre11a, widicus12}, indicating that this species arises in a cooler and less dense  environment than the HC.\\

\item  $\rm CH_3OCH_3$ (dimethyl ether) peaks towards mm3a and mm3b   and extends to the southern CR, similar to $\rm HCOOCH_3$.
Unlike the other COMs, its emission in mm2 is considerably stronger than towards the HC. This morphology has also been imaged with other interferometric studies  \citep{liu02, beuther05, guelin08,friedel08,favre11b,brouillet13}.\\

\item $\rm CH_3CH_2OH$ (ethanol),  the functional isomer\footnote{Two compounds with the same molecular formula but different functional groups (radicals) are called functional isomers.} of   $\rm CH_3OCH_3$, exhibits extended emission from the vicinity of HC to  mm2,  mm3a, and mm3b, indicating the same distinctive V-shaped molecular emission as the N-bearing COMs. However, unlike $\rm CH_3OCH_3$ and the other COMs, it peaks  in between the hotcore and  mm3a.  The difference in the location of their peaks is exactly the same as  reported by   \citet{guelin08,brouillet13} from PdBI and \citet{friedel08} with the EVLA and CARMA at high spatial resolution.  In the grain mantle chemistry models, both isomers  are produced from two radicals involved in the synthesis of $\rm CH_3OH$ ice \citep{garrod08}. Therefore, their spatial differentiation  may result from the chemical differences between the $\rm CH_2OH$ and $\rm CH_3O$ radicals across Orion-KL, e.g., variations in production rates or chemical evolution \citep{brouillet13}. \\

\item  $\rm CH_3COCH_3$ (acetone) is an O-bearing COM, though, it has a similar distribution to $\rm CH_3CH_2CN$, which peaks at SMA1, extends to mm2, and has no detection in the southern CR. Although its nature as a hot molecular core tracer has been confirmed by detections  in both Orion-KL \citep{friedel05, friedel08, widicus12,peng13} and Sagittarius B2 (N-LHM) \citep{snyder02}, whether this species mainly forms on dust grains or in the gas phase is still not clear.\\

\item $\rm CH_3CHO$ (acetaldehyde) is tentatively detected, as all lines except one  at 220.446 GHz (Table~\ref{tab:line}) are blended  with $\rm HCOOCH_3$. Morphologically, this line has a similar distribution to $\rm CH_2CO$, peaking at mm3a and mm3b, with extended emission to the NE and  SE tail. Although its origin is still unclear \citep{charnley04}, it has been detected in Orion-KL  with single-dish telescopes \citep{turner89}. Previous interferometric observations  failed to detect it because of filtering out the extended structure (e.g., \citealp{friedel08,widicus12}). However, its detection has been reported by recent CARMA and ALMA observations  (Loomis, priv. comm), which shows the same distribution as our single-dish complemented interferometric observation result.  \\

\item There are also several other organics, such as   $\rm CH_2OHCHO$ (glycolaldehyde), $\rm (CH_2OH)_2$ (ethylene glycol), and HCOOH (formic acid), which we do not detect in Orion-KL at 1.3\,mm in the SMA-30\,m combined data. Of the first two, although there are some candidate detections from our synthetic fitting, they are blended with other nearby lines, and so cannot be confirmed.  Although \citet{liu02} claimed to detect HCOOH in the CR, in their data the line at 1 mm is only a  $2\sigma$ feature. 
 \citet{widicus12} also did not detect any compact concentrations of HCOOH in our studied region either, and they conclude that the HCOOH emission must be extended on $\rm> 1\,700$\,AU scales.\\

\end{itemize}

In brief, from our detections, N- and O-bearing COMs not only show two segregated peaks, but also exhibit  more complicated spatial distributions.
Most of the N-bearing species (except $\rm CH_2CHCN$) are co-spatial and peak at the hotcore. While some O-bearing species ($\rm HCOOCH_3$, $\rm CH_3OCH_3$, and $\rm CH_3CHO$) peak at mm3a and mm3b, and others present more unusual
features (i.e., $\rm CH_3OH$ has velocity dependent peaks at both  hotcore and mm3a/mm3b; $\rm CH_3CH_2OH$ peaks in between mm3a and hotcore; and $\rm CH_3COCH_3$  peaks in a similar location to N-bearing COMs). This challenges  the traditional hypothesis that all of these molecules are forming in the same chemical and physical conditions \citep{garrod08, laas11}.\\

\item Unlike the saturated COMs,  long carbon chains  (organics with carbon-carbon multiple bonds)  are less likely to be detected in star-forming regions \citep{thorwirth01,rice13}.  

\begin{itemize}
\item  $\rm C_6H$ is the parent radical of $\rm C_6H^-$ (the first molecular anion  identified in the gas phase interstellar and circumstellar media; \citealt{mccarthy06}), after its presence was predicted by \citet{herbst81}.   It had not been detected in Orion-KL before.  In our dataset, there is one unblended line at 230.109 GHz with $\rm J=165/2\rightarrow163/2$, although most transitions identified by the synthetic fitting of $\rm C_6H$ are blended.  Emission from this tentative detected species appears to be confined to the HC.  \\

\item $\rm HC_3N$ and $\rm HC_7N$ are two cyanopolyynes ($\rm HC_{2n+1}N$, n is an integer).  We confirmed the detection of $\rm HC_3N (\nu_7=1)$ to the  north  of the hotcore and also tentatively detect $\rm HC_7N$.  Although only two out of four of the  $\rm HC_7N$  lines predicted by the synthetic fitting programme are detected in our observations\footnote{Lines with $\rm J=196\rightarrow195$ and $\rm J=205\rightarrow204$ are detected, but neither $\rm J=195\rightarrow194$   nor $\rm J=204\rightarrow203$  are detected, so these potential detected lines are marked with ``?" in Table~\ref{tab:line}.}, these lines are unblended and peak at a similar location to $\rm HC_3N$. The excited-vibrational state  $\rm (\nu_7=1)$ of  $\rm HC_3N$ shows a similar confined structure north-east of the HC in \citet{wright96,beuther05} and \citet{zapata11}, where it is said to form a shell around the HC. $\rm HC_3N$ is an especially useful dense gas tracer due to its relatively low optical depth transitions, large dipole moment, and simple linear structure \citep{bergin96}. The $\rm HC_3N (\nu_7=1)$ lines are believed to be primarily excited by a strong mid-infrared radiation field (e.g., \citealt{goldsmith82}).
A gas-phase chemical model of a hot molecular core with high but transitory abundances of cyanopolyynes has been presented in  \citet{chapman09},  where  these species are formed from reactions between the precursor $\rm C_{2n}H_2$ (e.g., $\rm C_2H_2$, acetylene) and N-bearing species via  gas phase chemistry (e.g., $\rm C_{2n}H_2 + CN \rightarrow HC_{2n+1}N + H$). \\

\end {itemize}

\end{enumerate}

%\begin{landscape}
\begin{table*} 
\small
\begin{center}
\begin{tabular}{c| cp{3cm}p{2cm} |p{5cm} }
\hline\hline

Species     &Identification  & Peak  & Column density @Peak $\rm (cm^{-2})$         &Other interferometric detections \\
\hline
$\rm CH_3CN$, $\rm CH_3^{13}CN$       &${\surd}$       &SMA1        &1.4(17), 9.4(14)       &a, b, e, i \\  
\hline
$\rm CH_2CHCN$      &${\surd}$       &North to hotcore       &$\ge$5.0(15)      &b, d, e \\
\hline
$\rm CH_3CH_2CN$      &${\surd}$       &hotcore        &$\ge$1.1(16)      &a, b, d, e, i, k  \\
\hline
\hline
$\rm CH_3OH$        &${\surd}$       &SMA1 \& mm3a       &$\ge$2.5(17)       &a, b, d, e, f, i, m\\ 
\hline
$\rm HCOOCH_3$        &${\surd}$       &mm3a \& mm3b         &$\ge$9.7(16)       &b, d, e, g, i, k\\ 
\hline
$\rm CH_3COCH_3$        &${\surd}$       &SMA1         &$\ge$1.5(16)       &c, i, k\\ 
\hline
$\rm CH_3OCH_3$           &${\surd}$     &mm3a \& mm3b     &$\ge$2.0(17)   &b, d, e, h, i, j\\
\hline
$\rm CH_3CH_2OH$        &${\surd}$       &near SMA1 \& mm3a    &$\ge1.5(16)$   &d ($\times$), e and j (only  1 line)\\
\hline
$\rm (CH_2OH)_2$           &$~\times$    &$-$    &$-$  &$-$\\
\hline
$\color{black}\rm CH_2OHCHO$   &$~\times$   &$-$   &$-$    &g. (some lines $\rm <3\sigma$) \\    
\hline
$\rm HCOOH$    &$~\times$   &$-$ &$-$   &d ($\times$), l ($\rm <3\sigma$), i (only from the extended configuration)\\
\hline
\hline
$\rm CH_3CHO$               &~*  &mm3a \& mm3b &$\ge$1.3(16)   &d ($\times$), i ($\times$), n (ALMA) \\
\hline
$\color{black}\rm C_6H$       &~*       &SMA1   &$\ge$1.0(15)   &$-$\\  
\hline    
$\color{black}\rm HC_7N$             &~*  &North to hotcore  &$\ge$1.0(16) &$-$\\
\hline

\hline
 \hline
\end{tabular}
\begin{tabular}{l}
\scriptsize{{\bf References.} a. \citet{wright96}; b. \citet{beuther05}; c. \citet{friedel05}; d. \citet{friedel08}; e. \citet{guelin08}; f. \citet{zapata11}; g. \citet{favre11a}; } \\
\scriptsize{h. \citet{favre11b}; i. \cite{widicus12}; j.  \citet{brouillet13}; k. \citet{peng13}; l.\citet{liu02}; m. \citet{peng12a}; n. R. Loomis, private communication.} \\

\end{tabular}\\
\caption{Identification of saturated COMs and complex carbon chains in Orion-KL.  Confirmed COMs are marked with ``${\surd}$",  tentative detections  with  ``*",   and nondetections with  ``$\times$".  At each of their main emission peak  (Figure~\ref{COMdis}), column density of $\rm CH_3CN$ is obtained with optical depth correction, while column densities of the rest molecules  are obtained at the temperature of an adjacent substructure from synthetic fittings, with an assumption that the species lines are optically thin (we denote them by ``$\ge$''), $x(y)=x\times10^y$.}
\label{COMcol}

\end{center}
\end{table*}

%stop here

\section{Rotational temperature and abundance estimates}\label{calcu}
 The above qualitative comparison of the morphology of different species show  the chemical variations in the above-mentioned substructures.  A more precise approach to studying these variations is to measure the quantitative differentiations in chemical parameters towards individual substructures, i.e., the temperatures, molecular column densities, and abundances.

\subsection{Temperature estimates}\label{tem}
Temperature is  commonly used as a benchmark for comparing model results and observations. Since N-, O-, and S-bearing species show different spatial distributions, we estimate their temperatures in each substructure separately.
\subsubsection{$\rm \bf CH_3CN$}\label{tem:ch3cn}
A good  tracer of temperature in  high-density gas ($\rm n\gtrsim10^5 cm^{-3}$) is $\rm CH_3CN$, which  is  a symmetric-top molecule with no dipole moment perpendicular to the molecular axis, and its K ladders in rotational levels are excited solely by collisions. Provided that the collision rate is sufficient to thermalize the rotational levels within each K ladder,  the rotational excitation temperature of $\rm CH_3CN$ is equal to  the gas kinetic temperature \citep{boucher80, wright95}.  Our observation band covers the $\rm J=12\rightarrow11$ transitions of $\rm CH_3CN$ around 220 GHz,  with upper state energy levels  from $\sim70$\,K to $\sim500$\,K,  yielding a wide range for determining temperatures.  If we assume that the emission from all of these transitions originate in the same parcels of gas along the line of sight in LTE,  having the same size,  velocity,  and a single temperature, then rotational temperatures can be derived from the rotational diagrams  (\citealt {hollis82,  loren84,  olmi93, goldsmith99}). Traditionally, in the { optically thin limit},  
the level populations  are then directly proportional to the line intensities of the K components.
The parameters used to estimate the level populations of  $\rm CH_3CN$ transitions (i.e., the line strength $S_{ul}\mu^2~\rm (Debye^2)$ and the lower state energy $\rm E_L~(cm^{-1})$) in our data are listed in Table~\ref{tab:rotch3cnline}.\\

\begin{figure*}[!ht]

  \small
\begin{center}
\begin{tabular}{ll}
\includegraphics[width=8cm]{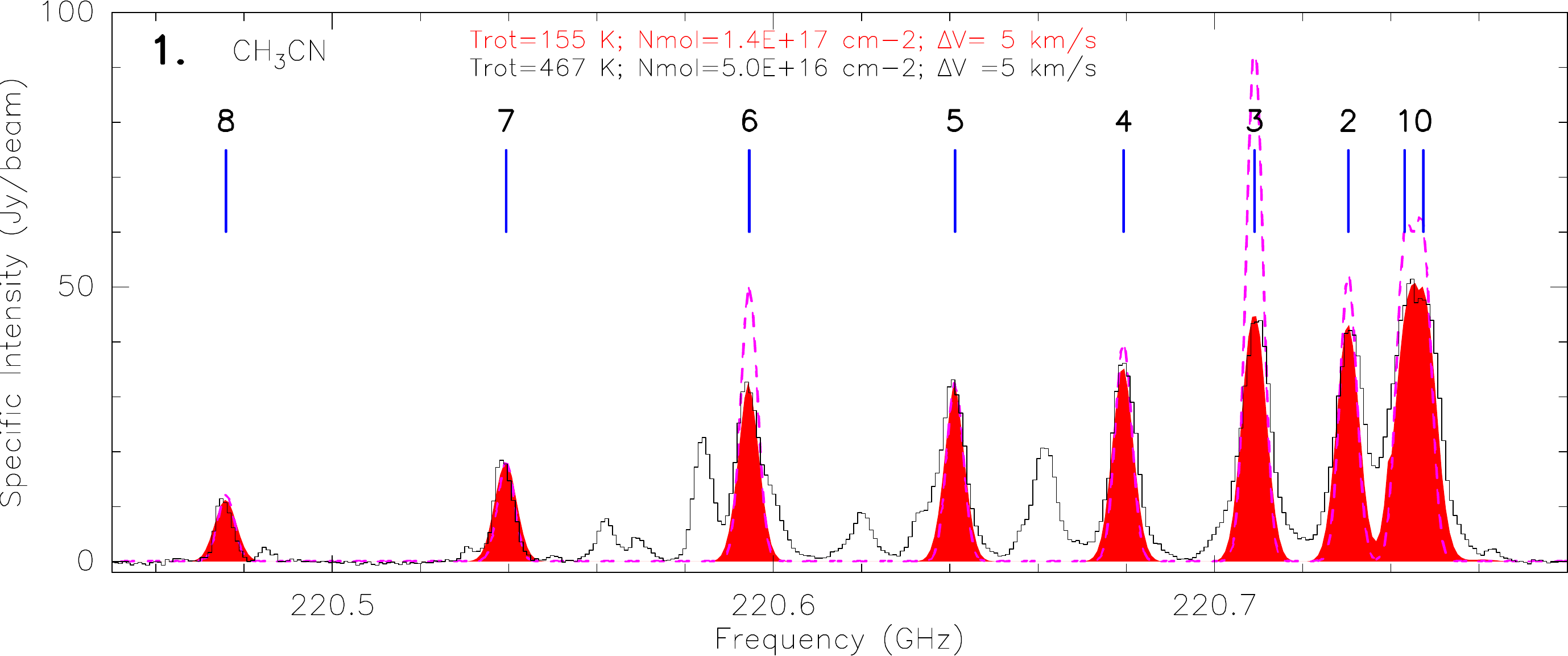}
&\includegraphics[width=5cm]{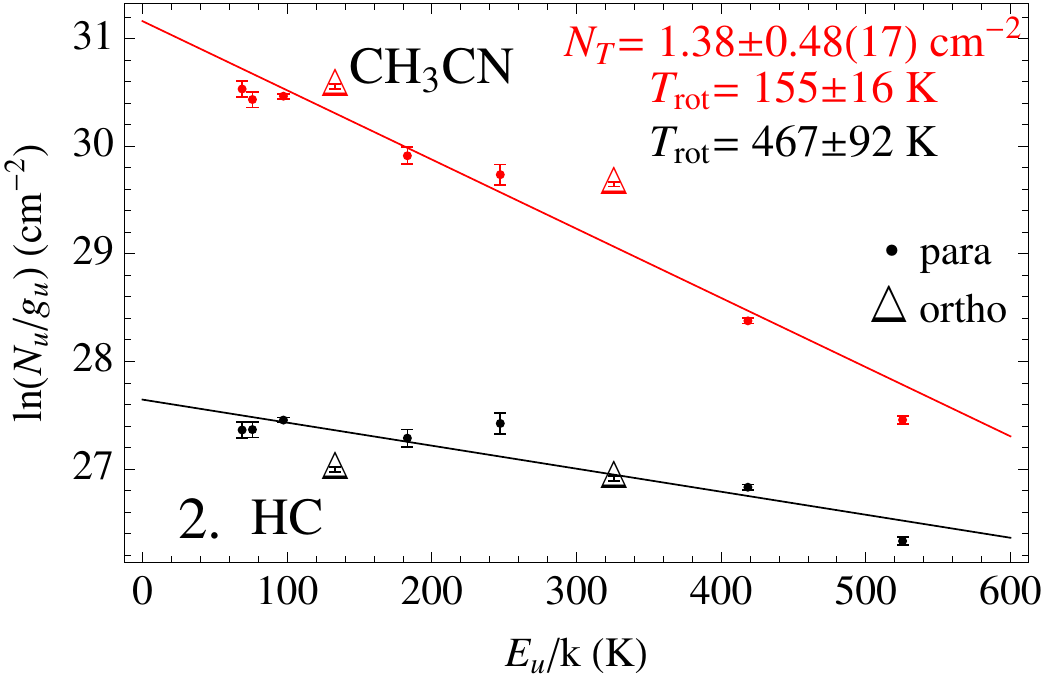}
\end{tabular}

\begin{tabular}{p{4.2cm}p{4.2cm}p{4.2cm}p{4.2cm}}
\includegraphics[width=4.5cm]{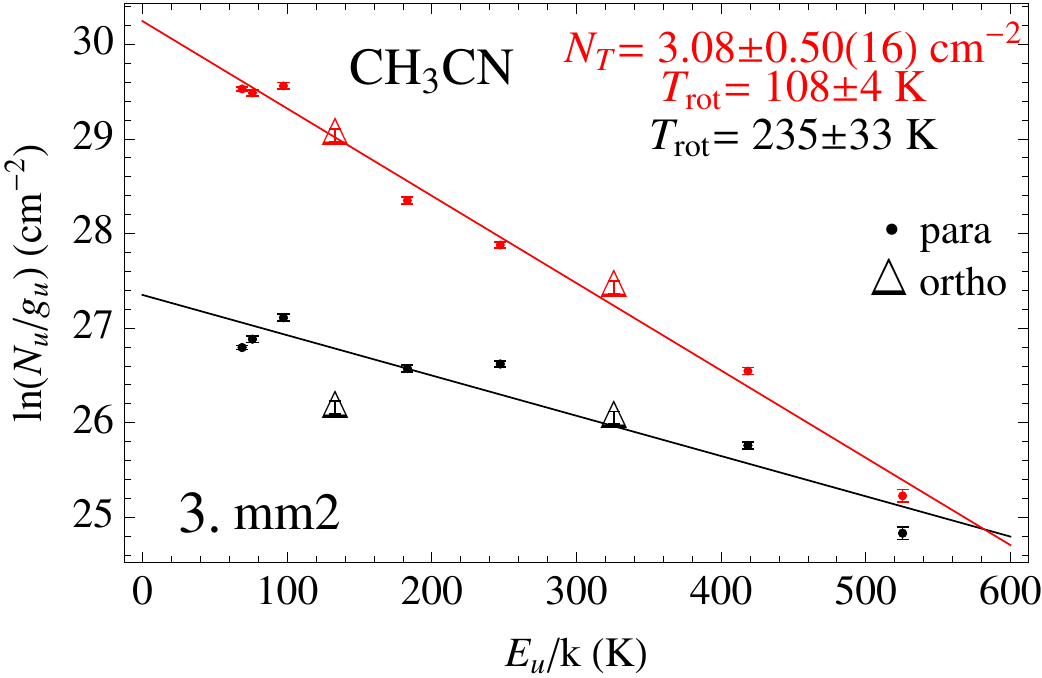}
&\includegraphics[width=4.5cm]{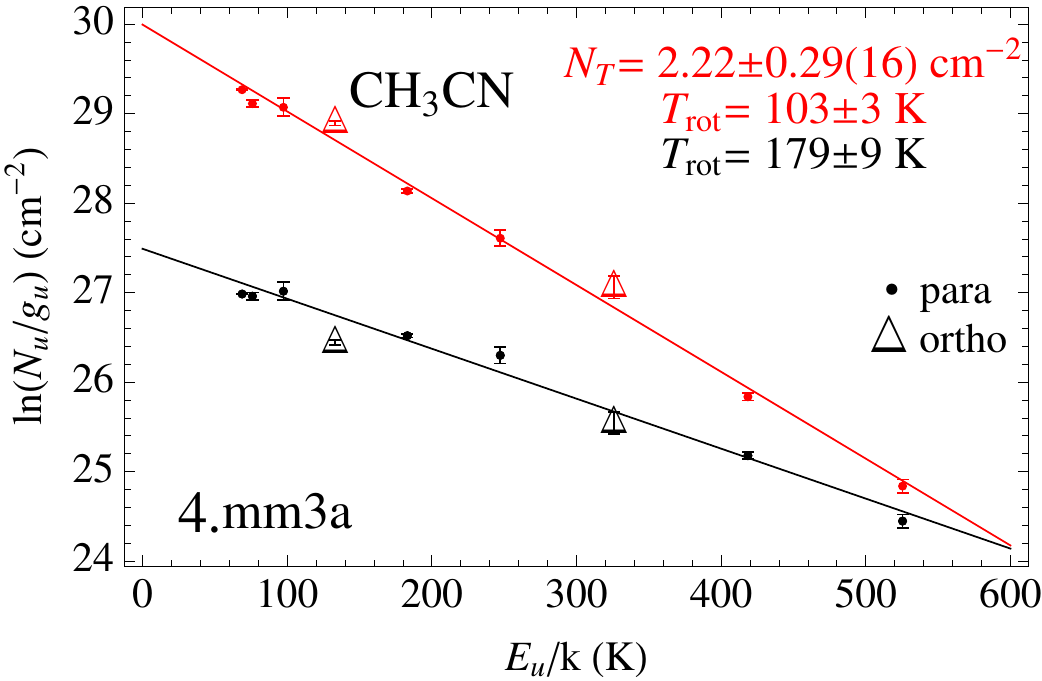}
&\includegraphics[width=4.5cm]{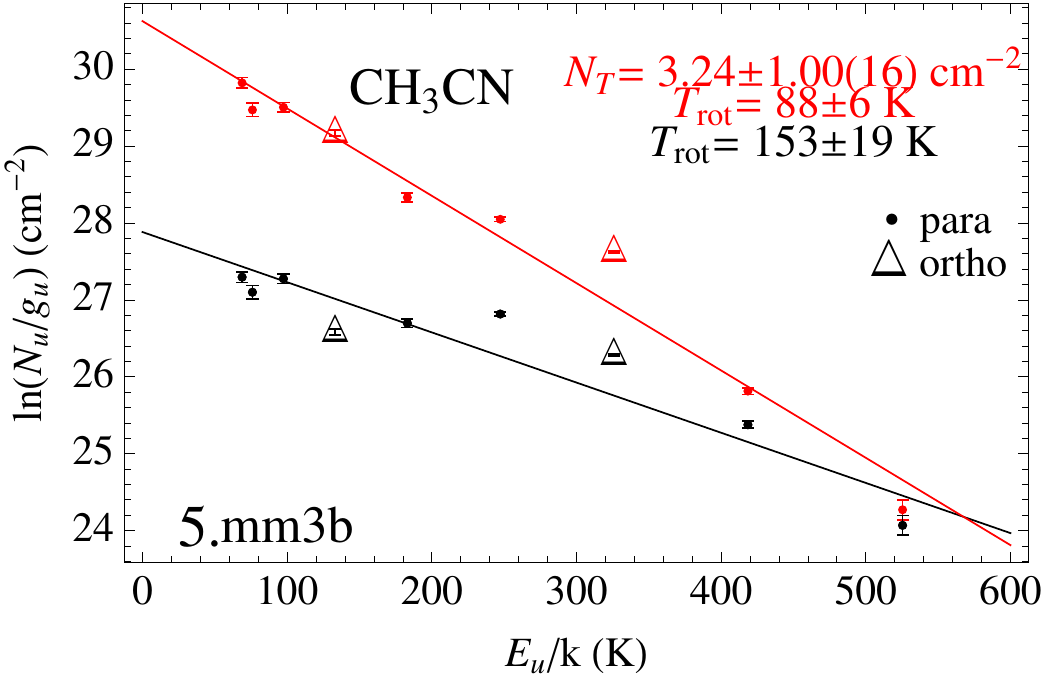}
&\includegraphics[width=4.5cm]{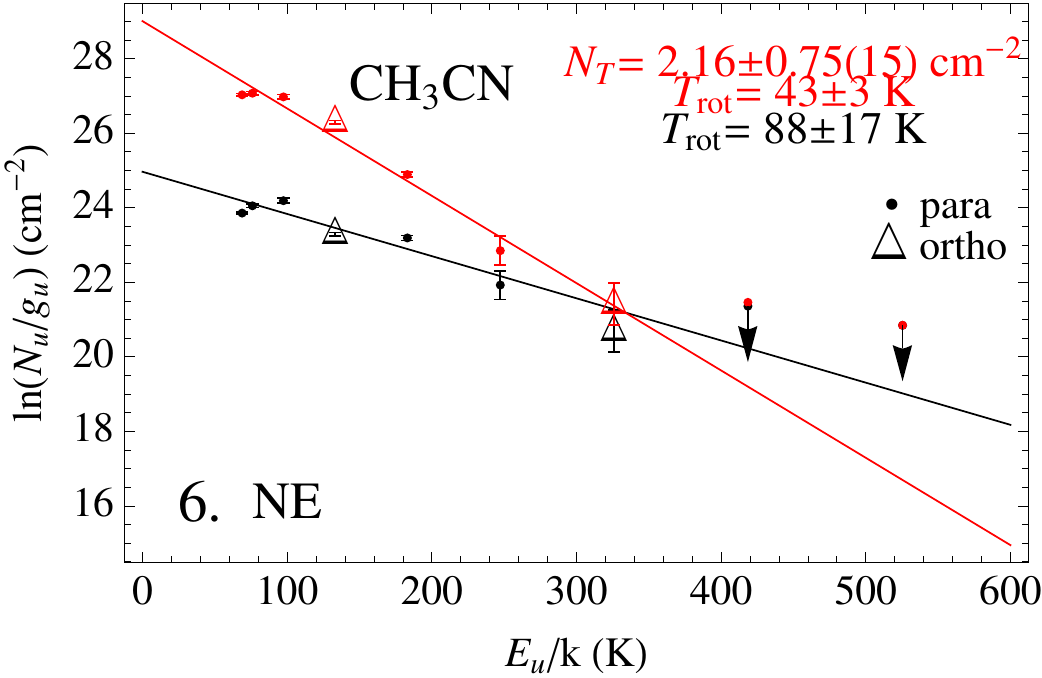}
\end{tabular}

\begin{tabular}{ll}
\includegraphics[width=8cm]{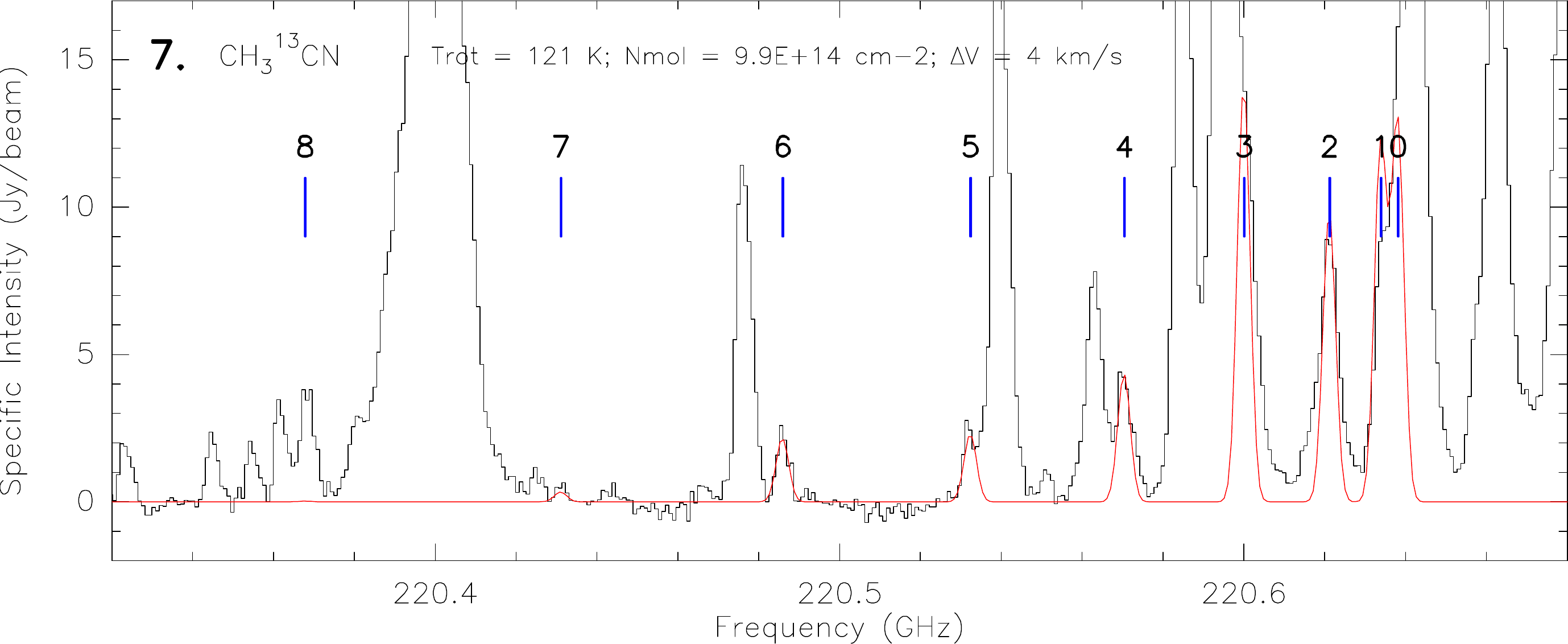}
&\includegraphics[width=5cm]{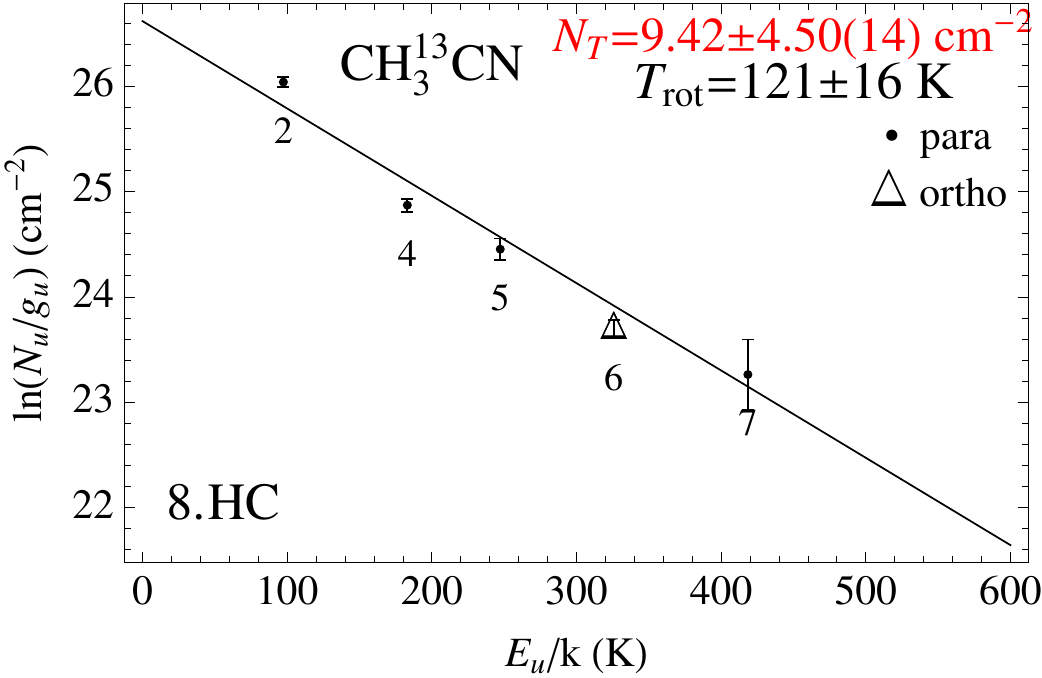}
\end{tabular}

\caption{Panel 1): $\rm CH_3CN$ spectrum towards the HC  in black,  LTE model fit in dashed purple, and fitting with optical depth correction in filled red.  Panels 2-6):  rotation diagram of $\rm CH_3CN$ derived for each substructure,  with dots for p-$\rm CH_3CN$ and triangles for o-$\rm CH_3CN$. Black fittings are only for the p-lines when assuming they are optically thin, and red fittings are for both o- and p- lines by optical depth correction. Panel 7):  $\rm CH_3^{13}CN$ spectrum towards the HC  in black and  the LTE model fit in red.  Panel 8:)  rotation diagram of  $\rm CH_3^{13}CN$ in HC with black dots (p-) and triangles (o-) from observation, and fitting is o-lines excluded by assuming lines are optically thin. For lines without detection,  an upper limit equal to the $\rm 3\sigma$ rms is shown as arrow. The estimated temperature and total column density from each fitting is  indicated in each panel.    Errors are derived from scatter in the data points. 
}\label{rotation}
\end{center}
\end{figure*}

The $\rm K=3,~6$ lines are from  ortho (o-) transitions, which have a statistical weight that is twice as strong as those of the other para (p-)  lines. Radiative transitions between o-/p-types are usually prohibited,  so we exclude these lines in the optically thin rotation diagram fitting (shown in black  lines in Figure~\ref{rotation}, also see \citealt{andersson84}). \\

Panel 1 of Figure~\ref{rotation}  shows the synthesised fitting under the assumption of optically thin emission in LTE for the HC with the dashed lines. To reduce the number of free parameters, we also assume that all the $\rm K=0-8$ components\footnote{The $\rm K=9$ line is blended by strong $\rm ^{13}CO~(2\rightarrow1)$ emission.} have the same line width. A satisfactory fit is obtained to all lines except $\rm K=0,~1,~2$. Since all transitions were observed simultaneously, calibration errors due to poor pointing or amplitude scaling are not responsible, and the remaining explanations  for the deviations are as follows: (1). The $\rm K=0, ~1$ lines are so close in frequency that  they are  blended in our fitting by the uniform line width we assumed.  
(2). It is also likely that lines at $\rm K=2$ are more optically thick.  \\

Compared to $\rm CH_3CN$, lines of its rare isotopologue $\rm CH_3^{13}CN$ should be optically thin. However,  $\rm CH_3^{13}CN~ (12_2\rightarrow11_2)$ is the only transition that is not blended with the broad line wings of $\rm CH_3CN$ in substructures other than HC. Even in the HC, only the $\rm K=2,~6$, and 7 lines are unblended (Panel 7 of Figure~\ref{rotation}). Excluding the o-type lines (K=3, 6), the rotation temperatures in HC we derive from the main and rare isotopologues of $\rm CH_3CN$ have large differences  (black fittings  in Panels 2 and 8 of Figure~\ref{rotation}).  Optical depth seems to be the most likely reason for such inconsistency, so to investigate this hypothesis, we use the following iterative approach to correct the optical depth of $\rm CH_3CN$ lines and derive  the opacity-corrected rotation temperature  $\rm T_{rot,\it p}$ ({\it p}=HC, mm2, mm3a, mm3b, NE) of each substructure (red  fittings in Panels 2-6 of Figure~\ref{rotation}): \\

\noindent (1) Obtain the rotation temperature $\rm T_{1,\it p}$ from weighted least squares fit to Boltzmann diagram, assuming that all the p-type $\rm CH_3CN$ lines are optically thin (see fit to the black dots in Panels 2-6 of Figure~\ref{rotation});\\
(2) Estimate the optical depth  of $\rm CH_3CN~ (12_2\rightarrow11_2)$ $\rm \tau_{K,1~ (K=2)}$ by comparing its main beam temperature with that of $\rm CH_3^{13}CN~ (12_2\rightarrow11_2)$, and calculate the column density of line $12_2\rightarrow11_2$ at $\rm T_{1,\it p}$ with the optical depth  correction. Then, estimate the total column density of $\rm CH_3CN$ as $\rm N_{1,\it p}$ assuming LTE (see Section~\ref{col} for more details about the correction);\\
(3) With $\rm T_{1,\it p}$ and $\rm N_{1,\it p}$ as trial input parameters, estimate the optical depth for the remaining transitions in the $\rm CH_3CN$ ladder $\rm \tau_{K,1~ (K=0-8)}$ with the RADEX code package\footnote{RADEX is a one-dimensional non-LTE radiative transfer code, providing an alternative to the widely used rotation diagram method. Without requiring the observation of  many optically thin emission lines, it can be used to roughly constrain the excitation temperature in addition to the column density.} \citep{vandertak07}, then correct their column densities line by line (dots and triangles in red in Figure~\ref{rotation}) and obtain a new rotation temperature $\rm T_{2,\it p}$;\\
(4) Repeat step (2)--(3) until $\rm (|T_{n, \it p}-T_{n-1,\it p}|)/T_{n-1,\it p}\le10\%$ (n=2,3,...; see fittings in red in Figure~\ref{rotation}).\\

The above approach\footnote{In reality, molecules do not distribute homogeneously in the clouds, and 
they are far from being homogeneous in density and temperature. The ultimate
best-fit  requires a  sophisticated, fully 3Dnd 
physico-chemo-LRT modelling, which is
beyond the scope of our paper. Our iterative approach is between realism and feasibility of the modelling, which fits both o-/p- $\rm CH_3CN$ well.}
can be applied to sources HC,  mm2 mm3a, mm3b, and NE, where the $\rm CH_3CN$ ladders and $\rm CH_3^{13}CN~(12_2\rightarrow11_2)$ can be clearly detected\footnote{The detection of $\rm CH_3^{13}CN~(12_2\rightarrow11_2)$ in NE is $4\sigma$.}. In SR and  OF1N(S),  the $\rm CH_3CN$  lines are below the  $\rm 3\sigma$ detection threshold, so we assume the rotation temperature in these  positions to be the same as that in NE. \\

\subsubsection{$\rm \bf HCOOCH_3$, $\rm \bf CH_3OH$, \bf and $\rm \bf ^{34}SO_2$}\label{tem:other}
$\rm HCOOCH_3$ has the largest number of lines detected in our dataset. From satisfactory synthetic fitting (especially for lines in the torsionally excited state $\nu=1$, Figure~\ref{COMspec}), we can reasonably assume that most of its lines are optically thin.  Line emissions from this species are not contaminated by N-bearing species owing to their different spatial distributions, so we use them to derive a second set of temperatures.\\

On the other hand, $\rm CH_3OH$ and $\rm H_2^{13}CO$ have an unusual spatial distribution (dual peaks /extended emissions at both HC and mm3a(b), and tail towards the SE), which may be affected by their different temperatures. Since multiple transitions of $\rm CH_3OH$ are available in our data,  we can use them to derive a third temperature set.\\

For S-bearing species, their extended emission indicates they may be sensitive to the ambient shock environment, so their temperature in each substructure may be different from the other species. However, in our data, some of the S-bearing species have only a single transition ($\rm O^{13}CS$, $\rm ^{13}CS$), and other lines may not be optically thin ($\rm SO_2$, SO, and OCS). Because $\rm ^{34}SO_2$ is the only S-bearing rare isotopologue that has multiple transitions in our observations, we applied  temperatures derived from transitions of  $\rm ^{34}SO_2$ to the other S-bearing species.\\

Using the lines list in Table~\ref{tab:rotline}, we derived the rotation temperatures in each substructure from $\rm HCOOCH_3$, $\rm CH_3OH$, and $\rm ^{34}SO_2$  (Figures~\ref{fig:trot_other}I--III) and list these temperature sets in bold face in Table~\ref{tab:trotcom}. Temperatures derived from $\rm HCOOCH_3$ lines  in the ground and torsionally excited states separately have some differences, but they do not deviate significantly from temperatures derived from all of its lines. Therefore, we use the ``mixed" temperatures in the end.\\

\subsubsection{The other organics}\label{tem:organics}
\begin{figure*}[!ht]
  \small
\begin{center}

\begin{tabular}{p{4.2cm}p{4.2cm}p{4.2cm}p{4.2cm}}
I. $\rm HCOOCH_3$\\
\includegraphics[width=4.5cm]{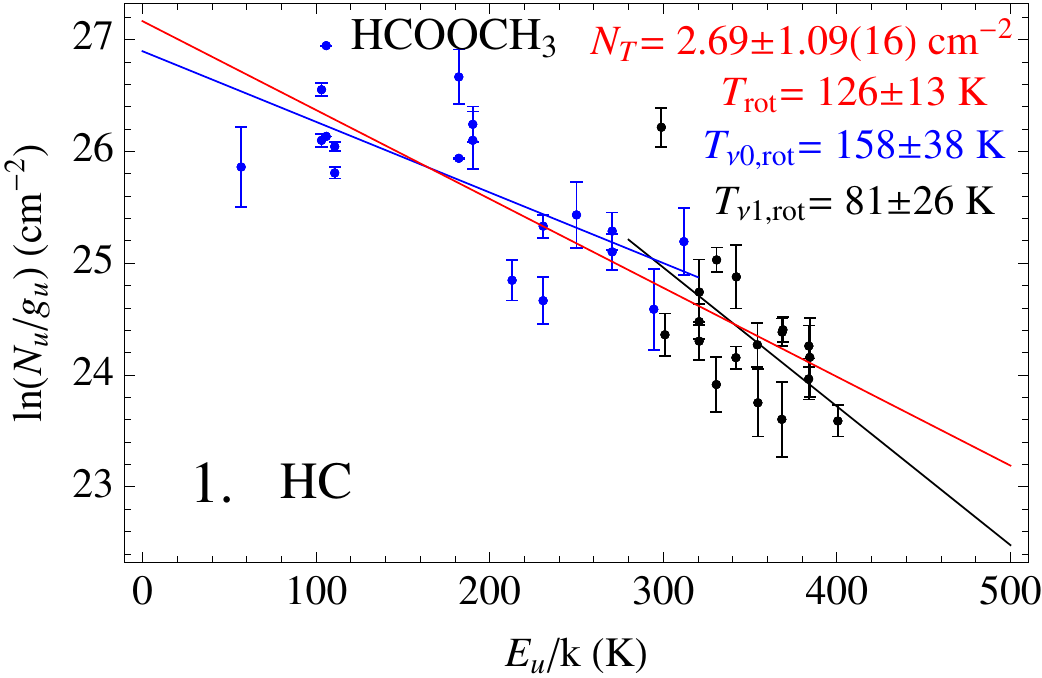}
&\includegraphics[width=4.5cm]{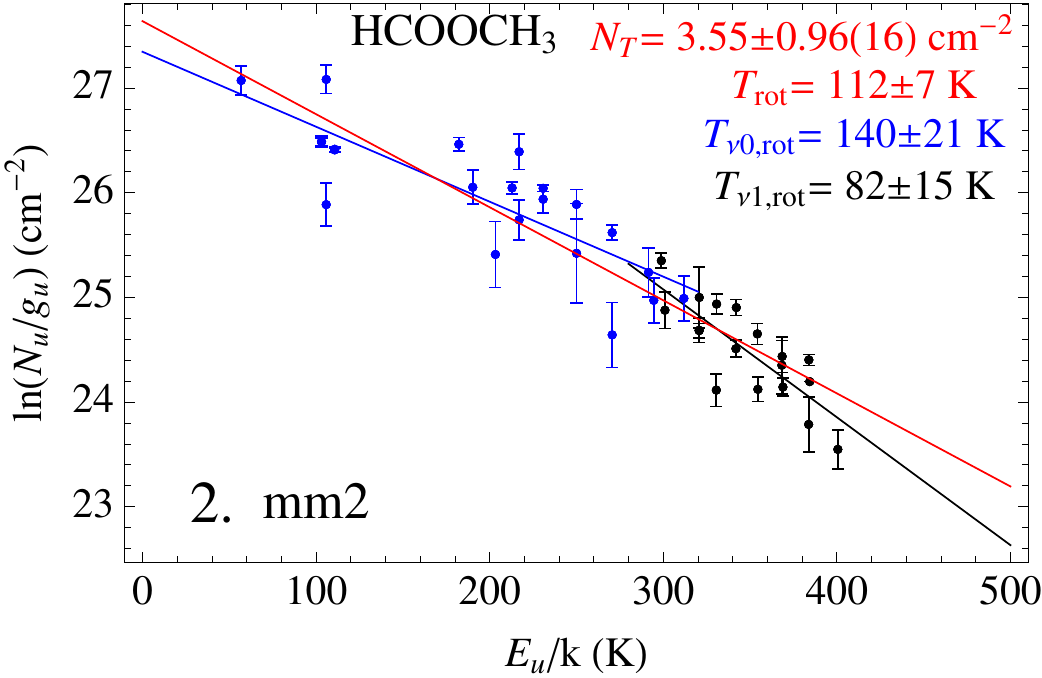}
&\includegraphics[width=4.5cm]{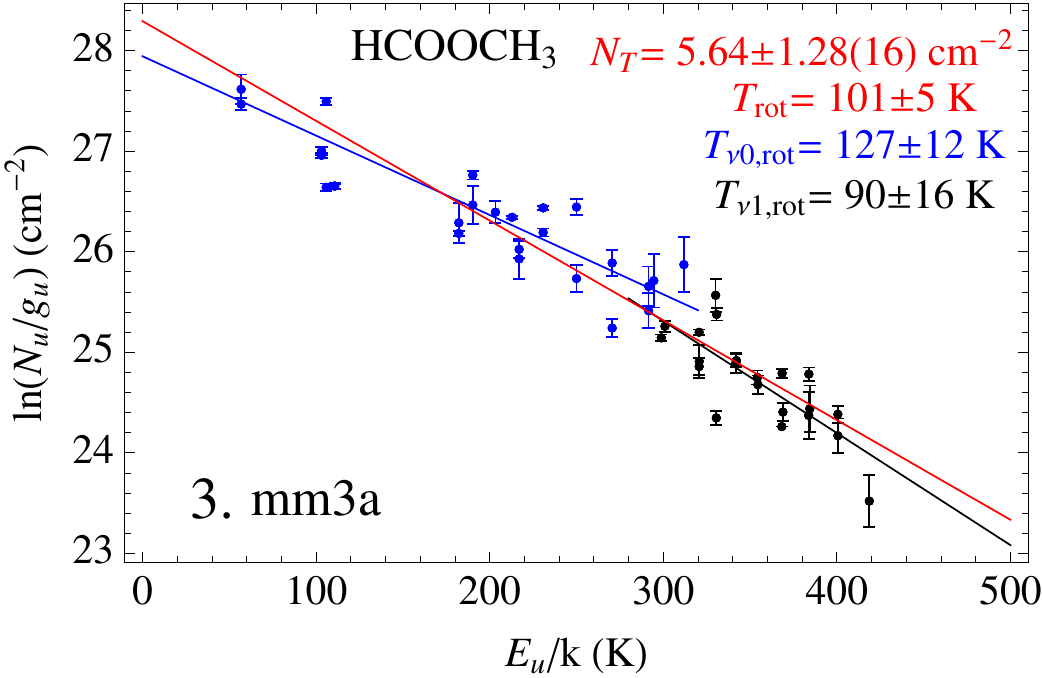}
&\includegraphics[width=4.5cm]{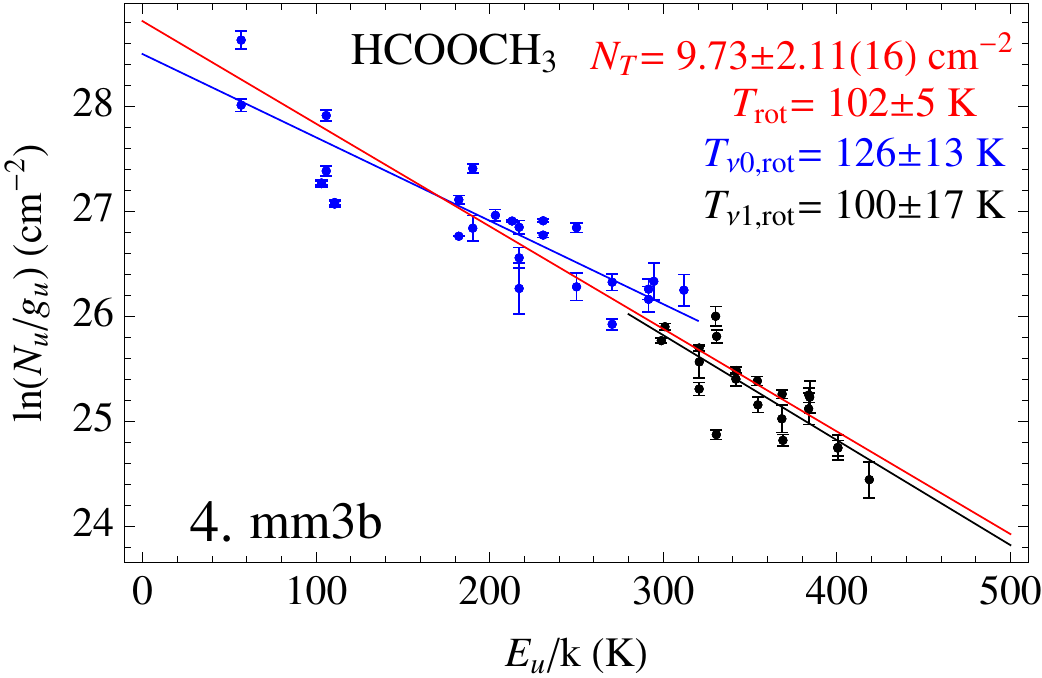}
\end{tabular}

\begin{tabular}{p{4.2cm}p{4.2cm}p{4.2cm}p{4.2cm}}
II. $\rm CH_3OH$\\
\includegraphics[width=4.5cm]{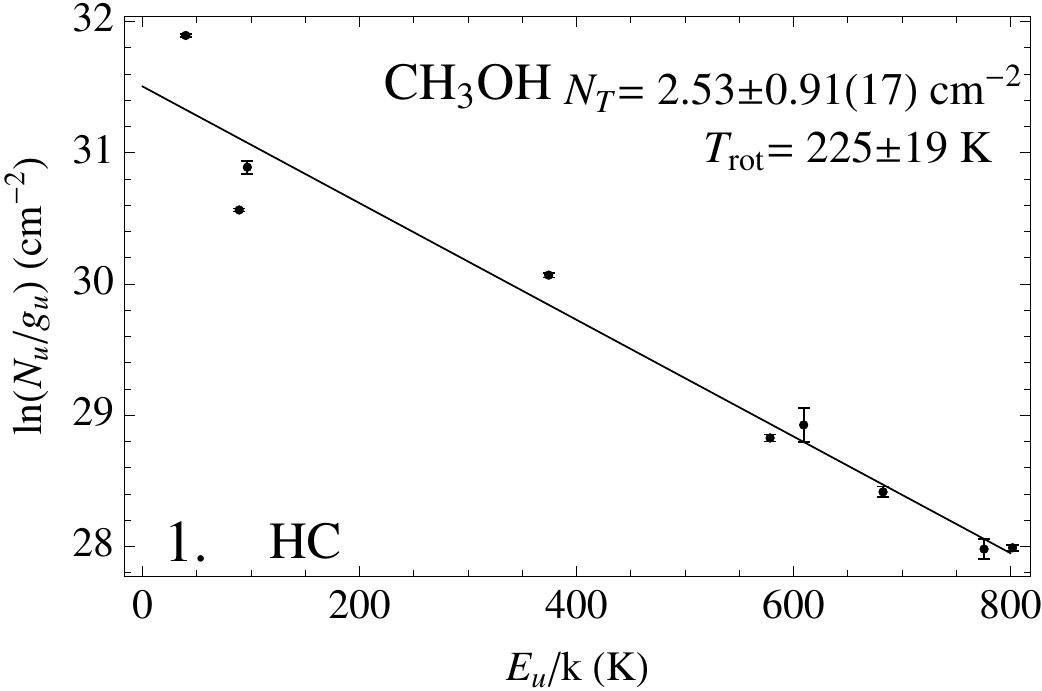}
&\includegraphics[width=4.5cm]{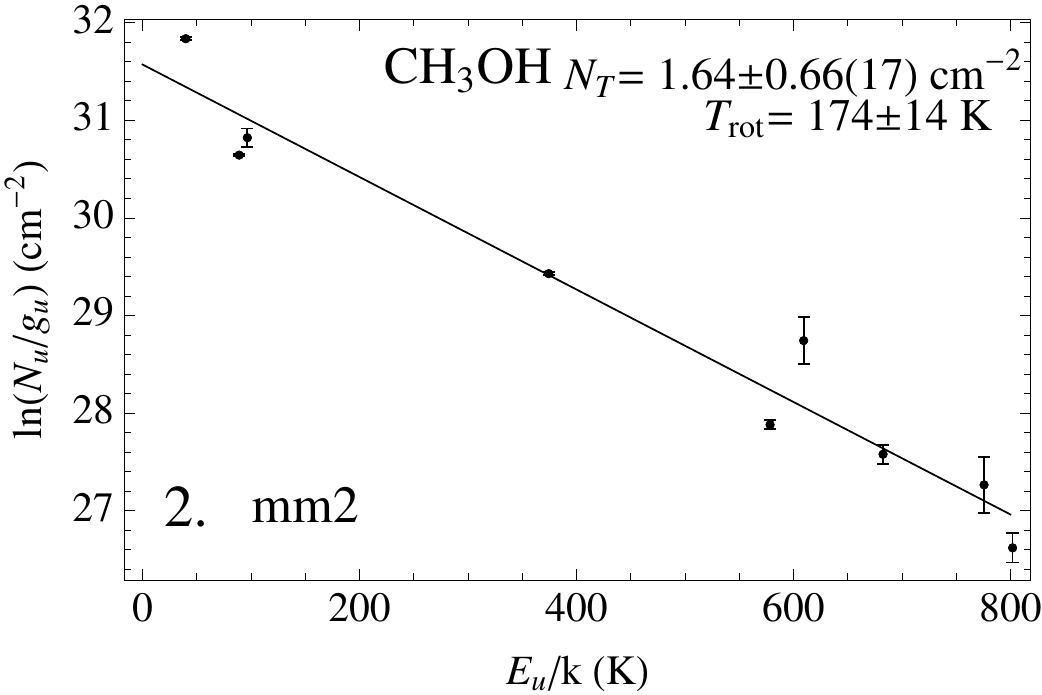}
&\includegraphics[width=4.5cm]{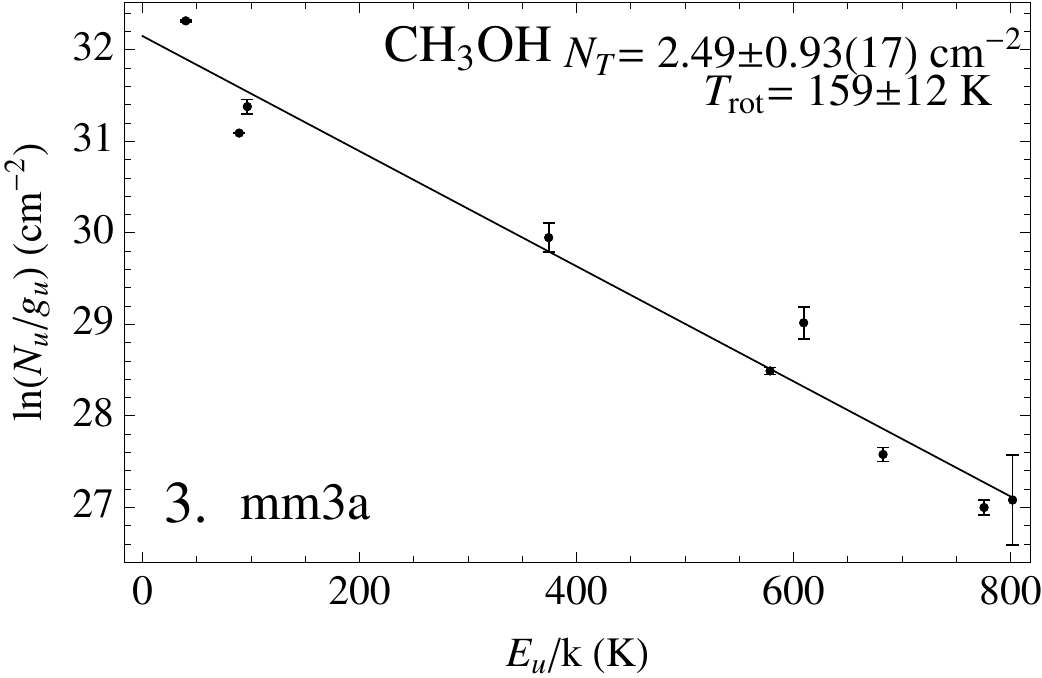}
&\includegraphics[width=4.5cm]{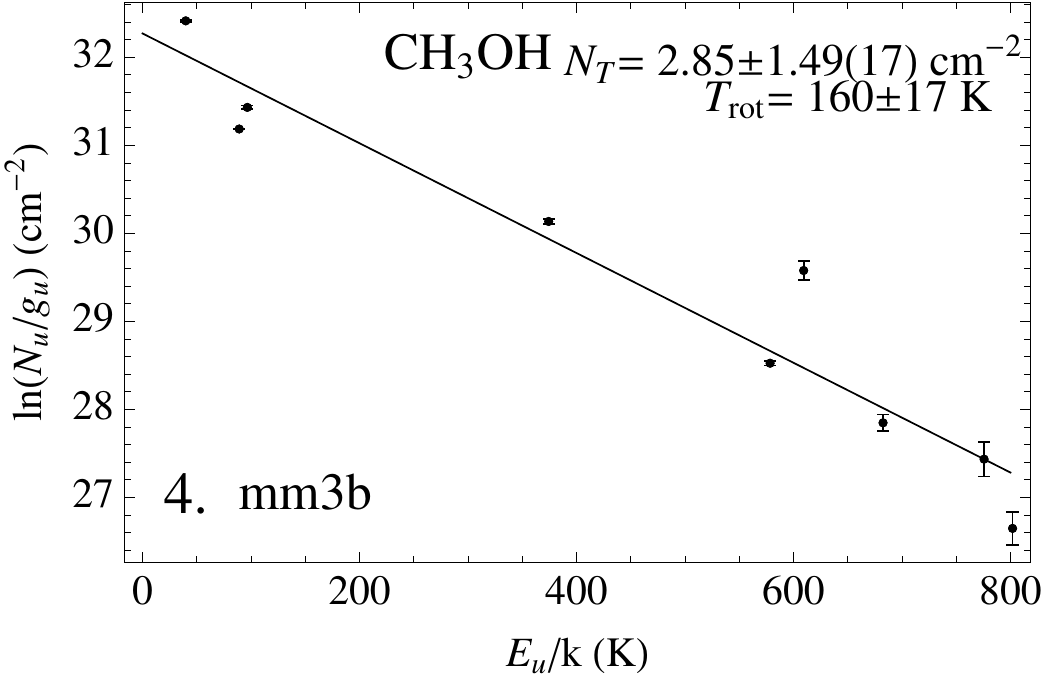}
\end{tabular}

\begin{tabular}{p{4.2cm}p{4.2cm}p{4.2cm}p{4.2cm}}
III. $\rm ^{34}SO_2$\\
\includegraphics[width=4.5cm]{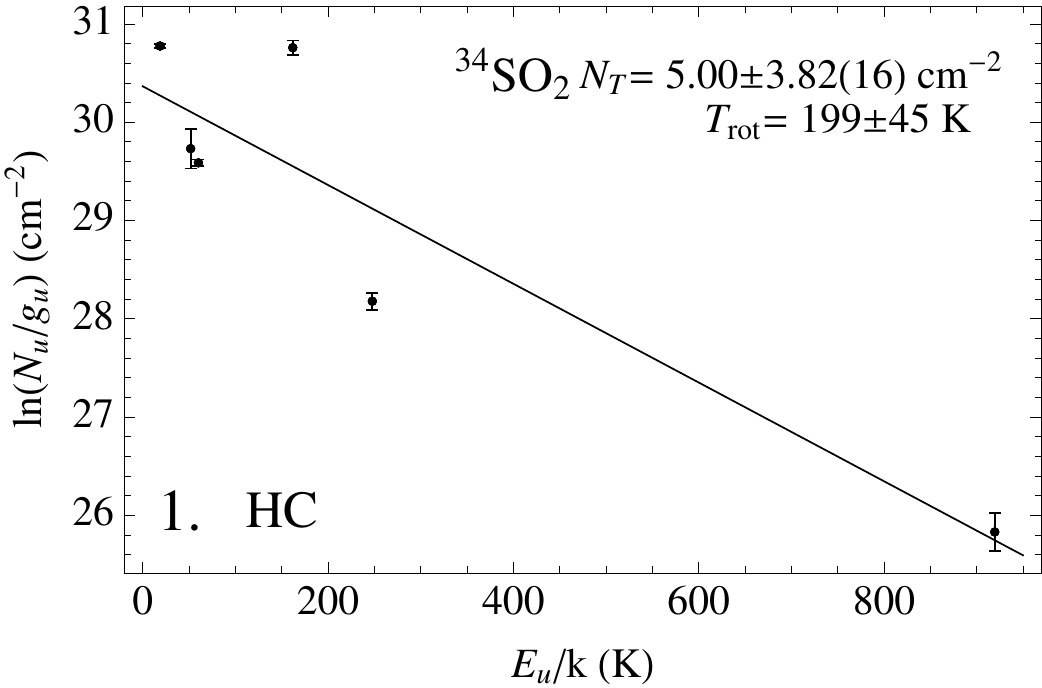}
&\includegraphics[width=4.5cm]{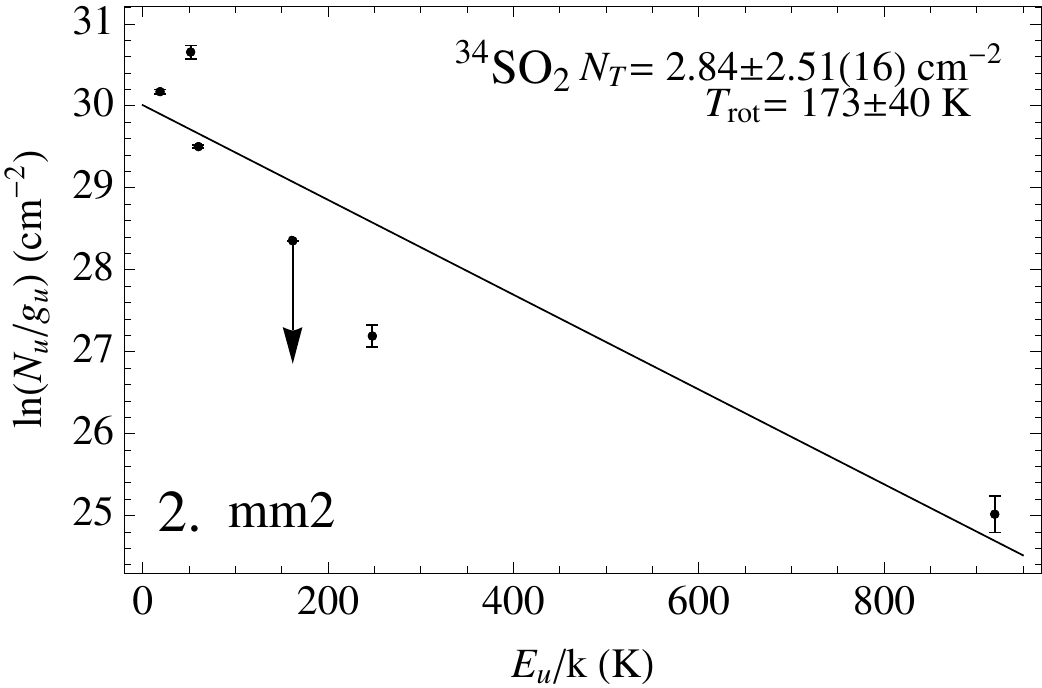}
&\includegraphics[width=4.5cm]{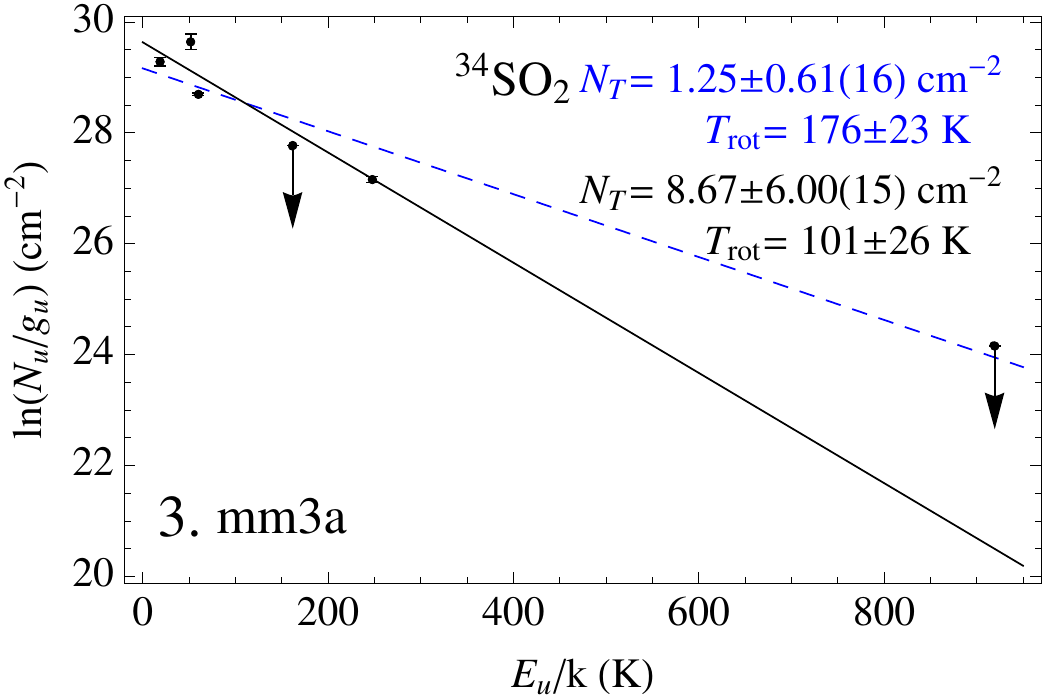}
&\includegraphics[width=4.5cm]{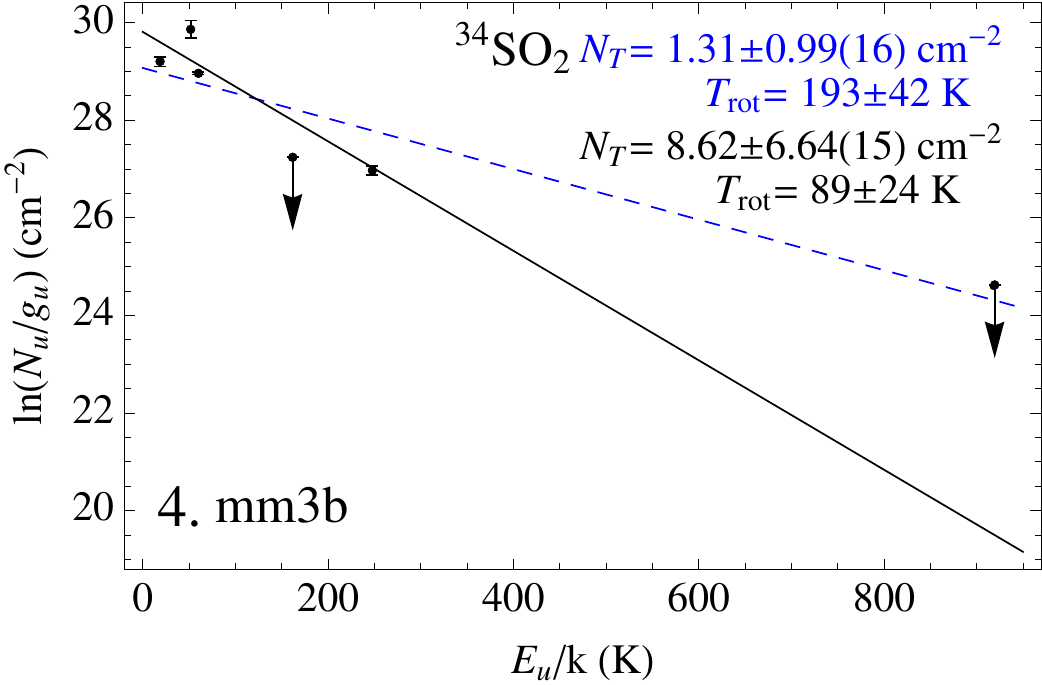}
\end{tabular}

\vspace{0.2cm}
\begin{tabular}{p{4.2cm}p{4.2cm}p{4.2cm}p{4.2cm}}
IV. HNCO\\
\includegraphics[width=4.5cm]{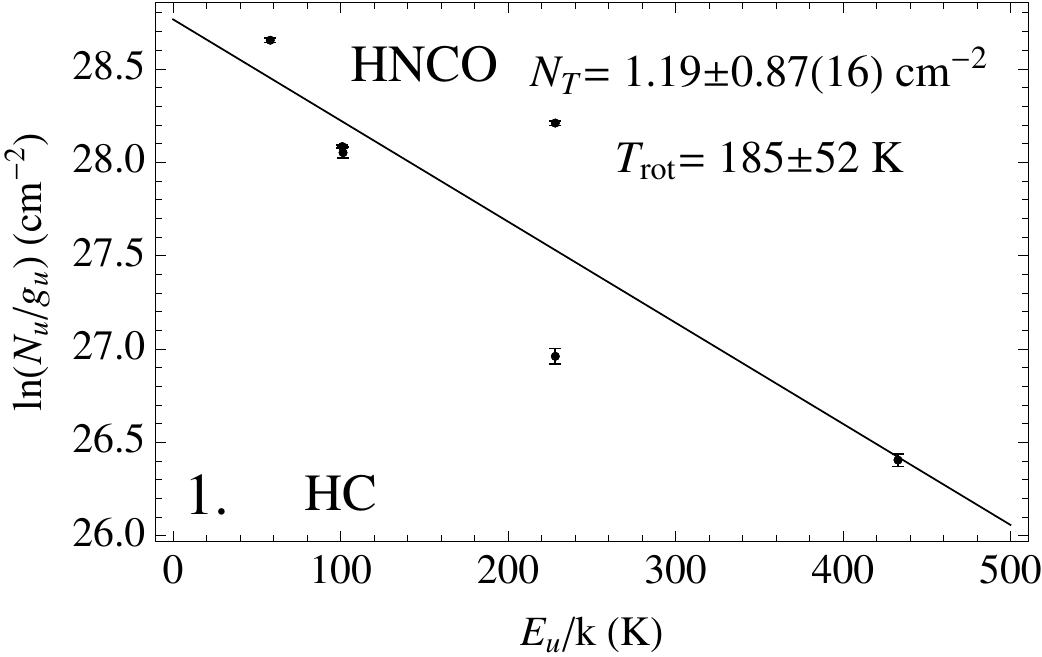}
&\includegraphics[width=4.5cm]{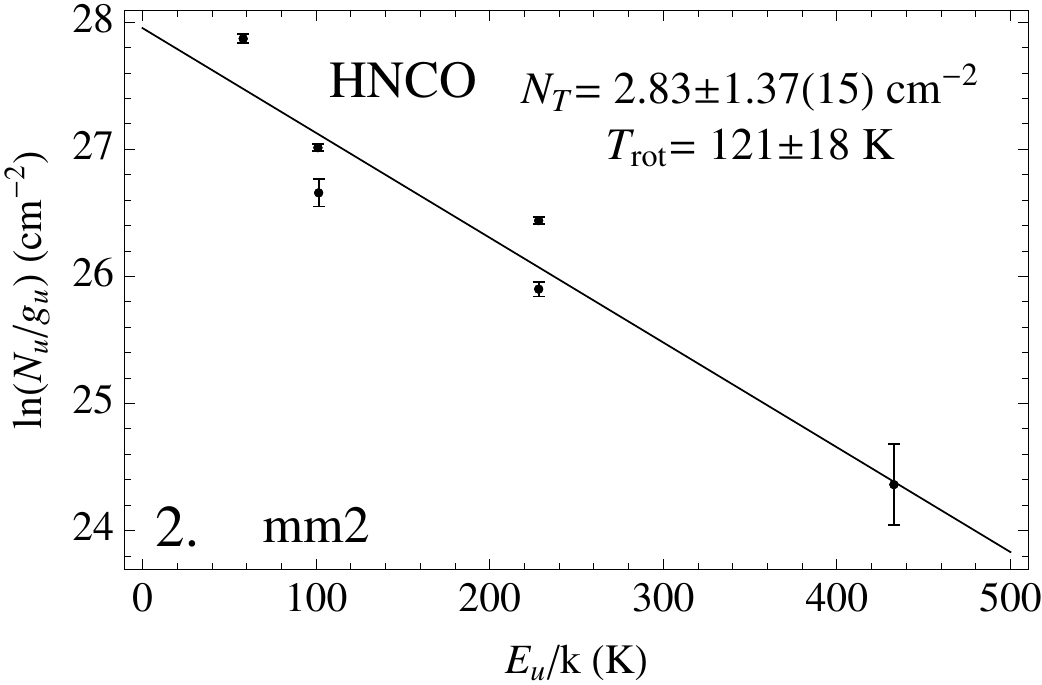}
&\includegraphics[width=4.5cm]{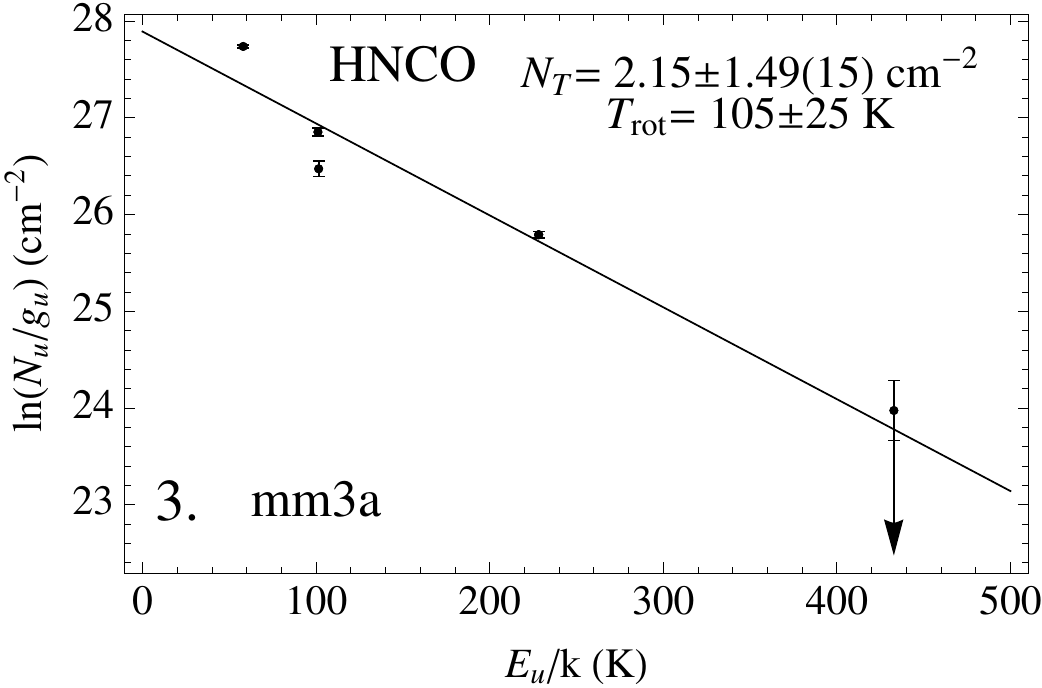}
&\includegraphics[width=4.5cm]{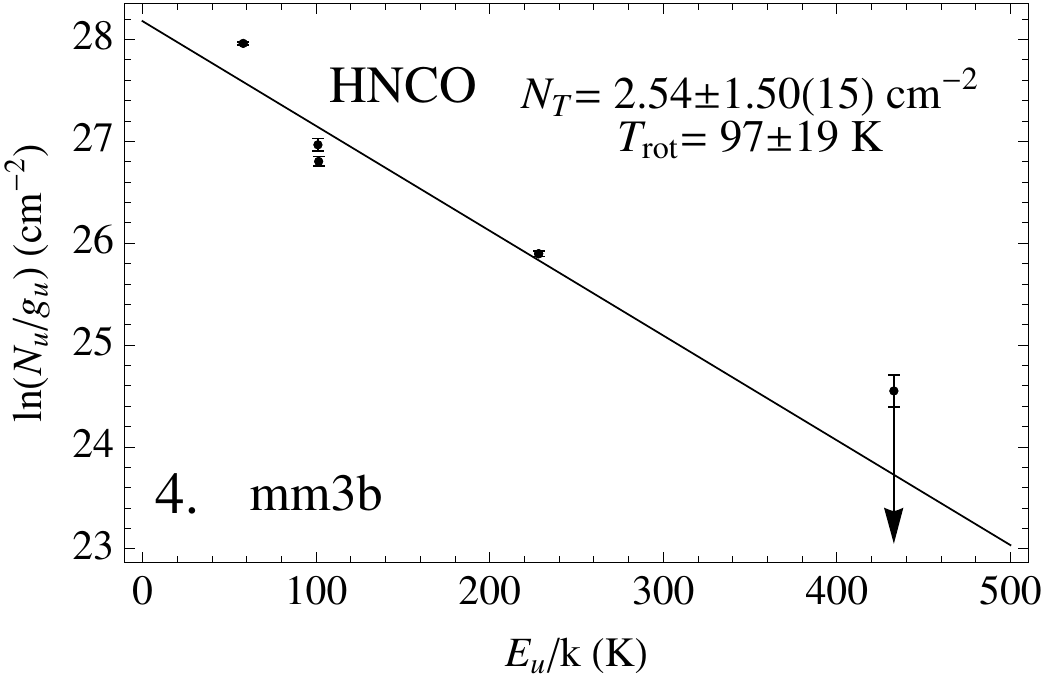}
\end{tabular}

\begin{tabular}{p{4.2cm}p{4.2cm}p{4.2cm}p{4.2cm}}
V. $\rm CH_3OCH_3$\\
\includegraphics[width=4.5cm]{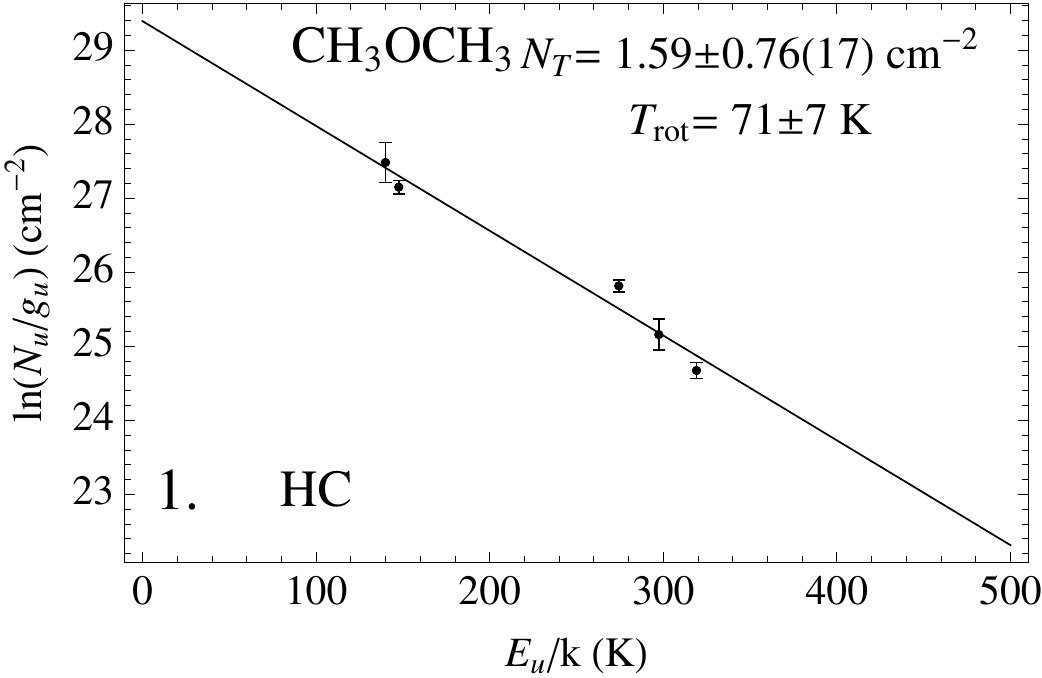}
&\includegraphics[width=4.5cm]{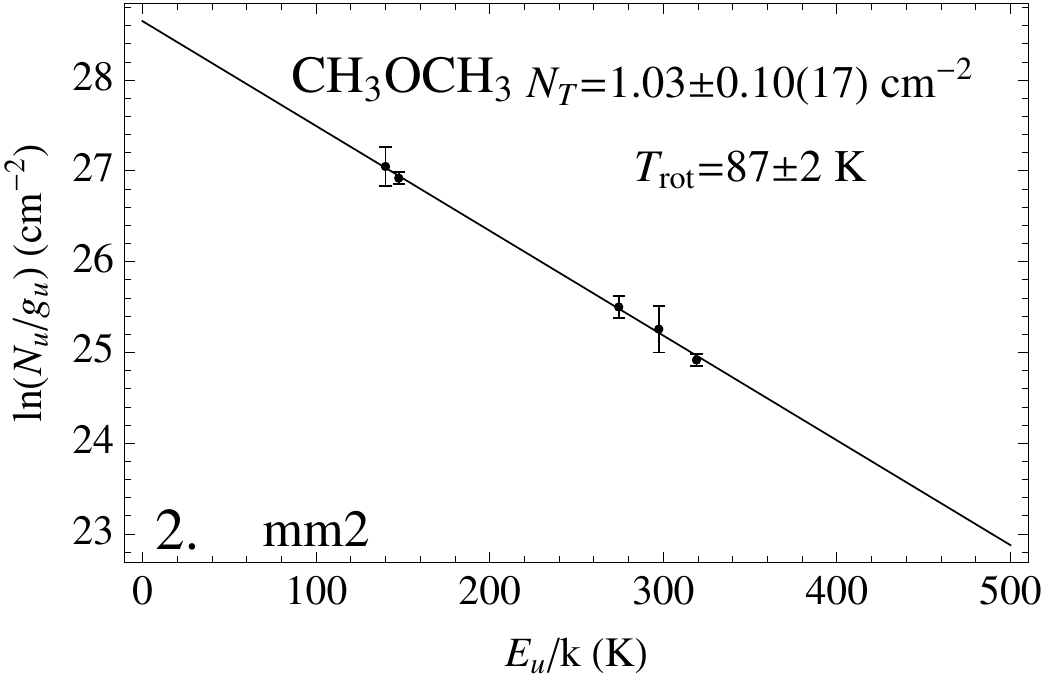}
&\includegraphics[width=4.5cm]{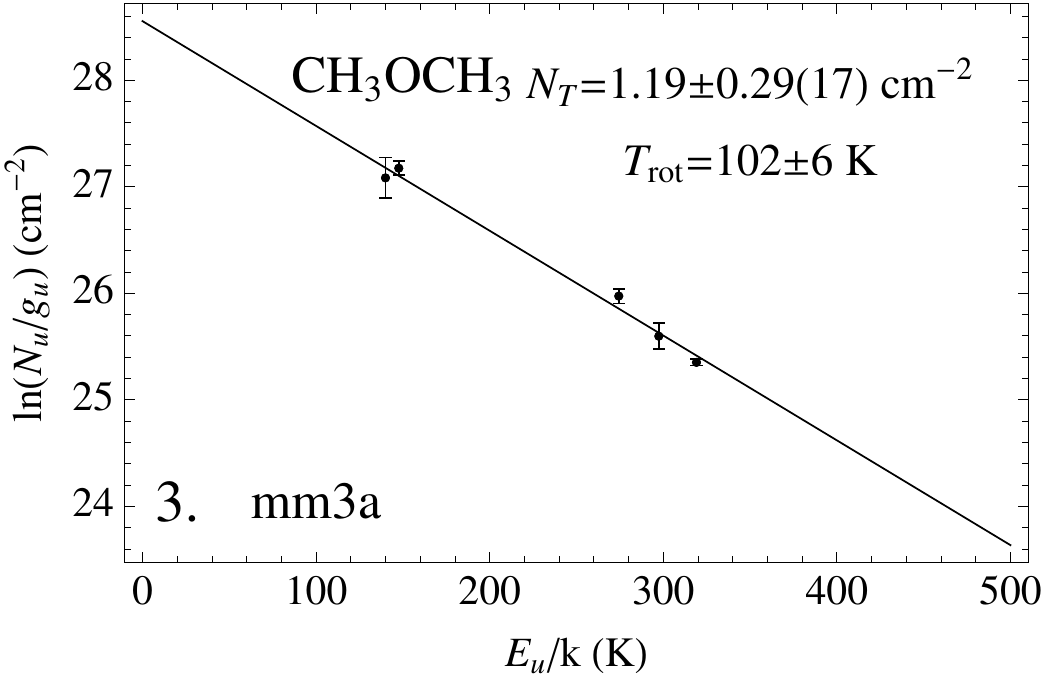}
&\includegraphics[width=4.5cm]{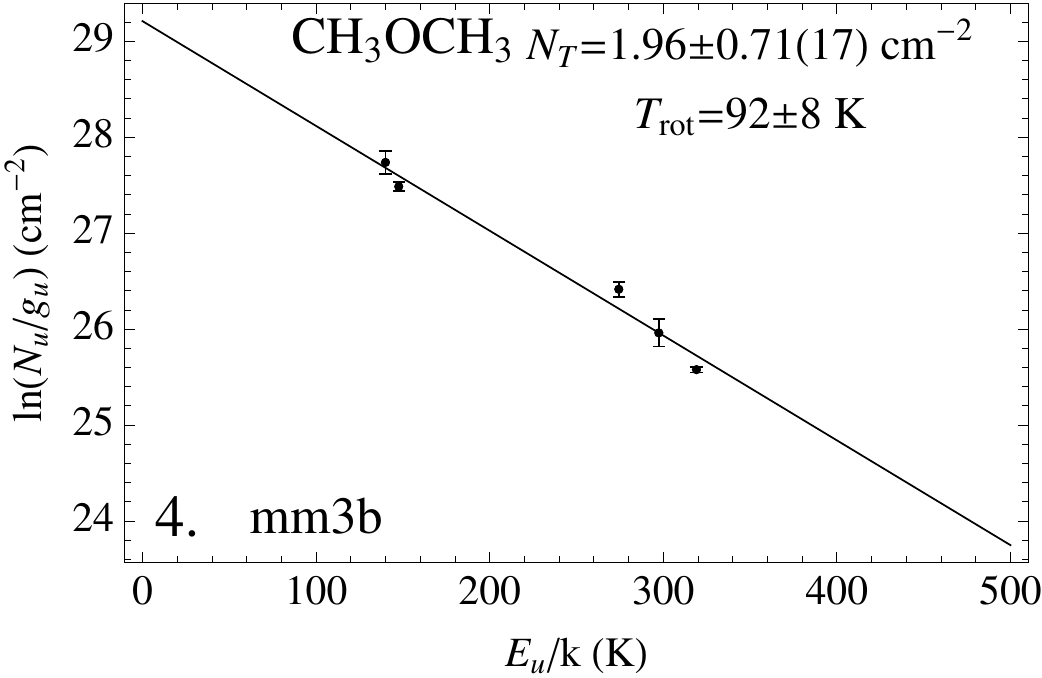}
\end{tabular}

\begin{tabular}{p{4.2cm}p{4.2cm}p{4.2cm}p{4.2cm}}
VI. $\rm CH_3CH_2OH$\\
\includegraphics[width=4.5cm]{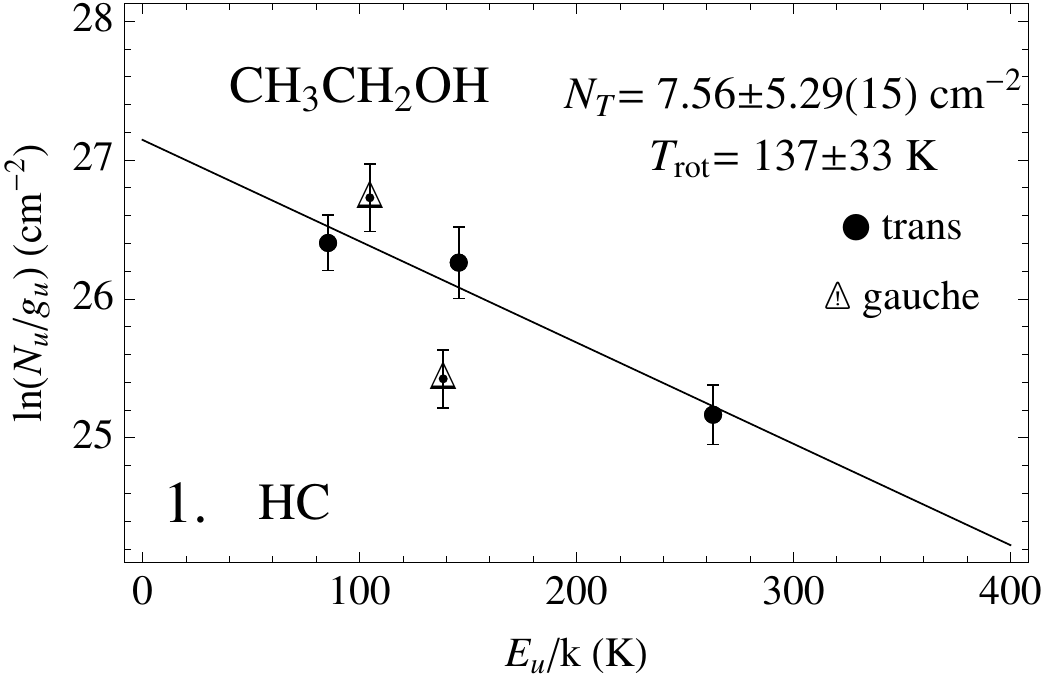}
&\includegraphics[width=4.5cm]{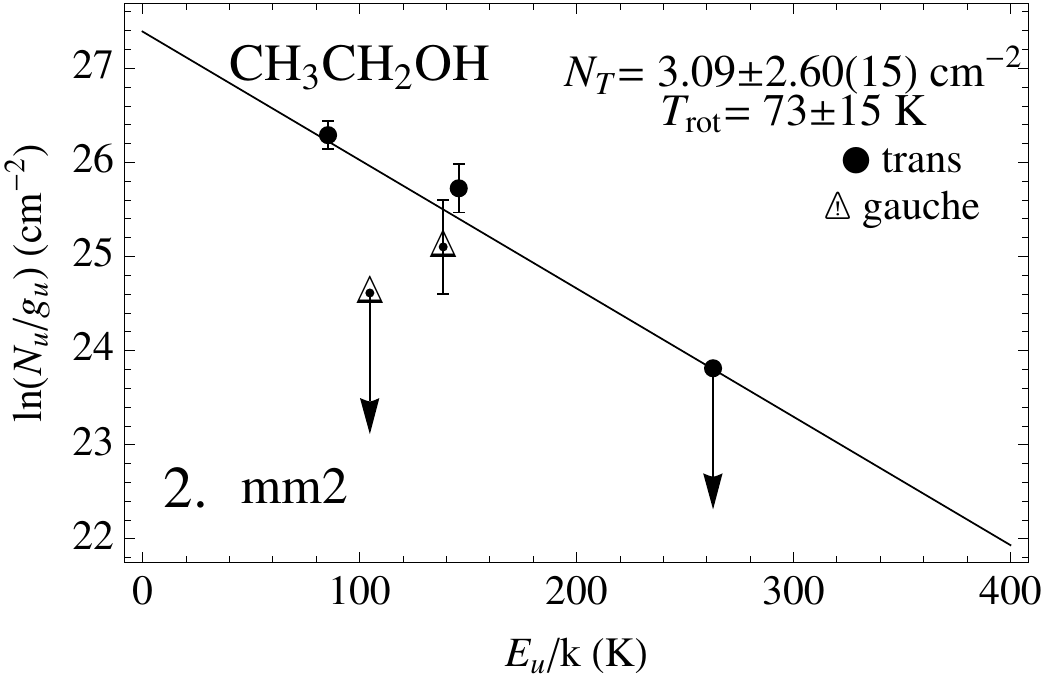}
&\includegraphics[width=4.5cm]{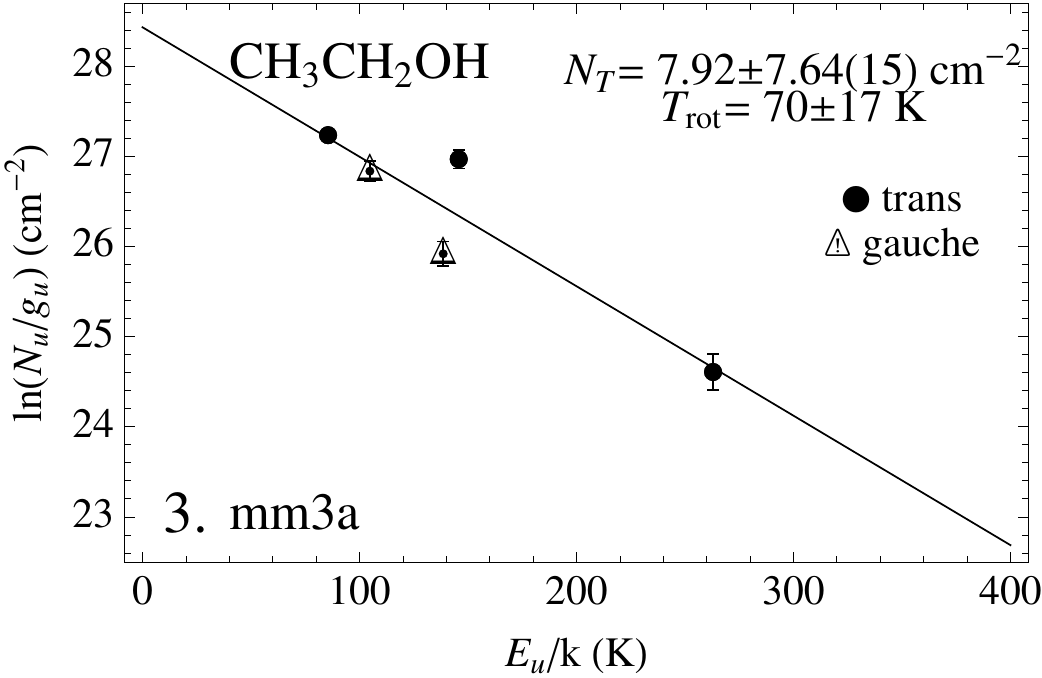}
&\includegraphics[width=4.5cm]{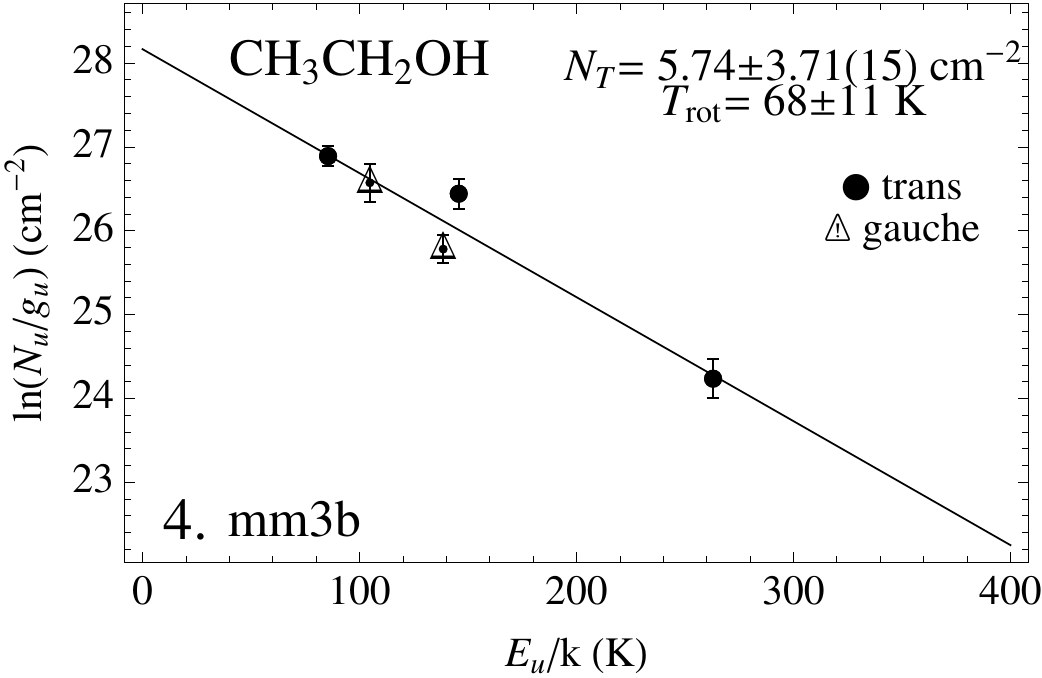}
\end{tabular}

\caption{ Rotation diagrams of the N-, O-, and S-bearing species with multiple unblended transitions detection in HC, mm2, mm3a, and mm3b. 
Panels in Row I show two states of $\rm HCOOCH_3$ lines, with the blue and black dots from observation  of ground state ($\rm \nu=0$) and torsionally excited state  ($\rm \nu=1$) lines, respectively. With the optically thin assumption,  fittings for each state  are plotted in blue or black, respectively, and fittings for all the transitions are plotted in red. Non-detection of $\rm HCOOCH_3$ lines are not shown.
Panels in Row II--VI show fittings with the optically thin assumption, with $\rm 3\sigma$ upper limits from non-detections plotted as arrows. 
Blue dashed lines in Row III are fittings with upper limits included.
The estimated temperature and total column density from each fitting is  indicated in each panel.   Errors are derived from scatters in the data points.
}\label{fig:trot_other}
\end{center}
\end{figure*}

In addition, since different organic molecules have specific spatial distributions, we also derive the rotation temperatures for the species that have multiple transitions with significantly different $\rm E_u/k_B$ detected in our dataset (Figures~\ref{fig:trot_other}IV--VI, listed in Table~\ref{tab:rotline}). In the condensations HC, mm2, mm3a, and mm3b, rotation temperatures of HNCO are similar to those derived from $\rm CH_3CN$ within uncertainties; and rotation temperatures derived from $\rm CH_3OCH_3$ and $\rm CH_3CH_2OH$ are close to $\rm HCOOCH_3~(\nu=1)$ within uncertainties. Therefore, temperature sets from $\rm CH_3CN$ and $\rm HCOOCH_3$ are applicable to the rest N- and O-bearing species  (including simple molecules and COMs) in estimating their column densities.\\

 Although  temperatures derived from different species at the same substructure vary, HC is consistently the warmest substructure, while mm2, mm3a, and mm3b are slightly cooler. In the outflow regions, there are no sufficiently strong lines for species other than $\rm CH_3CN$ to derive the temperature, which may suggest these regions are cooler, exhibiting similar chemical properties.

\begin{table*} 
\small
\begin{center}
\begin{tabular}{c|p{2cm}p{2cm} p{2cm}p{2cm} p{2cm}}
\hline\hline

Species                            &HC (K)                      & mm2 (K)               & mm3a (K)             &mm3b (K)            &NE (K) \\
\hline
$\rm \bf CH_3CN$                 &$ \bf 155\pm16$              &$\bf 108\pm4$            &$\bf 103\pm3$           &$\bf 88\pm6$         &$ \rm \bf 43\pm3$\\
$\rm CH_3^{13}CN$        &$\bf 121\pm16$                 &--                           &--                          &--                        &--\\
$\rm HNCO$          &$185\pm52$              &$121\pm18$            &$105\pm25$           &$97\pm19$           &--\\
\hline
$\rm \bf HCOOCH_3$          &$\bf 126\pm13$              &$\bf 112\pm7$            &$\bf 101\pm5$           &$\bf 102\pm5$           &--\\
$\rm \bf CH_3OH$               &$\bf 225\pm19$              &$\bf 174\pm14$            &$\bf 159\pm12$           &$\bf 160\pm17$           &--\\
$\rm CH_3OCH_3$          &$71\pm7$              &$87\pm2$            &$102\pm6$           &$92\pm8$           &--\\
$\rm CH_3CH_2OH$          &$137\pm33$              &$73\pm15$            &$70\pm17$           &$68\pm11$           &--\\
\hline
$\rm \bf ^{34}SO_2$              &$\bf 199\pm45$              &$\bf 173\pm40$            &$\bf 101\pm26$           &$\bf 89\pm24$           &--\\
\hline

\end{tabular}
\caption{Rotation temperature sets for each substructure derived from species, each of which has $\rm >4$ transitions with significantly different $\rm E_u/k_B$ detected in our data. The substructures where there are not enough sufficient lines to derive the rotation temperatures are denoted by ``--". Temperatures in bold face are used to estimate the column densities and abundances of species whose rotation diagrams are unable to be derived.
\label{tab:trotcom}
}
\end{center}
\end{table*}

\subsection{Column densities}\label{col}
\subsubsection{Molecular column densities} \label{spl}   
For all the species we detected, we calculate their column densities (from both main and rare isotopologues) from the integrated spectral line intensities as below.\\

The optical depth $\rm \tau_\upsilon$  (in velocity) along the line of sight is (\citealt{zeng06},  Eq. A.25): 
\begin{equation}
\rm \tau_\upsilon=\frac{c^3}{8\pi \nu^3 } A_{ul} N_u (e^{\frac{h \nu}{k_BT_{rot}}}-1) \Phi(\upsilon). 
\end{equation}
where c is the speed of light, $\rm k_B$  the Boltzmann constant, h  the Planck constant, $\rm \nu$  the line rest frequency, $ \Phi(\upsilon)$  the distribution function of the line shape in terms of velocity $\upsilon$. 
Here, $\rm  A_{ul}$ is the average spontaneous emission rate from the upper state $\rm E_u$ into the lower state $\rm E_l$, and it is calculated from line strength $S_{ul}$ and dipole moment $\mu$ as \\
\begin{equation}
\rm  A_{ul}=\frac{64\pi^4\nu^3}{3hc^3}\frac{\it S_{ul}\mu^2}{g_u}
\end{equation}

After integrating the optical depth  over the observed  linewidth, $\rm \int \Phi(\upsilon)  d\upsilon=1$, and the column density  $\rm N_{u}$ of the line at $\rm \nu$ Hz is
\begin{equation}\label{up}
\rm N_u=\frac{8 \pi \nu^3}{c^3 A_{ul}}\frac{1}{exp(\frac{h \nu}{k_BT_{rot}})-1} \int \tau_\upsilon  d\upsilon. ~~~~~(cm^{-2})
\end{equation}

When assuming LTE,  the total column density of all transitions for a molecule $\alpha$ is 
\begin{equation}\label{NT}
\rm N_{T,\alpha}=\frac{N_u}{g_u} Q(T_{rot}) e^{\frac{E_u}{k_BT_{rot}}} ~~~~~(cm^{-2})
\end{equation}
where $\rm Q(T_{rot})$ is the partition function 
for the given  (rotational) excitation temperature of each sources, which is interpolated from a table of partition functions for fixed temperatures,  obtained from CDMS/JPL.\\

Here,  $S_{ul}\mu^2$ can be  calculated from CDMS/JPL line intensity $\rm  \ell(T_{rot})$  \citep{pickett98},

 \begin{eqnarray}
  S_{ul}\mu^2&=&\rm \frac{3hc}{8\pi^3\nu}\ell(T_{rot})Q(T_{rot})(e^{-\frac{E_l}{k_BT_{rot}}}-e^{-\frac{E_u}{k_BT_{rot}}})^{-1}\\
&=&\rm 2.4025\times10^{10}[\frac{\nu}{Hz}]^{-1}[\frac{\ell(300\,K)}{nm^2MHz}]Q(300K) \nonumber\\
&&\rm \times[e^{-\frac{E_l}{k_B300K}}-e^{-\frac{E_u}{k_B300K}}]^{-1} Debye^2\label{sul}
 \end{eqnarray}

For the emission lines, we assume the Rayleigh-Jeans approximation ($\rm \frac{h\nu}{k_BT_{rot}}\ll1$) is valid, therefore Eqs.~\ref{up} and \ref{NT} can be simplified as  (\citealt{zeng06},  Eq.  4.26, also see \citealt{wilson09}),
 \begin{align}\label{emi}
 \rm \frac{N_u}{g_u}\cong \frac{3k_B}{8\pi^3\nu \it S_{ul}\mu^2} T_{rot} \int \tau_\upsilon  d\upsilon  &~~~~~~~~~~\rm(cm^{-2})\\
\rm N_{T,\alpha}=
\rm \frac{k_B}{h c} \frac{(e^\frac{h\nu}{k_BT_{rot}}-1)}{\ell(T_{rot})}T_{rot} \int \tau_\upsilon  d\upsilon  &~~~~~~~~~~\rm(cm^{-2})
 \end{align}

Assuming that  the observed emission in each substructure is homogeneous and fills the combined beam, 
the integration of measured  main beam brightness temperature within the  velocity range $\rm \int T_{B}(\upsilon)  d\upsilon$  can be  substituted for the last term in the above equation: 
\begin{equation}\label{eq:correc}
\rm T_{rot} \int \tau_\upsilon  d\upsilon \cong \frac{\tau_{\alpha,0}}{1-e^{-\tau_{\alpha,0}}}\int T_{B}(\upsilon)  d\upsilon~~~~~~~~~~~~(K~cm\,s^{-1})\\ 
\end{equation}
where $\rm \tau_{\alpha,0}$ is the optical depth at line centre, and $\rm \int T_{B}(\upsilon)  d\upsilon$ is measured from Gaussian/hyperfine structure (HFS) fitting by using Gildas software package.\\%\footnote{http://www.iram.fr/IRAMFR/GILDAS}.\\

The column densities and related uncertainties of species listed in Table~\ref{tab:trotcom} ($\rm CH_3CN$ ($\rm CH_3^{13}CN$), HNCO, $\rm HCOOCH_3$, $\rm CH_3OH$, $\rm CH_3OCH_3$, $\rm CH_3CH_2OH$, and $\rm ^{34}SO_2$) can be estimated directly from their rotation diagrams and the scatterings of the data points (Figure~\ref{fig:trot_other}).
For the species whose transitions are not sufficient to derive the rotation diagrams, we estimate their column densities by assuming that transitions of related species  have the same $\rm T_{rot}$:\\
$\bullet$ N-bearing species have  similar temperatures to those derived from the opacity-corrected $\rm CH_3CN$,\\
$\bullet$ S-bearing species have similar temperatures to $\rm ^{34}SO_2$,\\
$\bullet$ O-bearing species have  similar temperatures to $\rm HCOOCH_3$,\\
$\bullet$ Given the similarly unusual spatial distribution between $\rm H_2^{13}CO$ and $\rm CH_3OH$, we also estimate the column densities of $\rm H_2^{13}CO$ at temperatures derived from $\rm CH_3OH$. \\

In general, we assume that the observed transitions are optically thin ($\tau_{\alpha,0}\le1, \frac{\tau_{\alpha,0}}{1-e^{-\tau_{\alpha,0}}} \approx1$). However, for several molecules with large abundances, we  observe the corresponding transitions of  their main and rare isotopologues, e.g.,  $\rm OCS ~(J=19\rightarrow18)$-$\rm O^{13}CS ~(19\rightarrow18)$,  $\rm ^{13}CO ~(2\rightarrow1)$-$\rm CO ~(2\rightarrow1$)-$\rm C^{18}O ~(2\rightarrow1$) and  $\rm CH_3^{13}CN ~(12_2\rightarrow11_2)$-$\rm CH_3CN ~(12_2\rightarrow11_2)$. By measuring the ratio between main beam brightness temperature of the main line $\rm T_{B,~\alpha,0}$ and its rare isotopologue $\rm T_{B,~\beta,0}$, we can estimate the optical depth at line centre of a given transition $\rm \tau_{\alpha,0}$ \citep{myers83}:\\ 

  \begin{equation}\label{eq:tau}
\rm \frac{1- exp(-\tau_{\alpha,0}/\Re^\alpha)}{1-exp(-\tau_{\alpha,0})}\approx \frac{T_{B, ~\beta,0}}{T_{B, ~\alpha,0}}\\
  \end{equation}
where $\Re_\alpha$ is the intrinsic abundance of the main isotope (e.g., $\rm ^{12}C$) compared to its rare  isotope (e.g., $\rm ^{13}C$) in the ISM (Table \ref{tab:correction}, e.g., \citealt{wilson94,chin96}). \\

Using the above method, we calculate the column densities of 25 isotopologues (for both optically thin and thick cases) in each substructure, and present them in Tables \ref{col-Obearing}I,  \ref{col-Nbearing}I and  \ref{col-Sbearing}I. In Table \ref{tab:correction}, we list the transitions for which we corrected their optical depths. We found that $\rm CO ~(2\rightarrow1$),  $\rm OCS ~(19\rightarrow18)$, and $\rm CH_3CN ~(12_2\rightarrow11_2)$ are significantly optically thick ($\rm \tau=6-117$) in all substructures where they are detected. The validity of the assumption that the remaining molecular lines are optically thin  is discussed in Section \ref{error}.\\

In addition, the main emission peak(s) of some COMs ($\rm CH_3COCH_3$, $\rm CH_3CH_2OH$, $\rm CH_2CHCN$, $\rm C_6H$, and $\rm HC_7N$) do not strictly coincide with either of the continuum peaks. In determining their column densities at their main emission peaks, we use their strongest transitions  (shown in Figure~\ref{COMdis}) and  the temperature from an adjacent substructure.   Lines are assumed to be optically thin, and molecular column densities could be underestimated if this assumption is incorrect, so the lower limits are given in Table~\ref{COMcol}.

\subsubsection{$\rm H_2$ column densities}\label{continu}
In Orion-KL, at 1.3\,mm wavelength, the continuum emission is dominated by dust-neutral gas emission \citep{beuther04}. 
Assuming  the dust emission at mm wavelengths is optically thin and coupled with the gas, which  has a single excitation temperature,  $\rm T_{ex}\sim T_{rot}=T_{dust}$ \citep{hildebrand83},  we use the SMA-only data to calculate the $\rm H_2$ column densities for different substructures, according to \citet{schuller09} :
 \begin{equation}\label{gas}
 \rm N_{H_2}=\frac{I_{\nu}R}{B_\nu(T_{dust})\Omega_a \kappa_{\nu}\mu m_H}~~~~~~~~~(cm^{-2})  \\
    \end{equation}
    where $\rm I_\nu$ is the peak specific intensity for each substructure on the continuum  (in $\rm Jy~beam^{-1}$, taken to be the $3\sigma$ upper limit in the outflow regions);  R is the isothermal gas to dust mass ratio (taken to be 150, from \citealt{draine11});  $\rm B_\nu(T_{dust})$ is the Planck function for a dust temperature $\rm T_{dust}$;  $\rm \Omega_a$ is the solid angle of the SMA beam  (in $\rm rad^2$);  {\color{black} $\rm \kappa_{\nu}=1.0~\rm cm^2g^{-1} $} is the dust absorption coefficient at 1.3\,mm (assuming a model of agglomerated grains with thin ice mantles\footnote{ At around 100--150 K, water ice starts to be evaporated. However, in the line of sight, the envelope of the protostar may not have the same temperature at different radii, so ice may not be completely evaporated. Therefore, without knowing the real geometry of the source, we assume the dust absorption coefficient according to the ``thin ice mantle" case. Moreover, this assumption is consistent with the modelling result  in \citet{gerner14}. In the extremely case, if we assume the ice is completely evaporated, the $\rm H_2$ column density derived from continuum will be 2--4 times lower than our current estimation. } for densities {\color{black} $\rm 10^7~cm^{-3}$} \citep{blake87}, extrapolated from \citet{ossenkopf94};  $\mu$ is the mean molecular weight of the ISM, which is assumed to be 2.33;  and $\rm m_H$ is the mass of an hydrogen atom  ($\rm 1.67 \times 10^{-24}$ g).  The results are reported in column  $\rm N_{H_2,1}$ of Table~\ref{source}.\\

From the SMA-only observations, the total continuum flux  at 1.3\,mm is 22.6\,Jy. However, extrapolating the flux $\it S_\nu$ from existing SCUBA %(Submillimetre Common-User Bolometric Array)  
data  \citep{di08} at 450 $\mu m$ (28\,700\,Jy) and 850 $\mu m$ (2\,090\,Jy), the total flux should be closer to  $\rm \sim400$\,Jy at 1.3\,mm (assuming $\it S_\nu\propto\rm \nu^4$, \citealt{johnstone99}). The difference implies that $\rm >90\%$ of the flux has been filtered out by the SMA observations. 
In contrast, $\rm C^{18}O~(2\rightarrow1)$   from a combination of 30\,m and SMA observations is detected towards all the substructures, and it should not suffer from the same spatial filtering as the SMA-only continuum data. As a result, it is a more reliable probe of the total column density of gas in the cloud at all spatial scales. Moreover, as a rare isotopologue line, it can be reasonably assumed to be optically thin, so we use it to make a second independent estimation of the $\rm H_2$ column density in all the substructures. By assuming a constant abundance in all substructures $\rm \frac{N_{C^{18}O}}{N_{H_2}} =\frac{N_{C^{18}O}}{N_{CO}}\frac{N_{CO}}{N_{H_2}} \sim \frac{10^{-4}}{560}\sim1. 79\times10^{-7}$ \citep{wilson94},  $\rm H_2$ column densities can be calculated from the column densities of optically thin  $\rm C^{18}O$ emission (reported in the column $\rm N_{H_2,2}$ of Table~\ref{source}). 

\subsection{Molecular abundances} \label{molabun}   
Subsequently, to study the chemical properties on the small scale of Orion-KL, we derived the molecular abundances with respect to $\rm H_2$ (converted from $\rm C^{18}O$) in each substructure (Tables \ref{col-Obearing}II,  \ref{col-Nbearing}II, and  \ref{col-Sbearing}II).\\

\begin{figure*}
\centering
\scalebox{0.95}{
\begin{tabular}{ll}
\includegraphics[width=5.5cm, angle=-90]{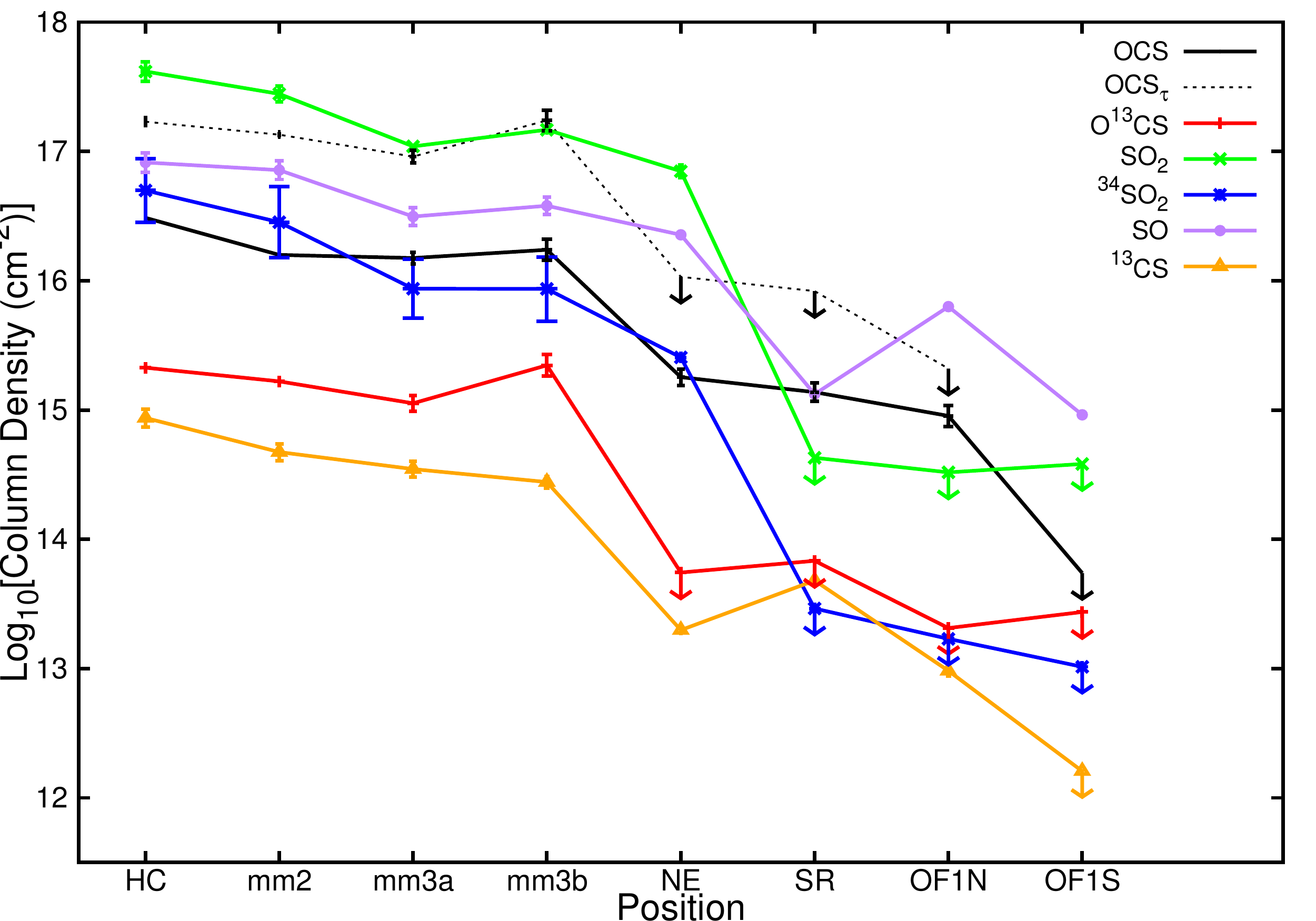}
&\includegraphics[width=5.5cm, angle=-90]{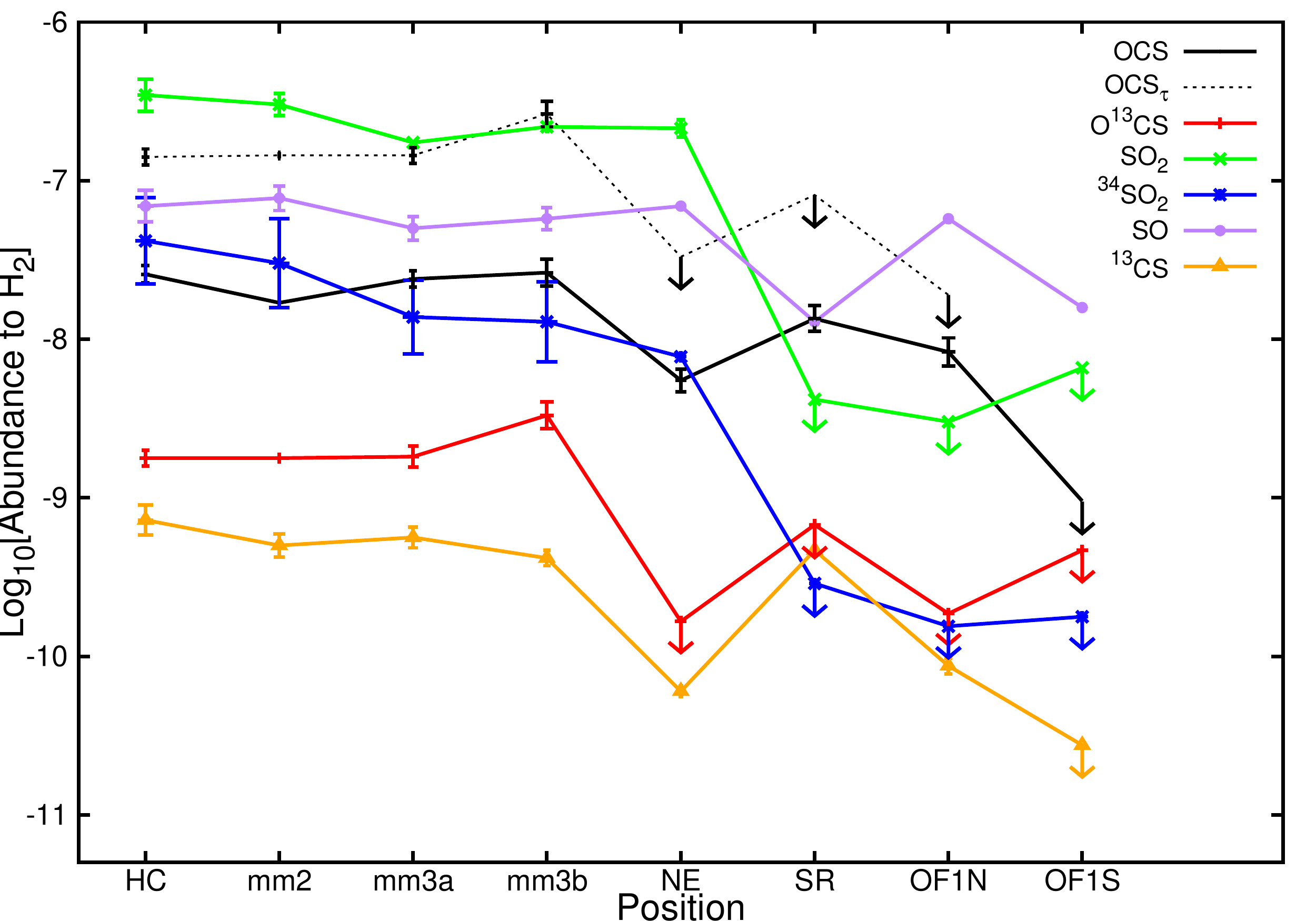}\\
\includegraphics[width=5.5cm, angle=-90]{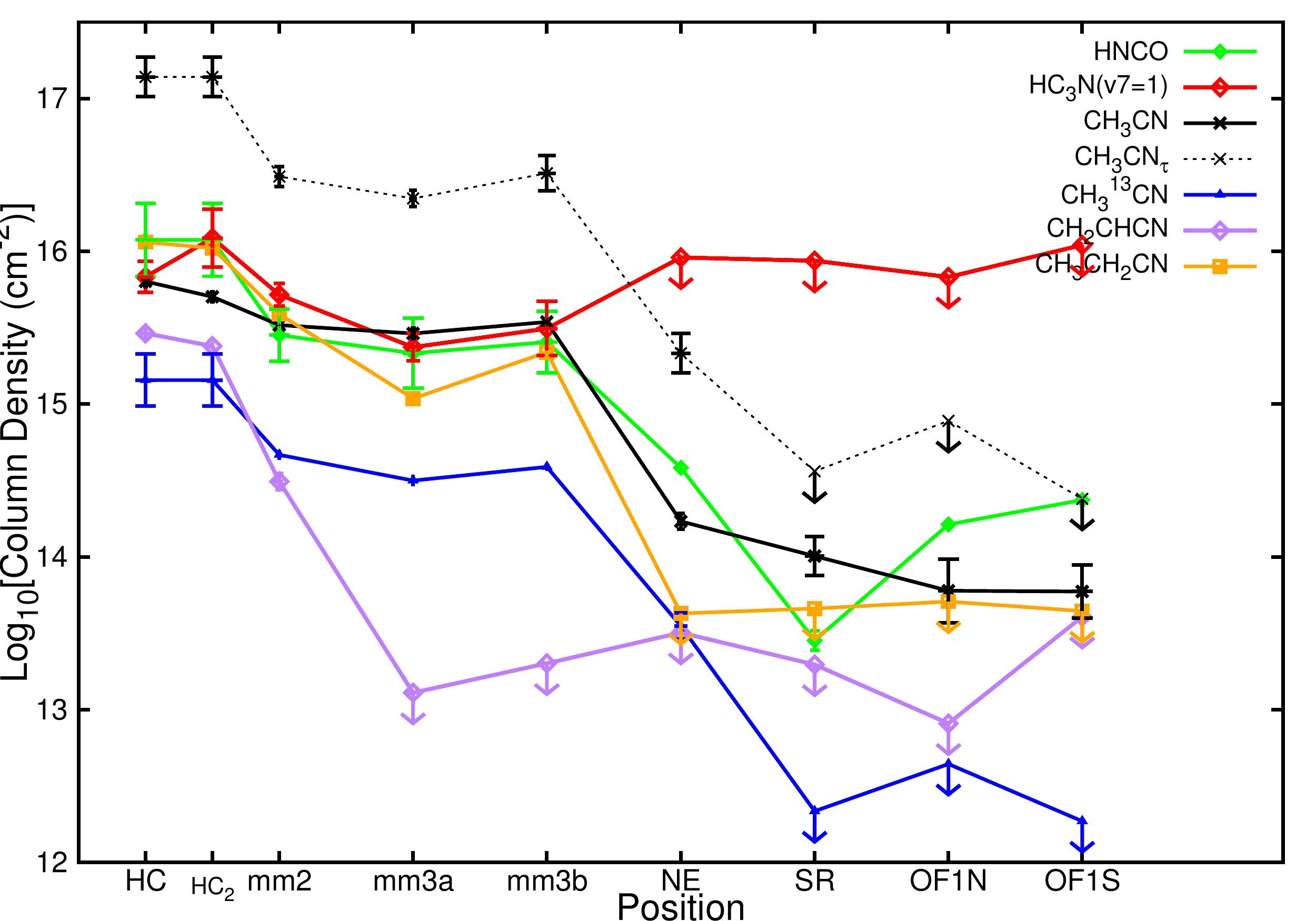}
&\includegraphics[width=5.5cm, angle=-90]{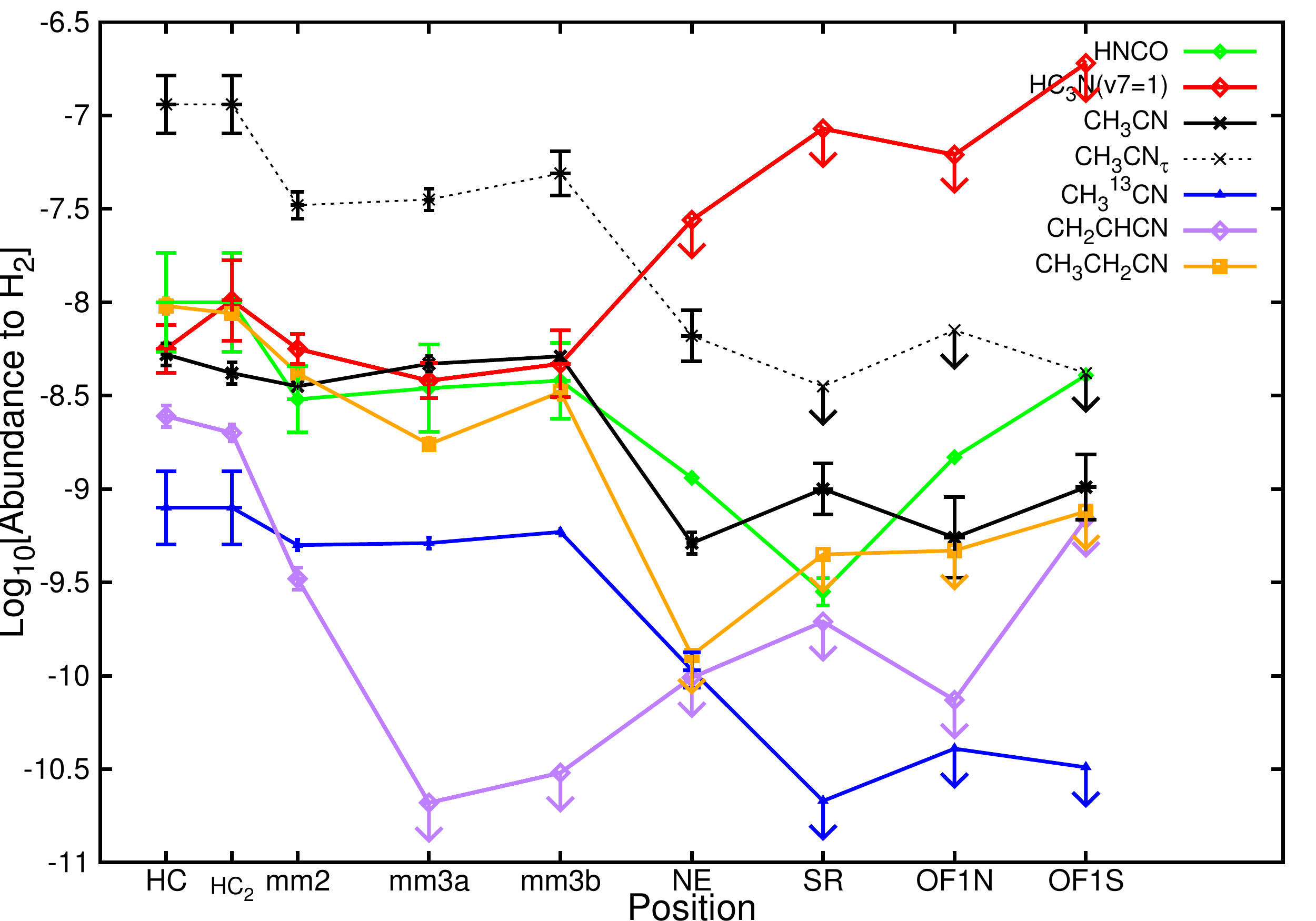}\\
\multicolumn{2}{l}{\scriptsize * $\rm HC_7N$ is not given because of large uncertainties of its tentative detection.}\\
\includegraphics[width=5.5cm, angle=-90]{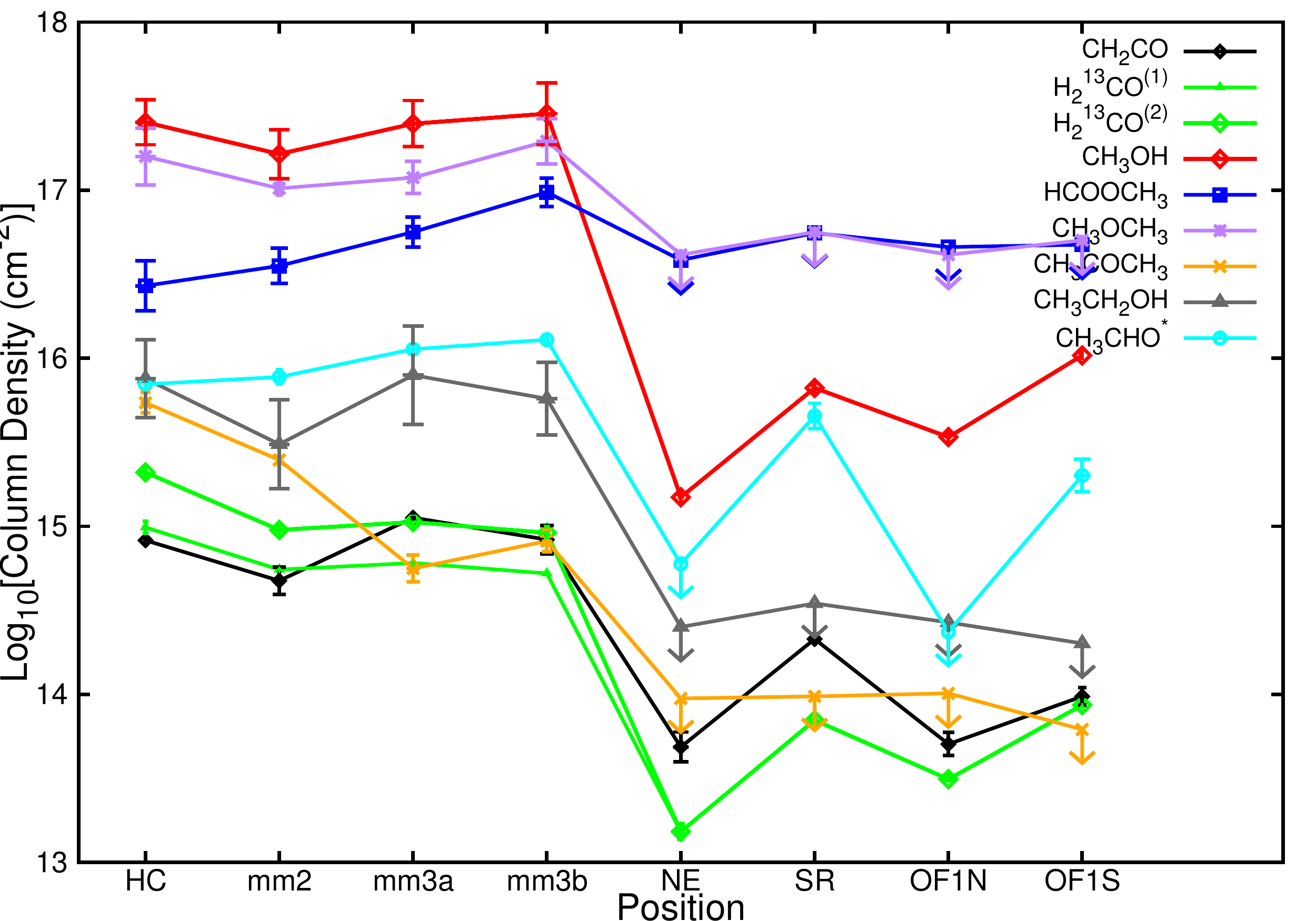}
&\includegraphics[width=5.5cm, angle=-90]{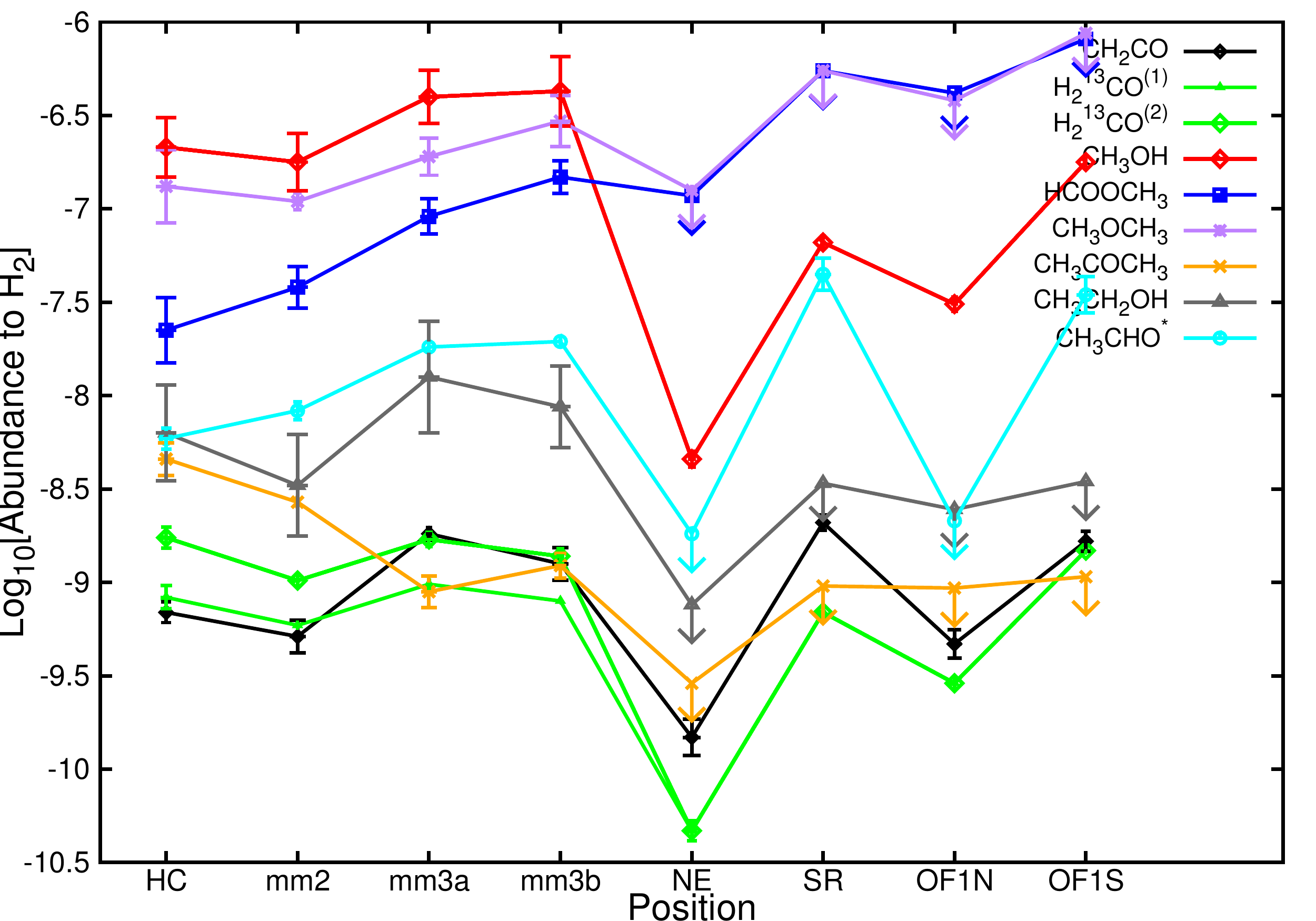}\\
\multicolumn{2}{l}{\scriptsize * Values of $\rm H_2^{13}CO^{(1)}$ are derived from the temperatures the same as $\rm HCOOCH_3$; Values of $\rm H_2^{13}CO^{(2)}$ are derived from the temperatures the same as $\rm CH_3OH$.}\\
\multicolumn{2}{l}{\scriptsize * $\rm C_6H$ is not given because of large uncertainties of its tentative detection.}\\
\includegraphics[width=5.5cm, angle=-90]{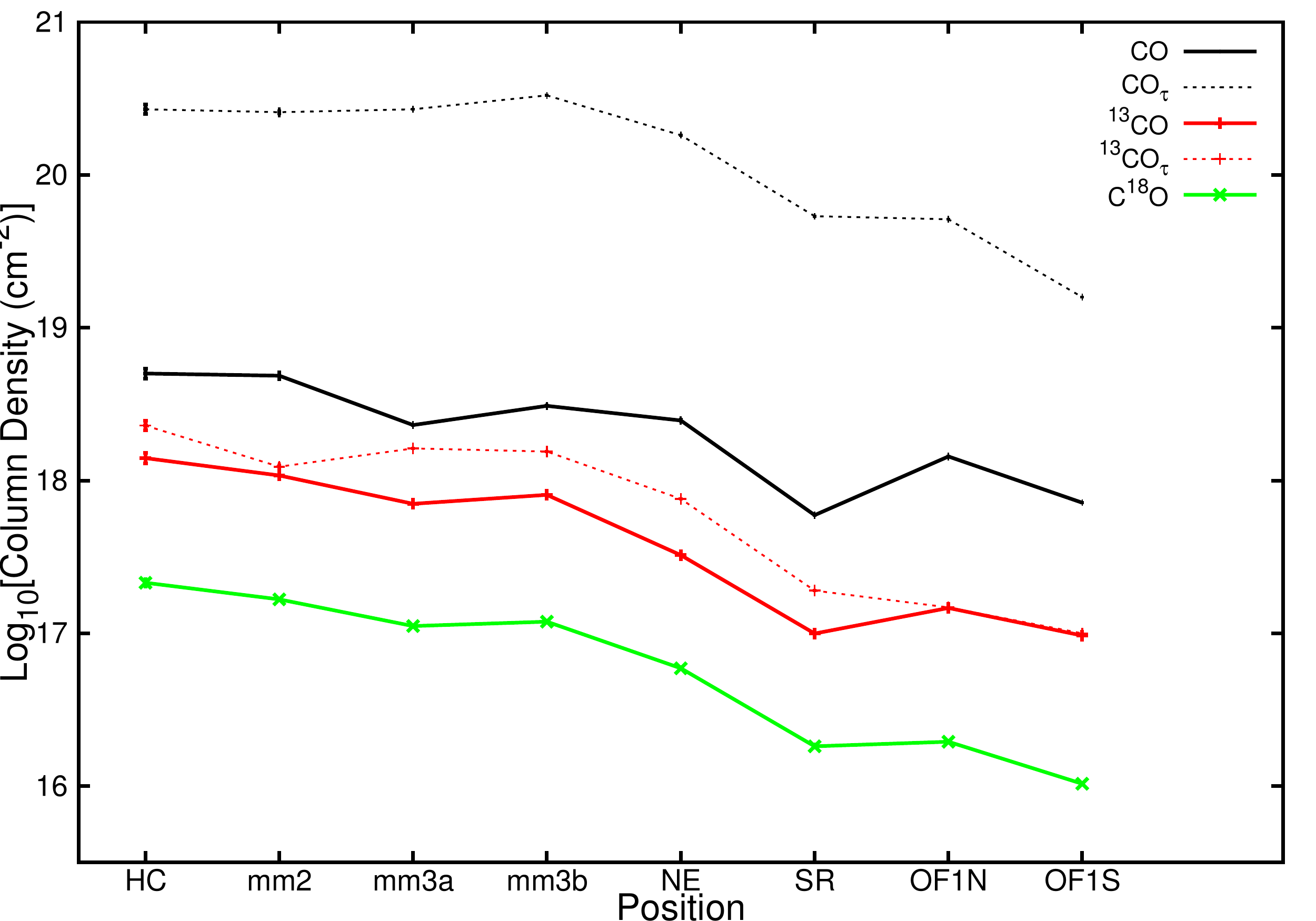}
&\includegraphics[width=5.5cm, angle=-90]{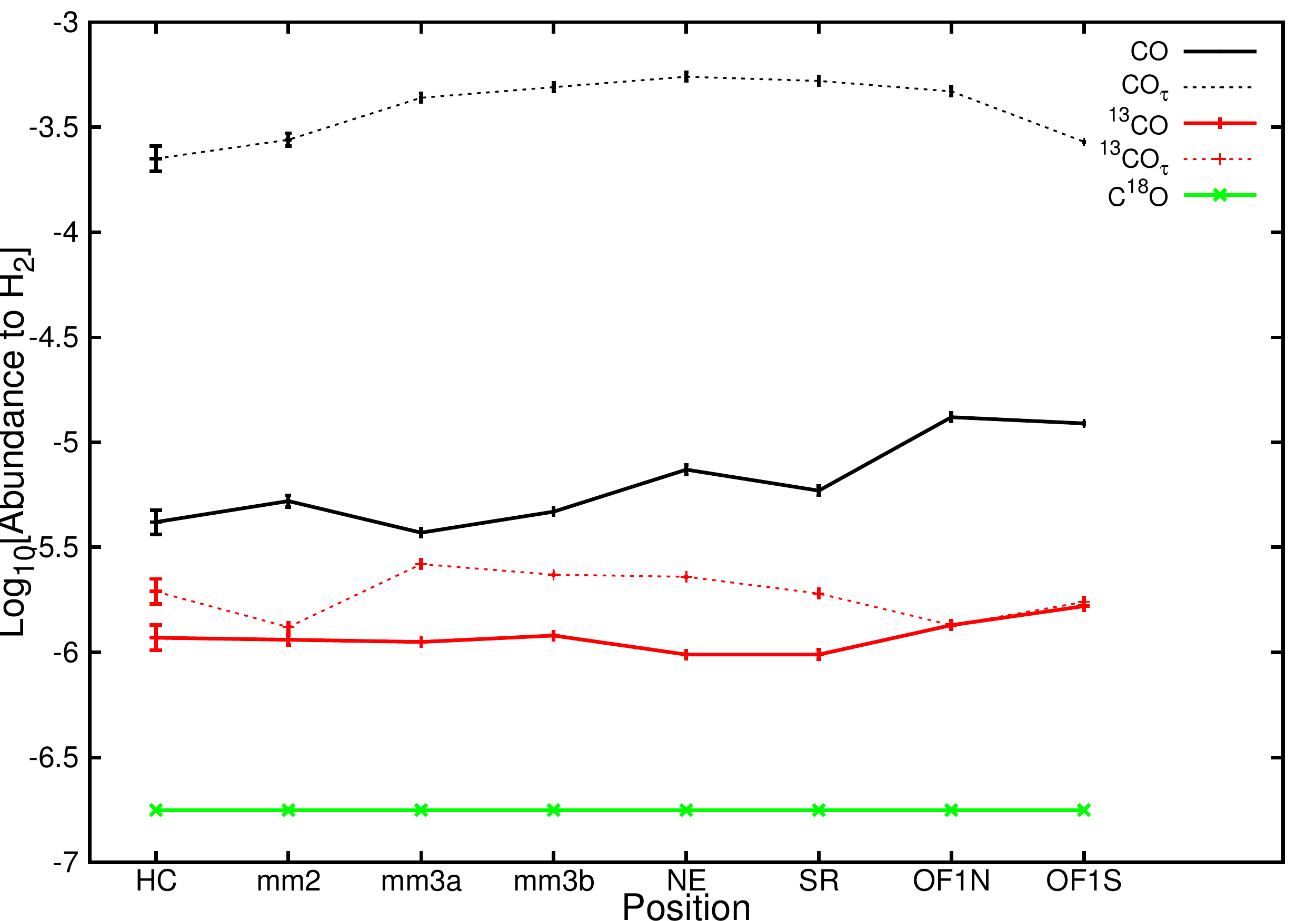}\\

\end{tabular}
}
\caption{\tiny{ Molecular column densities (panels in the left column) and abundances  (panels in the right column) towards 8 positions in Orion-KL.  The molecular abundances of each species with respect to $\rm H_2$ are converted from $\rm C^{18}O$, assuming a constant $\rm N_{C^{18}O}/N_{H_2} \sim1. 79\times10^{-7}$ ratio. Column densities and abundances with the uncertainties are measured as mentioned in Section~\ref{error} and Tables \ref{col-Obearing}--\ref{col-Sbearing}.  Coloured solid lines show the differentiation for each isotopologue calculated from optically thin assumption (arrows highlight the upper limits), while dashed lines show the  differentiation calculated with optical depth correction from Eq.~\ref{eq:tau}.  For abundant molecules (e.g.,  HNCO, SO, $\rm SO_2$, $\rm CH_3OH$)  which we have not measured their line optical depths, their densities could be underestimated by a factor of 7-10.}
 }\label{abun_line}
\end{figure*}

Figure~\ref{abun_line} presents the molecular column densitiy and the abundance of each species with respect to $\rm N_{H_2,2}$.  Generally, all the molecules have higher column densities in the central condensations (HC, mm2, mm3a, and mm3b) than the extended regions in the outflows (NE, SR, OF1N, and OF1S). While the abundances of some species
have relatively small variations among the substructures (e.g., SO, CO, $\rm C^{18}O$, and $\rm ^{13}CO$), some vary substantially. Specifically, N-bearing species are significantly underabundant in the SR and the  outflow regions; O-bearing organic molecules containing $\le6$ atoms (e.g., $\rm H_2^{13}CO$, $\rm CH_2CO$, $\rm CH_3OH$) have more extended emission across all the substructures, with nearly constant large abundances in all the substructures except for NE. The S-bearing molecules are an intermediate case:  most of them are not detected in the high-velocity outflow, but  are clearly detected in NE (OCS, $\rm SO_2$, $\rm ^{34}SO_2$, $\rm ^{13}CS$) and (or) the SR (SO), with lower abundance. Such abundance differentiation among species is consistent with the spatial  differentiation shown in Figure~\ref{into}. \\

 \subsection{Error budget}\label{error}
In this section, we consider the effect that the various assumptions that we made in our quantitative analysis have on the results presented in this paper. 

\begin{enumerate}

\item $\rm T_{rot}$ (rotation temperature):  The traditional rotation diagram method is subject to a number of uncertainties owing to various assumptions, e.g,  line emission is assumed to be optically thin and LTE is assumed to apply (see discussion in \citealt{snyder05, herbst09}). If these assumptions are valid, 
the  points on the population diagram will be well-fit by a straight line,  with the  slope proportional to the negative reciprocal of the temperature.  However, this is not the case (see Figure~\ref{rotation}).  According to Table~\ref{conti}, the volumn density in each substructure is $\rm \sim10^7~cm^{-3}$ (assuming a source size $\sim10$ times of the beam size, see also \citealt{blake87, plume12}),  so LTE can be a reasonable assumption (except for such extreme cases as masing and sub-/super thermal excited conditions), and the main reason for the fittings with uncertainties is the optical depth of the used lines.  \\

(1) In Section ~\ref{tem:ch3cn}, we obtain the optical depth of $\rm CH_3CN~(12_2\rightarrow11_2)$ by comparing its observed main beam brightness temperature to that of $\rm CH_3^{13}CN~(12_2\rightarrow11_2)$.
Then, we use this optical depth as a standard in non-LTE radiative transfer fitting and derive a lower $\rm T_{rot}$ in each substructure. 
 In the HC, the rotational temperatures derived from $\rm CH_3CN~ (155\pm16~ K)$ and $\rm CH_3^{13}CN~ (121\pm16~ K)$ are slightly different, but they agree within the uncertainties, and  column densities of N-bearing species are insensitive to temperature uncertainties of this magnitude  (Table~\ref{col-Nbearing}I).
  On the other hand, temperatures derived from HNCO agree with the above $\rm T_{rot}$ within the  uncertainties, so our approach should be applicable.\\

(2) In Section ~\ref{tem:other}, $\rm T_{rot}$ for most O-bearing species are assumed to be similar to those derived from $\rm HCOOCH_3$. In general, the temperatures derived from two different states of $\rm HCOOCH_3$ lines are consistent, while the torsionally excited $\rm HCOOCH_3$ ($\rm \nu=1$) lines appear to trace lower temperatures than lines in the ground state ($\rm \nu=0$), because LTE assumption may not be applied in sub-thermal conditions (see also \citealt{demyk08}). In addition, we derived a third temperature set from $\rm CH_3OH$, which is in general higher than those derived from $\rm HCOOCH_3$, $\rm CH_3OCH_3$, and $\rm CH_3CH_2OH$.  Synthetic spectrum fitting (Figure~\ref{COMspec}) indicates that some lines of $\rm CH_3OH$ may be optically thick, which may explain the higher measured  $\rm T_{rot}$. On the other hand, $\rm CH_3OH$ lines have broader line widths ($\rm \sim5-7~km\,s^{-1}$) than $\rm HCOOCH_3$ lines ($\rm \sim3-5~km\,s^{-1}$) in the central condensations, implying that they trace gas with different temperatures.  \\

(3) For S-bearing species, $\rm T_{rot}$ derived from $\rm ^{34}SO_2$ varies from HC through mm2 to mm3a and mm3b. However, detected lines in mm3a and mm3b are not sufficient to derive a reliable temperature, because we lose the constraint of fittings from the non-detection line with high $\rm E_u/k_B$ at 229.987 GHz (Figure~\ref{fig:trot_other}III). Nevertheless, we estimate the temperature upper limits for both substructures by including the non-detected lines (dashed blue fittings), and find the differences for column densities derived at temperatures from 89-101 K (non-detection are excluded in fittings) to 176-193 K (upper limits, columns in italic font in Table~\ref{col-Sbearing}I) are by a factor of $<3$. The insensitivity of S-bearing column densities to this temperature regime may also explain the slightly higher $\rm T_{rot}$ with large uncertainties in HC and mm2.\\

\item $\rm N_{T}$ (molecular column density): In Section~\ref{col}, we first assume that all the identified molecular lines are optically thin. Then, we use Eq.~\ref{eq:tau} to obtain the optical depth of those main isotopologue lines whose rare isotopologue lines with the same transition are detected. Finally, using Eq.~\ref{eq:correc}, we  correct the column density of the main isotopologue. In this approach, by assuming the rare isotopologue lines are optically thin, we list the optical  depths and abundance ratios for $\rm ^{13}CO~(2\rightarrow1)$, $\rm CO~(2\rightarrow1)$, $\rm OCS ~(19\rightarrow18)$, and  $\rm CH_3CN ~(12_2\rightarrow11_2)$ with respect to the corresponding rare isotopologue lines  with and without optical depth correction in Table~\ref{tab:correction}. \\

\begin{table*} 
\tiny
\begin{center}
\begin{tabular}{c p{0.01cm}p{1cm} | p{2.5cm}p{1.2cm} p{1.2cm}p{1.2cm} p{1.2cm} p{1.2cm} p{1.2cm} p{1.2cm} p{1.2cm}}\hline\hline

Species  &    &$\Re_\alpha$    & $\rm HC$         & $\rm mm2$    & $\rm mm3a$   &
$\rm mm3b$     & $\rm NE$  & $\rm S$      & $\rm OF1N$      & $\rm OF1S$       \\

\hline 
$\rm \tau_{\rm CO ~(2\rightarrow1)}$      &  &          &54         &53         &117         &107         &74         &90         &36         &22\\
$\rm N_{CO_\tau}/N_{C^{18}O}$  &$^\dag$    &$560\pm25$             &$\rm 1263.0_{\pm21.8}$       &$\rm 1541.0_{\pm57.1}$    &$\rm 2422.0_{\pm46.6}$    &$\rm 2765.0_{\pm72.3}$    &$\rm 3100.0_{\pm52.9}$    &$\rm 2934.0_{\pm18.8}$    &$\rm 2640.0_{\pm33.7}$    &$\rm 1517.0_{\pm29.8}$\\
$\rm N_{CO}/N_{C^{18}O}$  &$^\star$     &            &$\rm 23.4_{\pm0.4}$       &$\rm 29.1_{\pm1.1}$    &$\rm 20.7_{\pm0.4}$    &$\rm 25.8_{\pm0.7}$    &$\rm 41.9_{\pm0.7}$    &$\rm 32.6_{\pm0.2}$    &$\rm 73.3_{\pm0.9}$    &$\rm 68.9_{\pm1.4}$\\

\hline
$\rm \tau_{\rm ^{13}CO~(2\rightarrow1)}$   &    &    &1.1         &0.25         &2         &1.5         &2         &1.5         &0.02         &0.08\\
$\rm N_{^{13}CO_\tau}/N_{C^{18}O}$  &$^\dag$ &$7.2\pm0.9$           &$\rm 10.8_{\pm0.2}$        &$\rm 7.3_{\pm0.2}$    &$\rm 14.6_{\pm0.2}$    &$\rm 13.1_{\pm0.2}$    &$\rm 12.8_{\pm0.0}$    &$\rm 10.6_{\pm0.0}$    &$\rm 7.6_{\pm0.0}$    &$\rm 9.7_{\pm0.3}$\\
$\rm N_{^{13}CO}/N_{C^{18}O}$  &$^\star$ &           &$\rm 6.5_{\pm0.1}$    &$\rm 6.5_{\pm0.2}$    &$\rm 6.3_{\pm0.1}$    &$\rm 6.8_{\pm0.1}$    &$\rm 5.5_{\pm0.0}$    &$\rm 5.5_{\pm0.0}$    &$\rm 7.5_{\pm0.0}$    &$\rm 9.3_{\pm0.3}$\\

\hline 

$\rm \tau_{\rm OCS~(19\rightarrow18)}$          &     &      &5.5     &8.5    &6    &10    &$\le$6      &$\le$6        &$\le$2         &$--$\\
$\rm N_{OCS_\tau}/N_{O^{13}CS}$  &$^\dag$    &$77\pm7$        &$\rm 79.7_{\pm0.7}$    &$\rm 81.1_{\pm2.0}$    &$\rm 80.1_{\pm3.0}$    &$\rm 78.2_{\pm0.0}$    &$\rm \le195.2$    &$\rm \le121.3$    &$\rm \le101.1$    &$--$\\
$\rm N_{OCS}/N_{O^{13}CS}$  &$^\star$    &           &$\rm 14.4_{\pm0.1}$    &$\rm 9.5_{\pm0.2}$    &$\rm 13.3_{\pm0.5}$    &$\rm 7.8_{\pm0.0}$    &$\rm \ge 32.4$    &$\rm \ge 20.2$    &$\rm \ge 43.7$    &$--$\\          

\hline 
$\rm \tau_{\rm CH_3CN ~(12_2\rightarrow11_2)}$        &         &          &18       &11         &8.5         &9         &17         &$\le$3.5         &$\le$13        &$\le$4\\
$\rm N_{CH_3CN_\tau}/N_{CH_3^{13}CN}$  &$^\dag$    &$77\pm7$      
   &$\rm 79.6_{\pm3.0}~ (76.6_{\pm2.4})^*$    &$\rm 77.7_{\pm3.1}$    &$\rm 78.1_{\pm1.9}$    &$\rm 79.8_{\pm0.3}$    &$\rm 82.3_{\pm6.7}$    &$\rm \le168.3$    &$\rm \le 176.3$    &$\rm \le 130.0$\\
   
$\rm N_{CH_3CN}/N_{CH_3^{13}CN}$  &$^\star$    &            
 &$\rm 4.4_{\pm0.2}~(4.3_{\pm0.1})^*$    &$\rm 7.1_{\pm0.3}$    &$\rm 9.2_{\pm0.2}$    &$\rm 8.9_{\pm0.0}$    &$\rm 4.8_{\pm0.4}$   &$\rm \ge 46.6$    &$\rm \ge 13.6$    &$\rm \ge  31.9$\\

\hline
 \multicolumn{8}{l}{* ~Ratio at temperature derived from $\rm CH_3^{13}CN$.}
\end{tabular}

\caption{Optical depth and abundance ratios for $\rm CO$,  $\rm ^{13}CO$,  OCS, and $\rm \rm CH_3CN$  with respect to their rare isotopologues in each substructure of  Orion KL. The values are derived from the single strongest transition of each species using Eqs.~\ref{up} and ~\ref{NT}. $\dag$ or $\star$ mark the estimations with or without optical depth correction. ``$\ge$ ($\le$)" come from the $\rm 3\sigma$ limit of the non-detection of $\rm O^{13}CS$ (OCS) and $\rm CH_3^{13}CN$ . The ratios between isotopes $\Re_\alpha$  of $\rm ^{12}C/^{13}C$ and $\rm ^{16}O/^{18}O$ are provided from the local ISM ratio measured in \citet{wilson94}. Uncertainties on the measured values are typically $\le10\%$ (written as the subscript), as determined from $\rm T_{rot}$, partition function $\rm Q(T_{rot})$, and  $\rm \int T_B(\upsilon)d\upsilon$ of the transition.
  \label{tab:correction}}
\end{center}
\end{table*}

 However, Eq.~\ref{eq:correc}  is based on the assumption that the main and rare isotopologue lines have the same excitation temperature, which is often not the case. 
Therefore, a more accurate calculation requires the large velocity gradient (LVG) approach  and proper chemical modelling. 
Nevertheless, our analytic calculations suggest that,  the optical depth is the main situation driving the error in the estimated molecular column density. For abundant molecules such as $\rm CH_3OH$, HNCO, SO, and $\rm SO_2$, where we do not observe a corresponding transition of their rare isotopologues, our assumption that their lines are optically thin may 
result in  underestimating their column densities by a factor of 7-10  (coming from the average differences for  OCS, CO, and $\rm CH_3CN$ column density values before and after optical depth correction in Figure~\ref{abun_line}).\\

Aside from the optical depth, the errors also come from our assumption of LTE. Assuming that the transitions of one species have ideal Boltzmann distributions at the same $\rm T_{rot}$, molecular total column density can be derived from one transition. However, as represented by points on their rotation diagrams, multiple transitions of certain species   (e.g., $\rm ^{34}SO_2$, HNCO, and $\rm CH_3CH_2OH$, Figure~\ref{fig:trot_other}) show wide scatter, indicating that uncertainties in the estimation of $\rm T_{rot}$ lead to the over-/under-estimation of the total column densities  by a factor of 0.5--2 (bold face in Tables \ref{col-Obearing}I,  \ref{col-Nbearing}I, and  \ref{col-Sbearing}I). For the species whose total column densities are not able to be estimated from rotation diagram fittings owing to the lack of  sufficient transitions, our LTE assumption (using Eq.~\ref{NT}) may also result in the  under/over-estimation of their total column densities of the same magnitude. \\ 

In addition,  $\rm \int T_{B}({\upsilon})  d{\upsilon}$ measured from Gaussian/HFS fittings brings  on average $<10\%$ uncertainties (Table  \ref{tab:lineprofile}), which is negligible compared to the chemical differentiation amongst the  substructures  (written as the subscript in Tables \ref{col-Obearing}I, \ref{col-Nbearing}I,  and  \ref{col-Sbearing}I).\\

\item $\rm N_{H_2}$ and molecular abundance: Because no continuum information is available from the combined data, we used two methods to calculate the column density of $\rm H_2$ (Table~\ref{source}): 
from SMA-only continuum ($\rm N_{H_2,1}$) and from SMA-30\,m $\rm C^{18}O$ ($\rm N_{H_2,2}$).  The molecular abundances were derived with respect to $\rm H_{2,2}$. Comparing the  column density of $\rm H_{2,1}$ and $\rm H_{2,2}$, we found  
$\rm N_{H_2,2}/ N_{H_2,1}$  is $\rm \sim0.9$ in  mm2, $\rm \sim0.5$ in mm3a and mm3b, $\rm \sim0.4$  in HC\footnote{ In the extreme case, the ratio is 0.8--1, if assuming the ice mantle is completely evaporated.}, $\rm \sim0.2$ in NE, but 0.1 in SR.  Several reasons may lead to such difference: (1) the specific intensity ($\rm Jy~beam^{-1}$) used to calculate the column density depends on the beam. The combined beam is larger than the SMA-only beam by a factor of $\rm\sim2.5$, so it is reasonable that for larger beam size, we obtain lower $\rm H_2$ column density (similar case in \citealt{beuther07b}). (2) In producing the continuum, we select the ``line-free" part, but there are still lines with low emission (Figure~\ref{COMspec}), which may contaminate the continuum;  (3) we simply assume that the the gas-to-dust ratio and the volume densities in all substructures are similar within an order of magnitude ($\rm \sim 150$, $\rm \sim10^7~cm^{-3}$), hence also the dust opacity. Since we do not know the size of each substructure, gas is possibly much denser in a warmer source (HC, mm2, mm3a, and mm3b) than  in the other cooler substructures (NE, SR), so the homogeneous volume density assumption may  not be accurate in the whole region.  {\color{black} (4) Grain growth in the envelopes can  also be a reason for difference in dust emissivity from centre to extended substructures at (sub-)mm \citep{miotello14}}. (5) We also note that  $\rm C^{18}O$ column density  in the central region is a factor of 30 higher than  in the outflow.  Judging that $\rm N_{H_2,2}$ is derived with the assumption that $ \rm N_{C^{18}O}/N_{H_2}=const.$  in all the studied substructures, the  variation of  $ \rm N_{C^{18}O}/N_{H_2}$ ratio in different  substructures 
can be another factor that leads to error in our estimation. \\

\item In addition,  another simplification in our calculation is an assumed constant beam-filing factor. 
Determination of molecular abundances in Orion-KL substructures relies upon an accurate source structure model for each molecule.  When the source is smaller than the beam size, the main beam brightness temperature is diluted, leading to  an underestimated brightness temperature for more compact structures. To correct this, an appropriate correction factor 
needs to be applied.  Since we do not know the accurate sizes of the substructures, we simply assume that the filling factor is unity in all the calculations. \\

\end{enumerate}

\section{Discussion}\label{dis}
These observations comprise a large set of molecular lines (over {\color{black}160} lines from 20 species, including 25 isotopologues) from combined SMA and IRAM 30\,m data towards Orion-KL. Resolving Orion-KL into  eight substructures and comparing  S-, N-,  and  O-bearing molecules,  we find complicated spatial morphologies for COMs and strong chemical differentiation over the observed substructures. \\

%%%%%%%%%%%
 
\subsection{Comparison with the other results}
Previous quantitative studies of Orion-KL mainly focused on the HC, CR (with mm3a and mm3b  treated as a single substructure), and plateau (OF1N, OF1S), therefore 
comparison of our estimated molecular abundances with previous results  is limited to these substructures. 

\begin{enumerate}
\item Since different molecules have different chemistry, the rotation temperatures derived using different tracers in the same substructure vary. 
Comparing with  the results of others studies, the variation in $\rm T_{rot}$ for different species can be explained. \\
\begin{itemize}

\item Our opacity-corrected temperature in HC derived from $\rm CH_3CN$ is 120--160 K, which is a factor of 1.5 lower than  reported by  \citet{goddi11b} from $\rm NH_3$ at a higher angular resolution ($\rm \sim 0.8\arcsec$). Given that HNCO in our dataset probes the similar temperatures, our slightly lower temperatures for N-bearing molecules may trace the envelope region of the HC.\\

\item Our adopted temperatures from $\rm HCOOCH_3$  agree with the temperatures estimated by \citet{favre11a} from observations of this species at similar spatial resolution\footnote{MF1--5 are the $\rm HCOOCH_3$ emission peaks in \citet{favre11a}, which coincide with mm3b (MF1), HC (MF2), mm3a (MF3), and mm2 (MF4--5) in our continuum.}.  However, they are lower than those reported in \citet{brouillet13} owing to our spectral resolution. Multiplets of  $\rm CH_3OCH_3$ and $\rm CH_3CH_2OH$ are not resolved in this paper, so we treat each detected line of these species as a single component, and temperatures obtained from rotation diagrams are potentially underestimated\footnote{ \citet{brouillet13} resolves the $\rm CH_3OCH_3$ multiplets and uses them to estimate temperatures as well as column densities. However, if each multiplet is treated as a single component, it would derive temperatures consistent with those we estimate.}.\\

\item Temperatures probed by $\rm CH_3OH$ ($\rm >200 $ K in HC, $\rm \sim170$ in mm2, and $\rm \sim160$ K in CR) are higher than those probed by the other molecules, probably because of   the lack of optical depth corrections, or because they trace gas with different temperatures (discussed in Section \ref{error} part 1 (2)). Nevertheless,  these temperatures  agree with those derived from the same species by  \citet{sutton95} ($\rm ^{13}CH_3OH$) and \citet{friedel12} in HC and CR, and by \citet{beuther05} in HC and mm2. \\

\item  We note that S-bearing species may be affected by shocks in the  environment.  Although they were not often derived previously, and the high $\rm E_u/k_B$ transition leads to large uncertainties in the rotation temperatures derived from $\rm ^{34}SO_2$,  our estimated temperatures are within the uncertainty of those derived by \citet{esplugues13} from $\rm SO_2$ in HC ($\rm 190\pm60$ K) and CR ($\rm 80\pm30$ K).\\

\end{itemize}

\item Due to the compactness of the detected substructures, column densities of CO, $\rm ^{13}CO$, and $\rm C^{18}O$  are ten times higher than those from lower resolution observations in single-dish observations of \citet{wilson11} (FWHP $ \sim13\arcsec$)  and in \citet{plume12} (Herschel/HIFI beam ranges from 13\arcsec to $\rm 40\arcsec$). If taking the beam ratio into account, our results are consistent with previous measurement.\\

\item In both HC and CR, the abundances of SO, $\rm SO_2$, HNCO, $\rm HC_3N$, $\rm CH_3OH$, $\rm CH_2CHCN$, $\rm CH_3CH_2CN$, and CO are comparable to the values in \citet{blake87,wright96}, and \citet{esplugues13}, within our factor of 1.5-2 uncertainties. $\rm H_2^{13}CO$ is lower than the value derived   from $\rm H_2CO$ in \citet{blake87}, assuming a standard $\rm ^{12}C/^{13}C$ ratio, but is comparable to the value in \citet{wright96}. After the optical depth corrections, our OCS and $\rm CH_3CN$ abundances are ten times higher than those in \citet{blake87}, because the latter estimation was made assuming lines of these species  to be  optically thin. However, our abundances of  OCS and $\rm ^{13}CS$ are comparable to  those in \citet{tercero10}, within our factor of 2 uncertainty.\\ 

\item  Similar to the gradient in \citet{sutton95} and \citet{wright96}, CO and SO have stronger emission and higher abundances in the plateau of both the high- and low-velocity outflows, which  has been suggested to be due to shock enhancement in these regions.  Although  molecules  in general have  around 1-2 magnitudes lower abundances in our low-velocity outflow than the values reported in the NE--SW plateau from previous interferometric study \citep{wright96}, their abundances in our high-velocity outflow are comparable to the single dish results in the NW--SE plateau \citep{sutton95}.\\

\item The estimated column densities of COMs ($\rm CH_3CH_2CN$, $\rm CH_3COCH_3$, $\rm CH_3OCH_3$, and $\rm HCOOCH_3$) in HC are found to be lower than the values in \citet{friedel08} by a factor of 10, because their observations have a smaller synthetic beam. However, $\rm HCOOCH_3$ has  comparable column densities to \citet{favre11a} in mm2 (MF4--5), mm3a (MF3), and mm3b (MF1), as does $\rm CH_3OCH_3$ compared to \citet{brouillet13} at all the central condensations (HC, mm2, mm3a, and mm3b),  and  the abundance of $\rm CH_3COCH_3$ compared to \citet{peng13} at HC and mm2 (Ace-3). 
\end{enumerate}

To sum up,  the abundance of most species  from our study are comparable to previous single-dish studies especially in the central condensations. 
Therefore,  molecular abundance estimation  from  single-dish observations for the farther sources (e.g., ten times farther than Orion-KL) should be trustworthy at least in the source centre after optical depth correction.

\subsection{Chemistry in the hot molecular core}\label{hotcore}
At a spatial resolution of $\sim1200$ AU, our data resolves Orion-KL into several condensations.
 The  specific intensities ($\rm Jy~beam^{-1}$) of the dust continuum peaks exhibit a gradient from the central condensations to the  extended structures, and the gas temperatures exhibit the same gradient: 
 The HC is the strongest continuum substructure, because it has  both the highest gas temperature and column density; mm2, mm3a, and mm3b are slightly fainter and less dense; while NE and SR have weaker continuum and line emission than the former sources, but stronger emission than OF1N(S). \\ 
 
Differentiation in the spatial distribution of S-, N-, and O-bearing species clearly shows that  chemistry in the central condensations of Orion-KL is not  homogeneous.  Simple N- and O-bearing molecules show the strongest emission towards the central condensations. However, unlike most of the N-bearing species (including N-bearing COMs) which are not detected in SR, NE, or the outflow,  simple organic molecules ($\rm H_2^{13}CO$, $\rm CH_2CO$) and CO ($\rm ^{13}CO$, $\rm C^{18}O$) have bright extended emission even towards the cooler and less dense regions (SR, OF1N, and OF1S). Compared to the above extreme cases, the S-bearing molecules are  widely distributed over the central condensations, NE and (or) SR. A possible reason for this is that the S-bearing molecules are mainly formed via warm gas-phase chemistry or shock chemistry (e.g.,  \citealt{pineau93, charnley97}),  and then coincide with both N- and O-bearing molecules in the plateau (NE) of BN/KL (e.g., \citealt{chandler97, schreyer99}, see discussion in Section \ref{outflow}).  \\

For the COMs,  it has been widely suggested that ``N-/O- chemical differentiation" can be caused by lower or higher temperatures during the cold pre-hot molecular core phase and  different paces of warm-up in these regions \citep{caselli93, garrod08, neill11, laas11}.  Besides confirming the HC peak of N-bearing COMs ($\rm CH_2CHCN$,  $\rm CH_3CH_2CN$,  $\rm CH_3CN$, and $\rm CH_3^{13}CN$) and the  mm3a (or/and  mm3b) peak of several O-bearing COMs (e.g., $\rm HCOOCH_3$ and $\rm CH_3OCH_3$),  we found several other COMs which cannot be classified within these dichotomous emission peaks groups (also see \citealt{guelin08,friedel12,peng13}).  For example,  the peak of $\rm CH_3OH$ is velocity dependent, shifting from the HC (3.6$\rm ~km\,s^{-1}$) through mm3a (6$\rm ~km\,s^{-1}$) to mm3b (7.2$\rm ~km\,s^{-1}$); the peak of $\rm CH_3CH_2OH$ is in between the HC and mm3a. In contrast,  $\rm CH_3COCH_3$ shares a morphology with N-bearing molecules -- peaking towards the HC with a V-shaped distribution. This similarity in morphologies may indicate a link between the HC-peaking N-bearing species and the  mm3a (or/and mm3b)-peaking species, 
therefore their  formation pathways likely involve  $\rm NH_3$ (ammonia) or another major N-bearing molecule  (e.g.,  \citealt{chen11}), or need  similar physical conditions appropriate to their production and/or sublimation, such as shocks  \citep{peng13}.   Moreover, although $\rm CH_3CHO$ is only tentatively detected, it peaks in a similar location to $\rm HCOOCH_3$. Furthermore, like $\rm CH_2CO$, it exhibits  extended emission from the NE to the SE tail, suggesting these three species may share a common or at least a related  formation path.   \\

The favoured model of hot molecular core chemistry so far is the gas-grain model, where frozen radicals, partly produced by photolysis,  become mobile during the warm-up phase, and their reactions produce COMs
 \citep{markwick00, garrod08}. Three phases are usually modelled,  the cold,  warm-up, and hot molecular core stages.  During these stages, different species are defined as zeroth-generation (e.g.,   $\rm CH_3OH$),  first-generation (e.g., $\rm HCOOCH_3$),  and second-generation organic molecules (see \citealt{herbst09}).  In such a scenario,  since the critical densities are fairly similar for N- and O-bearing species, the distribution diversity of species in Orion-KL may also be explained by different initial compositions or different chemical ``ages"  (evolutionary stage,  the time that the gas phase species has had to evolve from an initial composition after being released from ice mantles, \citealp{friedel08}).
 If that is the case,  COMs like $\rm CH_3OCH_3/CH_3OH$ can act as chemical clocks,  implying that the different substructures in BN/KL could have different chemical ages (e.g., \citealp{charnley97, wakelam04}).  \\

In addition to grain-surface and gas-phase chemical processing, 
 it is also possible that varying physical conditions,  including temperature,  density,  and kinematics cause the differentiation in the spatial distributions of molecular line emission seen in the Orion-KL region.   Probably due to the difference in the central protostars, the warm-up paces of different substructures vary. The modelling result from \citet{garrod08} (especially their Figure~8) is an example. In a particular substructure during  the  warm-up process (e.g., from 100 K to 200 K), the abundances of some molecules increase (e.g., HNCO, $\rm CH_3CN$, $\rm CH_3OH$) because of their evaporation from the grain surface to the gas phase, {\color{black}  while the abundances of  some large molecules cannot be synthesized on grains any longer because their precursor species have desorbed to the gas phase. They decrease (e.g., $\rm HCOOCH_3$, $\rm CH_3OCH_3$) or start to decrease ($\rm CH_3OH$). Another possibility is that these species are  destroyed by ion-molecule chemistry (e.g., by protonation via $\rm XH^+$ molecules),  followed by dissociative recombination into multiple small fragments.}  
 In this scenario, the different emission peaks of different species can be explained by the higher temperature in HC (120--160 K) than in mm3a and mm3b (80--100 K), meaning that N-bearing species are still evaporating in HC but formation of molecules such as $\rm HCOOCH_3$ is stopped or it is  dissociating there. \\

\subsection{Chemistry in the outflow}\label{outflow}

Outflows from young stars  can destroy grain material and/or liberate ices when they impact the surrounding  cold matter,  meaning that chemical reactions can be driven by the associated shocks even in cold gas \citep{herbst09}.  For years,  the outflows in  Orion-KL  have been the main subject of chemical studies of this region  (\citealt{wright96} and references therein).  Although COMs are   less abundant or not detected in these  outflows, the chemistry of simple molecules especially  shock tracers can provide an alternative way to investigate the gas structure of the outflows.\\ 

Two outflows have been reported in Orion-KL,  almost perpendicular to each other.  In addition,  along the NE--SW direction, a large-scale ``integral-shaped" filament has been detected aligned with the low-velocity outflow \citep{wright92,lis98,di08}.  To confirm whether the NE clump  is associated with the filament or the outflow, we selected several widespread species, and use a Gaussian function to fit their line profile extracted from NE (Figure~\ref{wing}). Broad line wings ($\rm C^{18}O$) and a second component (HNCO,  $\rm SO_2$, and $\rm ^{34}SO$) with $\rm V_{peak}\sim20-23~km\,s^{-1}$ indicate that line emission in NE contains significant contribution from the outflow and is part of the plateau.\\

To study the chemistry in the outflows, we image the  blue- and red-shifted gas of several shock-tracing molecules along the outflow direction (Figure~\ref{shift}). To avoid contamination from blending,  we image  as broad a velocity range as possible (labelled in each panel),  though,  $\rm -20\sim+30 ~km\,s^{-1}$ is the limit for most lines other than SO, CO, and $\rm ^{13}CO$. Because of the range limit and  the limits of  the 52\arcsec ~ primary beam,  the high-velocity outflow ($\rm 30-100~km\,s^{-1}$) cannot be imaged in its entirety.  However,  the low-velocity outflow ($\rm \sim 18~km\,s^{-1}$) with a NE$-$SW elongated structure is clearly detected, showing bipolar structure 
in the intensity-integrated map of  $\rm H_2^{13}CO$. 
 In addition, intensity-integrated maps of  such molecules  as OCS, $\rm ^{34}SO_2$, SO, HNCO, $\rm CH_2CO$, $\rm CO$, and $\rm ^{13}CO$ all have a ``butterfly" morphology,  with blue-shifted radial velocities of the moving gas towards the NW  and the red-shifted velocities in the SE. This morphology is exactly the same as found in SiO  \citep{wright95, plambeck09, zapata12, niederhofer12} and $\rm H_2O$ maser observations \citep{greenhill13} on scales of tens of AU. 
Whether this morphology is due to the bipolar lobes of the high-velocity outflow or to the expanding low-velocity outflow is still not clear. 
Possibilities are discussed  (e.g., two bipolar outflows from a precessing  binary, ballistic ejecta from a rotating disk, a single expanding wide-angle NE$-$SW outflow; \citealt{zapata12}). The asymmetry shape of the lobes (e.g., ``U" shape blue-shifted and half ``U" red-shifted lobes in the map of  SO) is said to come from some density inhomogeneities in the surroundings  \citep{zapata12} or the interaction between outflow and the environment \citep{niederhofer12}.  \\

Therefore, it can be suggested that a picture of explosive  NW--SE outflow explains the excitation temperature gradient from the central condensations to the extended outflows in Orion-KL \citep{zapata12}: the decelerated bullets can be absorbed by the BN/KL cloud,  and their kinetic energy can be transferred to thermal energy \citep{peng12b};  the closer a gas parcel is to the explosion centre (e.g., the closer encounter of BN, source I and maybe also source N, suggested by proper motion studies from \citealt{rodrquez05,gomez05,gomez08} and 3D modelling from \citealt{nissen12}), the higher the thermal energy available to affect the chemical  properties of the gas in the HC and mm2. \\

\begin{figure}
 \centering
\includegraphics[width=7cm, angle=0]{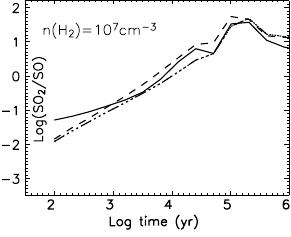}\\
\includegraphics[width=5cm, angle=270]{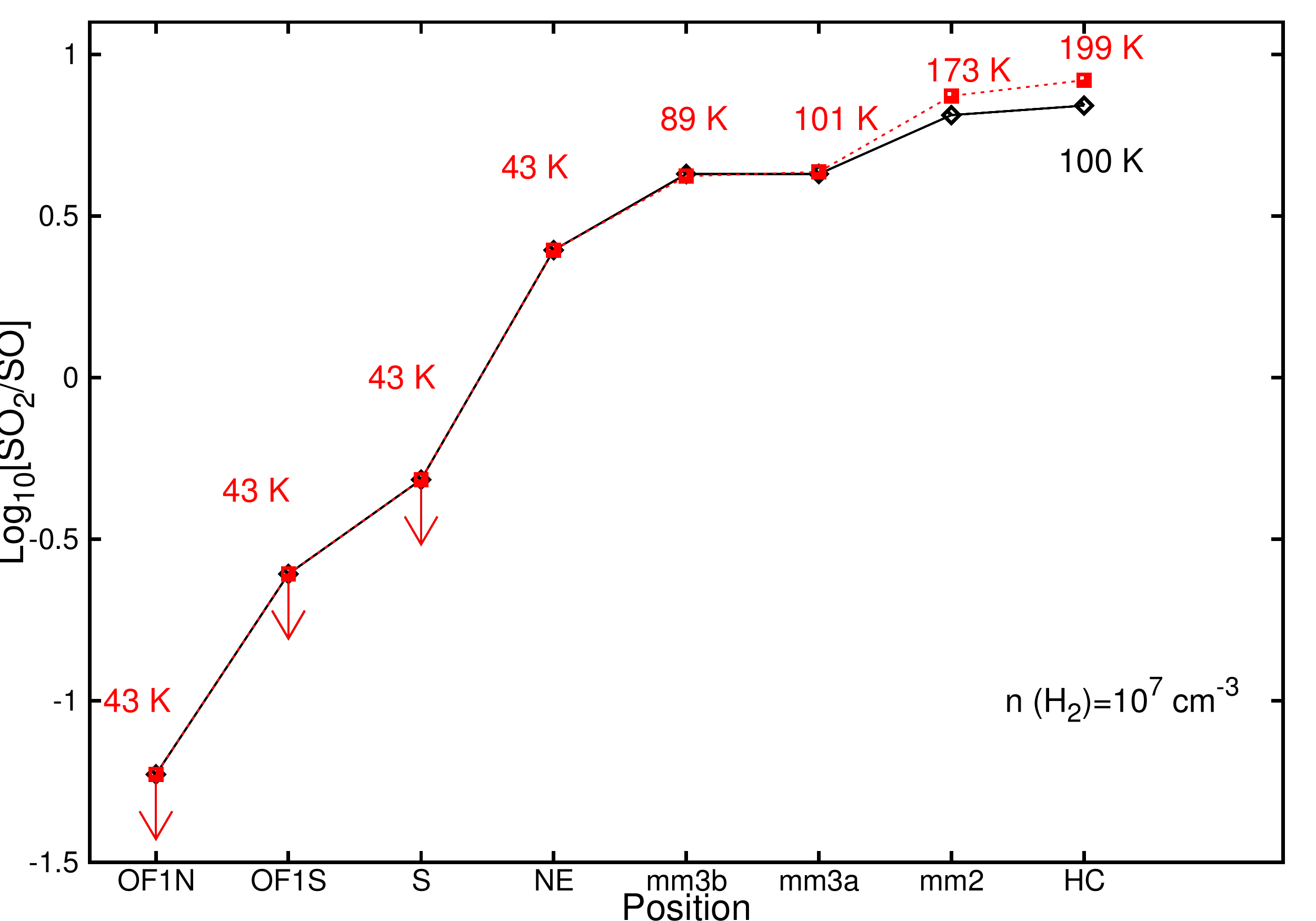}\\
\caption{Observed $\rm SO_2/SO$ ratio compared to the model prediction. The upper panel is adopted from Figure~5 of \citet{wakelam04} (100 K, $\rm 10^7\, cm^{-3}$),  where different line types are from different grain component  models. The lower panel is from our observed  $\rm SO_2/SO$ ratio by assuming (1) all substructures have similar  density $\rm 10^7\, cm^{-3}$ and (2) $\rm N_{SO_2}/ N_{^{34}SO_2}=22$ \citep{wilson94}. The black solid line comes from the assumption that all substructures have a temperature close to 100 K, and the  red dotted line comes from different temperatures derived in Section \ref{tem} (labelled in red). }\label{wakelam}
\end{figure}

Of all the shock-related molecules, S-bearing species are of special interest, not only because their abundance may be enhanced by shocks \citep{martin92, pineau93, chernin94, bachiller97, schilke97, charnley97,  bachiller01,  viti04,  garrod13}, but also because they have been proposed as  potential chemical clocks to date outflows (and hence their protostellar driving source; e.g.,  \citealt{charnley97, wakelam04}). 
The angular resolution of our study does not allow us to investigate the post/pre-shocked chemistry.
 In spite of that,  we do find that some  S-bearing molecules (e.g.,  $\rm SO_2$, $\rm ^{34}SO_2$, and SO) have almost the same abundances in NE as in the central sources, which are higher than  those in OF1N(S), even though we assume the temperatures in  NE and OF1N(S) are equally low%\footnote{CO and SO are more abundant in the plateau, which is suggested by  \citet{sutton95} and \citet{wright96} to be a result of shock enhancement.}
. The enhancement of their abundance  may indicate stronger shocks in this direction than the rest  of Orion-KL. On the other hand, although we cannot tell the precise ages of different substructures,  we can compare the variations in abundance ratios $\rm SO_2/SO$ in our observations to the gas-grain modelling result. To simplify the case, we assume that the temperatures in all substructures are 100 K and that all substructures have the same density $\rm \sim 10^7\,cm^{-3}$; and we estimate the value of $\rm SO_2$ (lines are possibly optically thick) by using $\rm ^{34}SO_2$ and a ratio of $\rm ^{32}S/^{34}S=22$  \citep{wilson94}.   Comparing the observed $\rm SO_2/SO$ ratio to modelling of \citet{wakelam04}\footnote{In \citet{wakelam04}, SO/$\rm H_2$ ratio is sensitive to the grain components. However, $\rm SO_2/SO$ ratios from different grain component models are similar at a particular time (yr), so we compare this ratio to the observations.} in all substructures (Figure~\ref{wakelam}), we find that  the HC has the highest  $\rm SO_2/SO$ ratio, which suggests it is the most evolved substructure. Also, mm2, mm3a, and mm3b have lower ratios and are younger. The lowest ratios are in the outflows, which indicate the age of the low-velocity outflow should be $\rm \ll1\times10^4$ years, while the high-velocity outflow should be $\rm \ll1\times10^3$ years \footnote{NE is part of the  filamentary structure, so the age of the low-velocity outflow is implied from the upper limit ratio in SR. Due to  the non-detection of $\rm SO_2~(^{34}SO_2)$ in OF1N(S), the age of the high-velocity outflow is implied from the upper limit ratio in OF1N.}.\\

The shocks could also be responsible for the observed chemical differentiation in COMs  (e.g., by stripping ices from grains).  The CR (together with mm3a and mm3b) is thought to be the place where icy grains are released into the gas phase by an outflow from the HC \citep{blake87, liu02}. {\color{black}Therefore, based on the reported association  between the emissions of $\rm HCOOCH_3$ and the $\rm 2.12 ~\mu m$ vibrationally excited $\rm H_2$ \citep{lacombe04}, it has been suggested that shocks between the ISM and high-velocity gas ``bullets" from the outflow \citep{zapata09, bally11}   explains the production of COMs, which has the same morphology as $\rm HCOOCH_3$ \citep{favre11b}}.  In a similar scenario,  N-bearing  COMs may be released to the gas phase  north-east  of the HC (e.g., $\rm HC_3N$, $\rm CH_2CHCN$),   where the low-velocity outflow is impacting the ambient dusty material, and the weak emission /nondetection of  these molecules along the high-velocity  outflow may be due to lack of dusty material in the NW--SE direction \citep{friedel12}. \\

%%%%%%%%%%%%%%%%%%%

\subsection{Search for COMs}\label{search}
The chemistry of Orion-KL is rich,  not only in high-density tracers and shock tracers,  but also with numerous transitions of COMs,  which are less abundant than the other tracers and which have a less well understood formation mechanism. \\

{\color{black}
 Three possibilities have been discussed for the formation of the COMs:  (1) ion-molecule chemistry involving large radicals  \citep{ehrenfreund00}; (2) formation on dust grains and then released to the gas at higher temperatures in the later stages of star formation \citep{garrod08, herbst09}; (3) a hybrid approach that COMs are ejected into the gas via efficient reactive desorption, after a sequence of gas-phase reactions between precursor species formed on grain surfaces \citep{vasyunin13}. 
}
In our dataset,  we search for several COMs (Table~\ref{COMcol})  by using the synthetic fitting programme described in Section \ref{image}. $\rm HC_7N$, $\rm C_6H$, and  $\rm CH_3CHO$ are  tentatively detected because of missing predicted lines. For the remaining species, if not taking  the blending problem into account, all the predicted transitions are robustly detected.  In addition, we  find that  saturated molecules  are likely to be more abundant than unsaturated molecules (e.g.,   $\rm HC_3N$, $\rm C_6H$,  $\rm HC_7N$). This  indicates that these saturated species may be  formed on dust grain surfaces by subsequent hydrogenation
of their precursor species (such as CO, O, N or their more complex
combinations).  \\

Unfortunately, we lack the sensitivity to detect HCOOH  (formic acid),  which has been previously reported in Orion-KL by \citet{liu02}. Line blending also prevents us from robustly detecting   
$\rm CH_2OHCHO$ (glycolaldehyde) and $\rm (CH_2OH)_2$ (ethylene glycol) in our data. We also fail to detect other prebiotic molecules,  e.g., $\rm NH_2CH_2COOH$ (glycine; \citealt{garrod13}),  $\rm CH_3COOH$ (acetic acid),  $\rm NH_2CH_2CN$  (amino acetonitrile; \citealt{belloche08}),  $\rm CH_3CONH_2$ (acetamide; \citealt{vD09}),  and the ring molecules. 
More sensitive observations (e.g., with ALMA) are required to determine whether these molecules exist in the Orion-KL environment.\\

\section{Conclusions}
Based on the analysis of combined SMA and 30\,m observations of Orion-KL, we have drawn the following conclusions:\\

\begin{enumerate}
\item From  interferometric observations with the SMA, Orion-KL has been resolved into several continuum substructures with different peak flux densities, at the angular resolution of 3\arcsec (1\,200\,AU at a distance of 414\,pc). In addition to the previous well-known compact continuum substructures like the HC and mm2, our data resolves the clump south-west of HC into two cores: mm3a and mm3b. In addition, we confirmed the detection of strong continuum emissions ($>5\sigma$) from two resolved clumps  (SR and NE), which lie along the axis of both the low-velocity outflow and the large-scale dusty filament. \\
\item  For the first time, we have combined  interferometric and single-dish observations of Orion-KL, to correct  missing short spacings, and achieved high sensitivity to all spatial scales.
Covering a field of view of $\rm 52\arcsec$, we mapped the distribution of  all the identified species. In particular, we obtained the first  maps of  molecules such $\rm ^{34}SO_2$, $\rm O^{13}CS$,  HNCO, $\rm H_2^{13}CO$, $\rm ^{13}CO$, and $\rm CH_2CO$  with both high resolution and sensitivity to spatial information on all scales. 
 We also  identified several transitions of low abundance  ($\rm <10^{-8}$ with respect to $\rm H_2$) COMs, such as  $\rm CH_3COCH_3$ and $\rm CH_3CH_2OH$, as well as made the tentative detection of  $\rm CH_3CHO$ and
long carbon chains  like $\rm C_6H$ and $\rm  HC_7N$.   \\
\item The Orion KL region exhibits clear chemical differentiation on the scales probed by the SMA-30\,m observations.  The emission from S-bearing molecules, CO isotopologues, and simple organic molecules are extended, covering not only the central condensations  (HC, mm2, mm3a, and mm3b), but also the NE clump, the SR, and even the outflow regions. In comparison, the distributions of COMs are more concentrated and  hard to group. The segregated emission peaks at either the HC (e.g., N-bearing) or CR (together with mm3a and mm3b, e.g., $\rm HCOOCH_3$,  $\rm CH_3OCH_3$)  indicates  different formation pathways for O-bearing and N-bearing molecules. However,  some COMs peak in-between  the HC  and mm3a (e.g., $\rm CH_3CH_2OH$), or have dual velocity dependent peaks ($\rm CH_3OH$), or share the same morphology as N-bearing species ($\rm CH_3COCH3$), which may indicate the linking of some formation pathways between O-bearing and N-bearing molecules. Moreover, line widths of the N-bearing, S-bearing species, $\rm HCOOCH_3$ and $\rm CH_3OH$ vary, and so do the derived rotation temperatures, indicating they trace different gas.\\
\item  By studying the chemistry within outflows of Orion-KL, we found   higher abundances of such shock tracers as  $\rm ^{34}SO_2$, $\rm SO_2$, and $\rm SO$ in the plateau of the low-velocity outflow (NE), indicating stronger shocks in this direction than the rest of Orion-KL.\\

\item  Molecular rotation temperatures and abundances show large gradience between central condensations and the outflow regions, indicating difference between hot molecular core and shock chemistry. \\

\end{enumerate}

\acknowledgements
{We would like to thank the SMA staff and IRAM 30\,m staff for
their helpful support during the reduction of the SMA data, the performance of the 
IRAM 30~m observations in service mode,  Z. Y. Zhang and Y. Wang for helping with code development, for useful discussion, and the referee for constructive suggestions. \\
This research made use of NASA's Astrophysics Data System.\\
The Submillimetre Array is a joint project between the Smithsonian Astrophysical Observatory and the Academia Sinica Institute of Astronomy and Astrophysics, and it is funded by the Smithsonian Institution and the Academia Sinica.\\
 S. F acknowledges
financial support by  the European Community Seventh Framework Programme [FP7/2007-2013] under grant agreement no. 238258.\\
D. S acknowledges support by the {\it Deutsche Forschungsgemeinschaft} through
SPP~1385: ``The first ten million years of the solar system - a
planetary materials approach'' (SE 1962/1-2). }\\

\bibliographystyle{aa}

%\bibliography{../..//HMSFR}
\bibliography{orion_accepted_proof.bbl}
%%%%%%%%%%%%%%%%%%%%%%%%%%%%%%%%%%%%%%%
%%%%%%%%%%%%%%%%%%%%%%%%%%%%%%%%%%%%%%%%
%%%%%%%%%%%%%%%%%%%%%%%%%%%%%%%%%%%%%%%%
\appendix
\section{}
\setcounter{table}{0}
\renewcommand{\thetable}{A\arabic{table}}

\onecolumn

\begin{longtable}{lrrr|lrrr}
\caption{Identified  emission lines from the SMA-30\,m combined dataset. 
}\label{tab:line}\\
%\begin{tabular}
\hline \hline
Freq.  & Mol.  & $E_u/k_B$   &Note      &Freq.  & Mol.  (candidates)    &$E_u/k_B$    &Note\\
(GHz) &          &  (K)         &       & (GHz) &                                & (K)           & \\
\hline
218.903          &$\rm OCS_{(\nu=0)} (18\rightarrow17)$           &100               &\\
%{\color{blue} 218.968}        &\color{blue} $\rm CH_2OHCHO  (36_{10, 27}\rightarrow36_{9, 28})$    &\color{blue} 434         &\color{blue} $*$   
218.966         &$\rm HCOOCH_{3 (\nu=1)} (18_{12, 6}\rightarrow17_{12, 5})E$           &384       &\\
218.981          &$\rm HNCO_{(\nu=0)}  (10_{1, 10}\rightarrow9_{1, 9})$           &101       &\\
219.068         &$\rm HCOOCH_{3 (\nu=1)} (18_{17, 2}\rightarrow17_{17, 1})E$           &481      & \\
219.079         &$\rm HCOOCH_{3 (\nu=1)} (28_{3, 25}\rightarrow28_{2, 26})E$           &434      & \\

%2	(CH3)2CO v=0		Acetone	219.07685 (8.05E-5)	 	34( 9,25)-34( 9,26) EA	-5.72740	 	397.82146	JPL
%3	(CH3)2CO v=0		Acetone	219.07685 (8.05E-5)	 	34(10,25)-34( 8,26) EA	-5.72740	 	397.82146	JPL
%4	(CH3)2CO v=0		Acetone	219.07700 (8.05E-5)	 	34(10,25)-34( 9,26) AE	-5.55130	 	397.82149	JPL

219.090         &$\rm HCOOCH_{3 (\nu=0)} (34_{7, 28}\rightarrow34_{5, 29})E$           &387      &*
&219.090         &$\rm HCOOCH_{3 (\nu=0)} (34_{7, 28}\rightarrow34_{5, 29})A$           &387      & \\
219.154         &$\rm HCOOCH_{3 (\nu=1)} (18_{11, 7}\rightarrow17_{11, 6})E$           &369      &*
&219.153         &$\rm HCOOCH_{3 (\nu=1)} (10_{4, 6}\rightarrow9_{3, 6})E$           &230      & \\
219.174         &$\rm HC_3N_{(\nu_7=1)} (24\rightarrow23)$         &452            &$\dag$\\
219.195         &$\rm HCOOCH_{3 (\nu=1)} (18_{16, 3}\rightarrow17_{16, 2})E$           &459      &* 
&219.196         &$\rm HCOOCH_{3 (\nu=0)} (36_{6, 30}\rightarrow36_{6, 31})E$           &429      &\\
219.220        &$\rm CH_3COCH_{3 (\nu=0)}  (21_{1, 20}\rightarrow20_{2, 19})AE$    &122           &*
&219.220        &$\rm CH_3COCH_{3 (\nu=0)}  (21_{2, 20}\rightarrow20_{1, 19})AE$    &122           &\\
~~~& & & &219.220        &$\rm CH_3COCH_{3 (\nu=0)}  (21_{2, 20}\rightarrow20_{2, 19})EA$    &122           &\\
~~~& & & &219.220        &$\rm CH_3COCH_{3 (\nu=0)}  (21_{1, 20}\rightarrow20_{1, 19})EA$    &122           &\\
219.242        &$\rm CH_3COCH_{3 (\nu=0)}  (21_{1, 20}\rightarrow20_{2, 19})EE$    &122           &*
&219.242        &$\rm CH_3COCH_{3 (\nu=0)}  (21_{2, 20}\rightarrow20_{1, 19})EE$    &122           &\\
~~~& & & &219.242        &$\rm CH_3COCH_{3 (\nu=0)}  (21_{2, 20}\rightarrow20_{2, 19})EE$    &122           &\\
~~~& & & &219.220        &$\rm CH_3COCH_{3 (\nu=0)}  (21_{1, 20}\rightarrow20_{1, 19})EE$    &122           &\\
219.264        &$\rm CH_3COCH_{3 (\nu=0)}  (21_{1, 20}\rightarrow20_{2, 19})AA$    &122           &*
&219.264        &$\rm CH_3COCH_{3 (\nu=0)}  (21_{2, 20}\rightarrow20_{1, 19})AA$    &122           &\\
219.276         &$\rm SO_{2 (\nu=0)}   (22_{7, 15}\rightarrow23_{6, 18}) $            &352             & \\
219.311        &$\rm CH_3COCH_{3 (\nu=0)}  (12_{9, 4}\rightarrow11_{8, 3})AA$    &66           &\\
219.325         &UL           &           &$?$\\
219.331       &$\rm HCOOCH_{3 (\nu=1)} (18_{15, 4}\rightarrow17_{15, 3)})E$           &438      &* 
&219.328        &$\rm HCOOCH_{3 (\nu=1)} (36_{9, 28}\rightarrow36_{8, 29})E$           &635      &\\
219.355         &$\rm ^{34} SO_{2 (\nu=0)}    (11_{1, 11}\rightarrow10_{0, 10})$           &60           &$\dag$\\
219.401         &$\rm CH_{2}CHCN_{(\nu=0)}  (23_{3, 20}\rightarrow22_{3, 19})$             &145       &\\
219.412         &$\rm HCOOCH_{3 (\nu=1)}  (18_{10, 8}\rightarrow17_{10, 7})E$           &355       &\\
219.417         &$\rm HCOOCH_{3 (\nu=0)} (30_{5, 26}\rightarrow30_{4, 27})E$           &291       &\\
219.463        &$\rm CH_3CH_2CN  (22_{2, 21}\rightarrow21_{1, 20})$    &112         &\\
%&219.469        &$\rm CH_3COCH_{3}  (33_{8, 25}\rightarrow33_{7, 26})$    &368          &AE,EA\\
219.479         &$\rm HCOOCH_{3 (\nu=1)} (18_{14, 5}\rightarrow17_{14, 4})E$           &419       &\\
219.484         &$\rm HCOOCH_{3 (\nu=0)} (30_{5, 26}\rightarrow30_{4, 27})A$           &291       &\\

219.506         &$\rm CH_{3}CH_{2}CN_{(\nu=0)}  (24_{2, 22}\rightarrow23_{2, 21})$             &136       &\\
%{\color{blue}219.540}        &\color{blue}$\rm (CH_2OH)_2  (22_{2, 21}\nu=1\rightarrow21_{2, 20}\nu=0)$    &\color{blue}122           &\color{blue}$*$\\
219.547          &$\rm HNCO_{(\nu=0)} (10_{4, 6}\rightarrow9_{4, 5})$             &750          &*
&219.547          &$\rm HNCO_{(\nu=0)} (10_{4, 7}\rightarrow9_{4, 6})$             &750          &\\
219.560          &$\rm C^{18}O (2\rightarrow1)$             &16          &$\dag$\\
219.566         &$\rm HCOOCH_{3 (\nu=1)} (18_{15, 4}\rightarrow17_{15, 3})A$           &438       &*
&219.566         &$\rm HCOOCH_{3 (\nu=1)} (18_{15, 3}\rightarrow17_{15, 2})A$           &438       &\\
219.568         &$\rm HCOOCH_{3 (\nu=1)} (18_{14, 5}\rightarrow17_{14, 4})A$           &419       &*
&219.568         &$\rm HCOOCH_{3 (\nu=1)} (18_{14, 4}\rightarrow17_{14, 3})A$           &419       &\\
219.571         &$\rm HCOOCH_{3 (\nu=1)} (18_{16, 3}\rightarrow17_{16, 2})A$           &458       &*
&219.571         &$\rm HCOOCH_{3 (\nu=1)} (18_{16, 2}\rightarrow17_{16, 1})A$           &458       &\\
219.579         &$\rm HCOOCH_{3 (\nu=0)} (28_{9, 19}\rightarrow28_{8, 20})A$           &294       &*
&219.579       &$\rm HCOOCH_{3 (\nu=1)} (18_{17, 2}\rightarrow17_{17, 1})A$           &481       &\\ 
~~~& & & &219.579       &$\rm HCOOCH_{3 (\nu=1)} (18_{17, 1}\rightarrow17_{17, 0})A$           &481       & \\
~~~& & & &219.578         &$\rm HCOOCH_{3 (\nu=1)} (29_{9, 21}\rightarrow29_{8, 21})E$           &498       &\\
219.584         &$\rm HCOOCH_{3 (\nu=1)} (18_{13, 6}\rightarrow17_{13, 5})A$           &401      &*
&219.584       &$\rm HCOOCH_{3 (\nu=1)} (18_{13, 5}\rightarrow17_{13, 4})A$           &401      & \\
~~~& & & &219.586         &$\rm HCOOCH_{3 (\nu=0} (30_{9, 22}\rightarrow30_{8, 23})A$           &330       &\\
219.592         &$\rm HCOOCH_{3 (\nu=0} (28_{9, 19}\rightarrow28_{8, 20})E$           &294       &\\
219.600         &$\rm HCOOCH_{3 (\nu=0} (30_{9, 22}\rightarrow30_{8, 23})E$           &330       &\\
219.607         &$\rm HCOOCH_{3 (\nu=0} (30_{5, 26}\rightarrow30_{3, 27})E$           &291       &*
&219.610        &$\rm CH_3COCH_{3 (\nu=0)}  (33_{8, 25}\rightarrow33_{8, 26})EE$    &368           &\\
~~~& & & &219.610         &$\rm CH_3COCH_{3 (\nu=0)}  (33_{9, 25}\rightarrow33_{7, 26})EE$    &368           &\\
219.622        &$\rm HCOOCH_{3 (\nu=1)}  (18_{12, 7}\rightarrow17_{12, 6})A$           &384        &*
&219.622        &$\rm HCOOCH_{3 (\nu=1)}  (18_{12, 6}\rightarrow17_{12, 5})A$           &384        & \\
219.642        &$\rm HCOOCH_{3 (\nu=1)}  (18_{13, 6}\rightarrow17_{13, 5})E$           &401          &\\
219.657          &$\rm HNCO (10_{3, 8}\rightarrow9_{3, 7})$            &447    &*
&219.657          &$\rm HNCO (10_{3, 7}\rightarrow9_{3, 6})$            &447    &\\
219.696        &$\rm HCOOCH_{3 (\nu=1)}  (18_{11, 8}\rightarrow17_{11, 7})A$           &369         &*
&219.696        &$\rm HCOOCH_{3 (\nu=1)}  (18_{11, 7}\rightarrow17_{11, 6})A$           &369         & \\
219.705        &$\rm HCOOCH_{3 (\nu=1)}  (18_{4, 15}\rightarrow17_{4, 14})A$           &299         & \\
219.734         &$\rm HNCO_{(\nu=0)} (10_{2, 9}\rightarrow9_{2, 8})$            &231       &\\
219.737          &$\rm HNCO_{(\nu=0)} (10_{2, 8}\rightarrow9_{2, 7})$            &231       &\\
219.764        &$\rm HCOOCH_{3 (\nu=1)}  (18_{9, 9}\rightarrow17_{9, 8})E$           &342          &\\  
%219.765         &$\rm CH_{3}CH_{2}CN_{(\nu=0)}  (35_{3, 32}\rightarrow34_{4, 31})$             &285     &   \\
%{\color{blue}219.780}        &\color{blue}$\rm CH_3CHO  (11_{1, 10}\rightarrow10_{1, 9}A)$    &\color{blue}435              &  \\
219.798          &$\rm HNCO_{(\nu=0)} (10_{0, 10}\rightarrow9_{0, 9})$            &58            &$\dag$\\
219.822         &$\rm HCOOCH_{3 (\nu=1)} (18_{10, 9}\rightarrow17_{10,8})A$           &355        &* 
&219.822         &$\rm HCOOCH_{3 (\nu=1)} (18_{10,8}\rightarrow17_{10,7})A$           &355        &\\
219.827         &$\rm HCOOCH_{3 (\nu=1)} (18_{12, 7}\rightarrow17_{12,6})E$           &384       &\\ 
219.909         &$\rm H_{2} ^{13}CO (3_{1, 2}\rightarrow2_{1, 1})$          &33      &$\dag$\\
219.949         &$\rm SO_{(\nu=0)} (6_{5}\rightarrow5_{4})$           &35                 &$\dag$\\
219.984         &$\rm CH_{3}OH_{(\nu_t=0)} (25_{3, 22}\rightarrow24_{4, 20})E$           &802    & \\
219.994         &$\rm CH_{3}OH_{(\nu_t=0)}  (23_{5, 19}\rightarrow22_{6, 17})E$           &776     & \\
220.030        &$\rm HCOOCH_{3 (\nu=1)} (18_{9, 10}\rightarrow17_{9, 9})A$           &342       &*
&220.030        &$\rm HCOOCH_{3 (\nu=1)} (18_{9, 9}\rightarrow17_{9, 8})A$           &342       &\\
220.043        &$\rm HCOOCH_{3 (\nu=1)} (18_{11, 8}\rightarrow17_{11, 7})E$           &368       &\\
220.078        &$\rm CH_{3}OH_{(\nu_t=0)}  (8_{0, 8}\rightarrow7_{1, 6})E$           &97       &\\
220.154        &t-$\rm CH_{3}CH_2OH  (24_{3, 22}\rightarrow24_{2, 23})$           &263       &\\
220.167         &$\rm HCOOCH_{3 (\nu=0)} (17_{4, 13}\rightarrow16_{4, 12})E$           &103    &$\ddag$\\
220.178        &$\rm CH_{2}CO (11_{1, 11}\rightarrow10_{1, 10})$           &76         &$\dag$\\
220.190         &$\rm HCOOCH_{3 (\nu=0)} (17_{4, 13}\rightarrow16_{4, 12})A$           &103    &\\
220.258         &$\rm HCOOCH_{3 (\nu=1)} (18_{8, 10}\rightarrow17_{8, 9})E$           &330   &$\ddag$\\
220.296         &$\rm CH_{3}^{13}CN (12_{9}\rightarrow11_{9})$           &646      & \\
220.307         &$\rm HCOOCH_{3 (\nu=1)} (18_{10, 9}\rightarrow17_{10, 8})E$           &354       &\\
220.324         &$\rm CH_{3}CN_{(\nu=0)} (12_{10}\rightarrow11_{10})$           &782       &\\
220.355        &$\rm CH_3COCH_{3 (\nu=0)} (22_{0,22}\rightarrow21_{0,21})$   &124        & AE,EA*
&220.355        &$\rm CH_3COCH_{3 (\nu=0)} (22_{1,22}\rightarrow21_{1,21})$   &124        & AE,EA\\
~~~& & & &220.355        &$\rm C_6H  (J=159/2$-157/2,$\rm \Omega$=3/2, l=e)    &425           &? \\
220.362        &$\rm CH_3COCH_{3 (\nu=0)} (22_{0,22}\rightarrow21_{1,21})EE$   &124        &*
&220.362        &$\rm CH_3COCH_{3 (\nu=0)} (22_{1,22}\rightarrow21_{1,21})EE$   &124        &\\
~~~& & & &220.362        &$\rm CH_3COCH_{3 (\nu=0)} (22_{1,22}\rightarrow21_{0,21})EE$   &124        &\\
~~~& & & &220.362        &$\rm CH_3COCH_{3 (\nu=0)} (22_{0,22}\rightarrow21_{0,21})EE$   &124        &\\
220.368        &$\rm HCOOCH_{3 (\nu=1)} (18_{8, 11}\rightarrow17_{8, 10})A$           &331       &* 
&220.370        &$\rm HCOOCH_{3 (\nu=1)} (18_{8, 10}\rightarrow17_{8, 9})A$           &331       &\\
~~~&& & & 220.368        &$\rm CH_3COCH_{3 (\nu=0)} (22_{1,22}\rightarrow21_{1,21})AA$   &124        &\\
~~~&& & & 220.368        &$\rm CH_3COCH_{3 (\nu=0)} (22_{0,22}\rightarrow21_{0,21})AA$   &124        &\\
~~~&& & & 220.368         &$\rm CH_{3}^{13}CN (12_{8}\rightarrow11_{8})$           &526      & \\
220.399         &$\rm ^{13}CO (2\rightarrow1)$           &16        &$\dag$ \\
220.404         &$\rm CH_{3}CN_{(\nu=0)} (12_{9}\rightarrow11_{9})$           &646      & \\
220.416        &$\rm HCOOCH_{3 (\nu=1)} (18_{3, 16}\rightarrow17_{2, 15})A$           &293       & \\
220.431         &$\rm CH_{3}^{13}CN (12_{7}\rightarrow11_{7})$           &417      &* 	
&220.433        &$\rm HCOOCH_{3 (\nu=0)} (35_{5, 28}\rightarrow33_{5, 29})A$           &357       & \\
220.446        &$\rm CH_3CHO  (13_{3, 10}\rightarrow13_{2, 11}\nu=0)E$    &105           &?$\ddag$\\
%220.444        &g-$\rm CH_3CH_2OH  (28_{7, 22}\rightarrow28_{6, 22},\nu_t=0-1)$    &455           &*
%&{\color{red}?}220.443        &g-$\rm CH_3CH_2OH  (30_{7, 23}\rightarrow30_{6, 25},\nu_t=0-1)$    &503           &\\
%{\color{blue}220.464}        &\color{blue}$\rm CH_2OHCHO  (20_{2, 18}\rightarrow19_{3, 17})$    &\color{blue}120     \color{blue}& $*$$\ddag$           
220.465        &$\rm CH_3COCH_{3 (\nu=0)}  (11_{11, 1}\rightarrow10_{10, 0})AE$    &63          &*
&220.466        &$\rm CH_3COCH_{3 (\nu=0)}  (11_{11, 0}\rightarrow10_{10, 1})AE$    &63          &\\
220.476        &$\rm CH_{3}CN_{(\nu=0)}  (12_{8}\rightarrow11_{8})$            &526        &\\
220.486        &$\rm CH_{3}^{13}CN  (12_{6}\rightarrow11_{6})$            &326        &*
&220.487        &$\rm C_6H  (J=159/2$-157/2, $\rm \Omega$=3/2,, l=f)    &426               &? \\	
220.525        &$\rm HCOOCH_{3 (\nu=1)} (10_{4, 6}\rightarrow9_{3, 7})A$           &231       & \\
220.533        &$\rm CH_{3}^{13}CN  (12_{5}\rightarrow11_{5})$            &247        &\\
220.539         &$\rm CH_{3}CN_{(\nu=0)} (12_{7}\rightarrow11_{7})$           &419       &\\
220.561       &$\rm CH_{2}CHCN  (24_{1, 24}\rightarrow23_{1, 23})$            &134       &\\
220.570        &$\rm CH_{3}^{13}CN  (12_{4}\rightarrow11_{4})$            &183        &\\
220.585        &$\rm HNCO_{(\nu=0)}  (10_{1, 9}\rightarrow9_{1, 8})$            &101     &\\
220.594         &$\rm CH_{3}CN_{(\nu=0)} (12_{6}\rightarrow11_{6})$           &326       &\\
220.621         &$\rm CH_{3}^{13}CN (12_{2}\rightarrow11_{2})$            &97           &$\ddag$\\

220.641        &$\rm CH_{3}CN_{(\nu=0)}  (12_{5}\rightarrow11_{5})$            &248    &\\
220.661        &$\rm CH_{3}CH_{2}CN_{(\nu=0)}  (25_{2, 24}\rightarrow24_{2, 23})$            &143      &\\
220.679        &$\rm CH_{3}CN_{(\nu=0)}  (12_{4}\rightarrow11_{4})$            &183     &\\
220.709       &$\rm CH_{3}CN_{(\nu=0)}  (12_{3}\rightarrow11_{3})$            &133       &\\
220.730        &$\rm CH_{3}CN_{(\nu=0)}  (12_{2}\rightarrow11_{2})$            &98     &$\ddag$\\
220.743       &$\rm CH_{3}CN_{(\nu=0)}  (12_{1}\rightarrow11_{1})$            &76     &\\
220.747        &$\rm CH_{3}CN_{(\nu=0)}  (12_{0}\rightarrow11_{0})$            &69      &\\
220.764        &$\rm CH_3COCH_{3 (\nu=0)}  (11_{11, 0}\rightarrow10_{10, 0})EE$    &63           &\\
220.786        &$\rm HCOOCH_{3 (\nu=0)} (28_{3, 25}\rightarrow28_{3, 26})E$           &248       & \\
220.812        &$\rm HCOOCH_{3 (\nu=0)}  (18_{3, 16}\rightarrow17_{2, 15})E$           &105       & \\
220.815        &$\rm HCOOCH_{3 (\nu=0)}  (18_{3, 16}\rightarrow17_{2, 15})A$           &105       & \\
220.820        &$\rm HCOOCH_{3 (\nu=0)}  (24_{2, 23}\rightarrow24_{1, 24})A$           &169       & \\
220.848        &$\rm CH_3OCH_{3}  (24_{4, 20}\rightarrow23_{5, 19})$    &297          &A(E)A(E)\\
220.866        &$\rm HCOOCH_{3 (\nu=0)} (28_{3, 25}\rightarrow28_{3, 26})A$           &248        &*
&220.866        &$\rm HCOOCH_{3 (\nu=1)} (19_{2, 17}\rightarrow18_{3, 16})A$           &303        &\\
220.889        &$\rm HCOOCH_{3 (\nu=0)} (18_{17, 1}\rightarrow17_{17, 0})A$           &293        &*
&220.889        &$\rm HCOOCH_{3 (\nu=0)} (18_{17, 2}\rightarrow17_{17, 1})A$           &293        &\\
220.893         &$\rm CH_3OCH_{3} (23_{4,20}\rightarrow23_{3,21})$           &274      &A(E)A(E) \\
220.901        &$\rm HCOOCH_{3 (\nu=0)} (18_{17, 1}\rightarrow17_{17, 0})E$           &293        &\\
220.913        &$\rm HCOOCH_{3 (\nu=1)} (18_{7, 12}\rightarrow17_{7, 11})A$           &320        &*
&220.910        &$\rm HCOOCH_{3 (\nu=0)} (18_{17, 2}\rightarrow17_{17, 1})E$           &293        &\\
220.926        &$\rm HCOOCH_{3 (\nu=0)} (18_{16, 2}\rightarrow17_{16, 1})A$           &271        &*
&220.926        &$\rm HCOOCH_{3 (\nu=0)} (18_{16, 3}\rightarrow17_{16, 2})A$           &271        &\\
220.935        &$\rm HCOOCH_{3 (\nu=0)} (18_{16, 2}\rightarrow17_{16, 1})E$           &271        &\\
220.946        &$\rm HCOOCH_{3 (\nu=1)} (18_{7, 11}\rightarrow17_{7, 10})A$           &321        &*
&220.947        &$\rm HCOOCH_{3 (\nu=0)} (18_{16, 3}\rightarrow17_{16, 2})E$           &271        &\\
%220.962        &$\rm CH_{3}CH_{2}CN_{(\nu=0)}  (46_{6, 40}\rightarrow46_{5, 41})$            &506      & \\
220.967        &$\rm HC_7N_{(\nu=0)}  (J=196-195)$    &1045           &?      \\
220.978       &$\rm HCOOCH_{3 (\nu=0)}  (18_{15, 4}\rightarrow17_{15, 3})A$           &250       &*
&220.978       &$\rm HCOOCH_{3 (\nu=0)}  (18_{15, 3}\rightarrow17_{15, 2})A$           &250       & \\
220.985      &$\rm HCOOCH_{3 (\nu=1)}  (18_{7, 11}\rightarrow17_{7, 10})E$           &320       &*
&220.984      &$\rm HCOOCH_{3 (\nu=0)}  (18_{15, 3}\rightarrow17_{15, 2})E$           &250       &\\
%{\color{blue}220.999}        &\color{blue}$\rm CH_3CH_2OH  (13_{0, 13}\rightarrow12_{0, 12},\nu_t=0-1)$    &\color{blue}136           &\color{blue}$*$
220.998       &$\rm HCOOCH_{3 (\nu=0)}  (18_{15, 4}\rightarrow17_{15, 3})E$           &250        &*
&220.999        &g-$\rm CH_3CH_2OH  (13_{0, 13}\rightarrow12_{0, 12},\nu_t=0-1)$    &136           &\\
%{\color{blue}221.008}        &\color{blue}$\rm (CH_2OH)_2  (21_{4, 18}\nu=1\rightarrow20_{4, 17}\nu=0)$    &\color{blue}122          &\color{blue}$\ddag$\\
221.048       &$\rm HCOOCH_{3 (\nu=0)}  (18_{14, 5}\rightarrow17_{14, 4})A$           &231         &*
&221.048       &$\rm HCOOCH_{3 (\nu=0)}  (18_{14, 4}\rightarrow17_{14, 3})A$           &231         &\\
~~~& & & &221.048       &$\rm HCOOCH_{3 (\nu=0)}  (18_{14, 4}\rightarrow17_{14, 3})E$           &231         &\\
221.067       &$\rm HCOOCH_{3 (\nu=0)}  (18_{14, 5}\rightarrow17_{14, 4})A$           &231         &\\
221.076       &$\rm HCOOCH_{3 (\nu=0)}  (29_{9, 21}\rightarrow29_{8, 22})A$           &312        & \\
221.086       &$\rm HCOOCH_{3 (\nu=0)}  (29_{9, 21}\rightarrow29_{8, 22})E$           &312        & \\
221.111      &$\rm HCOOCH_{3 (\nu=1)}  (18_{8, 11}\rightarrow17_{8, 10})E$           &330        & \\
221.115     &$\rm ^{34}SO_{2(\nu=0)} (22_{2, 20}\rightarrow22_{1, 21})$           &248     &\\
221.124       &$\rm CH_{2}CHCN_{(\nu=0)}  (23_{1, 22}\rightarrow22_{1, 21})$            &130       &\\
221.140      &$\rm HCOOCH_{3 (\nu=0)}  (18_{13, 5}\rightarrow17_{13, 4})E$           &213        &* 
&221.141      &$\rm HCOOCH_{3 (\nu=0)}  (18_{13, 5}\rightarrow17_{13, 4})A$           &213        &\\
~~~& & & &221.141      &$\rm HCOOCH_{3 (\nu=0)}  (18_{13,6}\rightarrow17_{13, 5})A$           &213        &\\

229.056        &$\rm CH_3COCH_{3 (\nu=0)}  (22_{1, 21}\rightarrow21_{2, 20})EE$    &133          &*
&229.056        &$\rm CH_3COCH_{3 (\nu=0)}  (22_{1, 21}\rightarrow21_{1, 20})EE$    &133          &\\
~~~& & & &229.056        &$\rm CH_3COCH_{3 (\nu=0)}  (22_{2, 21}\rightarrow21_{2, 20})EE$    &133          &\\
~~~& & & &229.056        &$\rm CH_3COCH_{3 (\nu=0)}  (22_{2, 21}\rightarrow21_{1, 20})EE$    &133          &\\
~~~& & & &229.058        &$\rm CH_3COCH_{3 (\nu=0)}  (14_{9,6}\rightarrow13_{8, 5})EE$    &85          &\\
229.078      &$\rm CH_3COCH_{3 (\nu=0)}  (22_{1, 21}\rightarrow21_{2, 20})AA$    &133          &*
&229.078        &$\rm CH_3COCH_{3 (\nu=0)}  (22_{2, 21}\rightarrow21_{1, 20})AA$    &133          &\\
229.087      &$\rm CH_{2}CHCN_{(\nu=0)}  (24_{3, 21}\rightarrow23_{3, 20})$            &156       &\\
229.117      &$\rm HCOOCH_{3 (\nu=1)}  (19_{3, 17}\rightarrow18_{2, 16})E$           &303        & \\
229.127        &$\rm CH_3COCH_{3 (\nu=0)}  (12_{10, 2}\rightarrow11_{9, 2})EE$    &68          &\\
229.134       &$\rm CH_3CHO  (9_{3, 7}\rightarrow9_{2, 7})$    &62          &\\
229.203         &UL           &           &$?$\\
229.224        &$\rm HCOOCH_{3 (\nu=0)}  (23_{9, 14}\rightarrow23_{8, 15})A$    &217          &\\
%229.227         &UL           &           &$?$
%&{\color{blue}229.227}        &\color{blue}$\rm CH_3CH_2OH (12_{5, 8}\rightarrow12_{4, 9})$    &\color{blue}157           &\color{blue}$?$\\
%~~~& & & &{\color{blue}229.228}        &\color{blue}$\rm CH_3OCHO  (14_{5, 9}\rightarrow14_{3, 12}A)$    &\color{blue}266           &\color{blue}$?$\\
%{\color{blue}229.234}        &\color{blue}$\rm (CH_2OH)_2  (23_{11, 13}\nu=0\rightarrow22_{11, 12}\nu=1)$    &\color{blue}195         & \\
229.259       &$\rm HCOOCH_{3 (\nu=0)}  (23_{9, 14}\rightarrow23_{8, 15})E$    &217          &*
&229.258        &$\rm CH_3CHO  (10_{3, 7}\rightarrow10_{2, 8})$    &71     &\\
229.265      &$\rm CH_{3}CH_{2}CN_{(\nu=0)}  (26_{2, 25}\rightarrow25_{5, 24})$            &154          &$\ddag$\\
229.320        &$\rm HCOOCH_{3 (\nu=0)}  (23_{9, 15}\rightarrow23_{8, 16})E$    &217         & \\
229.348      &$\rm SO_{2(\nu=0)}  (11_{5, 7}\rightarrow12_{4, 8})$            &122        &$\dag$\\
229.389        &$\rm HCOOCH_{3 (\nu=0)}  (23_{9, 15}\rightarrow23_{8, 16})A$    &217         & \\
229.405      &$\rm HCOOCH_{3 (\nu=0)} (18_{3, 15}\rightarrow17_{3, 14})E$           &111      &\\
229.420      &$\rm HCOOCH_{3 (\nu=0)} (18_{3, 15}\rightarrow17_{3, 14})A$           &111     &\\
229.474      &$\rm HCOOCH_{3 (\nu=0)} (20_{3, 17}\rightarrow19_{4, 16})E$           &134        &\\
229.491        &t-$\rm CH_3CH_2OH  (17_{5, 12}\rightarrow17_{4, 13})$    &160     &*
&229.498        &g-$\rm CH_3CH_2OH  (41_{11, 31}\rightarrow40_{12, 29},\nu_t=1-0)$    &925     &\\
~~~& & & &229.498        &g-$\rm CH_3CH_2OH  (41_{11, 30}\rightarrow40_{12, 28},\nu_t=1-0)$    &925     &\\
229.505      &$\rm HCOOCH_{3 (\nu=0)} (20_{3, 17}\rightarrow19_{4, 16})A$           &134       &\\
229.539      &$\rm HCOOCH_{3 (\nu=0)} (31_{5, 27}\rightarrow31_{3, 28})E$           &309       &\\
229.589      &$\rm CH_3OH_{(\nu_t=0)} (15_{4, 11}\rightarrow16_{3, 13})E$           &374      &\\
229.607      &$\rm HCOOCH_{3 (\nu=0)} (31_{5, 27}\rightarrow31_{3, 28})A$           &309       &\\
229.648      &$\rm CH_{2}CHCN_{(\nu=0)}  (25_{1, 25}\rightarrow24_{1, 24})$            &146        &\\
229.661      &$\rm HCOOCH_{3 (\nu=1)} (25_{9, 16}\rightarrow25_{8, 17})A$           &432       &\\
%{\color{blue}229.747}        &\color{blue}$\rm CH_2OHCHO  (29_{10, 20}\rightarrow29_{9, 21})$    &\color{blue}303       &\color{blue}$*$           \\
229.759     &$\rm CH_{3}OH_{(\nu_t=0)}  (8_{-1, 8}\rightarrow7_{0, 7})E$           &89            &$\ddag$ \\
%{\color{blue}229.775}        &\color{blue}$\rm CH_3CHO  (11_{1, 11}\rightarrow10_{0, 10})$    &\color{blue}61     &\color{blue}$*$           \\
229.858     &$\rm ^{34}SO_{2(\nu=0)} (4_{2, 2}\rightarrow3_{1, 3})$           &19      & \\
229.864     &$\rm CH_{3}OH_{(\nu_t=0)}  (19_{5, 15}\rightarrow20_{4, 16}) A++$           &578        & \\
229.888     &$\rm ^{34}SO_{2(\nu=0)} (17_{3, 15}\rightarrow18_{0, 18})$           &162       & \\
229.939     &$\rm CH_{3}OH_{(\nu_t=0)}  (19_{5, 14}\rightarrow20_{4, 17})A--$           &578        &\\
229.987     &$\rm ^{34}SO_{2(\nu=0)} (42_{6, 36}\rightarrow41_{7, 35})$           &921      &  \\
230.027     &$\rm CH_{3}OH_{(\nu_t=0)}  (3_{-2, 2}\rightarrow4_{-1, 4})E$           &40     & \\
230.109        &$\rm C_6H  (J=165/2$-163/2, $\rm \Omega$=1/2,  l=f)    &482           &? $\ddag$     \\
230.140        &$\rm CH_3OCH_3  (25_{4, 22}\rightarrow25_{3, 22})EA$    &319         &A(E)A(E)$\ddag$\\
230.170        &$\rm CH_3COCH_{3(\nu=0)}  (23_{0, 23}\rightarrow22_{1, 22})AE$    &135         &*
&230.170        &$\rm CH_3COCH_{3(\nu=0)}  (23_{1, 23}\rightarrow22_{0, 22})AE$    &135         &\\
~~~& & & &230.170        &$\rm CH_3COCH_{3(\nu=0)}  (23_{0, 23}\rightarrow22_{0, 22})EA$    &135         &\\
~~~& & & &230.170        &$\rm CH_3COCH_{3(\nu=0)}  (23_{1, 23}\rightarrow22_{1, 22})EA$    &135         &\\
230.177        &$\rm CH_3COCH_{3(\nu=0)}  (23_{0, 23}\rightarrow22_{1, 22})EE$    &135         &*
&230.177        &$\rm CH_3COCH_{3(\nu=0)}  (23_{1, 23}\rightarrow22_{0, 22})EE$    &135         &\\
~~~& & & &230.177        &$\rm CH_3COCH_{3(\nu=0)}  (23_{0, 23}\rightarrow22_{0, 22})EE$    &135         &\\
~~~& & & &230.177        &$\rm CH_3COCH_{3(\nu=0)}  (23_{1, 23}\rightarrow22_{1, 22})EE$    &135         &\\
230.183        &$\rm CH_3COCH_{3(\nu=0)}  (23_{0, 23}\rightarrow22_{1, 22})AA$    &135        &$\ddag$*
&230.183        &$\rm CH_3COCH_{3(\nu=0)}  (23_{1, 23}\rightarrow22_{0, 22})AA$    &135        &\\
230.234        &$\rm CH_3OCH_3  (17_{2, 15}\rightarrow16_{3, 14})$    &148         &A(E)A(E)
&230.234        &$\rm C_6H  (J=165/2$-163/2, $\rm \Omega$=1/2,  l=e)    &483           &?      \\
230.283        &$\rm CH_3COCH_{3(\nu=0)}  (16_{8, 8}\rightarrow15_{9, 7})EE$    &107        &\\
230.293      &$\rm CH_3OH_{(\nu_t=0)}  (22_{2, 20}\rightarrow21_{-3, 19})E$           &609        &\\
%{\color{blue}230.302}        &\color{blue}$\rm CH_3CHO  (12_{2, 11}\rightarrow11_{2, 10}A)$    &\color{blue}81              & \\
230.318     &$\rm O^{13}CS (19\rightarrow18)$           &111          &$\dag$ \\
%&{\color{blue}230.316}        &\color{blue}$\rm CH_3CHO  (12_{2, 11}\rightarrow11_{2, 10}E)$    &\color{blue}81               &\\
230.369     &$\rm CH_{3}OH_{(\nu_t=0)}  (22_{4, 18}\rightarrow21_{5, 17})E$           &683          &\\
%{\color{blue}230.395}        &\color{blue}$\rm CH_3CHO  (12_{2, 11}\rightarrow11_{2, 10}A--)$    &\color{blue}287             &\color{blue}$\ddag$\\
%{\color{blue}230.439}        &\color{blue}$\rm CH_3CHO  (12_{2, 11}\rightarrow11_{2, 10}A++)$    &\color{blue}440              & \\
230.377      &$\rm HCOOCH_{3 (\nu=0)} (22_{9, 14}\rightarrow22_{8, 15})A$           &203       &\\
%230.412        &$\rm CH_3COCH_{3(\nu=0)}  (27_{1, 26}\rightarrow22_{1, 27})EE$    &195         &*
%&230.412        &$\rm CH_3COCH_{3(\nu=0)}  (27_{2, 26}\rightarrow27_{1, 27})EE$    &195         &\\
%~~~& & & &230.412        &$\rm CH_3COCH_{3(\nu=0)}  (27_{1, 26}\rightarrow27_{0, 27})EE$    &195         &\\
%~~~& & & &230.412        &$\rm CH_3COCH_{3(\nu=0)}  (27_{2, 26}\rightarrow27_{0, 27})EE$    &195         &\\
230.436      &$\rm HCOOCH_{3 (\nu=0)} (25_{1, 24}\rightarrow25_{1, 25})E$           &182       &*
&230.436      &$\rm HCOOCH_{3 (\nu=0)} (25_{1, 24}\rightarrow25_{0, 25})E$           &182       &\\
230.441      &$\rm HCOOCH_{3 (\nu=0)} (25_{2, 24}\rightarrow25_{1, 25})E$           &182       &*
&230.441      &$\rm HCOOCH_{3 (\nu=0)} (25_{2, 24}\rightarrow25_{0, 25})E$           &182       &\\
230.481        &$\rm CH_3OCH_3  (10_{8, 2}\rightarrow11_{7, 5})EE$    &140         &*
&230.474        &$\rm CH_3OCH_3  (10_{8, 3}\rightarrow11_{7, 4})EA$    &140         &\\
~~~& & & &230.476        &$\rm CH_3OCH_3  (10_{8, 3}\rightarrow11_{7, 4})EE$    &140         &\\
~~~& & & &230.478        &$\rm CH_3OCH_3  (10_{8, 3}\rightarrow11_{7, 4})AA$    &140         &\\
~~~& & & &230.478        &$\rm CH_3OCH_3  (10_{8, 2}\rightarrow11_{7, 5})AA$    &140         &\\
~~~& & & &230.479        &$\rm CH_3OCH_3  (10_{8, 3}\rightarrow11_{7, 4})AE$    &140         &\\
~~~& & & &230.479        &$\rm CH_3OCH_3  (10_{8, 2}\rightarrow11_{7, 5})AE$    &140         &\\
~~~& & & &230.483        &$\rm CH_3OCH_3  (10_{8, 2}\rightarrow11_{7, 5})EA$    &140         &\\
230.488     &$\rm CH_{2}CHCN_{(\nu=0)} (24_{1, 23}\rightarrow23_{1, 22})$           &141        &\\
230.505        &$\rm CH_3OCH_3  (26_{4, 23}\rightarrow25_{5, 20})$    &343         &A(E)A(E)\\
230.538     &$\rm CO_{(\nu=0)} (2\rightarrow1)$           &17           &$\dag$ \\
230.673        &g-$\rm CH_3CH_2OH  (13_{2, 11}\rightarrow12_{2, 10},\nu_t=0-0)$    &139                &\\
230.697        &g-$\rm CH_3CH_2OH  (8_{3, 5}\rightarrow8_{2, 7},\nu_t=1-0)$    &103               & \\
230.739     &$\rm CH_{2}CHCN_{(\nu=0)}  (25_{0, 25}\rightarrow24_{0, 24})$            &146     &$\ddag$\\
%{\color{blue}230.791}        &\color{blue}$\rm CH_3CHO  (12_{9, 3}\rightarrow11_{9, 2}A)$    &\color{blue}460              &\color{blue}$*$ 
230.794        &g-$\rm CH_3CH_2OH  (6_{5, 1}\rightarrow5_{4, 1},\nu_t=0-1)$    &105                &*
&230.794        &g-$\rm CH_3CH_2OH  (6_{5, 2}\rightarrow5_{4, 2},\nu_t=0-1)$    &105                &\\
230.844      &$\rm HCOOCH_{3 (\nu=1)} (19_{17, 2}\rightarrow18_{17, 1})E$           &493       &\\
230.852      &$\rm HCOOCH_{3 (\nu=1)} (19_{16, 3}\rightarrow18_{16, 2})E$          &470      &\\
230.879        &$\rm HCOOCH_{3 (\nu=1)}  (18_{4, 14}\rightarrow17_{4, 13})A$    &301         &\\
230.889        &$\rm HCOOCH_{3 (\nu=1)}  (19_{15, 4}\rightarrow18_{15, 3})E$    &450          &\\
%{\color{blue}230.934}        &\color{blue}$\rm (CH_2OH)_2  (23_{3, 20}\nu=0\rightarrow22_{3, 19}\nu=1)$    &\color{blue}143          &\color{blue}$*$  
230.933     &$\rm ^{34}SO_{2(\nu=0)} (5_{4, 2}\rightarrow6_{3, 3})$           &52      &* 
&230.936        &$\rm HCOOCH_{3 (\nu=0)}  (29_{3, 26}\rightarrow29_{2, 27})E$    &265          &\\
230.954        &t-$\rm CH_3CH_2OH  (16_{5, 11}\rightarrow16_{4, 12})$    &145              &\\
230.960        &$\rm HCOOCH_{3 (\nu=1)}  (19_{14, 5}\rightarrow18_{14, 4})E$    &430          &\\
%{\color{blue}230.966}        &\color{blue}$\rm (CH_2OH)_2  (23_{7, 17}\nu=0\rightarrow22_{7, 16}\nu=1)$    &\color{blue}160         &\\
%{\color{blue}230.967}        &\color{blue}$\rm (CH_2OH)_2  (24_{1, 23}\nu=0\rightarrow23_{2, 22}\nu=1)$    &\color{blue}144         & \\
230.991    &t-$\rm CH_3CH_2OH  (14_{0, 14}\rightarrow13_{1, 13})$      &86  &*$\ddag$
&230.992        &g-$\rm CH_3CH_2OH  (6_{5, 1}\rightarrow6_{4, 2},\nu_t=1-1)$    &110               &\\
~~~& & & &230.993        &g-$\rm CH_3CH_2OH  (6_{5, 2}\rightarrow6_{4, 3},\nu_t=1-1)$    &110               &\\
%{\color{blue}231.016}        &\color{blue}$\rm (CH_2OH)_2  (24_{2, 23}\nu=0\rightarrow23_{1, 22}\nu=1)$    &\color{blue}143           &\color{blue}$*$          \\
231.019        &$\rm HCOOCH_{3 (\nu=0)}   (12_{4, 9}\rightarrow11_{3, 8})E$    &57        &*   
&231.020       &$\rm HCOOCH_{3 (\nu=0)}  (29_{3, 26}\rightarrow29_{2, 27})A$    &264        & \\
231.046       &$\rm HCOOCH_{3 (\nu=0)}   (12_{4, 9}\rightarrow11_{3, 8})A$    &57        &\\  
%~~~& & & &{\color{blue}231.021}        &$\rm \color{blue}(CH_2OH)_2  (24_{2, 23}\nu=1\rightarrow23_{1, 22}\nu=1)$    &\color{blue}144         &$\color{blue}*$    \\
231.061     &$\rm OCS_{(\nu=0)} (19\rightarrow18)$           &111        &$\dag$\\
231.102        &$\rm HC_7N  (J=205-204)$    &1142           &?$\ddag$      \\
231.188        &$\rm HCOOCH_{3 (\nu=0)}   (21_{9, 13}\rightarrow21_{8, 14})E$    &190          &\\
%{\color{blue}231.200}        &\color{blue}$\rm CH_2OHCHO  (28_{10, 19}\rightarrow28_{9, 20})$    &\color{blue}287    &\color{blue}$*$        
%&{\color{blue}231.200}        &\color{blue}$\rm CH_3CHO  (12_{8, 4}\rightarrow11_{8, 3}E)$    &\color{blue}421               &\color{blue}$*$ \\
231.199        &$\rm HCOOCH_{3 (\nu=0)}   (21_{9, 12}\rightarrow21_{8, 13})A$    &190          &*
&231.199        &$\rm HCOOCH_{3 (\nu=0)}   (21_{9, 12}\rightarrow21_{8, 13})E$    &190          &\\
231.221     &$\rm ^{13}CS_{(\nu=0)} (5\rightarrow4)$           &33           &$\dag$\\
231.231       &$\rm HCOOCH_{3 (\nu=1)}   (19_{12, 7}\rightarrow18_{12, 6})E$    &396          &
&231.231        &g-$\rm  CH_3CHO  (12_{8, 5}\rightarrow11_{8, 4}E)$    &216               &\\

\hline \hline
%\end{tabular}
%~\\

\multicolumn{8}{l}{{\bf Note.} 1.~Due to the frequency resolution (0.8125 MHz, $\rm \sim 1.2 km\,s^{-1}$), lines with broad FWHM width can be attributed to different species, so  lines with }\\
\multicolumn{8}{l}{~ ~~~~~~~~~~ stronger CDMS/JPL intensity are listed here in the left column with ``*", and the potential blended weaker  transitions are listed in the right column.}\\
\multicolumn{8}{l}{~ ~~~~~~~~2.~ Tentative detections and unidentified  lines are marked with ``?". }\\
\multicolumn{8}{l}{~ ~~~~~~~~3.~ Lines with  ``$\dag$" are imaged in Figure~\ref{into},  with ``$\ddag$" are imaged in Figure~\ref{dis}. }\\
\multicolumn{4}{l}{~ ~~~~~~~~4.~``A(E)A(E)" represents 4 types of transitions are possible: AA, EE, AE, EA.}\\
\end{longtable}

%%%%%%%%%%%%%%%%%%%%%%%%rotation
%%%%%%%%%%%%%%%%%%%%%%%%rotation
%%%%%%%%%%%%%%%%%%%%%%%%rotation

\newpage
\begin{landscape}
\begin{table}
\caption{Laboratory parameters of $\rm CH_3CN$ and $\rm CH_3^{13}CN$ and the intensity integrated  over the width of each line $\rm \int T_{B}({\upsilon})  d{\upsilon}$, which is used for rotation diagrams in  Figure~\ref{rotation}. Uncertainties on the measured intensities are typically $\le10\%$  (written as the subscript), as determined from Gaussian fitting. 
 %For lines of $\rm CH_3CN$,   $\rm CH_3OH$ and  $\rm ^{34}SO_2$ which are not detected, an upper limit equal to the $\rm 3\sigma$ rms is given. 
  \label{tab:rotch3cnline} 
 }
\small
\begin{center}
%I(a). Laboratory parameters and integrated intensities $\rm \int T_B(\upsilon)d\upsilon$ of  $\rm CH_3CN~(J=12\rightarrow11)^*$\\

\begin{tabu}{cc| p{0.2cm} p{1cm} p{1cm} p{1cm}p{0.7cm}|p{2cm} p{2cm}p{2cm} p{2cm} p{2cm}|p{1cm}}\hline\hline

Mol. &Freq    &K  &$\rm I(300K)$    &$\rm E_{L}$    &$S_{ul}\mu^2$       &o-/p-  &$\rm HC$                      & $\rm mm2$                     & $\rm mm3a$                 &$\rm mm3b$                    &$\rm NE$   &Note\\
       &(GHz)    &  &$\rm Log_{10}$    &$\rm (cm^{-1})$     &$\rm (D^2)$                       &     &$\rm (K~km~s^{-1})$     &$\rm (K~km~s^{-1})$      &$\rm (K~km~s^{-1})$    &$\rm (K~km~s^{-1})$       &$\rm (K~km~s^{-1})$       \\
\hline
\multirow{9}*{$\rm CH_3CN~(J=12\rightarrow11)$}
    &220.476   &         8   &    -3.098  &   357.938  &   205.033 &p-     &$\rm    73.64 _{\pm2.81  }$      &$\rm    16.46_{\pm1.14  }$      &$\rm    11.22 _{\pm0.85  }$      &$\rm    7.67_{\pm1.06   }$      &$\rm     \le0.31$    &\multirow{9}*{\it 1, 2, 3, 4}\\

   &220.539   &         7   &    -2.868  &   283.608  &   243.499 &p-     &$\rm    144.31 _{\pm3.70  }$      &$\rm     49.48_{\pm1.90  }$      &$\rm     27.68 _{\pm1.16  }$      &$\rm     33.72_{\pm1.52   }$      &$\rm      \le0.61$\\

   &220.594   &         6   &    -2.377  &   219.154  &   553.690 &o-     &$\rm    356.10 _{\pm7.70  }$      &$\rm    150.47_{\pm10.38  }$      &$\rm     90.69_{\pm11.98  }$      &$\rm    190.53_{\pm2.74   }$      &$\rm      0.71_{\pm0.53 }$\\

   &220.641   &         5   &    -2.522  &   164.592  &   305.047 &p-     &$\rm    327.23_{\pm32.55  }$      &$\rm    146.62_{\pm4.76  }$      &$\rm    106.37_{\pm10.00  }$      &$\rm    178.31_{\pm5.00   }$      &$\rm      1.34_{\pm0.63 }$\\

   &220.679   &         4   &    -2.397  &   119.933  &   328.188 &p-     &$\rm    307.21_{\pm25.39  }$      &$\rm    150.29_{\pm5.73  }$      &$\rm    142.61 _{\pm2.77  }$      &$\rm    170.24_{\pm10.00   }$      &$\rm      5.11_{\pm0.36 }$\\

   &220.709   &         3   &    -2.001  &    85.187  &   692.234 &o-     &$\rm    484.10_{\pm10.61  }$      &$\rm    209.82_{\pm15.40  }$      &$\rm    278.38 _{\pm7.16  }$      &$\rm    319.87_{\pm13.08   }$      &$\rm     11.92_{\pm0.56 }$\\

   &220.730   &         2   &    -2.234  &    60.363  &   358.988 &p-     &$\rm    397.54 _{\pm9.19  }$      &$\rm    281.55_{\pm10.27  }$      &$\rm    256.41_{\pm27.26  }$      &$\rm    331.98_{\pm21.20   }$      &$\rm     15.17_{\pm1.00 }$\\

   &220.743   &         1   &    -2.194  &    45.467  &   366.614 &p-     &$\rm    371.68_{\pm28.18  }$      &$\rm    229.32_{\pm8.17  }$      &$\rm    247.64_{\pm10.46  }$      &$\rm    284.65_{\pm25.77   }$      &$\rm     13.52_{\pm0.66 }$\\

  & 220.747   &         0   &    -2.180  &    40.501  &   369.277 &p-     &$\rm    372.96_{\pm28.18  }$      &$\rm    211.34_{\pm4.55  }$      &$\rm    255.62 _{\pm1.40  }$      &$\rm    348.84_{\pm25.77   }$      &$\rm     11.19_{\pm0.37 }$\\
\hline

\multirow{5}*{$\rm CH_3^{13}CN~(J=12\rightarrow11)$}
&220.431   &7   &-2.707   &283.596   &243.492   &p-   &$\rm4.07_{\pm1.61}$   &&&&&\multirow{5}*{\it 2, 3, 5}\\
&220.485   &6   &-2.216   &219.141   &553.669   &o-   &$\rm14.29_{\pm1.28}$  &&&&& \\
&220.532   &5   &-2.361   &164.575   &305.101   &p-   &$\rm16.78_{\pm1.80}$   &&&&&\\
&220.570   &4   &-2.236   &119.915   &328.169   &p-   &$\rm27.41_{\pm1.75}$  &&&&& \\
&220.621   &2   &-2.073   &60.344     &358.964   &p-   &$\rm96.44_{\pm4.41}$   &&&&&\\

\hline
 \hline
%    \multicolumn{7}{l}{* None detection are not given}\\
\multicolumn{13}{l}{ {\bf Notes:} {\it 1.} None detection is given by an upper limit of $\rm 3\sigma$ rms.}\\
\multicolumn{13}{l}{~~~~~~~~~~~~{\it 2.}  Integrated intensities are  obtained by Gildas ``Gaussian" fittings.}\\
\multicolumn{13}{l}{~~~~~~~~~~~~{\it 3.}  Laboratory parameters are given from CDMS. }\\
\multicolumn{13}{l}{~~~~~~~~~~~~{\it 4.}  Q(300K, {\scriptsize{$\rm CH_3CN$}})=14683.6324 } \\
\multicolumn{13}{l}{~~~~~~~~~~~~{\it 5.}  Q(300K, {\scriptsize{$\rm CH_3^{13}CN$}})=199602.70 } \\
\end{tabu}

\end{center}
\end{table}

\end{landscape}

%%%%%%%%%%%%%%%%%%%%%%%%%%%%HNCO
\begin{landscape}
\begin{table}
\caption{The laboratory parameters and the intensity integrated over the width of each line $\rm \int T_{B}({\upsilon})  d{\upsilon}$, which is used for rotation diagrams in Figure~\ref{fig:trot_other}. Uncertainties on the measured intensities are typically $\le10\%$  (written as the subscript), as determined from Gaussian or HFS fitting. 
 %For lines of $\rm CH_3CN$,   $\rm CH_3OH$ and  $\rm ^{34}SO_2$ which are not detected, an upper limit equal to the $\rm 3\sigma$ rms is given. 
  \label{tab:rotline} 
 }
\small
\begin{center}

%II. Laboratory parameters and integrated intensities $\rm \int T_B(\upsilon)d\upsilon$ of $\rm HNCO$]\\
\scalebox{0.95}{
\begin{tabu}{cc| p{1cm} p{1cm} p{1cm} |p{2cm} p{2cm} p{2cm} p{2cm}|p{2cm}}\hline\hline

Mol.    &Freq      &$\rm I(300K)$    &$\rm E_{L}$    &$S_{ul}\mu^2$         &$\rm HC$                      & $\rm mm2$                     & $\rm mm3a$                 &$\rm mm3b$                       &Notes\\
          &(GHz)      &$\rm Log_{10}$    &$\rm (cm^{-1})$     &$\rm (D^2)$                            &$\rm (K~km~s^{-1})$     &$\rm (K~km~s^{-1})$      &$\rm (K~km~s^{-1})$    &$\rm (K~km~s^{-1})$           \\

%IV(a). Laboratory parameters and integrated intensities $\rm \int T_B(\upsilon)d\upsilon$ of $\rm HCOOCH_3~(\nu=1)^*$]\\
   \hline
   \multirow{21}*{$\rm HCOOCH_3~(\nu=1)$}
   &218.966   &    -4.619  &   259.918  &    53.255  &$\rm       2.17_{\pm0.92   }$      &$\rm      2.25_{\pm0.00  }$      &$\rm      2.87_{\pm0.75   }$      &$\rm      6.32_{\pm1.03  }$
         &\multirow{21}*{\it 1, 2, 3, 4}\\
   &219.155   &    -4.543  &   249.175  &    60.055  &$\rm       6.26_{\pm0.74   }$      &$\rm      4.81_{\pm0.43  }$      &$\rm      6.26_{\pm0.58   }$      &$\rm      9.44_{\pm0.55  }$\\
   &219.479   &    -4.813   &   283.781  &    38.005  &$\rm     --$      &$\rm      --$      &$\rm      0.82_{\pm0.24   }$      &$\rm      2.06_{\pm0.39  }$\\
   &219.584   &    -4.704  &   271.154  &    45.955  &$\rm       3.18_{\pm0.48   }$      &$\rm      3.05_{\pm0.63  }$      &$\rm      7.04_{\pm0.63   }$      &$\rm     10.08_{\pm0.77  }$\\
   &219.623   &    -4.615  &   259.539  &    53.333  &$\rm       4.81_{\pm0.98   }$      &$\rm      5.55_{\pm0.30  }$      &$\rm      8.12_{\pm0.56   }$      &$\rm     12.94_{\pm0.94  }$\\
   &219.642   &    -4.703   &   271.148  &    46.014  &$\rm      --$      &$\rm     --$      &$\rm      1.90_{\pm0.35   }$      &$\rm      3.39_{\pm0.42  }$\\
   &219.696   &    -4.540  &   248.860  &    60.113  &$\rm       6.14_{\pm0.79   }$      &$\rm      6.47_{\pm1.07  }$      &$\rm      9.24_{\pm0.43   }$      &$\rm     14.74_{\pm0.62  }$\\
   &219.705   &    -4.262  &   200.469  &    90.462  &$\rm      28.88_{\pm5.46   }$      &$\rm     12.16_{\pm0.89  }$      &$\rm      9.88_{\pm0.34   }$      &$\rm     18.45_{\pm0.46  }$\\
   &219.764   &    -4.424  &   230.499  &    71.907  &$\rm       6.03_{\pm1.97   }$      &$\rm      6.18_{\pm0.49  }$      &$\rm      6.27_{\pm0.40   }$      &$\rm     11.04_{\pm0.34  }$\\
   &219.822   &    -4.477  &   239.119  &    66.281  &$\rm       3.61_{\pm1.27   }$      &$\rm      5.22_{\pm0.64  }$      &$\rm      9.10_{\pm0.86   }$      &$\rm     14.66_{\pm1.09  }$\\
   &219.827   &    -4.613  &   259.457  &    53.422  &$\rm       1.80_{\pm0.36   }$      &$\rm      1.50_{\pm0.45  }$      &$\rm      2.70_{\pm0.71   }$      &$\rm      5.69_{\pm0.92  }$\\
   &220.030   &    -4.423  &   230.319  &    71.871  &$\rm       5.86_{\pm0.61   }$      &$\rm      8.35_{\pm0.70  }$      &$\rm     12.28_{\pm1.32   }$      &$\rm     20.35_{\pm1.32  }$\\
   &220.043   &    -4.537  &   248.708  &    60.241  &$\rm       1.41_{\pm0.56   }$      &$\rm      2.99_{\pm0.93  }$      &$\rm      2.73_{\pm0.00   }$      &$\rm      5.84_{\pm0.81  }$\\
   &220.258   &    -4.376  &   222.578  &    76.973  &$\rm       7.52_{\pm0.88   }$      &$\rm      6.85_{\pm0.70  }$      &$\rm     10.65_{\pm0.67   }$      &$\rm     16.40_{\pm1.02  }$\\
   &220.307   &    -4.473  &   238.906  &    66.458  &$\rm       3.04_{\pm0.66   }$      &$\rm      4.45_{\pm0.47  }$      &$\rm      4.91_{\pm0.35   }$      &$\rm      9.25_{\pm0.40  }$\\
   &220.368   &    -4.376  &   222.466  &    76.856  &$\rm      12.33_{\pm3.46   }$      &$\rm     15.04_{\pm2.53  }$      &$\rm     19.00_{\pm1.36   }$      &$\rm     32.09_{\pm1.55  }$\\
   &220.914   &    -4.335  &   215.570  &    81.244  &$\rm       7.70_{\pm1.43   }$      &$\rm     11.26_{\pm0.80  }$      &$\rm     13.42_{\pm1.22   }$      &$\rm     21.02_{\pm1.34  }$\\
   &220.946   &    -4.335  &   215.571  &    81.239  &$\rm       9.18_{\pm1.53   }$      &$\rm     11.30_{\pm1.40  }$      &$\rm     18.88_{\pm0.53   }$      &$\rm     31.07_{\pm0.91  }$\\
   &220.985   &    -4.334  &   215.616  &    81.434  &$\rm      11.98_{\pm4.04   }$      &$\rm     15.49_{\pm5.21  }$      &$\rm     14.17_{\pm2.53   }$      &$\rm     27.29_{\pm4.45  }$\\
   &221.111   &    -4.371   &   222.159  &    77.098  &$\rm     --$      &$\rm     --$      &$\rm     12.95_{\pm2.31   }$      &$\rm     19.97_{\pm1.96  }$\\
   &230.879   &    -4.218  &   201.531  &    91.238  &$\rm       4.79_{\pm1.00   }$      &$\rm      8.03_{\pm1.53  }$      &$\rm     11.76_{\pm0.68   }$      &$\rm     22.34_{\pm0.71  }$\\

%IV(b). Laboratory parameters and integrated intensities $\rm \int T_B(\upsilon)d\upsilon$ of $\rm HCOOCH_3~(\nu=0)^*$]\\

   \hline
      \multirow{27}*{$\rm HCOOCH_3~(\nu=0)$}
   &219.417   &    -5.168  &   195.355  &    10.989  &$\rm       --$      &$\rm     --$      &$\rm      1.57 _{\pm0.30   }$      &$\rm      3.66 _{\pm0.38  }$          
    &\multirow{21}*{\it 1, 2, 3, 4}\\
   &219.484   &    -5.168  &   195.354  &    10.980  &$\rm       --$      &$\rm      1.32 _{\pm0.35  }$      &$\rm      2.00 _{\pm0.44   }$      &$\rm      3.32 _{\pm0.43  }$\\
%   219.579   &    -5.053   &   197.441  s&    14.433  &$\rm      --$      &$\rm      2.10 _{\pm0.86  }$      &$\rm      --$      &$\rm      7.77 _{\pm0.00  }$\\
   &219.592   &    -5.055  &   197.439  &    14.378  &$\rm       0.90_{\pm0.39   }$      &$\rm      1.32 _{\pm0.32  }$      &$\rm      2.77 _{\pm0.85   }$      &$\rm      5.16 _{\pm1.00  }$\\
%   219.600   &    -5.063   &   221.999  s&    15.882  &$\rm      5.08 _{\pm4.63   }$      &$\rm     --$      &$\rm      3.00 _{\pm0.48   }$      &$\rm      3.17 _{\pm0.71  }$\\
   &220.167   &    -3.999  &    64.350  &    85.865  &$\rm      38.44_{\pm2.34   }$      &$\rm     36.01 _{\pm2.02  }$      &$\rm     57.74 _{\pm1.44   }$      &$\rm     78.96 _{\pm1.40  }$\\
   &220.190   &    -3.999  &    64.344  &    85.884  &$\rm      24.45_{\pm1.42   }$      &$\rm     35.99 _{\pm1.39  }$      &$\rm     60.42 _{\pm2.36   }$      &$\rm     78.54 _{\pm2.59  }$\\
   &220.812   &    -4.986  &    66.218  &     8.871  &$\rm       5.91_{\pm0.00   }$      &$\rm      6.77 _{\pm0.98  }$      &$\rm     10.22 _{\pm0.36   }$      &$\rm     15.57 _{\pm0.85  }$\\
   &220.815   &    -4.987  &    66.210  &     8.869  &$\rm       5.24_{\pm0.00   }$      &$\rm      4.09 _{\pm0.93  }$      &$\rm      8.71 _{\pm0.33   }$      &$\rm     18.34 _{\pm0.93  }$\\
   &220.926   &    -4.867  &   180.745  &    20.185  &$\rm       4.25_{\pm0.75   }$      &$\rm      7.14 _{\pm0.54  }$      &$\rm      9.34 _{\pm1.30   }$      &$\rm     14.46 _{\pm1.19  }$\\
   &220.935   &    -4.867  &   180.739  &    20.183  &$\rm       5.12_{\pm0.93   }$      &$\rm      2.69 _{\pm0.98  }$      &$\rm      4.89 _{\pm0.44   }$      &$\rm      9.68 _{\pm0.51  }$\\
   &220.978   &    -4.675  &   166.459  &    29.375  &$\rm       --$      &$\rm      8.52 _{\pm5.18  }$      &$\rm     11.64 _{\pm1.66   }$      &$\rm     20.14 _{\pm2.81  }$\\
   &220.998   &    -4.674  &   166.446  &    29.374  &$\rm       4.31_{\pm1.48   }$      &$\rm      6.80 _{\pm1.04  }$      &$\rm     11.86 _{\pm0.98   }$      &$\rm     17.70 _{\pm0.82  }$\\
   &221.048   &    -4.535  &   153.092  &    37.964  &$\rm      15.11_{\pm1.64   }$      &$\rm     30.75 _{\pm0.98  }$      &$\rm     45.62 _{\pm0.93   }$      &$\rm     73.34 _{\pm1.07  }$\\
   &221.067   &    -4.535  &   153.081  &    37.964  &$\rm       2.59_{\pm0.61   }$      &$\rm      9.24 _{\pm1.28  }$      &$\rm     11.91 _{\pm0.48   }$      &$\rm     21.32 _{\pm0.47  }$\\
   &221.086   &    -5.051  &   209.447  &    15.138  &$\rm       1.75_{\pm0.61   }$      &$\rm      1.43 _{\pm0.35  }$      &$\rm      3.45 _{\pm1.09   }$      &$\rm      5.03 _{\pm0.81  }$\\
   &221.140   &    -4.426  &   140.647  &    45.951  &$\rm      11.28_{\pm2.24   }$      &$\rm     37.37 _{\pm2.28  }$      &$\rm     50.30 _{\pm0.70   }$      &$\rm     88.58 _{\pm0.99  }$\\
   &229.224   &    -5.027  &   143.199  &    10.854  &$\rm       --$      &$\rm      2.25 _{\pm0.48  }$      &$\rm      2.71 _{\pm0.60   }$      &$\rm      5.09 _{\pm0.53  }$\\
   &229.259   &    -5.092  &   143.201  &     9.338  &$\rm       --$      &$\rm     --$      &$\rm      --$      &$\rm      3.27 _{\pm0.90  }$\\
%   229.320   &    -5.092   &   143.192  s&     9.344  &$\rm      1.25 _{\pm0.58   }$      &$\rm      6.92 _{\pm0.34  }$      &$\rm     11.34 _{\pm0.61   }$      &$\rm     13.46 _{\pm1.09  }$\\
   &229.389   &    -5.026  &   143.193  &    10.851  &$\rm       8.04_{\pm0.00   }$      &$\rm      4.31 _{\pm0.80  }$      &$\rm      2.99 _{\pm0.27   }$      &$\rm      6.82 _{\pm0.46  }$\\
   &229.405   &    -3.943  &    69.316  &    92.251  &$\rm      20.44_{\pm1.05   }$      &$\rm     37.31 _{\pm0.80  }$      &$\rm     47.60 _{\pm1.09   }$      &$\rm     72.27 _{\pm2.26  }$\\
   &229.420   &    -3.943  &    69.310  &    92.257  &$\rm      25.89_{\pm1.12   }$      &$\rm     37.54 _{\pm1.04  }$      &$\rm     47.68 _{\pm1.72   }$      &$\rm     73.39 _{\pm1.86  }$\\
%   229.474   &    -5.103   &    85.667  s&     6.903  &$\rm      7.20 _{\pm0.70   }$      &$\rm     14.72 _{\pm1.28  }$      &$\rm     11.30 _{\pm0.60   }$      &$\rm     17.01 _{\pm0.52  }$\\
%   229.505   &    -5.103   &    85.662  s&     6.903  &$\rm     55.84 _{\pm5.44   }$      &$\rm     10.00 _{\pm0.00  }$      &$\rm      4.99 _{\pm0.92   }$      &$\rm      9.50 _{\pm0.73  }$\\
%   229.539   &    -5.470   &   207.454  s&     5.308  &$\rm     19.43 _{\pm1.02   }$      &$\rm      5.16 _{\pm0.40  }$      &$\rm      4.82 _{\pm0.49   }$      &$\rm      7.59 _{\pm0.45  }$\\
   &230.377   &    -5.031  &   133.656  &    10.167  &$\rm       --$      &$\rm      1.52 _{\pm0.56  }$      &$\rm      4.07 _{\pm0.47   }$      &$\rm      7.19 _{\pm0.41  }$\\
   &230.436   &    -5.566  &   118.982  &     2.761  &$\rm       1.40_{\pm0.01   }$      &$\rm     --$      &$\rm      1.79 _{\pm0.05   }$      &$\rm      3.20 _{\pm0.01  }$\\
   &230.441   &    -5.566  &   118.983  &     2.761  &$\rm       2.90_{\pm0.81   }$      &$\rm      2.37 _{\pm0.16  }$      &$\rm      1.99 _{\pm0.43   }$      &$\rm      4.53 _{\pm0.19  }$\\
   &231.019   &    -5.268  &    31.810  &     3.599  &$\rm       1.69_{\pm0.73   }$      &$\rm      5.69 _{\pm0.83  }$      &$\rm      8.45 _{\pm0.51   }$      &$\rm     14.60 _{\pm0.91  }$\\
   &231.046   &    -5.268  &    31.801  &     3.599  &$\rm       --$      &$\rm     --$      &$\rm      4.90 _{\pm0.78   }$      &$\rm     13.58 _{\pm1.24  }$\\
   &231.188   &    -5.043  &   124.543  &     9.415  &$\rm       2.81_{\pm0.83   }$      &$\rm     --$      &$\rm      4.06 _{\pm0.84   }$      &$\rm      5.91 _{\pm0.76  }$\\
   &231.199   &    -5.039  &   124.546  &     9.492  &$\rm       6.55_{\pm1.13   }$      &$\rm      5.42 _{\pm0.96  }$      &$\rm     11.01 _{\pm0.48   }$      &$\rm     21.05 _{\pm0.96  }$\\

\hline
%\hline
%(to be continued)\\

\end{tabu}
}
\end{center}
\end{table}

\end{landscape}

%%%%%%%%%%%%%%%%%%%%

%%%%%%%%%%%%%%%%%%%%%%%%%%%HCOOCH3

\begin{landscape}
\begin{table}
\ContinuedFloat
\caption{(continued)}
\small
\begin{center}

\begin{tabu}{cc| p{1cm} p{1cm} p{1cm} |p{2cm} p{2cm} p{2cm} p{2cm}|p{2cm}}\hline\hline

Mol.    &Freq      &$\rm I(300K)$    &$\rm E_{L}$    &$S_{ul}\mu^2$         &$\rm HC$                      & $\rm mm2$                     & $\rm mm3a$                 &$\rm mm3b$                       &Notes\\
          &(GHz)      &$\rm Log_{10}$    &$\rm (cm^{-1})$     &$\rm (D^2)$                            &$\rm (K~km~s^{-1})$     &$\rm (K~km~s^{-1})$      &$\rm (K~km~s^{-1})$    &$\rm (K~km~s^{-1})$           \\
\hline
%V. Laboratory parameters and integrated intensities $\rm \int T_B(\upsilon)d\upsilon$ of $\rm CH_3OH$]\\
\multirow{9}*{$\rm  CH_3OH$}
   &219.984   &    -4.527  &   550.204  &    12.441  &$\rm      23.44_{\pm0.53   }$      &$\rm      5.96 _{\pm0.96  }$      &$\rm      9.44 _{\pm6.00   }$      &$\rm      6.13 _{\pm1.28  }$
    &\multirow{9}*{\it  3, 5, 6}\\
   &219.994   &    -4.591  &   531.938  &     9.844  &$\rm      18.37_{\pm1.44   }$      &$\rm      8.99 _{\pm3.00  }$      &$\rm      6.88 _{\pm0.57   }$      &$\rm     10.65 _{\pm2.33  }$\\
   &220.078   &    -3.892  &    59.809  &     5.111  &$\rm     175.00_{\pm9.00   }$      &$\rm    163.50_{\pm16.00  }$      &$\rm    285.56_{\pm23.00   }$      &$\rm    300.80 _{\pm6.77  }$\\
   &229.589   &    -4.132  &   252.591  &     6.806  &$\rm     106.85_{\pm2.03   }$      &$\rm     56.47 _{\pm0.95  }$      &$\rm     94.80_{\pm16.00   }$      &$\rm    114.36 _{\pm3.63  }$\\
   &229.759   &    -3.678  &    54.266  &     7.483  &$\rm     193.14_{\pm2.06   }$      &$\rm    209.17 _{\pm2.45  }$      &$\rm    326.94 _{\pm1.27   }$      &$\rm    359.76 _{\pm1.47  }$\\
   &229.939   &    -4.332  &   394.477  &     8.463  &$\rm      38.47_{\pm1.03   }$      &$\rm     14.97 _{\pm0.68  }$      &$\rm     27.50 _{\pm1.10   }$      &$\rm     28.46 _{\pm0.71  }$\\
   &230.027   &    -4.442  &    20.009  &     1.089  &$\rm     106.32_{\pm1.44   }$      &$\rm    100.37 _{\pm1.57  }$      &$\rm    162.28 _{\pm1.84   }$      &$\rm    179.02 _{\pm2.06  }$\\
   &230.293   &    -5.130  &   415.980  &     1.487  &$\rm       7.48_{\pm1.02   }$      &$\rm      6.23 _{\pm1.70  }$      &$\rm      8.19 _{\pm1.58   }$      &$\rm     14.36 _{\pm1.60  }$\\
   &230.368   &    -4.418  &   466.848  &     9.772  &$\rm      29.52_{\pm1.24   }$      &$\rm     12.79 _{\pm1.32  }$      &$\rm     12.78 _{\pm1.02   }$      &$\rm     16.73 _{\pm1.69  }$\\

\hline
%III. Laboratory parameters and integrated intensities $\rm \int T_B(\upsilon)d\upsilon$ of $\rm ^{34}SO_2^*$]\\
\multirow{6}*{$\rm ^{34}SO_2$}
   &219.355   &    -3.036  &    34.438  &    20.782  &$\rm     192.87_{\pm6.47   }$      &$\rm    177.38_{\pm3.74  }$      &$\rm     78.92_{\pm1.48   }$      &$\rm    103.15_{\pm2.72  }$
      &\multirow{6}*{\it 5, 7, 8,  9}\\
   &221.115   &    -3.089  &   164.876  &    33.785  &$\rm      77.29_{\pm6.98   }$      &$\rm     28.80_{\pm4.13  }$      &$\rm     27.83_{\pm1.59   }$      &$\rm     23.11_{\pm2.11  }$\\
   &229.858   &    -3.595  &     5.311  &     4.548  &$\rm     145.14_{\pm3.13   }$      &$\rm     79.54_{\pm2.07  }$      &$\rm     32.57_{\pm2.65   }$      &$\rm     30.12_{\pm3.15  }$\\
   &229.888   &    -4.951  &   104.865  &     0.322  &$\rm      10.12_{\pm0.75   }$      &$\rm      \le0.91$      &$\rm      \le 0.51$      &$\rm      \le0.30$\\
   &229.987   &    -4.310  &   631.573  &    17.640  &$\rm       4.01_{\pm0.85   }$      &$\rm      1.78_{\pm0.44  }$      &$\rm      \le0.75$      &$\rm      \le1.20$\\
   &230.933   &    -4.466  &    28.220  &     0.676  &$\rm       7.63_{\pm1.72   }$      &$\rm     19.26_{\pm1.63  }$      &$\rm      6.97_{\pm1.02   }$      &$\rm      8.70_{\pm1.75  }$\\

\hline
%II. Laboratory parameters and integrated intensities $\rm \int T_B(\upsilon)d\upsilon$ of $\rm HNCO$]\\
\multirow{6}*{$\rm HNCO$}
   &218.981   &    -2.677  &    62.949  &    70.596  &$\rm     145.59_{\pm1.40   }$      &$\rm     49.86 _{\pm1.43  }$      &$\rm     42.52 _{\pm1.70   }$      &$\rm     47.57 _{\pm3.08  }$
   &\multirow{6}*{\it 5, 8, 10, 11}\\
   &219.657   &    -3.232  &   293.597  &    59.063  &$\rm      45.60_{\pm1.57   }$      &$\rm      5.91 _{\pm2.20  }$      &$\rm      4.01 _{\pm1.46   }~^*$      &$\rm      7.13 _{\pm1.24  }~^*$\\
   &219.734   &    -2.887  &   151.337  &    66.112  &$\rm     155.14_{\pm2.02   }$      &$\rm     26.39 _{\pm0.78  }$      &$\rm     13.81 _{\pm0.45   }$      &$\rm     15.35 _{\pm0.42  }$\\
   &219.737   &    -2.887  &   151.337  &    66.110  &$\rm      44.49_{\pm1.89   }$      &$\rm     15.39 _{\pm0.91  }$      &$\rm     13.81 _{\pm0.45   }$      &$\rm     15.35 _{\pm0.42  }$\\
   &219.798   &    -2.602  &    32.994  &    72.127  &$\rm     263.86_{\pm3.43   }$      &$\rm    120.56 _{\pm4.15  }$      &$\rm    105.66 _{\pm1.53   }$      &$\rm    131.98 _{\pm2.02  }$\\
   &220.585   &    -2.671  &    63.190  &    70.599  &$\rm     141.92_{\pm4.26   }$      &$\rm     35.20 _{\pm4.03  }$      &$\rm     29.29 _{\pm2.42   }$      &$\rm     40.71 _{\pm1.98  }$\\

\hline
%VI. Laboratory parameters and integrated intensities $\rm \int T_B(\upsilon)d\upsilon$ of $\rm CH_3OCH_3$]\\
\multirow{6}*{$\rm CH_3OCH_3$}
   &220.848   &    -4.939  &   199.405  &    21.418  &$\rm       4.77_{\pm1.13   }$      &$\rm      5.27 _{\pm1.55  }$      &$\rm      7.41 _{\pm0.97   }$      &$\rm     10.65 _{\pm1.63  }$
      &\multirow{6}*{\it 2, 8, 12}\\
   &220.893   &    -4.372  &   183.375  &    73.169  &$\rm      31.48_{\pm2.58   }$      &$\rm     22.98 _{\pm2.95  }$      &$\rm     36.87 _{\pm2.65   }$      &$\rm     57.36 _{\pm4.60  }$\\
   &230.141   &    -4.374  &   214.188  &    77.819  &$\rm      11.12_{\pm1.27   }$      &$\rm     14.20 _{\pm1.01  }$      &$\rm     21.92 _{\pm0.64   }$      &$\rm     27.49 _{\pm0.85  }$\\
   &230.234   &    -4.675  &    94.945  &    21.945  &$\rm      37.39_{\pm3.46   }$      &$\rm     29.69 _{\pm1.98  }$      &$\rm     38.29 _{\pm2.58   }$      &$\rm     52.30 _{\pm2.45  }$\\
   &230.476   &    -5.794  &    89.617  &     1.622  &$\rm       5.79_{\pm1.80   }$      &$\rm      3.75 _{\pm0.88  }$      &$\rm      3.88 _{\pm0.82   }$      &$\rm      7.47 _{\pm0.95  }$\\

\hline
%VII. Laboratory parameters and integrated intensities $\rm \int T_B(\upsilon)d\upsilon$ of $\rm CH_3CH_2OH$]\\
\multirow{5}*{$\rm CH_3CH_2OH$}
   &220.155 $^{t}$   &    -4.109  &   175.464  &    25.923      &$\rm      2.91_{\pm0.69  }$      &$\rm      \le0.75$      &$\rm      1.66_{\pm0.36  }$      &$\rm      1.15_{\pm0.30   }$
         &\multirow{6}*{\it 2, 3, 7, 13, 14}\\
   &230.673 $^{g}$    &    -3.995  &    88.651  &    20.283      &$\rm      3.09_{\pm0.71  }$      &$\rm      2.23_{\pm1.45  }$      &$\rm      5.06_{\pm0.74  }$      &$\rm      4.41_{\pm0.81   }$\\
   &230.794 $^{g}$    &    -4.508  &    65.142  &     5.557      &$\rm      6.23_{\pm1.71  }$      &$\rm      \le0.75$      &$\rm      6.93_{\pm0.82  }$      &$\rm      5.33_{\pm1.36   }$\\
   &230.954 $^{t}$    &    -4.098  &    93.618  &    16.339      &$\rm      5.74_{\pm1.69  }$      &$\rm      3.36_{\pm0.98  }$      &$\rm     11.65_{\pm1.27  }$      &$\rm      6.88_{\pm1.35   }$\\
   &230.991 $^{t}$    &    -4.081  &    51.739  &    13.896      &$\rm      5.63_{\pm1.24  }$      &$\rm      5.02_{\pm0.81  }$      &$\rm     12.93_{\pm0.86  }$      &$\rm      9.16_{\pm1.15   }$\\

 \hline
\multicolumn{7}{l}{{\bf Notes:} {\it 1.}   None detection is marked with ``$--$". }\\
\multicolumn{7}{l}{~~~~~~~~~~~~{\it 2.}  Integrated intensities of multiplet (blending) are obtained by Gildas ``HFS" fittings. }\\
\multicolumn{7}{l}{~~~~~~~~~~~~{\it 3.}   Laboratory parameters are given from JPL. }\\
\multicolumn{7}{l}{~~~~~~~~~~~~{\it 4.}  Q(300K, {\scriptsize{$\rm HCOOCH_3$}})=199602.70 } \\

\multicolumn{7}{l}{~~~~~~~~~~~~{\it 5.} Integrated intensities are  obtained by Gildas ``Gaussian" fittings.}\\
\multicolumn{7}{l}{~~~~~~~~~~~~{\it 6.} Q(300K, {\scriptsize{$\rm CH_3OH$}})=9473.1198 }\\
          
\multicolumn{7}{l}{~~~~~~~~~~~~{\it 7.}  None detection is given by an upper limit of $\rm 3\sigma$ rms.}\\
\multicolumn{7}{l}{~~~~~~~~~~~~{\it 8.}  Laboratory parameters are given from CDMS.}\\
\multicolumn{7}{l}{~~~~~~~~~~~~{\it 9.}  Q(300K, {\scriptsize{$\rm ^{34}SO_2$}})=6020.754 }\\

\multicolumn{7}{l}{~~~~~~~~~~~~{\it 10.}  Lines marked with ``*" are blended with $\rm HCOOCH_3$ lines.}\\
\multicolumn{7}{l}{~~~~~~~~~~~~{\it 11.}  Q(300K, {\scriptsize{$\rm HNCO$}})=7785.741 }\\

\multicolumn{7}{l}{~~~~~~~~~~~~{\it 12.} Q(300K, {\scriptsize{$\rm CH_3OCH_3$}})=228016}  \\
\multicolumn{7}{l}{~~~~~~~~~~~~{\it 13.} Lines marked with ``{\it t}" are trans-$\rm CH_3CH_2OH$, with ``{\it g}" are gauche-$\rm CH_3CH_2OH$.}  \\
\multicolumn{7}{l}{~~~~~~~~~~~~{\it 14.} Q(300K, {\scriptsize{$\rm CH_3CH_2OH$}})=45583.66}  
\end{tabu}

\end{center}

\end{table}
\end{landscape}

%\end{landscape}
%%%%%%%%%%%%%%%%%%%%%%%%intensity
%%%%%%%%%%%%%%%%%%%%%%%%intensity
%%%%%%%%%%%%%%%%%%%%%%%%intensity
\newpage

\begin{landscape}
\begin{table}
\small
\begin{center}
Integrated intensity $\rm \int T_B(\upsilon)d\upsilon$ ($\rm K~km\,s^{-1}$) \\
\begin{tabular}{c| p{1.6cm} |p{2cm}  p{2cm} p{2cm} p{2cm} p{2cm} p{2cm} p{2cm} p{2cm}}\hline\hline

Species   &Freq (GHz)   & $\rm HC$      & $\rm mm2$      & $\rm mm3a$    &$\rm mm3b$     
       & $\rm NE$           & $\rm SR$      & $\rm OF1N$      & $\rm OF1S$\\
\hline
$\rm CO~~~^\S$     & 230.538     &$\rm 3204.00_{\pm40.97}$     &$\rm 3411.50_{\pm8.27}$     &$\rm 1777.80_{\pm17.41}$     &$\rm 2346.00_{\pm15.90}$     &$\rm 3553.00_{\pm1.13}$     &$\rm 853.91_{\pm5.08}$     &$\rm 2062.80_{\pm10.77}$     &$\rm 1028.70_{\pm12.43}$\\
$\rm ^{13}CO~~~^\S$     & 220.399     &$\rm 828.40_{\pm7.77}$     &$\rm 702.33_{\pm9.91}$     &$\rm 502.14_{\pm6.32}$     &$\rm 568.77_{\pm9.99}$     &$\rm 437.89_{\pm9.19}$     &$\rm 134.30_{\pm1.22}$     &$\rm 197.20_{\pm2.72}$     &$\rm 129.92_{\pm0.56}$\\
$\rm C^{18}O~~~^\S$     & 219.560     &$\rm 126.00_{\pm4.23}$     &$\rm 108.00_{\pm4.61}$     &$\rm 79.10_{\pm2.41}$     &$\rm 83.60_{\pm4.07}$     &$\rm 79.00_{\pm1.55}$     &$\rm 24.40_{\pm0.34}$     &$\rm 26.20_{\pm0.52}$     &$\rm 13.90_{\pm0.47}$\\
$\rm H_2^{13}CO$     & 219.909     &$\rm 53.93_{\pm2.30}$     &$\rm 34.70_{\pm2.01}$     &$\rm 43.23_{\pm0.80}$     &$\rm 36.86_{\pm1.61}$     &$\rm 2.53_{\pm0.40}$     &$\rm 11.63_{\pm0.40}$     &$\rm 5.18_{\pm0.39}$     &$\rm 14.31_{\pm0.26}$\\
$\rm CH_2CO$     & 220.178     &$\rm 14.44_{\pm2.31}$     &$\rm 9.02_{\pm2.39}$     &$\rm 23.11_{\pm2.52}$     &$\rm 17.00_{\pm4.34}$     &$\rm 1.27_{\pm0.26}$     &$\rm 5.58_{\pm0.30}$     &$\rm 1.32_{\pm0.20}$     &$\rm 2.54_{\pm0.26}$\\

\hline
$\rm CH_3OH$     & 229.759     &$\rm 193.14_{\pm2.06}$     &$\rm 209.17_{\pm2.45}$     &$\rm 326.94_{\pm1.27}$     &$\rm 359.76_{\pm1.47}$     &$\rm 6.62_{\pm0.37}$     &$\rm 29.56_{\pm0.77}$     &$\rm 15.08_{\pm0.67}$     &$\rm 46.21_{\pm0.53}$\\
$\rm CH_3OH$     & 229.939     &$\rm 38.47_{\pm1.03}$     &$\rm 14.97_{\pm0.68}$     &$\rm 27.50_{\pm1.10}$     &$\rm 28.46_{\pm0.71}$     &$\le0.25$     &$\le0.32$     &$\le0.17$     &$\le0.28$\\
$\rm HCOOCH_3$     & 220.258     &$\rm 7.52_{\pm0.88}$     &$\rm 6.85_{\pm0.70}$     &$\rm 10.65_{\pm0.67}$     &$\rm 16.40_{\pm1.02}$     &$\le0.24$     &$\le0.34$     &$\le0.28$     &$\le0.29$\\
$\rm HCOOCH_3$     & 220.167     &$\rm 38.44_{\pm2.34}$     &$\rm 36.01_{\pm2.02}$     &$\rm 57.74_{\pm1.44}$     &$\rm 78.96_{\pm1.40}$     &$\le43.22$     &$\rm 3.97_{\pm0.30}$     &$\le0.34$     &$\le0.04$\\
$\rm CH_3OCH_3$     & 230.141     &$\rm 11.12_{\pm1.27}$     &$\rm 14.20_{\pm1.01}$     &$\rm 21.92_{\pm0.64}$     &$\rm 21.49_{\pm0.85}$     &$\le0.21$     &$\le0.28$     &$\le0.21$     &$\le0.25$\\

$\rm CH_3COCH_3$     & 230.183     &$\rm 6.72_{\pm0.86}$     &$\rm 4.12_{\pm0.51}$     &$\rm 1.20_{\pm0.40}$     &$\rm 1.70_{\pm0.49}$     &$\le0.15$     &$\le0.15$     &$\le0.16$     &$\le0.10$\\
$\rm CH_3CH_2OH$     & 230.991     &$\rm 5.63_{\pm1.24}$     &$\rm 5.02_{\pm0.81}$     &$\rm 12.93_{\pm0.86}$     &$\rm 9.16_{\pm1.15}$     &$\le0.47$     &$\le0.66$     &$\le0.51$     &$\le0.38$\\
$\rm CH_3CHO~~~^*$     & 220.446     &$\rm 3.50_{\pm0.68}$     &$\rm 4.33_{\pm0.72}$     &$\rm 6.93_{\pm0.53}$     &$\rm 7.82_{\pm0.77}$     &$\le0.37$     &$\rm 2.79_{\pm0.34}$     &$\le0.14$     &$\rm 1.24_{\pm0.22}$\\
$\rm C_6H~~~^*$     & 230.109     &$\rm 10.46_{\pm0.92}$     &$\rm 2.08_{\pm0.80}$     &$\le0.22$     &$\le0.44$     &$\le0.92$     &$\le0.34$     &$\le0.60$     &$\le0.08$\\
\hline

$\rm HNCO$     & 219.798     &$\rm 263.86_{\pm3.43}$     &$\rm 120.56_{\pm4.15}$     &$\rm 105.66_{\pm1.53}$     &$\rm 131.98_{\pm2.02}$     &$\rm 22.61_{\pm1.08}$     &$\rm 1.68_{\pm0.28}$     &$\rm 9.65_{\pm0.28}$     &$\rm 13.94_{\pm0.40}$\\
$\rm HC_3N(v_7=1)$     & 219.174     &$\rm 227.01_{\pm2.74}$     &$\rm 70.10_{\pm2.64}$     &$\rm 27.10_{\pm2.77}$     &$\rm 19.87_{\pm2.36}$     &$\le0.51$     &$\le0.48$     &$\le0.38$     &$\le0.61$\\
$\rm CH_3CN$     & 220.730     &$\rm 397.54_{\pm9.19}$     &$\rm 281.55_{\pm10.27}$     &$\rm 256.41_{\pm27.26}$     &$\rm 331.98_{\pm21.20}$     &$\rm 15.17_{\pm1.00}$     &$\rm 8.98_{\pm2.42}$     &$\rm 5.32_{\pm2.84}$     &$\rm 5.27_{\pm2.19}$\\
$\rm CH_3^{13}CN$     & 220.621     &$\rm 96.44_{\pm4.41}$     &$\rm 40.71_{\pm3.12}$     &$\rm 28.38_{\pm2.19}$     &$\rm 37.73_{\pm2.46}$     &$\rm 3.13_{\pm0.50}$     &$\le0.19$     &$\le0.39$     &$\le0.17$\\
$\rm CH_2CHCN$     & 230.739     &$\rm 48.99_{\pm0.90}$     &$\rm 6.68_{\pm0.96}$     &$\le0.28$     &$\le0.45$     &$\le0.37$     &$\le0.23$     &$\le0.10$     &$\le0.48$\\
$\rm CH_3CH_2CN$     & 229.265     &$\rm 160.64_{\pm1.52}$     &$\rm 60.39_{\pm1.36}$     &$\rm 16.99_{\pm1.18}$     &$\rm 33.60_{\pm2.38}$     &$\le0.30$     &$\le0.33$     &$\le0.36$     &$\le0.31$\\
$\rm HC_7N~~~^*$     & 231.102     &$\rm 4.39_{\pm0.46}$     &$\le0.58$     &$\le0.38$     &$\le0.77$     &$\le0.31$     &$\le0.48$     &$\le0.33$     &$\le0.63$\\

\hline
$\rm OCS$     & 231.061     &$\rm 345.79_{\pm8.67}$     &$\rm 189.26_{\pm6.48}$     &$\rm 194.18_{\pm4.19}$     &$\rm 219.78_{\pm9.42}$     &$\rm 12.25_{\pm0.54}$     &$\rm 9.38_{\pm0.57}$     &$\rm 6.15_{\pm0.53}$     &$\le0.37$\\
$\rm O^{13}CS$     & 230.318     &$\rm 23.84_{\pm0.84}$     &$\rm 19.77_{\pm1.21}$     &$\rm 14.54_{\pm0.91}$     &$\rm 28.06_{\pm1.26}$     &$\le0.38$     &$\le0.47$     &$\le0.14$     &$\le0.19$\\
$\rm SO_2$     & 229.348     &$\rm 295.36_{\pm3.65}$     &$\rm 225.41_{\pm3.96}$     &$\rm 120.92_{\pm3.94}$     &$\rm 166.97_{\pm3.86}$     &$\rm 54.45_{\pm1.15}$     &$\le0.33$     &$\le0.25$     &$\le0.30$\\
$\rm ^{34}SO_2$     & 219.355     &$\rm 192.87_{\pm6.47}$     &$\rm 177.38_{\pm3.74}$     &$\rm 78.92_{\pm1.48}$     &$\rm 103.15_{\pm2.72}$     &$\rm 52.70_{\pm1.24}$     &$\le0.60$     &$\le0.35$     &$\rm \le0.21$\\
$\rm SO$     & 219.949     &$\rm 2281.00_{\pm16.36}$     &$\rm 2244.90_{\pm24.92}$     &$\rm 1500.70_{\pm17.58}$     &$\rm 1981.60_{\pm19.50}$     &$\rm 1748.00_{\pm58.77}$     &$\rm 102.16_{\pm3.93}$     &$\rm 487.06_{\pm2.87}$     &$\rm 70.85_{\pm3.55}$\\
$\rm ^{13}CS$     & 231.221     &$\rm 108.47_{\pm1.87}$     &$\rm 66.13_{\pm1.40}$     &$\rm 73.18_{\pm1.25}$     &$\rm 62.71_{\pm3.86}$     &$\rm 6.23_{\pm0.45}$     &$\rm 15.05_{\pm0.33}$     &$\rm 3.01_{\pm0.35}$     &$\le0.51$\\

 \hline
\end{tabular}
\caption{Intensity integrated over the width of each line, corresponding to the transitions shown in Figure~\ref{velpro}. Most species are measured from Gaussian fittings, and uncertainties (written as the subscript) on the measured intensities are typically $\le10\%$. For those marked with  ``$\S$", integration is from the whole area because they have none-Gaussian profiles caused by strong self-absorption, and uncertainties  are from their difference to Gaussian fitting.  
 For species which are not detected, an upper limit equal to the $\rm 3\sigma$ rms is given. Species with ``*" are tentative detected due to only one unblended line which has not been confirmed previously. \label{tab:lineprofile} }
\end{center}
\end{table}
\end{landscape}

%%%%%%%%%%%%%%%%%%%%%%%%column density
%%%%%%%%%%%%%%%%%%%%%%%%column density
%%%%%%%%%%%%%%%%%%%%%%%%column density

\newpage
\renewcommand{\baselinestretch}{1.0} 
\begin{landscape}
\begin{table}
\caption{Column densities and abundances for O-bearing molecules and carbon chains from different substructures in Orion-KL denoted in Figures~\ref{into} and \ref{COMdis}.}
\label{col-Obearing} 
\small
\begin{center}
%I. Molecular Column density [$\rm x\pm y (z) =(x\pm y) \times 10^z cm^{-2}$]\\

%%%%%%%%%%%%%%%%%%%%!!!!!!!!!!!!!!!!!!!!!!!!!!!!!!!!!!
%%%%%%%%%%%%%%%%%%%%!!!!!!!!!!!!!!!!!!!!!!!!!!!!!!!!!!OOOOOOOOOOOOOOOOOOOOOOO
%%%%%%%%%%%%%%%%%%%%!!!!!!!!!!!!!!!!!!!!!!!!!!!!!!!!!!

\begin{tabu}{c|p{2.3cm} p{2.3cm} p{2.3cm} p{2.3cm} p{1.8cm} p{1.8cm} p{1.8cm} p{1.8cm}}
%  \multicolumn{4}{l}{A1.}\\
\hline\hline
Species      & $\rm HC$      & $\rm mm2$      & $\rm mm3a$    &$\rm mm3b$     
       & $\rm NE$           & $\rm SR$      & $\rm OF1N$      & $\rm OF1S$\\

&($126\pm13$~K)  &($112\pm7$~K)   &($101\pm5$~K)    &($102\pm5$~K)    &($43\pm3$~K)   &($43\pm3$~K)    &($43\pm3$~K)   &($43\pm3$~K)\\
\hline
\multicolumn{9}{c}{I. Column densities (in the form of $\rm x\pm y (z) =(x\pm y) \times 10^z  cm^{-2}$)}    \\
\hline
$\rm C^{18}O$      &$\rm 2.14_{\pm0.13}(17)$            &$\rm 1.67_{\pm0.02}(17)$      &$\rm 1.11_{\pm0.01}(17)$      &$\rm 1.19_{\pm0.01}(17)$      &$\rm 5.90_{\pm0.10}(16)$      &$\rm 1.82_{\pm0.04}(16)$      &$\rm 1.96_{\pm0.03}(16)$      &$\rm 1.04_{\pm0.00}(16)$\\
$\rm CO$      %&$\rm 5.01_{\pm0.38}(18)$           &$\rm 4.85_{\pm0.25}(18)$      &$\rm 2.31_{\pm0.07}(18)$      &$\rm 3.08_{\pm0.10}(18)$      &$\rm 2.47_{\pm0.09}(18)$      &$\rm 5.94_{\pm0.17}(17)$      &$\rm 1.43_{\pm0.04}(18)$      &$\rm 7.15_{\pm0.17}(17)$\\
&~\color{black}$\rm 2.71_{\pm0.21}(20)~~~^\dagger$        &~\color{black}$\rm 2.57_{\pm0.13}(20)~~~^\dagger$ &~\color{black}$\rm 2.70_{\pm0.08}(20)~~~^\dagger$    &~\color{black}$\rm 3.29_{\pm0.11}(20)~~~^\dagger$    &~\color{black}$\rm 1.83_{\pm0.06}(20)~~~^\dagger$    &~\color{black}$\rm 5.34_{\pm0.16}(19)~~~^\dagger$    &~\color{black}$\rm 5.16_{\pm0.15}(19)~~~^\dagger$    &~\color{black}$\rm 1.57_{\pm0.04}(20)~~~^\dagger$\\

$\rm ^{13}CO$     % &$\rm 1.40_{\pm0.11}(18)$            &$\rm 1.08_{\pm0.04}(18)$      &$\rm 7.03_{\pm0.20}(17)$      &$\rm 8.05_{\pm0.19}(17)$      &$\rm 3.25_{\pm0.05}(17)$      &$\rm 9.98_{\pm0.27}(16)$      &$\rm 1.46_{\pm0.03}(17)$      &$\rm 9.65_{\pm0.31}(16)$\\
&~\color{black}$\rm 2.31_{\pm0.19}(18)~~~^\dagger$       &~\color{black}$\rm 1.22_{\pm0.05}(18)~~~^\dagger$ &~\color{black}$\rm 1.63_{\pm0.05}(18)~~~^\dagger$    &~\color{black}$\rm 1.55_{\pm0.04}(18)~~~^\dagger$    &~\color{black}$\rm 7.52_{\pm0.12}(17)~~~^\dagger$    &~\color{black}$\rm 1.93_{\pm0.05}(17)~~~^\dagger$    &~\color{black}$\rm 1.48_{\pm0.03}(17)~~~^\dagger$    &~\color{black}$\rm 1.00_{\pm0.03}(18)~~~^\dagger$\\

$\rm CH_2CO$      &$\rm 8.26_{\pm0.57}(14)$            &$\rm 4.74_{\pm0.97}(14)$      &$\rm 1.12_{\pm0.08}(15)$      &$\rm 8.34_{\pm1.79}(14)$      &$\rm 4.86_{\pm1.11}(13)$      &$\rm 2.14_{\pm0.16}(14)$      &$\rm 5.07_{\pm0.87}(13)$      &$\rm 9.72_{\pm1.23}(13)$\\
$\rm H_2^{13}CO^{(1)}$      &$\rm 9.86_{\pm0.85}(14)$            &$\rm 5.54_{\pm0.13}(14)$      &$\rm 6.05_{\pm0.24}(14)$      &$\rm 5.24_{\pm0.08}(14)$      &$\rm 1.53_{\pm0.17}(13)$      &$\rm 7.04_{\pm0.07}(13)$      &$\rm 3.14_{\pm0.09}(13)$      &$\rm 8.66_{\pm0.22}(13)$\\
%$\rm CH_3CHO~~~^*$      &$\rm 1.75_{\pm0.13}(16)$          &$\rm 1.94_{\pm0.20}(16)$      &$\rm 2.83_{\pm0.11}(16)$      &$\rm 3.23_{\pm0.19}(16)$      &$\rm \le 1.50(15)$      &$\rm 1.14_{\pm0.22}(16)$      &$\rm \le 5.89(14)$      &$\rm 5.03_{\pm1.26}(15)$\\
%change
$\rm CH_3CHO~~~^*$       &$\rm 6.99_{\pm0.51}(15)$           &$\rm 7.74_{\pm0.79}(15)$       &$\rm 1.13_{\pm0.04}(16)$ &$\rm 1.29_{\pm0.08}(16)$  &$\rm \le 5.99(14)$          &$\rm 4.54_{\pm0.87}(15)$      &$\rm \le 2.35(14)$      &$\rm 2.01_{\pm0.50}(15)$\\
$\rm C_6H~~~^*$      &$\rm 9.74_{\pm4.88}(14)$          &$\rm 2.71_{\pm1.96}(14)$      &$\rm \le 4.26(13)$      &$\rm \le 7.89(13)$      &$\rm \le 4.46(16)$      &$\rm \le 1.66(16)$      &$\rm \le 2.92(16)$      &$\rm \le 4.09(15)$\\
$\rm CH_3COCH_3$      &$\rm 5.43_{\pm0.82}(15)$          &$\rm 2.48_{\pm0.10}(15)$      &$\rm 5.61_{\pm1.14}(14)$      &$\rm 8.16_{\pm1.30}(14)$      &$\rm \le 9.47(13)$      &$\rm \le 9.72(13)$      &$\rm \le 1.02(14)$      &$\rm \le 6.20(13)$\\
$\rm  HCOOCH_3$      &$\rm  \bf   2.69\pm1.09(16)$           &$\rm  \bf   3.55\pm0.96(16)$      &$\rm  \bf   5.64\pm1.28(16)$      &$\rm  \bf   9.73\pm2.11(16)$      &$\rm \le 3.84(16)$      &$\rm \le 5.56(16)$      &$\rm \le 4.59(16)$      &$\rm \le 4.74(16)$\\
%\hline\hline
%\end{tabu}
%\vspace{1em}

%\begin{tabu}{c|p{1.8cm} p{1.8cm} p{1.8cm} p{1.8cm} p{1.8cm} p{1.8cm} p{1.8cm} p{1.8cm}}
%  \multicolumn{4}{l}{A2.}\\
%\hline\hline
%\cline{2-5}
%Species      & $\rm HC$      & $\rm mm2$      & $\rm mm3a$    &$\rm mm3b$            & $\rm NE$           & $\rm S$      & $\rm OF1N$      & $\rm OF1S$\\
%&  &   &    &    &($43\pm3$~K)   &($43\pm3$~K)    &($43\pm3$~K)   &($43\pm3$~K)\\
$\rm  CH_3OCH_3$      &$\rm  \bf   1.59\pm0.76(17)~~^\sharp$            &$\rm  \bf   1.03\pm0.10(17)~~^\sharp$      &$\rm  \bf   1.19\pm0.29(17)~~^\sharp$      &$\rm  \bf   1.96\pm0.71(17)~~^\sharp$      &$\rm \le 4.13(16)$      &$\rm \le 5.63(16)$      &$\rm \le 4.13(16)$      &$\rm \le 5.04(16)$\\
$\rm CH_3CH_2OH$      &$\rm  \bf   7.56\pm5.29(15)~~^\sharp$          &$\rm  \bf   3.09\pm2.60(15)~~^\sharp$      &$\rm  \bf   7.92\pm7.64(15)~~^\sharp$      &$\rm  \bf   5.74\pm3.71(15)~~^\sharp$      &$\rm \le 2.52(14)$      &$\rm \le 3.48(14)$      &$\rm \le 2.69(14)$      &$\rm \le 2.01(14)$\\

%\hline\hline

%\end{tabu}
%\vspace{1em}

%%%%%%%%%%%%%%%%%%%%!!!!!!!!!!!!!!!!!!!!!!!!!!!!!!!!!!
%%%%%%%%%%%%%%%%%%%%!!!!!!!!!!!!!!!!!!!!!!!!!!!!!!!!!!22222222222222222OOOOOOOOOOOOOOOOOO
%%%%%%%%%%%%%%%%%%%%!!!!!!!!!!!!!!!!!!!!!!!!!!!!!!!!!!

%\begin{tabu}{c| p{1.8cm}p{1.8cm} p{1.8cm} p{1.8cm} p{1.8cm} p{1.8cm} p{1.8cm} p{1.8cm}}
%  \multicolumn{4}{l}{A3.}\\
%\hline\hline
%Species      & $\rm HC$      & $\rm mm2$      & $\rm mm3a$    &$\rm mm3b$            & $\rm NE$           & $\rm S$      & $\rm OF1N$      & $\rm OF1S$\\
%\hline
%&($225\pm19$~K)  &($174\pm14$~K)   &($159\pm12$~K)    &($160\pm17$~K)    &($43\pm3$~K)   &($43\pm3$~K)    &($43\pm3$~K)   &($43\pm3$~K)\\
%\hline
$\rm H_2^{13}CO^{(2)}$      &$\rm 2.09_{\pm0.16}(15)~~~~~^\sharp$           &$\rm 9.53_{\pm0.51}(14)~~~~~^\sharp$      &$\rm 1.06_{\pm0.08}(15)~~~~~^\sharp$      &$\rm 9.13_{\pm0.83}(14)~~~~~^\sharp$      &$\rm 1.53_{\pm0.17}(13)$      &$\rm 7.04_{\pm0.07}(13)$      &$\rm 3.14_{\pm0.09}(13)$      &$\rm 8.66_{\pm0.22}(13)$\\
$\rm CH_3OH$      &$\rm  \bf   2.53\pm0.91(17)~~^\sharp$            &$\rm  \bf   1.64\pm0.66(17)~~^\sharp$      &$\rm  \bf   2.49\pm0.93(17)~~^\sharp$      &$\rm  \bf   2.85\pm1.49(17)~~^\sharp$      &$\rm 1.49_{\pm0.13}(15)$      &$\rm 6.66_{\pm0.36}(15)$      &$\rm 3.40_{\pm0.25}(15)$      &$\rm 1.04_{\pm0.04}(16)$\\

\hline
\multicolumn{9}{c}{II. Abundances (in the form of $\rm x\pm y (z) =(x\pm y) \times 10^z $)}    \\
\hline
$\rm C^{18}O$     &$\rm 1.79_{}(-7)$          &$\rm 1.79_{}(-7)$     &$\rm 1.79_{}(-7)$     &$\rm 1.79_{}(-7)$     &$\rm 1.79_{}(-7)$     &$\rm 1.79_{}(-7)$     &$\rm 1.79_{}(-7)$     &$\rm 1.79_{}(-7)$\\
$\rm CO$   %  &$\rm 4.19_{\pm0.60}(-6)$          &$\rm 5.20_{\pm0.35}(-6)$     &$\rm 3.71_{\pm0.16}(-6)$     &$\rm 4.63_{\pm0.19}(-6)$     &$\rm 7.50_{\pm0.40}(-6)$     &$\rm 5.83_{\pm0.31}(-6)$     &$\rm 1.31_{\pm0.06}(-5)$     &$\rm 1.23_{\pm0.03}(-5)$\\
&~\color{black}$\rm 2.26_{\pm0.32}(-4)~~~^\dagger$       &~\color{black}$\rm 2.76_{\pm0.18}(-4)~~~^\dagger$    &~\color{black}$\rm 4.34_{\pm0.19}(-4)~~~^\dagger$   &~\color{black}$\rm 4.95_{\pm0.20}(-4)~~~^\dagger$    &~\color{black}$\rm 5.55_{\pm0.29}(-4)~~~^\dagger$   & ~\color{black}$\rm 5.25_{\pm0.28}(-4)~~~^\dagger$    &~\color{black}$\rm 4.73_{\pm0.22}(-4)~~~^\dagger$   & ~\color{black}$\rm 2.71_{\pm0.07}(-4)~~~^\dagger$\\

$\rm ^{13}CO$    % &$\rm 1.17_{\pm0.17}(-6)$        &$\rm 1.16_{\pm0.07}(-6)$     &$\rm 1.13_{\pm0.05}(-6)$     &$\rm 1.21_{\pm0.04}(-6)$     &$\rm 9.87_{\pm0.33}(-7)$     &$\rm 9.80_{\pm0.50}(-7)$     &$\rm 1.34_{\pm0.05}(-6)$     &$\rm 1.66_{\pm0.06}(-6)$\\
&~\color{black}$\rm 1.93_{\pm0.28}(-6)~~~^\dagger$     &~\color{black}$\rm 1.31_{\pm0.07}(-6)~~~^\dagger$   &~\color{black}$\rm 2.61_{\pm0.11}(-6)~~~^\dagger$   &~\color{black}$\rm 2.34_{\pm0.07}(-6)~~~^\dagger$   &~\color{black}$\rm 2.28_{\pm0.08}(-6)~~~^\dagger$   &~\color{black}$\rm 1.89_{\pm0.10}(-6)~~~^\dagger$   &~\color{black}$\rm 1.35_{\pm0.05}(-6)~~~^\dagger$   &~\color{black}$\rm 1.73_{\pm0.06}(-6)~~~^\dagger$\\

$\rm CH_2CO$     &$\rm 6.90_{\pm0.93}(-10)$         &$\rm 5.09_{\pm1.13}(-10)$     &$\rm 1.81_{\pm0.15}(-9)$     &$\rm 1.25_{\pm0.28}(-9)$     &$\rm 1.48_{\pm0.37}(-10)$     &$\rm 2.10_{\pm0.21}(-9)$     &$\rm 4.64_{\pm0.89}(-10)$     &$\rm 1.68_{\pm0.22}(-9)$\\
$\rm H_2^{13}CO^{(1)}$     &$\rm 8.24_{\pm1.27}(-10)$        &$\rm 5.94_{\pm0.22}(-10)$     &$\rm 9.73_{\pm0.51}(-10)$     &$\rm 7.89_{\pm0.17}(-10)$     &$\rm 4.64_{\pm0.60}(-11)$     &$\rm 6.92_{\pm0.23}(-10)$     &$\rm 2.87_{\pm0.13}(-10)$     &$\rm 1.49_{\pm0.04}(-9)$\\
%$\rm CH_3CHO~~~^*$     &$\rm 1.46_{\pm0.20}(-8)$         &$\rm 2.08_{\pm0.25}(-8)$     &$\rm 4.55_{\pm0.23}(-8)$     &$\rm 4.86_{\pm0.32}(-8)$     &$\rm \le 4.55(-9)$     &$\rm 1.12_{\pm0.24}(-7)$     &$\rm \le 5.39(-9)$     &$\rm 8.68_{\pm2.20}(-8)$\\
%change
$\rm CH_3CHO~~~^*$    &$\rm 5.84_{\pm0.82}(-9)$         &$\rm 8.31_{\pm0.98}(-9)$      &$\rm 1.82_{\pm0.09}(-8)$   &$\rm 1.94_{\pm0.13}(-8)$   &$\rm \le 1.82(-9)$         &$\rm 4.46_{\pm0.98}(-8)$     &$\rm \le 2.15(-9)$      &$\rm 3.47_{\pm0.88}(-8)$\\

$\rm C_6H~~~^*$     &$\rm 8.14_{\pm4.84}(-10)$        &$\rm 2.91_{\pm2.18}(-10)$     &$\rm \le 6.84(-11)$     &$\rm \le 1.19(-10)$     &$\rm \le 1.35(-7)$     &$\rm \le 1.63(-7)$     &$\rm \le 2.67(-7)$     &$\rm \le 7.05(-8)$\\
$\rm CH_3COCH_3$     &$\rm 4.54_{\pm1.01}(-9)$         &$\rm 2.66_{\pm0.15}(-9)$     &$\rm 9.02_{\pm1.96}(-10)$     &$\rm 1.23_{\pm0.21}(-9)$     &$\rm \le 2.87(-10)$     &$\rm \le 9.55(-10)$     &$\rm \le 9.29(-10)$     &$\rm \le 1.07(-9)$\\

$\rm  HCOOCH_3$     &$\rm  \bf   2.25\pm1.11(-8)$          &$\rm  \bf   3.81\pm1.10(-8)$     &$\rm  \bf   9.06\pm2.19(-8)$     &$\rm  \bf   1.46\pm0.33(-7)$     &$\rm \le 1.16(-7)$     &$\rm \le 5.46(-7)$     &$\rm \le 4.20(-7)$     &$\rm \le 8.18(-7)$\\

%\hline
$\rm  CH_3OCH_3$     &$\rm  \bf   1.33\pm0.76(-7)~~~^\sharp$         &$\rm  \bf   1.10\pm0.12(-7)~~~^\sharp$     &$\rm  \bf   1.91\pm0.50(-7)~~~^\sharp$     &$\rm  \bf   2.94\pm1.09(-7)~~~^\sharp$     &$\rm \le 1.25(-7)$     &$\rm \le 5.53(-7)$     &$\rm \le 3.78(-7)$     &$\rm \le 8.69(-7)$\\
$\rm  CH_3CH_2OH$     &$\rm  \bf   6.32\pm5.09(-9)~~~^\sharp$          &$\rm  \bf   3.31\pm2.88(-9)~~~^\sharp$     &$\rm  \bf   1.27\pm1.26(-8)~~~^\sharp$     &$\rm  \bf   8.63\pm5.67(-9)~~~^\sharp$     &$\rm \le 7.63(-10)$     &$\rm \le 3.42(-9)$     &$\rm \le 2.46(-9)$     &$\rm \le 3.47(-9)$\\

%\hline
$\rm H_2^{13}CO^{(2)}$      &$\rm 1.75_{\pm0.25}(-9)~~~~~^\sharp$     &$\rm 1.02_{\pm0.07}(-9)~~~~~^\sharp$     &$\rm 1.70_{\pm0.15}(-9)~~~~~^\sharp$     &$\rm 1.37_{\pm0.14}(-9)~~~~~^\sharp$     &$\rm 4.64_{\pm0.60}(-11)$     &$\rm 6.92_{\pm0.23}(-10)$     &$\rm 2.87_{\pm0.13}(-10)$     &$\rm 1.49_{\pm0.04}(-9)$\\
$\rm CH_3OH$      &$\rm  \bf   2.12\pm0.94(-7)~~~^\sharp$        &$\rm  \bf   1.76\pm0.74(-7)~~~^\sharp$     &$\rm  \bf   4.00\pm1.57(-7)~~~^\sharp$     &$\rm  \bf   4.29\pm2.29(-7)~~~^\sharp$     &$\rm 4.52_{\pm0.47}(-9)$     &$\rm 6.54_{\pm0.52}(-8)$     &$\rm 3.11_{\pm0.29}(-8)$     &$\rm 1.79_{\pm0.08}(-7)$\\

\hline\hline
\multicolumn{9}{l}{{\bf Notes:} 1. ``*" mark the tentative detected molecules.}\\
\multicolumn{9}{l}{~~~~~~~~~~~ 2. Species except for those marked with ``$\sharp$" are assumed to have the same temperature (listed in the table head)  in the same substructure.}\\
\multicolumn{9}{l}{~~~~~~~~~~~ 3. Values and uncertainties in bold face are obtained directly  from   the rotation diagram fittings in  Figure~\ref{fig:trot_other}.}\\
\multicolumn{9}{l}{~~~~~~~~~~~ 4. For  the species whose transitions are not sufficient to derive the rotation diagrams, their column densities are  obtained from one strongest transition}\\
\multicolumn{9}{l}{~~~~~~~~~~~~~~~ by assuming LTE with the  excitation temperatures list in the table head  (derived from $\rm HCOOCH_3$ for HC, mm2, mm3a, and mm3b).}\\
\multicolumn{9}{l}{~~~~~~~~~~~ 5. Values of $\rm H_2^{13}CO^{(2)}$ are derived from the temperatures the same as $\rm CH_3OH$ list in Table~\ref{tab:trotcom}.}\\
\multicolumn{9}{l}{~~~~~~~~~~~ 6. Values of all species in the outflow regions (NE, SR, OF1N, and OF1S) are derived from the temperatures the same as $\rm CH_3CN$ at NE.}\\
\multicolumn{9}{l}{~~~~~~~~~~~ 7. Uncertainties for the values derived from one transition   (written as the subscript) are determined from $\rm T_{rot}$, partition function $\rm Q(T_{rot})$, and  fitting to  $\rm \int T_B(\upsilon)d\upsilon$. }\\
\multicolumn{9}{l}{~~~~~~~~~~~ 8. For species which are not detected, an upper limit derived from  $\rm 3\sigma$ rms is given. }\\
\multicolumn{9}{l}{~~~~~~~~~~~ 9. $``\dagger"$  mark the values which are likely obtained from optically thick lines and  for those we did the optical depth correction.  }
\end{tabu}
\end{center}
\end{table}

\end{landscape}
%\vspace{1em}

%%%%%%%%%%%%%%%%%%%%!!!!!!!!!!!!!!!!!!!!!!!!!!!!!!!!!!
%%%%%%%%%%%%%%%%%%%%!!!!!!!!!!!!!!!!!!!!!!!!!!!!!!!!!!NNNNNNNNNNNNNNNNNNNNNN
%%%%%%%%%%%%%%%%%%%%!!!!!!!!!!!!!!!!!!!!!!!!!!!!!!!!!!
\begin{landscape}
\begin{table}
\caption{Column densities and abundances for N-bearing molecules  from different substructures in Orion-KL denoted in Figure~\ref{into} and  \ref{COMdis}.   }
\label{col-Nbearing} 
\small
\begin{center}
\begin{tabu}{c| p{2.3cm} p{2.3cm} p{2.3cm} p{2.3cm} p{2.3cm} p{2.3cm} p{1.8cm} p{1.8cm} p{1.8cm}}
%  \multicolumn{4}{l}{B1.}\\
\hline\hline

Species     & \multicolumn{2}{c} {$\rm HC$}      & $\rm mm2$      & $\rm mm3a$    &$\rm mm3b$     
       & $\rm NE$           & $\rm SR$      & $\rm OF1N$      & $\rm OF1S$\\
&($155\pm16$~K)  &($121\pm16$~K) &($108\pm4$~K)   &($103\pm3$~K)    &($88\pm6$~K)    &($43\pm3$~K)   &($43\pm3$~K)    &($43\pm3$~K)   &($43\pm3$~K)\\

\hline
\multicolumn{10}{c}{I. Column densities (in the form of $\rm x\pm y (z) =(x\pm y) \times 10^z  cm^{-2}$)}    \\
\hline

$\rm HC_3N(v_7=1)$      &$\rm 6.81_{\pm1.81}(15)$      &$\rm 1.22_{\pm0.67}(16)$      &$\rm 5.20_{\pm0.96}(15)$      &$\rm 2.36_{\pm0.54}(15)$      &$\rm 3.13_{\pm1.56}(15)$      &$\rm \le 9.13(15)$      &$\rm \le 8.67(15)$      &$\rm \le 6.79(15)$      &$\rm \le 1.10(16)$\\
$\rm CH_3CN$      &~\color{black}$\rm  \bf   1.38\pm0.48(17)~~~^\dagger$      &~\color{black}$--$      &~\color{black}$\rm  \bf   3.08\pm0.50(16)~~~^\dagger$      &~\color{black}$\rm  \bf   2.22\pm0.29(16)~~~^\dagger$      &~\color{black}$\rm  \bf   3.24\pm1.00(16)~~~^\dagger$       &~\color{black}$\rm \bf  2.16\pm0.75(15)~~~^\dagger$    &~\color{black}$\rm \le 3.65(14)~~~^\dagger$    &~\color{black}$\rm \le 7.79(14)~~~^\dagger$    &~\color{black}$\rm \le 2.42(16)~~~^\dagger$\\

$\rm CH_3^{13}CN$      &$--$      &$\rm  \bf   9.42\pm4.50(14)$            &$\rm 4.66_{\pm0.24}(14)$      &$\rm 3.16_{\pm0.18}(14)$      &$\rm 3.88_{\pm0.15}(14)$      &$\rm 3.53_{\pm0.78}(13)$      &$\rm \le 2.17(12)$      &$\rm \le 4.42(12)$      &$\rm \le 1.86(12)$\\
$\rm CH_2CHCN$      &$\rm 2.91_{\pm0.21}(15)$      &$\rm 2.40_{\pm0.10}(15)$      &$\rm 3.11_{\pm0.40}(14)$      &$\rm \le 1.29(13)$      &$\rm \le 2.01(13)$      &$\rm \le 3.19(13)$      &$\rm \le 1.97(13)$      &$\rm \le 8.13(12)$      &$\rm \le 4.05(13)$\\
$\rm CH_3CH_2CN$      &$\rm 1.15_{\pm0.04}(16)$      &$\rm 1.05_{\pm0.01}(16)$      &$\rm 3.88_{\pm0.08}(15)$      &$\rm 1.09_{\pm0.08}(15)$      &$\rm 2.19_{\pm0.20}(15)$      &$\rm \le 4.26(13)$      &$\rm \le 4.59(13)$      &$\rm \le 5.10(13)$      &$\rm \le 4.41(13)$\\
$\rm HC_7N~~~^*$      &$\rm 8.25_{\pm11.50}(15)$      &$\rm 5.83_{\pm18.30}(16)$      &$\rm \le 2.10(16)$      &$\rm \le 2.24(16)$      &$\rm \le 2.62(17)$      &$\rm \le 5.28(22)$      &$\rm \le 8.12(22)$      &$\rm \le 5.58(22)$      &$\rm \le 1.06(23)$\\
%\hline\hline
%\end{tabu}
%\vspace{1em}

%\begin{tabu}{c| p{1.8cm}  p{1.8cm} p{1.8cm} p{1.8cm} p{1.8cm} p{1.8cm} p{1.8cm} p{1.8cm}}
%  \multicolumn{4}{l}{B2.}\\
%\hline\hline

%Species     &$\rm HC$      & $\rm mm2$      & $\rm mm3a$    &$\rm mm3b$           & $\rm NE$           & $\rm S$      & $\rm OF1N$      & $\rm OF1S$\\
% & &   &    &   &($43\pm3$~K)   &($43\pm3$~K)    &($43\pm3$~K)   &($43\pm3$~K)\\
% \cline{2-6}
$\rm HNCO$      &$\rm  \bf   1.19\pm0.87(16)~~^\sharp$      &$\rm     --$     &$\rm  \bf   2.83\pm1.37(15)~~^\sharp$      &$\rm  \bf   2.15\pm1.49(15)~~^\sharp$      &$\rm  \bf   2.54\pm1.50(15)~~^\sharp$      &$\rm 3.82_{\pm0.15}(14)$      &$\rm 2.84_{\pm0.44}(13)$      &$\rm 1.63_{\pm0.04}(14)$      &$\rm 2.36_{\pm0.05}(14)$\\

\hline
\multicolumn{10}{c}{II. Abundances (in the form of $\rm x\pm y (z) =(x\pm y) \times 10^z $)}    \\
\hline
$\rm HC_3N(v_7=1)$     &$\rm 5.69_{\pm1.96}(-9)$     &$\rm 1.02_{\pm0.66}(-8)$     &$\rm 5.58_{\pm1.12}(-9)$     &$\rm 3.79_{\pm0.92}(-9)$     &$\rm 4.70_{\pm2.40}(-9)$     &$\rm \le 2.77(-8)$     &$\rm \le 8.52(-8)$     &$\rm \le 6.21(-8)$     &$\rm \le 1.90(-7)$\\
$\rm  CH_3CN$    &~\color{black}$\rm  \bf   1.16\pm0.50(-7)~~^\dagger$     &$--$    &~\color{black}$\rm  \bf   3.31\pm0.60(-8)~~^\dagger$    &~\color{black}$\rm  \bf   3.56\pm0.52(-8)~~^\dagger$    &~\color{black}$\rm  \bf   4.88\pm1.54(-8)~~^\dagger$    &~\color{black}$\rm  \bf  6.54\pm2.42(-9)~~^\dagger$   &~\color{black}$\rm \le 3.59(-9)~~^\dagger$   &~\color{black}$\rm \le 7.13(-9)~~^\dagger$   &~\color{black}$\rm \le 4.17(-9)~~^\dagger$\\
$\rm  CH_3^{13}CN$     &$--$     &$\rm  \bf   7.87\pm4.48(-10)$         &$\rm 5.00_{\pm0.33}(-10)$     &$\rm 5.07_{\pm0.36}(-10)$     &$\rm 5.84_{\pm0.27}(-10)$     &$\rm 1.07_{\pm0.26}(-10)$     &$\rm \le 2.13(-11)$     &$\rm \le 4.04(-11)$     &$\rm \le 3.21(-11)$\\

$\rm CH_2CHCN$     &$\rm 2.43_{\pm0.34}(-9)$     &$\rm 2.01_{\pm0.21}(-9)$     &$\rm 3.34_{\pm0.48}(-10)$     &$\rm \le 2.07(-11)$     &$\rm \le 3.02(-11)$     &$\rm \le 9.68(-11)$     &$\rm \le 1.94(-10)$     &$\rm \le 7.44(-11)$     &$\rm \le 6.99(-10)$\\
$\rm CH_3CH_2CN$     &$\rm 9.59_{\pm0.97}(-9)$     &$\rm 8.77_{\pm0.63}(-9)$     &$\rm 4.16_{\pm0.15}(-9)$     &$\rm 1.75_{\pm0.14}(-9)$     &$\rm 3.29_{\pm0.33}(-9)$     &$\rm \le 1.29(-10)$     &$\rm \le 4.51(-10)$     &$\rm \le 4.67(-10)$     &$\rm \le 7.61(-10)$\\
$\rm HC_7N~~~^*$     &$\rm 6.89_{\pm10.60}(-9)$     &$\rm 4.87_{\pm16.50}(-8)$     &$\rm \le 2.25(-8)$     &$\rm \le 3.60(-8)$     &$\rm \le 3.94(-7)$     &$\rm \le 1.60(-1)$     &$\rm \le 7.98(-1)$     &$\rm \le 5.10(-1)$     &$\rm \le 18.30(-1)$\\
$\rm  HNCO$     &$\rm  \bf   9.93\pm8.33(-9)~~^\sharp$       &$\rm     --$    &$\rm  \bf   3.04\pm1.53(-9)~~^\sharp$     &$\rm  \bf   3.46\pm2.47(-9)~~^\sharp$     &$\rm  \bf   3.83\pm2.30(-9)~~^\sharp$     &$\rm 1.16_{\pm0.07}(-9)$     &$\rm 2.79_{\pm0.51}(-10)$     &$\rm 1.49_{\pm0.06}(-9)$     &$\rm 4.07_{\pm0.10}(-9)$\\

\hline\hline
\multicolumn{10}{l}{{\bf Notes:} 1. ``*" marks the tentative detected molecule.}\\
\multicolumn{10}{l}{~~~~~~~~~~~ 2. Species except for those marked with ``$\sharp$" are assumed to have the same temperature (listed in the table head)  in the same substructure.}\\
\multicolumn{10}{l}{~~~~~~~~~~~ 3. Values and uncertainties in bold face are obtained directly  from   the rotation diagram fittings in Figures~\ref{rotation} and ~\ref{fig:trot_other}.}\\
\multicolumn{10}{l}{~~~~~~~~~~~ 4. For  the species whose transitions are not sufficient to derive the rotation diagrams, their column densities are  obtained from one strongest transition}\\
\multicolumn{10}{l}{~~~~~~~~~~~~~~~ by assuming LTE with the  excitation temperatures list in the table head  (derived from $\rm CH_3CN$ for HC, mm2, mm3a and mm3b).}\\
\multicolumn{10}{l}{~~~~~~~~~~~ 5. Values of all species in the outflow regions (NE, SR, OF1N, and OF1S) are derived from the temperatures the same as $\rm CH_3CN$ at NE.}\\
\multicolumn{10}{l}{~~~~~~~~~~~ 6. Uncertainties for the values derived from one transition   (written as the subscript) are determined from $\rm T_{rot}$, partition function $\rm Q(T_{rot})$, and  fitting to  $\rm \int T_B(\upsilon)d\upsilon$. }\\
\multicolumn{10}{l}{~~~~~~~~~~~ 7. For species which are not detected, an upper limit derived from  $\rm 3\sigma$ rms is given. }\\
\multicolumn{10}{l}{~~~~~~~~~~~ 8. $``\dagger"$  mark the values which are likely obtained from optically thick lines and  for those we did the optical depth correction.  }
\end{tabu}
\end{center}
\end{table}
\end{landscape}

%%%%%%%%%%%%%%%%%%%%!!!!!!!!!!!!!!!!!!!!!!!!!!!!!!!!!!
%%%%%%%%%%%%%%%%%%%%!!!!!!!!!!!!!!!!!!!!!!!!!!!!!!!!!!SSSSSSSSSSSSSSSSSSSSSSSS
%%%%%%%%%%%%%%%%%%%%!!!!!!!!!!!!!!!!!!!!!!!!!!!!!!!!!!
\begin{landscape}
\begin{table}
\caption{Column densities and abundances for S-bearing molecules  from different substructures in Orion-KL denoted in Figure~\ref{into}.   }
\label{col-Sbearing} 
\footnotesize
\begin{center}

\begin{tabu}{c| p{2cm} p{2cm} p{2cm} p{2cm} p{2cm}p{2cm} p{1.8cm} p{1.8cm} p{1.8cm} p{1.8cm}}
%  \multicolumn{4}{l}{C.}\\
\hline\hline

Species      & $\rm HC$      & $\rm mm2$      & \multicolumn{2}{c} {$\rm mm3a$}    & \multicolumn{2}{c} {$\rm mm3b$}      
       & $\rm NE$           & $\rm SR$      & $\rm OF1N$      & $\rm OF1S$\\
&($199\pm45$~K)  &($173\pm40$~K)   &($101\pm26$~K)    &~\color{black}($\it 176\pm23$~K)   &($89\pm24$~K)    &~\color{black}($\it 193\pm42$~K)   &($43\pm3$~K)   &($43\pm3$~K)    &($43\pm3$~K)   &($43\pm3$~K)\\

\hline
\multicolumn{11}{c}{I. Column densities (in the form of $\rm x\pm y (z) =(x\pm y) \times 10^z  cm^{-2}$)}    \\
\hline

$\rm OCS$      %&$\rm 3.06_{\pm0.20}(16)$            &$\rm 1.58_{\pm0.06}(16)$      &$\rm 1.50_{\pm0.17}(16)$   &   &$\rm 1.73_{\pm0.36}(16)$   &   &$\rm 1.79_{\pm0.28}(15)$      &$\rm 1.37_{\pm0.24}(15)$      &$\rm 9.00_{\pm1.85}(14)$      &$\rm \le 5.49(13)$\\
 &~\color{black}$\rm 1.69_{\pm0.11}(17)~~~^\dagger$       &~\color{black}$\rm 1.35_{\pm0.05}(17)~~~^\dagger$ &~\color{black}$\rm 9.02_{\pm1.00}(16)~~~^\dagger$  &~\color{black}$\it 9.84_{\pm0.23}(16)~~~^\dagger$  &~\color{black}$\rm 1.73_{\pm0.36}(17)~~~^\dagger$  &~\color{black}$\it 1.92_{\pm0.08}(17)~~~^\dagger$  &~\color{black}$\rm \le 1.08(16)~~~^\dagger$    &~\color{black}$\rm \le 8.26(15)~~~^\dagger$    &~\color{black}$\rm \le2.08(15)~~~^\dagger$     &~\color{black}$--~~~^\dagger$\\
 
$\rm O^{13}CS$      &$\rm 2.12_{\pm0.12}(15)$           &$\rm 1.66_{\pm0.02}(15)$      &$\rm 1.13_{\pm0.17}(15)$   &~\color{black}$\it 1.23_{\pm0.02}(15)$   &$\rm 2.22_{\pm0.46}(15)$   &~\color{black}$\it 2.46_{\pm0.10}(15)$        &$\rm \le 5.53(13)$      &$\rm \le 6.81(13)$      &$\rm \le 2.06(13)$      &$\rm \le 2.73(13)$\\

$\rm SO_2$      &$\rm 4.16_{\pm0.78}(17)$            &$\rm 2.79_{\pm0.43}(17)$      &$\rm 1.09_{\pm0.00}(17)$  &~\color{black}$\it 1.52_{\pm0.12}(17) $   &$\rm 1.47_{\pm0.09}(17)$   &~\color{black}$\it 2.28_{\pm0.38}(17)$        &$\rm 7.03_{\pm0.81}(16)$      &$\rm \le 4.26(14)$      &$\rm \le 3.29(14)$      &$\rm \le 3.82(14)$\\

$\rm ^{34}SO_2$      &$\rm  \bf   5.00\pm3.82(16)$           &$\rm  \bf   2.84\pm2.51(16)$      &$\rm  \bf   8.67\pm6.00(15)$   &~\color{black}$\it {\bf 1.25\pm 0.61 (16)}$   &$\rm  \bf   8.62\pm6.64(15)$  &~\color{black}$\it  {\bf 1.31\pm 0.99 (16)}$    &$\rm 2.55_{\pm0.05}(15)$      &$\rm \le 2.91(13)$      &$\rm \le 1.70(13)$      &$\rm \le 1.03(13)$\\

$\rm SO$      &$\rm 8.20_{\pm1.53}(16)$            &$\rm 7.16_{\pm1.29}(16)$      &$\rm 3.13_{\pm0.55}(16)$   &~\color{black}$\it 4.86_{\pm0.48}(16)$    &$\rm 3.79_{\pm0.63}(16)$      &~\color{black}$\it 6.94_{\pm1.23}(16)$   &$\rm 2.26_{\pm0.03}(16)$      &$\rm 1.32_{\pm0.02}(15)$      &$\rm 6.31_{\pm0.09}(15)$      &$\rm 9.18_{\pm0.26}(14)$\\

%$\rm ^{13}CS$      &$\rm 6.42_{\pm2.05}(14)$         &$\rm 3.13_{\pm1.00}(14)$      &$\rm 1.51_{\pm0.47}(14)$    &~\color{black}$\it 3.56_{\pm0.65}(14)$    &$\rm 1.10_{\pm0.31}(14)$       &~\color{black}$\it 3.53_{\pm0.99}(14)$   &$\rm 4.92_{\pm0.12}(12)$      &$\rm 1.19_{\pm0.03}(13)$      &$\rm 2.38_{\pm0.15}(12)$      &$\rm \le 3.99(11)$\\

%change
$\rm ^{13}CS$        &$\rm 8.69_{\pm1.49}(14)$        &$\rm 4.72_{\pm0.78}(14)$      &$\rm 3.49_{\pm0.53}(14)$ &~\color{black}$\it 5.30_{\pm0.48}(14)$    &$\rm 2.77_{\pm0.30}(14)$       &~\color{black}$\it 4.89_{\pm0.63}(14)$    &$\rm 1.99_{\pm0.12}(13)$      &$\rm 4.80_{\pm0.05}(13)$      &$\rm 9.61_{\pm0.98}(12)$        &$\rm \le 1.61(12)$\\

\hline
\multicolumn{11}{c}{II. Abundances (in the form of $\rm x\pm y (z) =(x\pm y) \times 10^z $)}    \\
\hline
$\rm OCS$     %&$\rm 2.56_{\pm0.34}(-8)$         &$\rm 1.70_{\pm0.09}(-8)$     &$\rm 2.41_{\pm0.30}(-8)$     &$\rm 2.61_{\pm0.56}(-8)$     &$\rm 5.44_{\pm0.97}(-9)$     &$\rm 1.35_{\pm0.28}(-8)$     &$\rm 8.23_{\pm1.86}(-9)$     &$\rm \le 9.46(-10)$\\
&~\color{black}$\rm 1.41_{\pm0.18}(-7)~~~^\dagger$   &~\color{black}$\rm 1.45_{\pm0.07}(-7)~~~^\dagger$   &~\color{black}$\rm 1.45_{\pm0.18}(-7)~~~^\dagger$  &~\color{black}$\it 1.58_{\pm0.06}(-7)~~~^\dagger$ &~\color{black}$\rm 2.61_{\pm0.56}(-7)~~~^\dagger$ &~\color{black}$\it  2.89_{\pm0.14}(-7)~~~^\dagger$  &~\color{black}$\rm \le 3.27(-8)~~~^\dagger$   &~\color{black}$\rm \le 8.12(-8)~~~^\dagger$   &~\color{black}$\rm \le 1.90(-8)~~~^\dagger$  &~\color{black}$--~~~^\dagger$\\

$\rm O^{13}CS$         &$\rm 1.77_{\pm0.22}(-9)$     &$\rm 1.78_{\pm0.04}(-9)$     &$\rm 1.81_{\pm0.30}(-9)$    &~\color{black}$\it 1.98_{\pm0.05}(-9)$      &$\rm 3.34_{\pm0.72}(-9)$      &~\color{black}$\it 3.70_{\pm0.18}(-9)$   &$\rm \le 1.68(-10)$     &$\rm \le 6.69(-10)$     &$\rm \le 1.88(-10)$     &$\rm \le 4.72(-10)$\\

$\rm SO_2$          &$\rm 3.47_{\pm0.91}(-7)$     &$\rm 2.99_{\pm0.51}(-7)$     &$\rm 1.76_{\pm0.02}(-7)$  &~\color{black}$\it 2.44_{\pm0.22}(-7)$    &$\rm 2.21_{\pm0.15}(-7)$    &~\color{black}$\it 3.43_{\pm0.60}(-7)$       &$\rm 2.13_{\pm0.29}(-7)$     &$\rm \le 4.19(-9)$     &$\rm \le 3.01(-9)$     &$\rm \le 6.58(-9)$\\

$\rm  ^{34}SO_2$         &$\rm  \bf   4.18\pm3.65(-8)$     &$\rm  \bf   3.04\pm2.78(-8)$     &$\rm  \bf   1.39\pm0.99(-8)$  &~\color{black}$\it \bf 2.01\pm 1.01 (-8)$   &$\rm  \bf   1.30\pm1.01(-8)$   &~\color{black}$\it \bf 1.97\pm 1.51 (-8)$  &$\rm 7.75_{\pm0.29}(-9)$     &$\rm \le 2.86(-10)$     &$\rm \le 1.55(-10)$     &$\rm \le 1.78(-10)$\\

$\rm SO$         &$\rm 6.85_{\pm1.79}(-8)$     &$\rm 7.69_{\pm1.51}(-8)$     &$\rm 5.04_{\pm0.95}(-8)$  &~\color{black}$\it 7.81_{\pm0.88}(-8)$    &$\rm 5.71_{\pm1.00}(-8)$  &~\color{black}$\it 1.04_{\pm0.19}(-7)$     &$\rm 6.87_{\pm0.21}(-8)$     &$\rm 1.30_{\pm0.05}(-8)$     &$\rm 5.77_{\pm0.19}(-8)$     &$\rm 1.58_{\pm0.05}(-8)$\\

%$\rm ^{13}CS$          &$\rm 5.36_{\pm2.15}(-10)$     &$\rm 3.36_{\pm1.14}(-10)$     &$\rm 2.43_{\pm0.80}(-10)$   &~\color{black}$\it 5.72_{\pm1.13}(-10)$      &$\rm 1.65_{\pm0.48}(-10)$   &~\color{black}$\it 5.31_{\pm1.53}(-10)$  &$\rm 1.49_{\pm0.06}(-11)$     &$\rm 1.17_{\pm0.06}(-10)$     &$\rm 2.18_{\pm0.18}(-11)$     &$\rm \le 6.88(-12)$\\
%change
$\rm ^{13}CS$          &$\rm 7.26_{\pm1.77}(-10)$    &$\rm 5.06_{\pm0.92}(-10)$      &$\rm 5.62_{\pm0.92}(-10)$  
&~\color{black}$\it 8.51_{\pm0.88}(-10)$       &$\rm 4.16_{\pm0.49}(-10)$       &~\color{black}$\it 7.36_{\pm1.00}(-10)$    &$\rm 6.03_{\pm0.48}(-11)$     &$\rm 4.71_{\pm0.16}(-10)$     &$\rm 8.80_{\pm1.06}(-11)$      &$\rm \le 2.78(-11)$\\

\hline
 \hline
\multicolumn{11}{l}{{\bf Notes:} 1. Values and uncertainties in bold face are obtained directly  from   the rotation diagram fittings in  Figure~\ref{fig:trot_other}.}\\
\multicolumn{11}{l}{~~~~~~~~~~~ 2. Values  in italic font are derived from upper limit of $\rm T_{rot}$ in mm3a and mm3b.}\\
\multicolumn{11}{l}{~~~~~~~~~~~ 3. For  the species whose transitions are not sufficient to derive the rotation diagrams, their column densities are  obtained from one strongest transition}\\
\multicolumn{11}{l}{~~~~~~~~~~~~~~~ by assuming LTE with the  excitation temperatures list in the table head  (derived from $\rm ^{34}SO_2$ for HC, mm2, mm3a and mm3b).}\\
\multicolumn{11}{l}{~~~~~~~~~~~ 4. Values of all species in the outflow regions (NE, SR, OF1N, and OF1S) are derived from the temperatures the same as $\rm CH_3CN$ at NE.}\\
\multicolumn{11}{l}{~~~~~~~~~~~ 5. Uncertainties for the values derived from one transition   (written as the subscript) are determined from $\rm T_{rot}$, partition function $\rm Q(T_{rot})$, and  fitting to  $\rm \int T_B(\upsilon)d\upsilon$. }\\
\multicolumn{11}{l}{~~~~~~~~~~~ 6. For species which are not detected, an upper limit derived from  $\rm 3\sigma$ rms is given. }\\
\multicolumn{11}{l}{~~~~~~~~~~~ 7. $``\dagger"$  mark the values which are likely obtained from optically thick lines and  for those we did the optical depth correction.  } 

\end{tabu}
\end{center}

\end{table}
\end{landscape}

%%%%%%%%%%%%%%%%%%%%%%%%%%
%%%%%%%%%%%%%%%%%%%%%%%%%% 
\newpage
\setcounter{figure}{0}
\renewcommand{\thefigure}{A\arabic{figure}}

\begin{figure*}[!ht]
\begin{center}
\includegraphics[width=18cm]{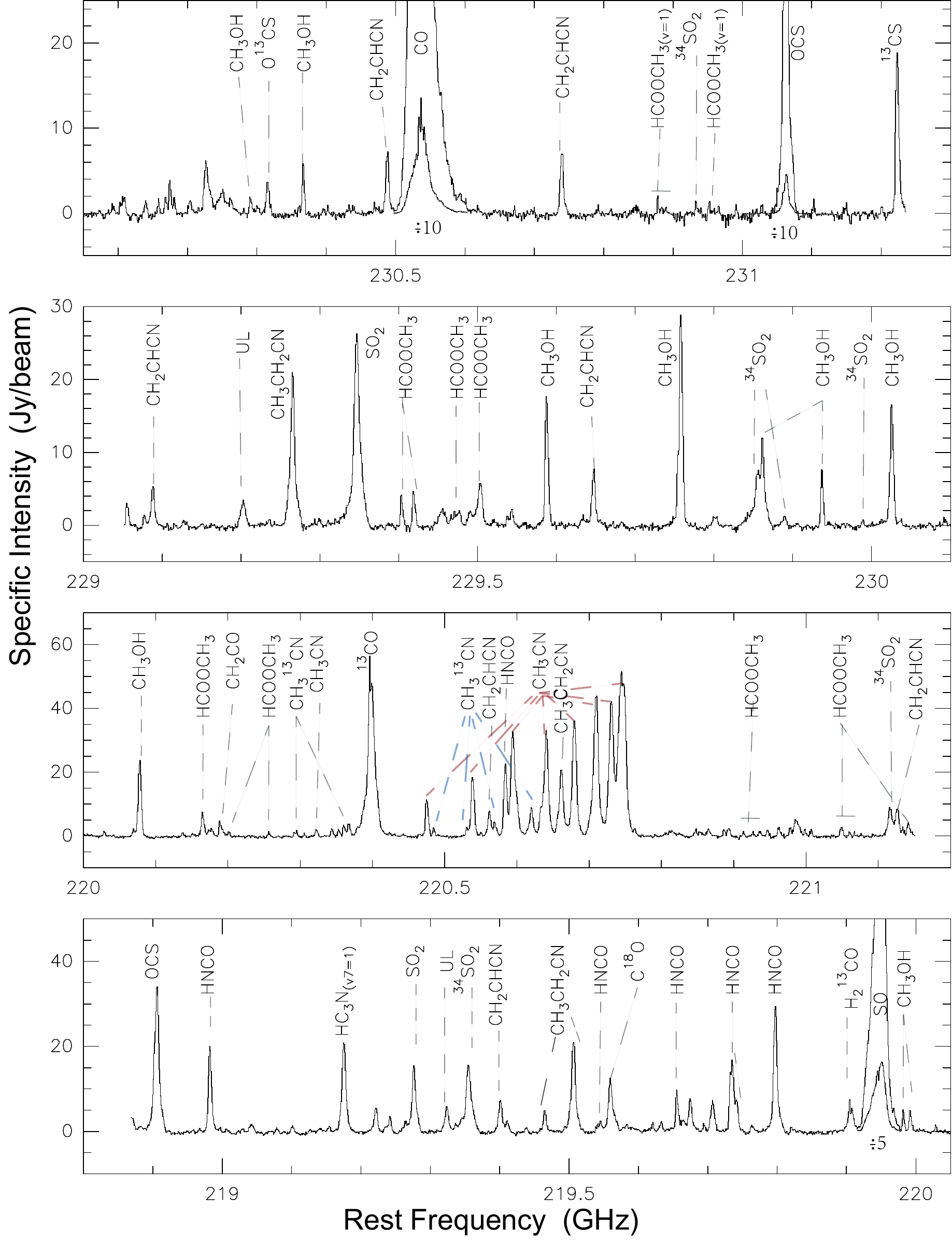}
\end{center}
\caption{Spectrum from the HC, with identified strong lines marked. From 220.5 GHz to 220.8 GHz, $\rm CH_3CN$ and $\rm CH_3^{13}CN$ are labelled in red and blue, respectively. COMs containing only C, H, O elements and other tentatively  detections are labelled in Figure~\ref{COMspec}.
%{\color{red}remove the /10, /5 that are appearing in the spectral boxes}
}\label{blow}
%\end{figure}
\end{figure*}

\newpage
\begin{figure*}[htb]
\begin{center}
\begin{tabular}{p{9cm} p{9cm}}
\multicolumn{2}{l}{$\rm CH_3OH$ @ mm3a  ($\rm T_{rot}$=160 K; $\rm N_{mol}=2.5\times10^{17}~cm^{-2}$; $\rm V_{lsr}=\rm 7.6~km\,s^{-1}$; $\rm \Delta V=5~km\,s^{-1}$)}\\
\includegraphics[width=9cm]{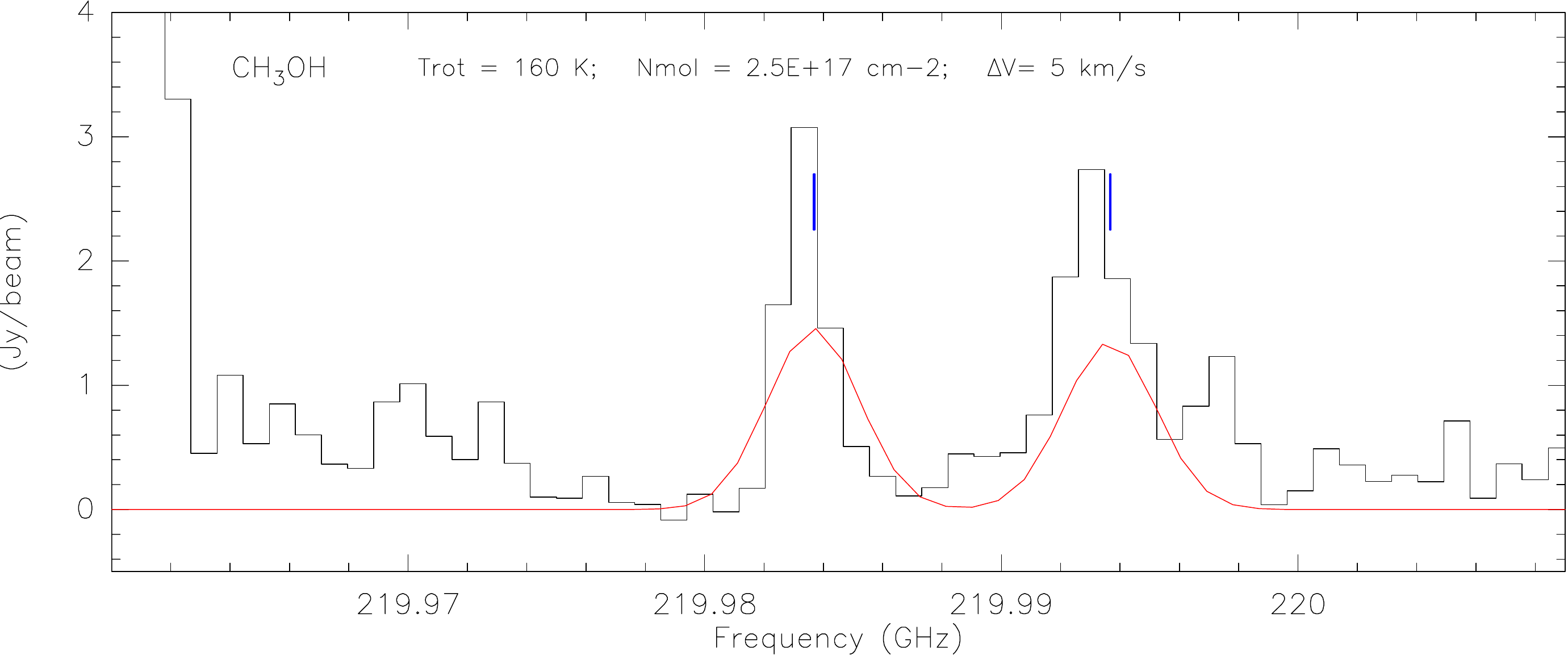}
&\includegraphics[width=9cm]{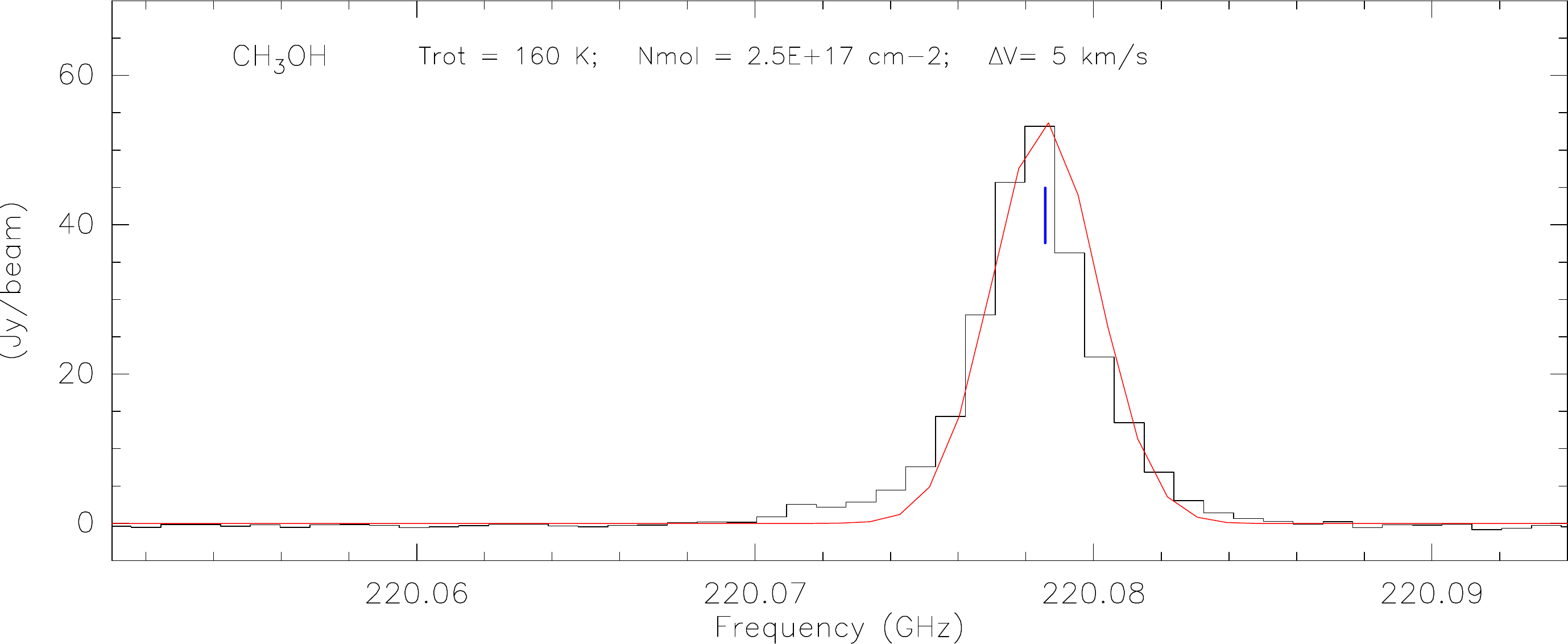}\\
\includegraphics[width=9cm]{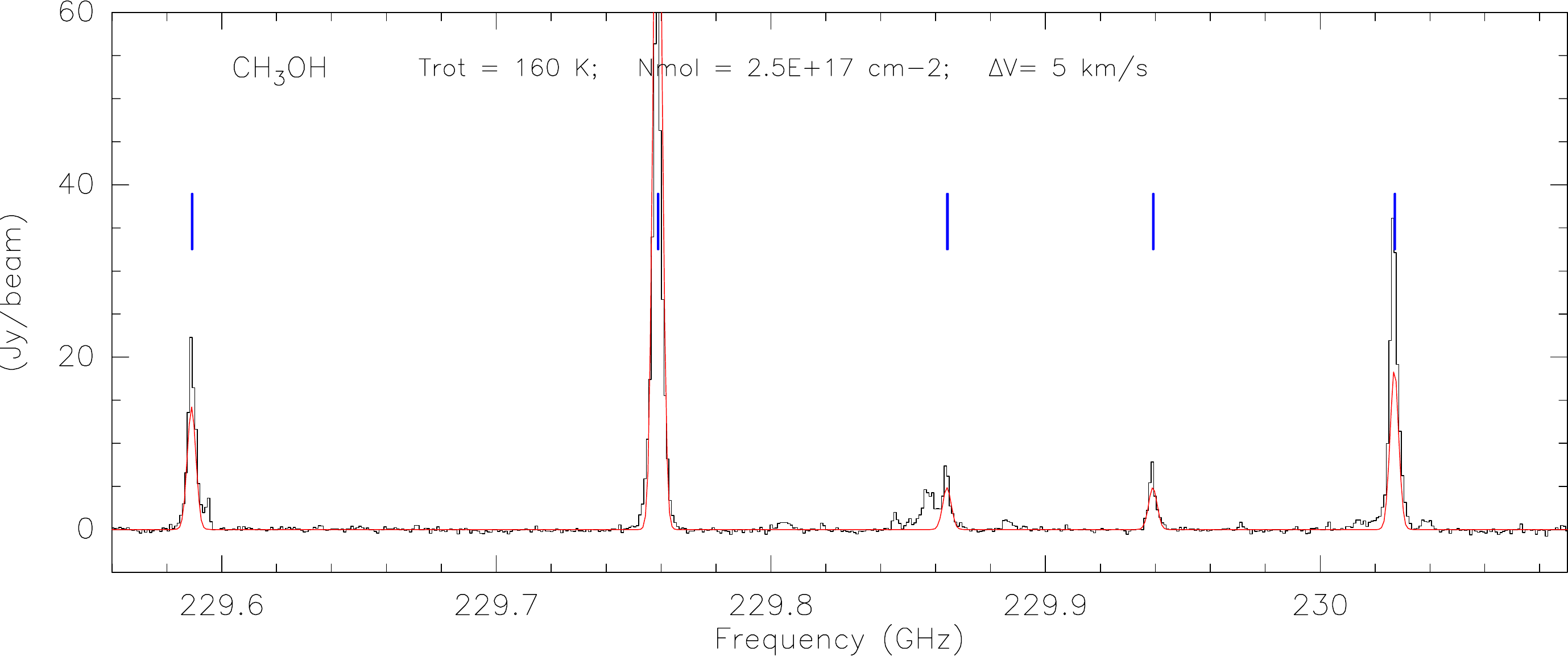}
&\includegraphics[width=9cm]{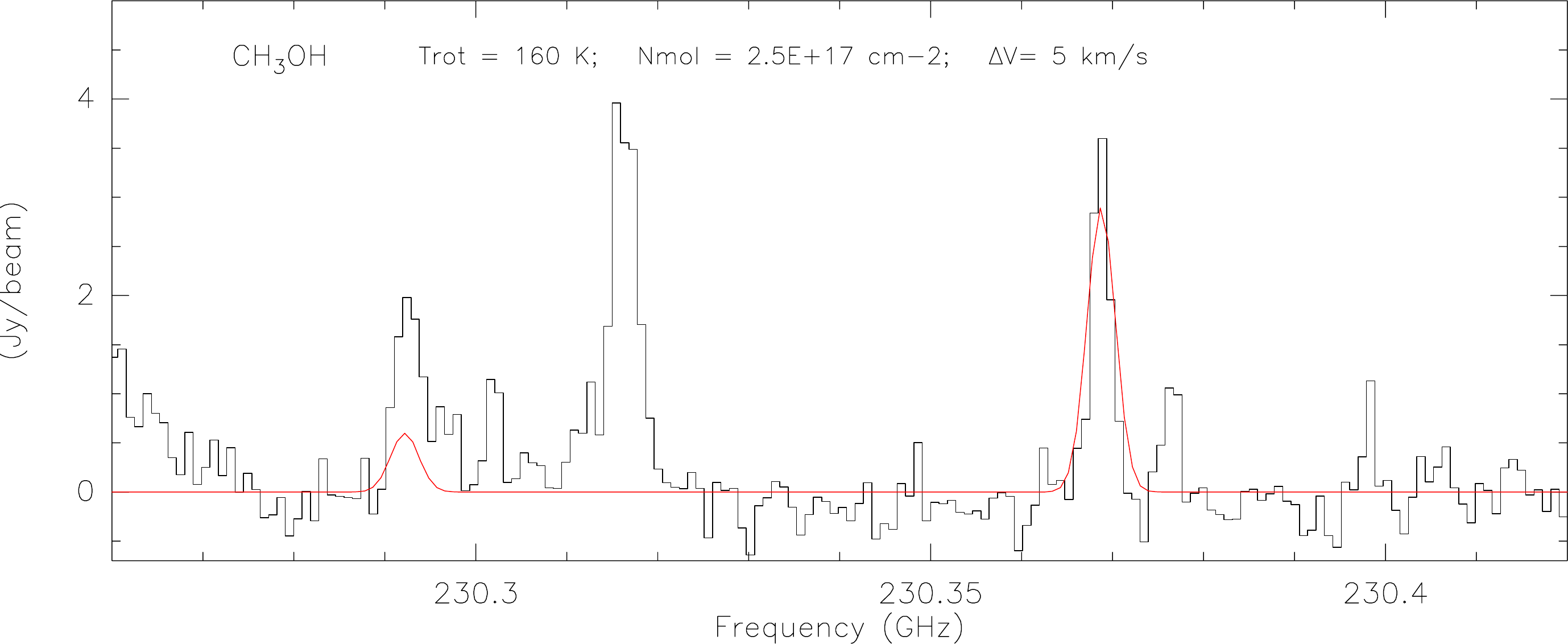}\\

\multicolumn{2}{l}{$\rm HCOOCH_3$ @ mm3b  ($\rm T_{rot}$=102 K; $\rm N_{mol}=1.0\times10^{17}~cm^{-2}$; $\rm V_{lsr}=\rm 7.8~km\,s^{-1}$; $\rm \Delta V=3~km\,s^{-1}$)}\\
\includegraphics[width=9cm]{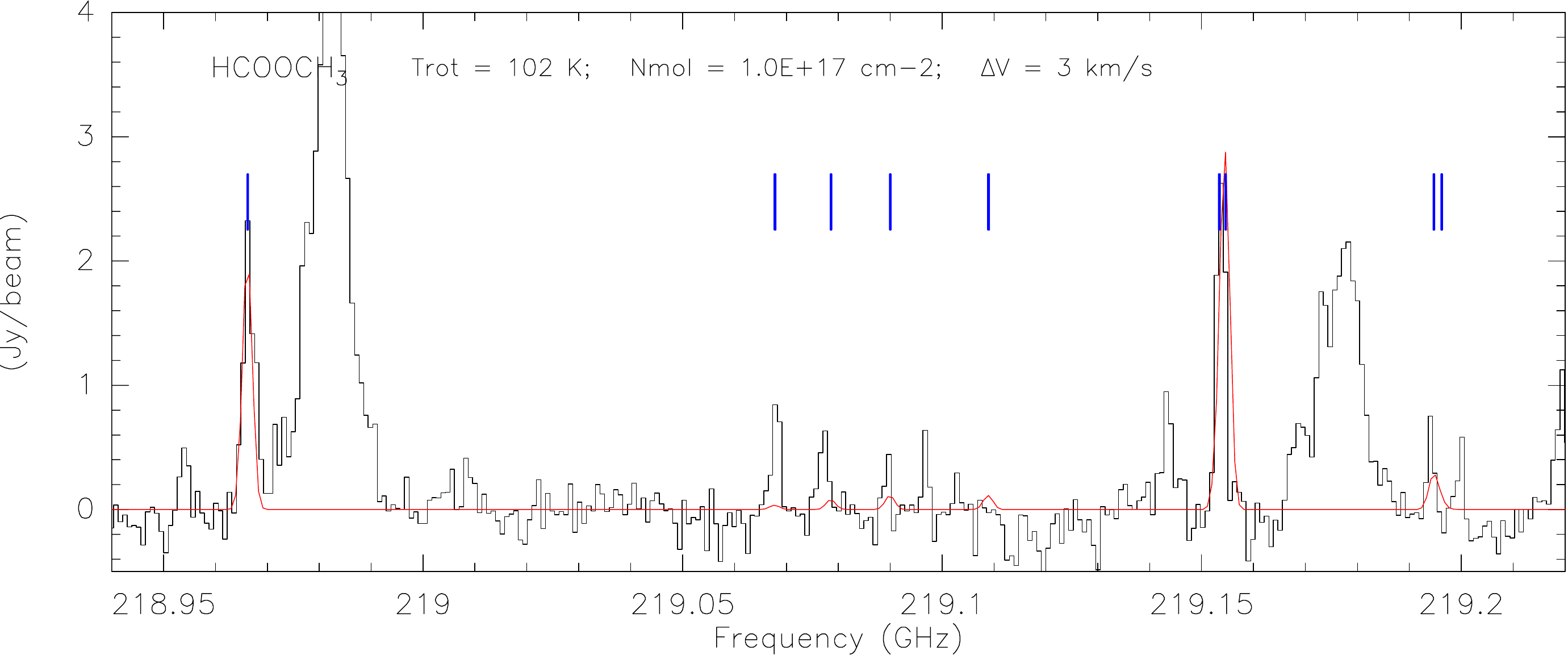}
&\includegraphics[width=9cm]{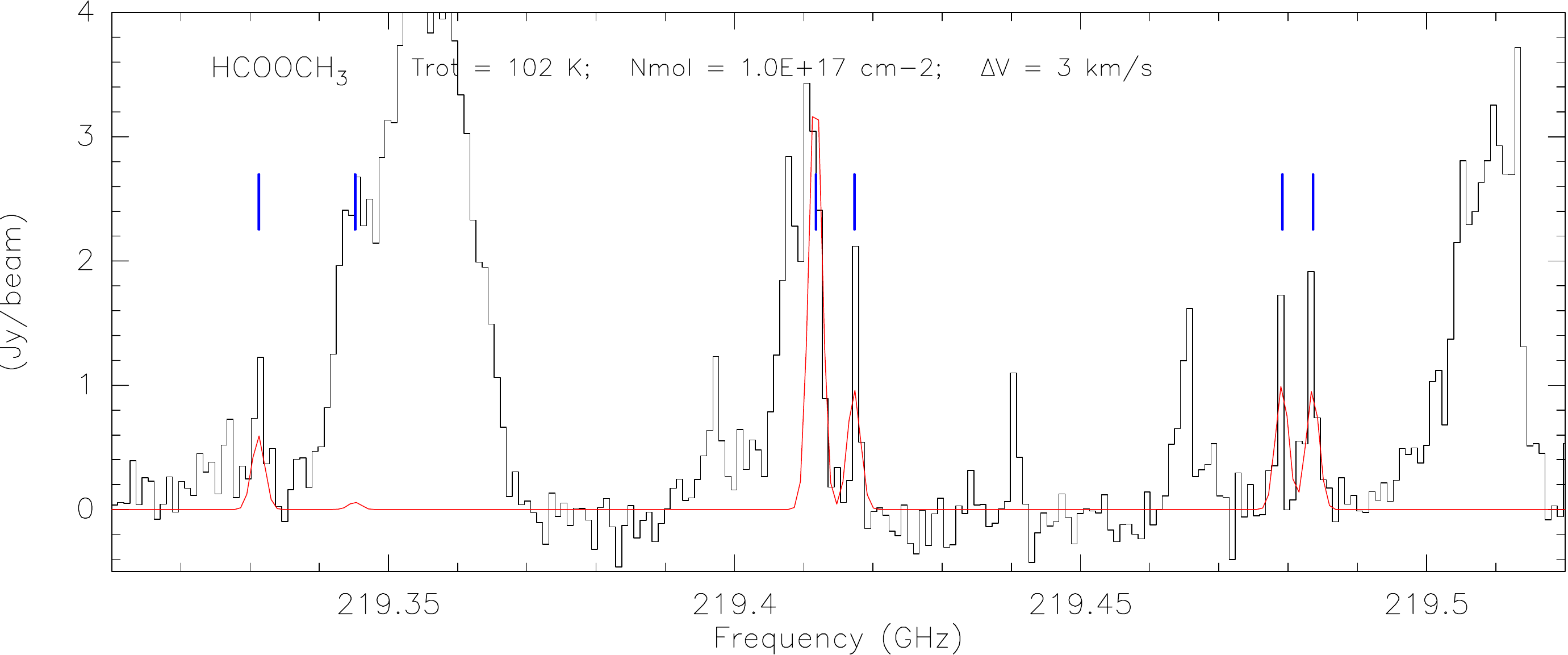}\\
\includegraphics[width=9cm]{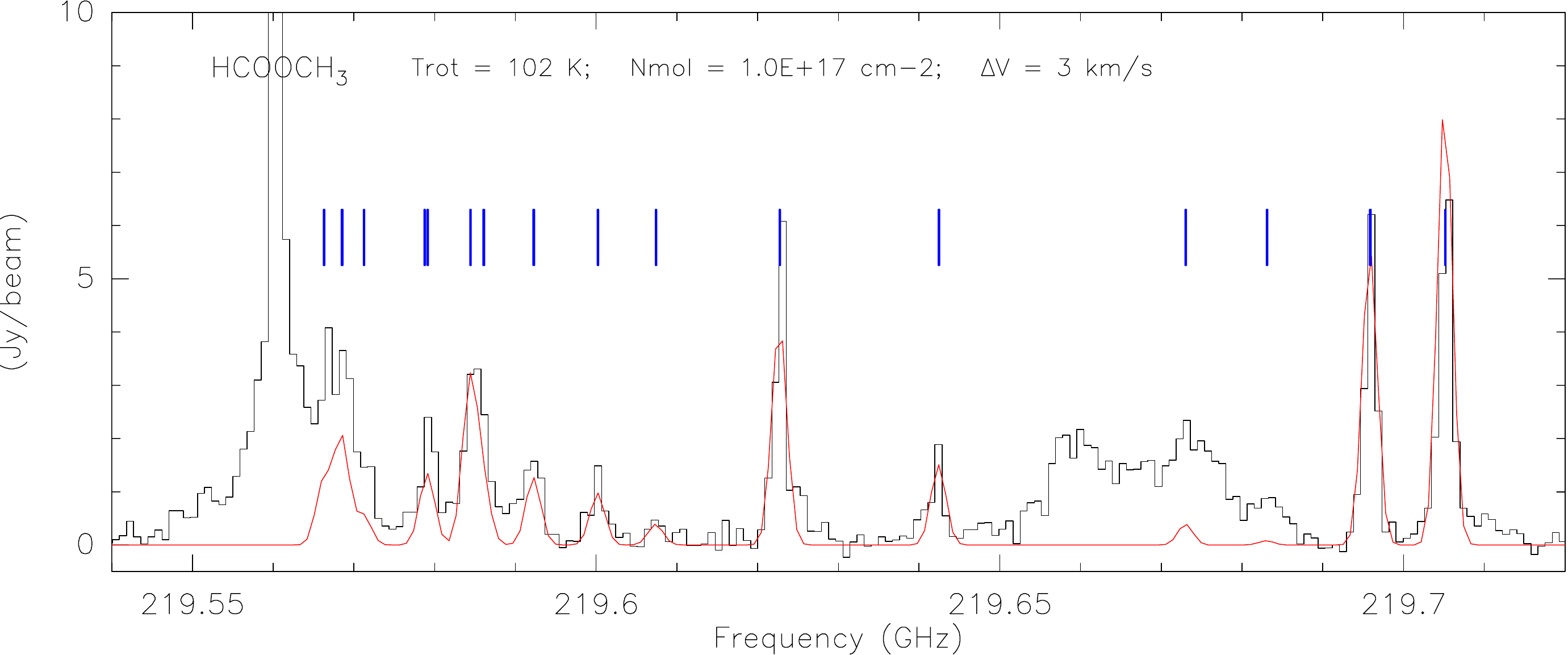}
&\includegraphics[width=9cm]{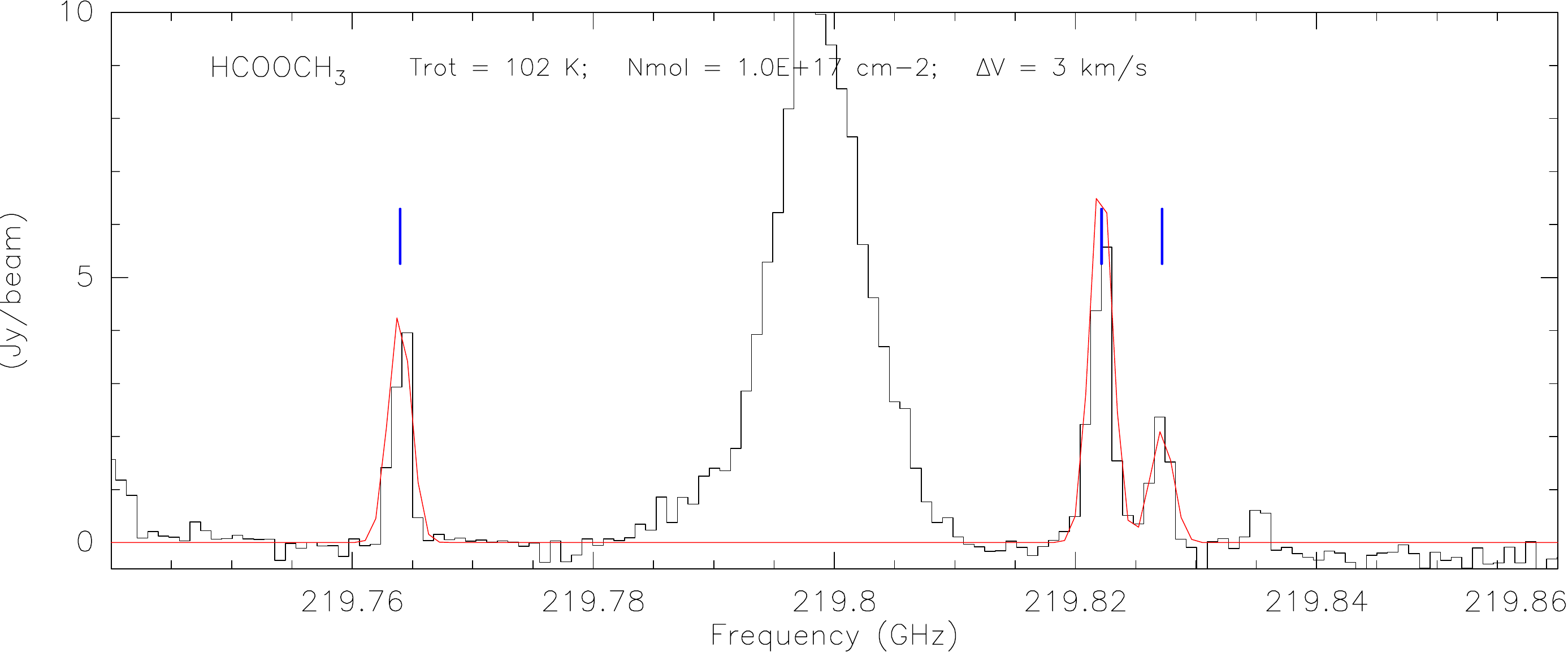}\\
\includegraphics[width=9cm]{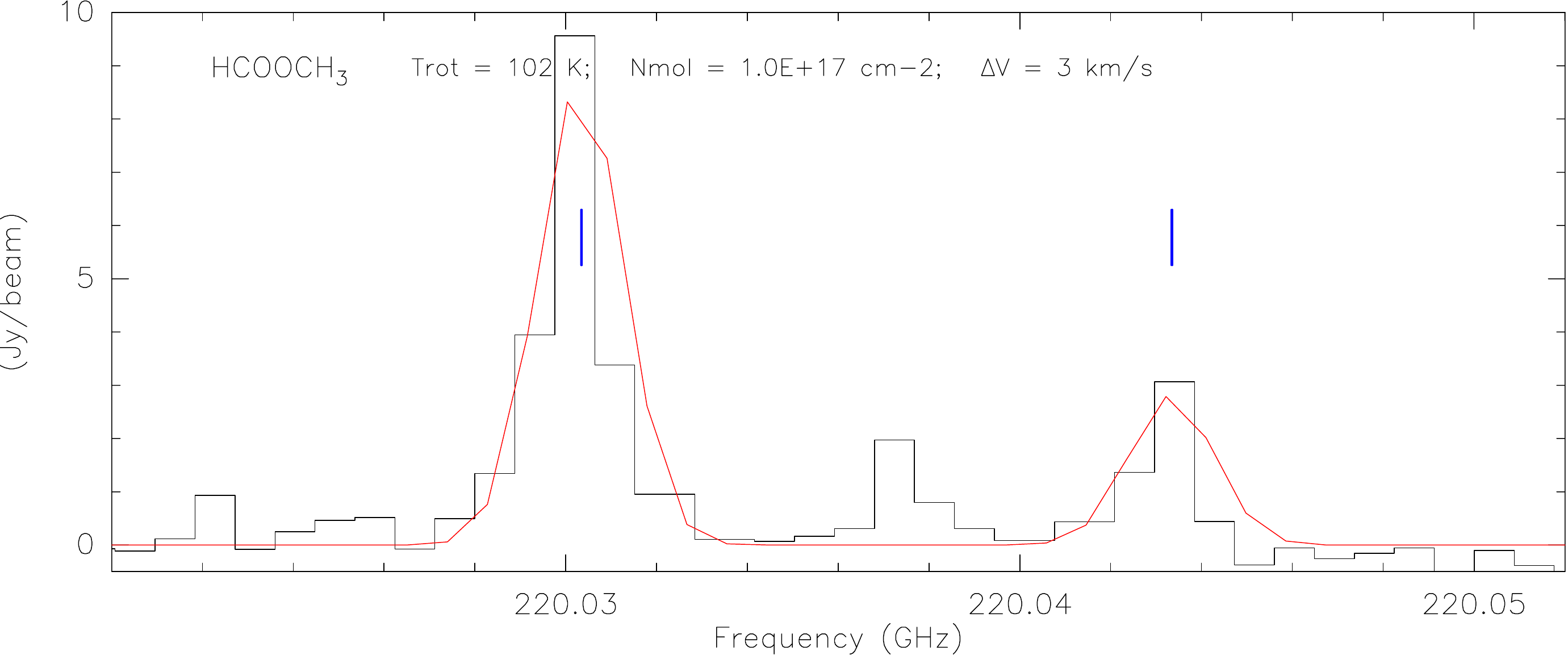}
&\includegraphics[width=9cm]{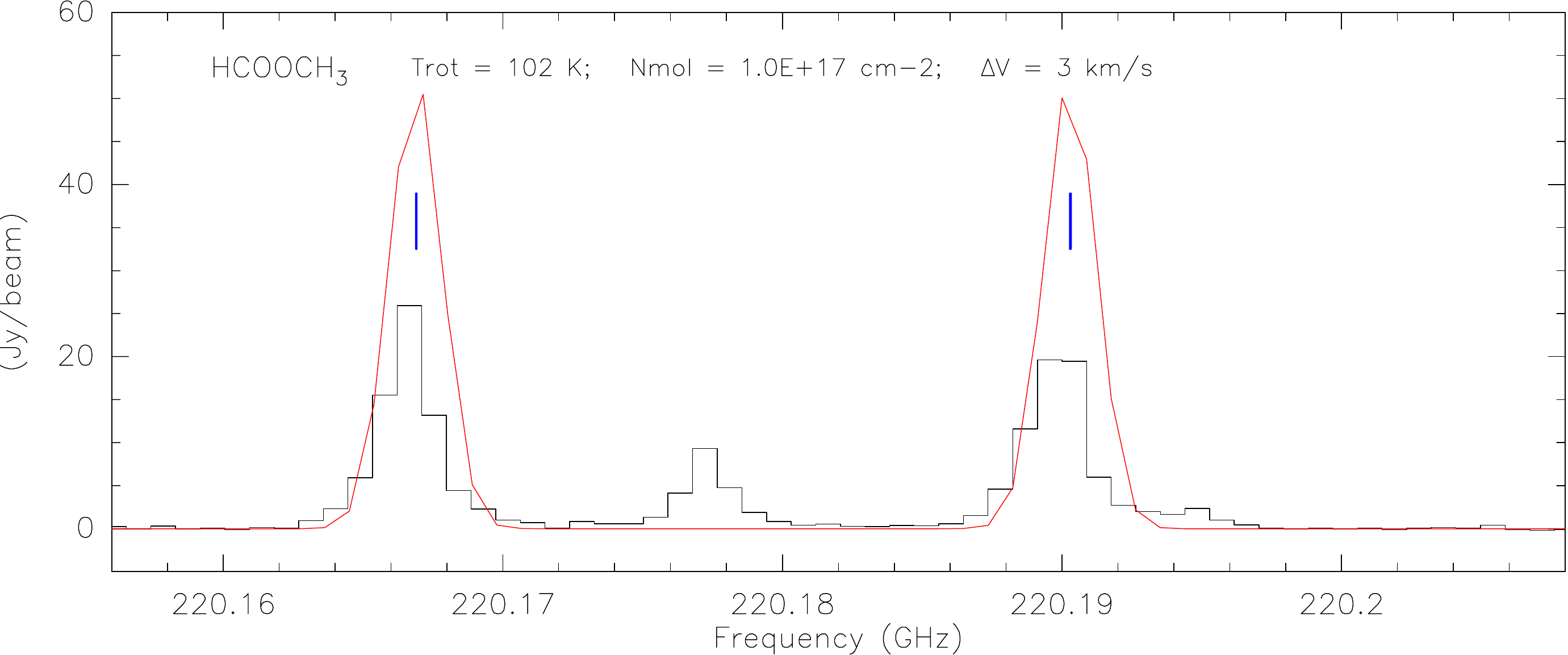}\\
\end{tabular}
\end{center}
\caption{Synthetic spectrum fitting of COMs towards their emission peaks ($\rm V_{lsr}$ of each species are corrected according to $\rm V_{peak}$  from Figure~\ref{velpro}).  The black histograms are the observed spectra,  overlaid with a best fit model spectrum for  given uniform FWHM linewidth ($\rm \triangle V$) (fit from Figure~\ref{velpro}),  $\rm T_{rot}$ (assumed from an adjacent substructure), and $\rm N_{mol}$ (adjusted based on  $\rm N_{T}$ in Tables~\ref{col-Obearing}I, \ref{col-Nbearing}I, and \ref{col-Sbearing}I) listed on top of each panel.  Blue line mark the central frequency of each transition.  Uncertainties of the fitting to observations come from  the assumption that under LTE, the candidate molecule lines are optically thin,  and  each species has the same line width at different transitions.  Referenced molecular data are from CDMS/JPL \citep{sanchez11, palau11}.
  }\label{COMspec}
\end{figure*}

\begin{figure*}[htb]
\ContinuedFloat
\begin{center}
\begin{tabular}{p{9cm} p{9cm}}
\includegraphics[width=9cm]{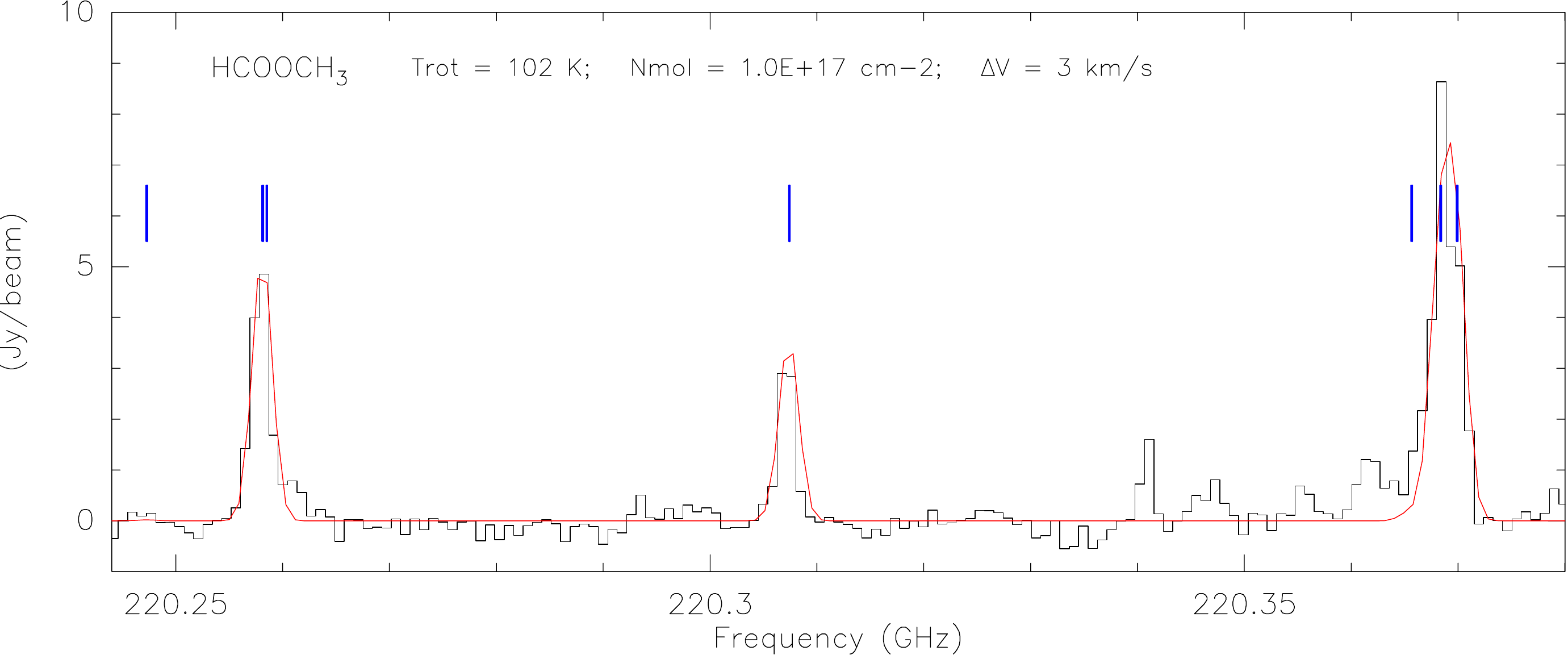}
&\includegraphics[width=9cm]{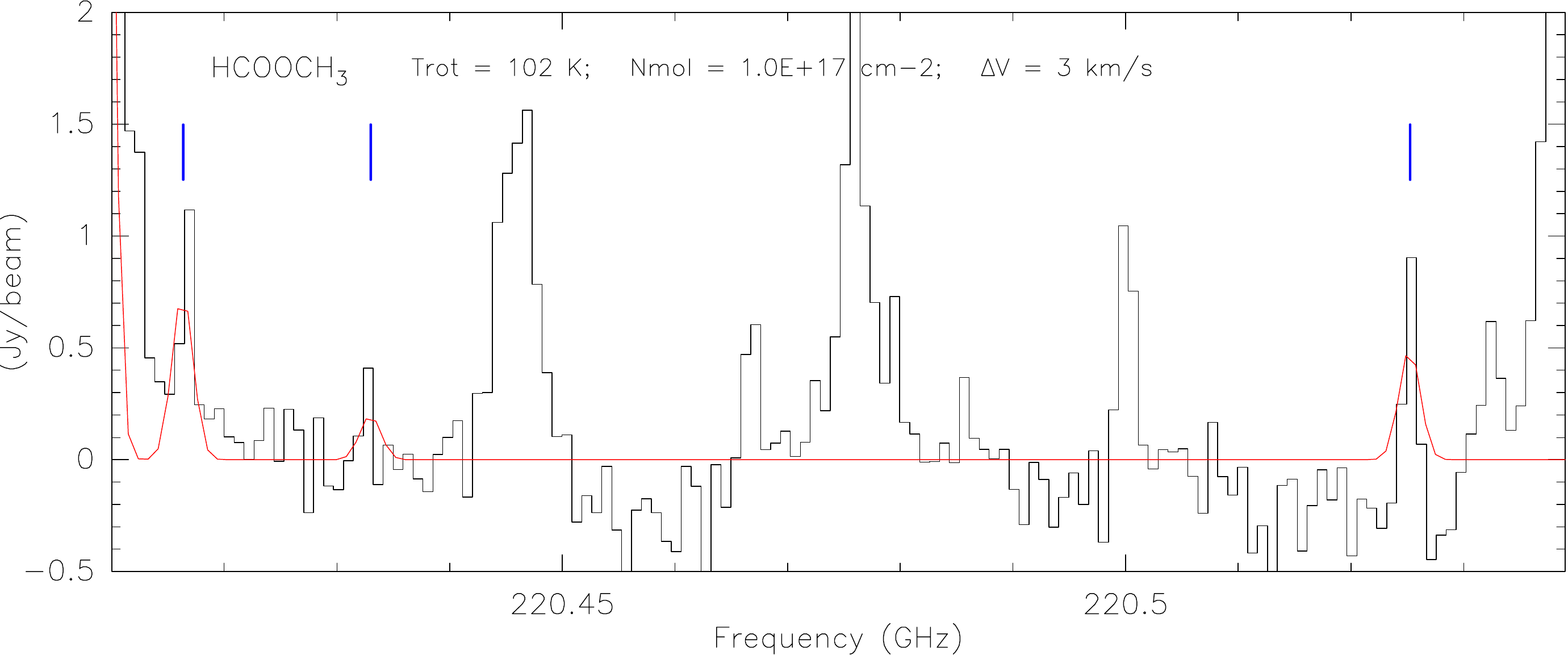}\\
\includegraphics[width=9cm]{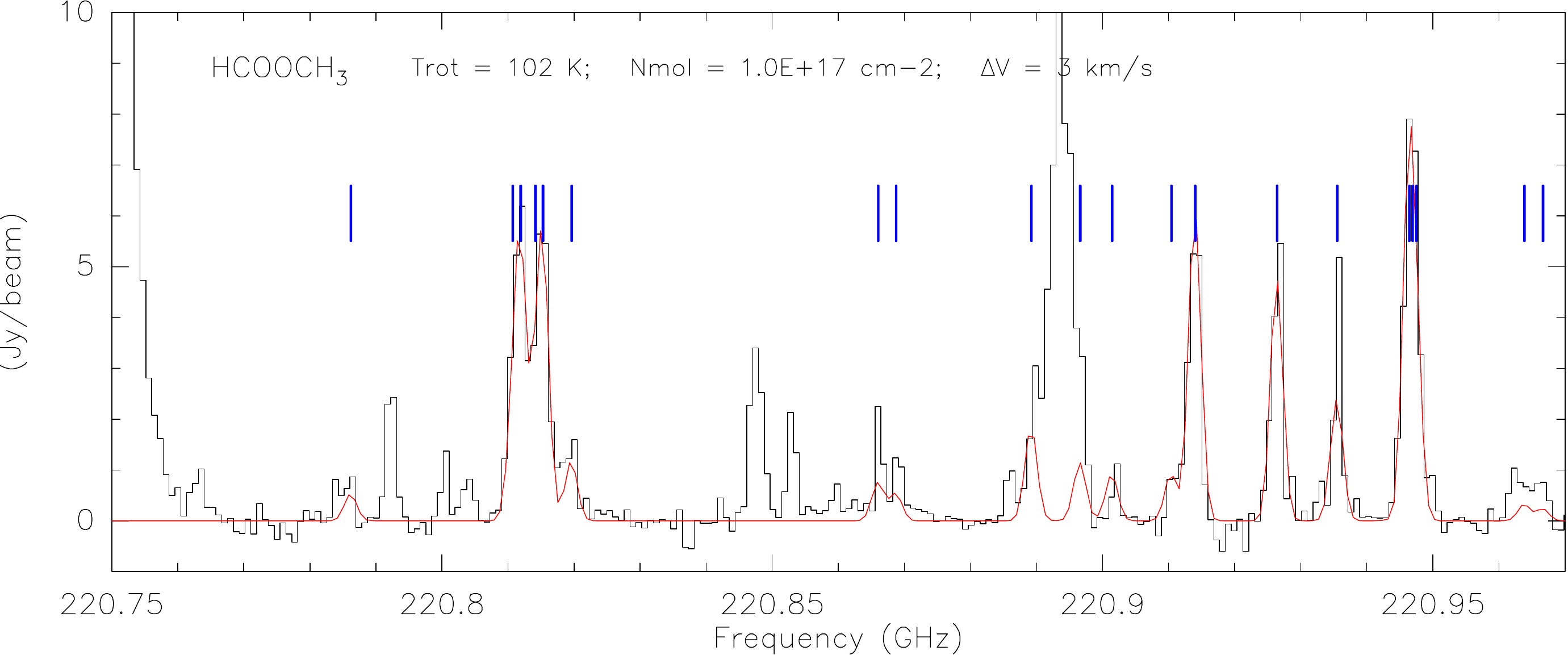}
&\includegraphics[width=9cm]{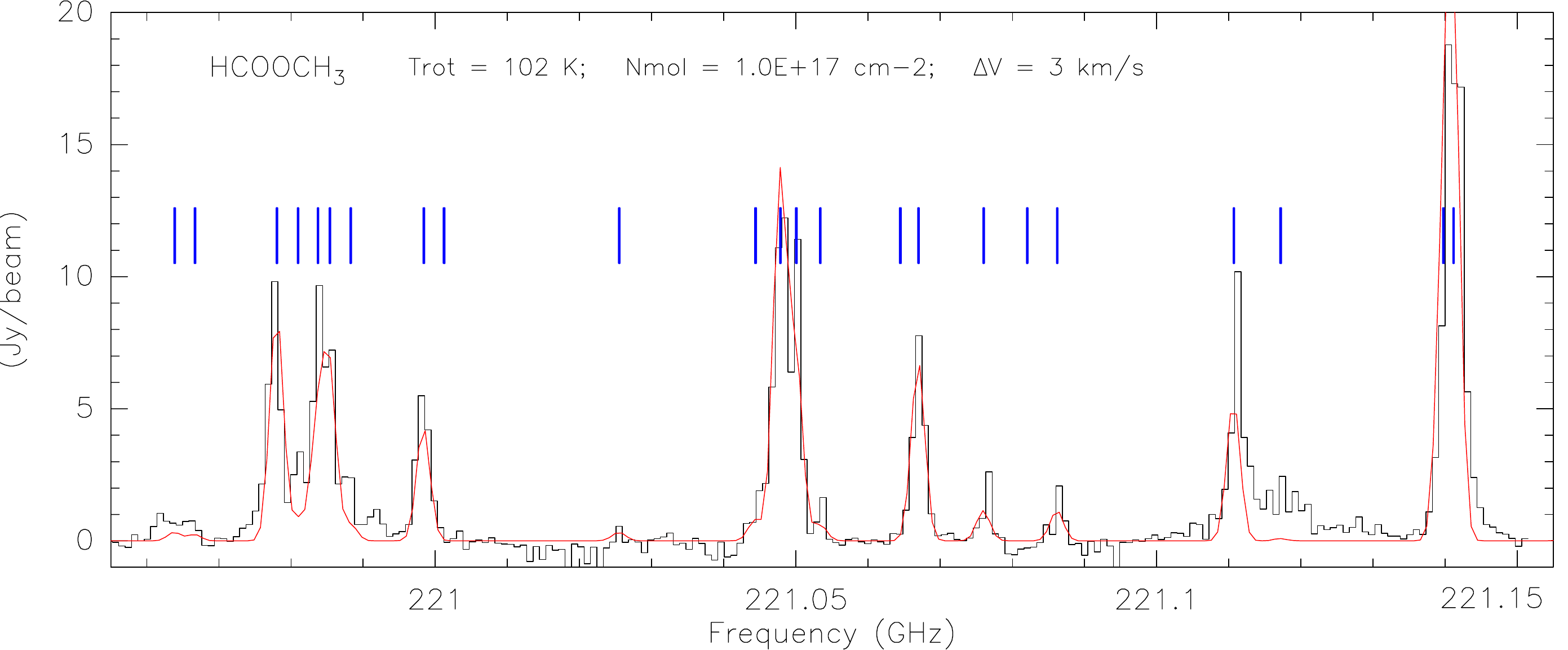}\\
\includegraphics[width=9cm]{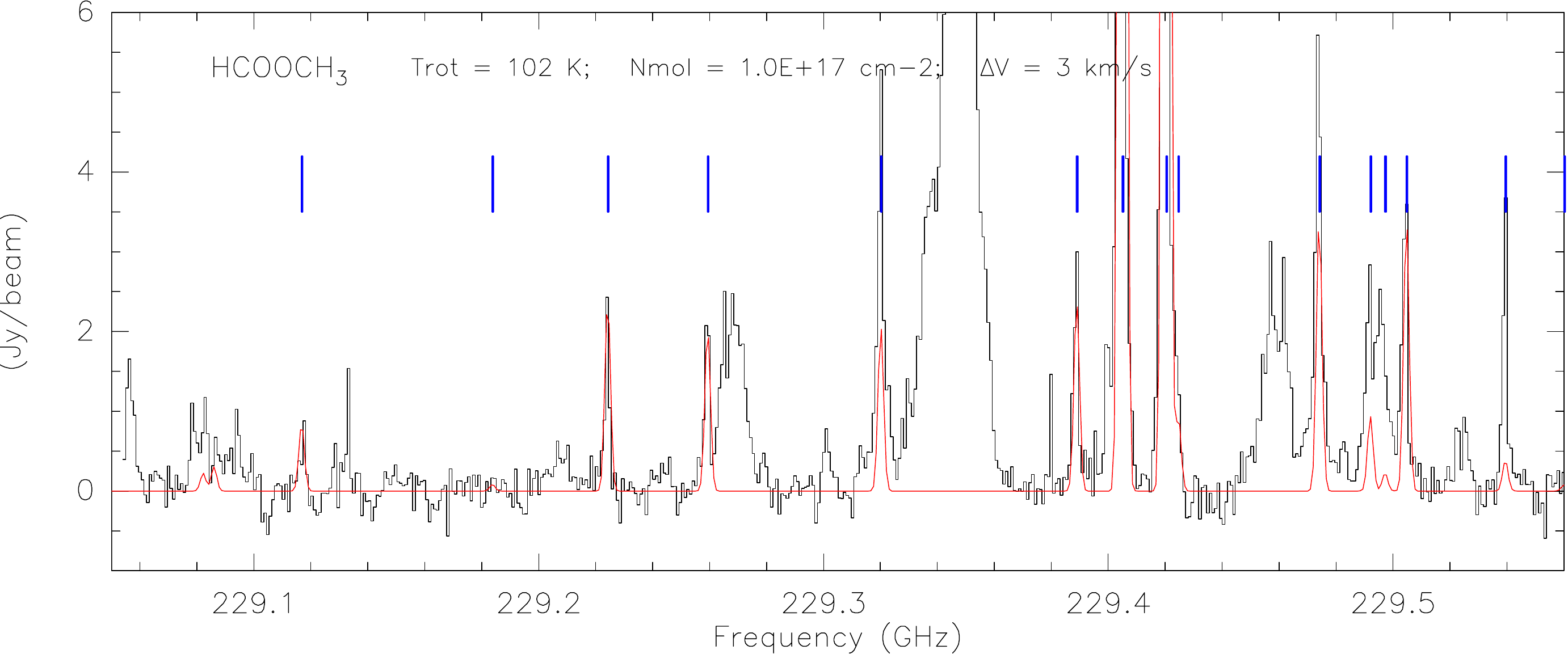}
&\includegraphics[width=9cm]{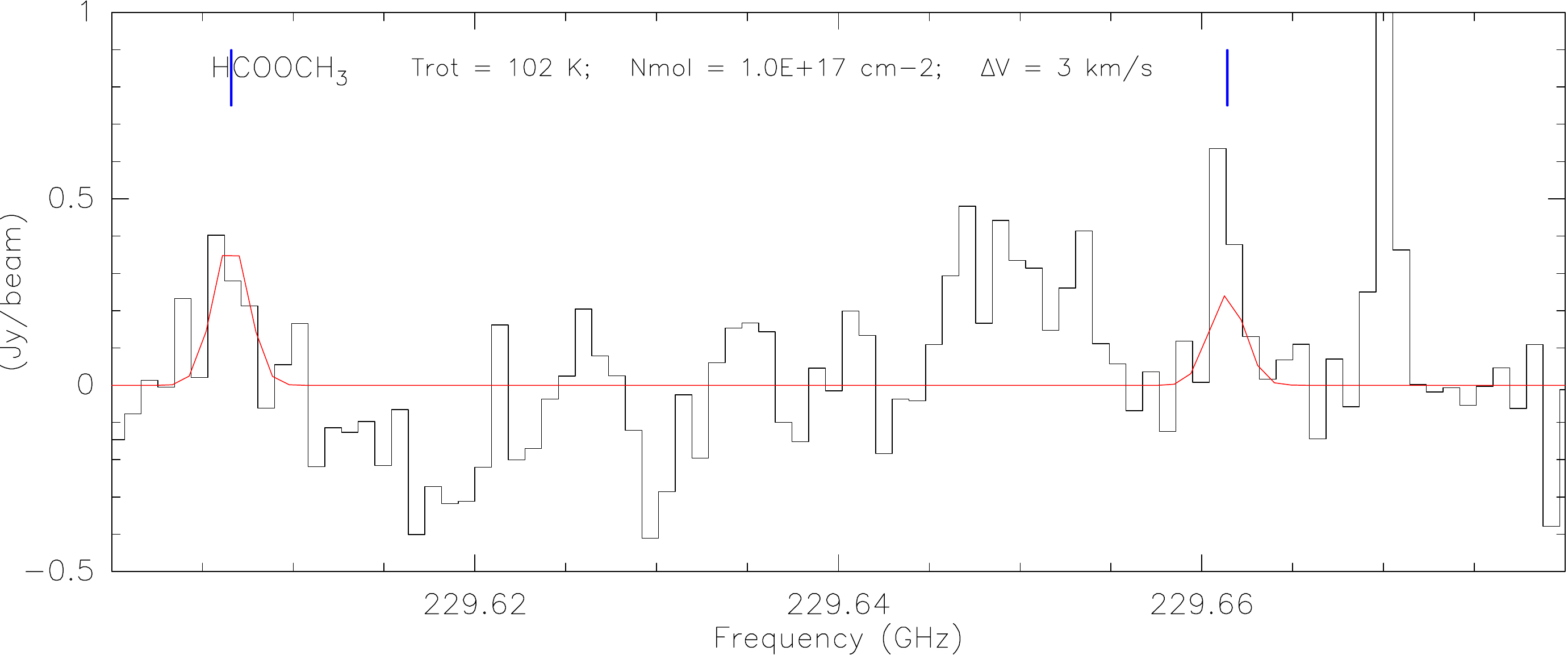}\\
\includegraphics[width=9cm]{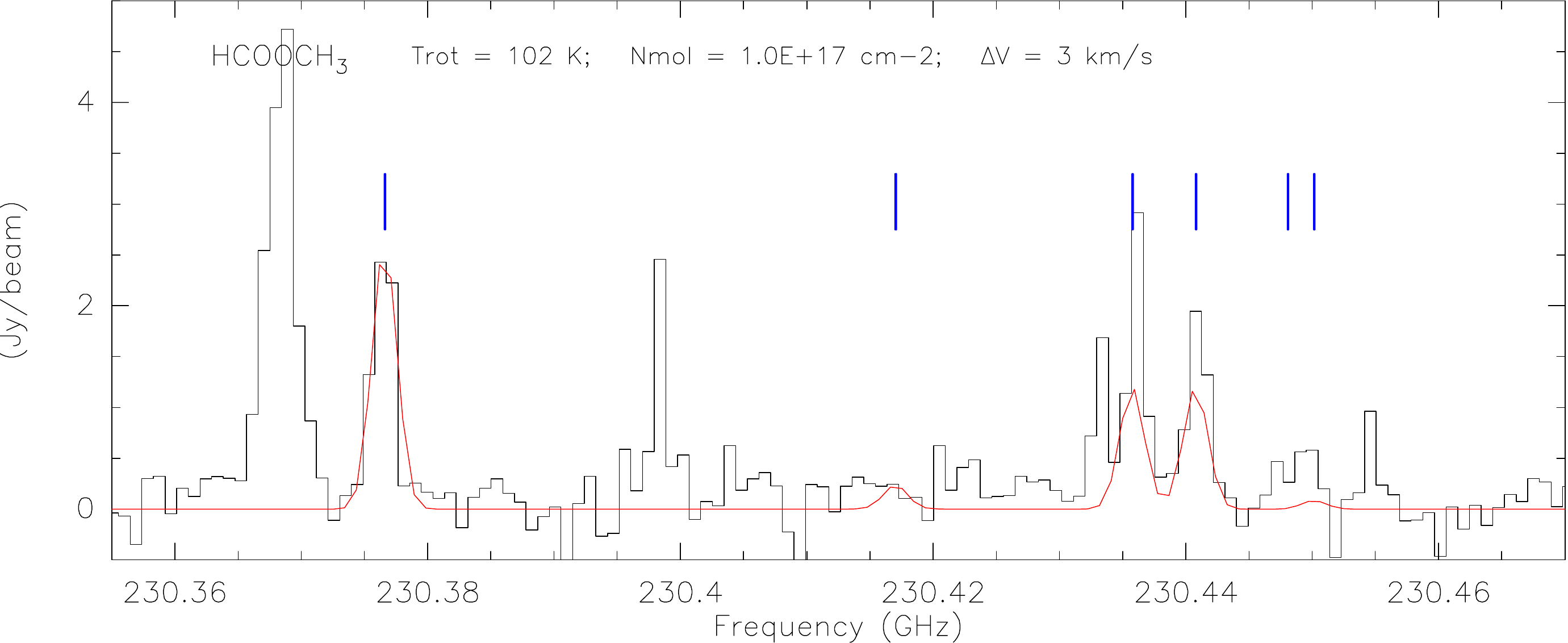}
&\includegraphics[width=9cm]{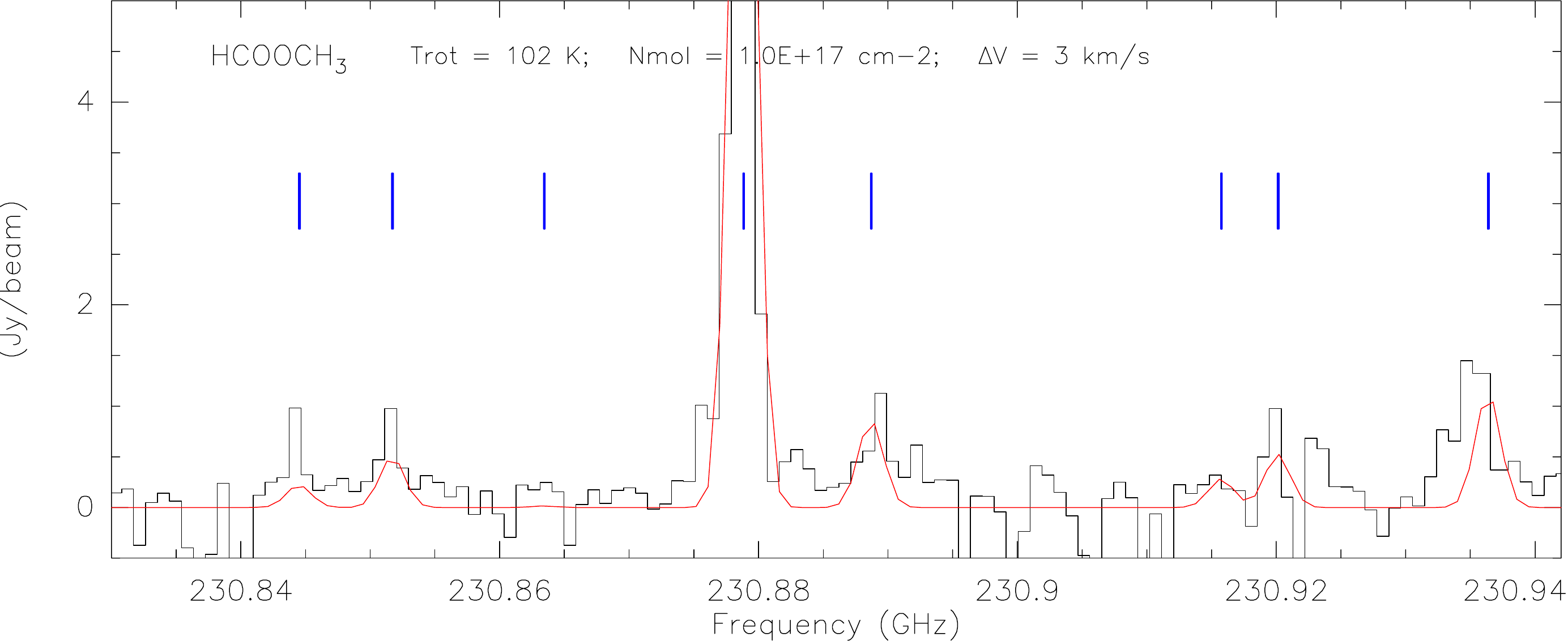}\\
\includegraphics[width=9cm]{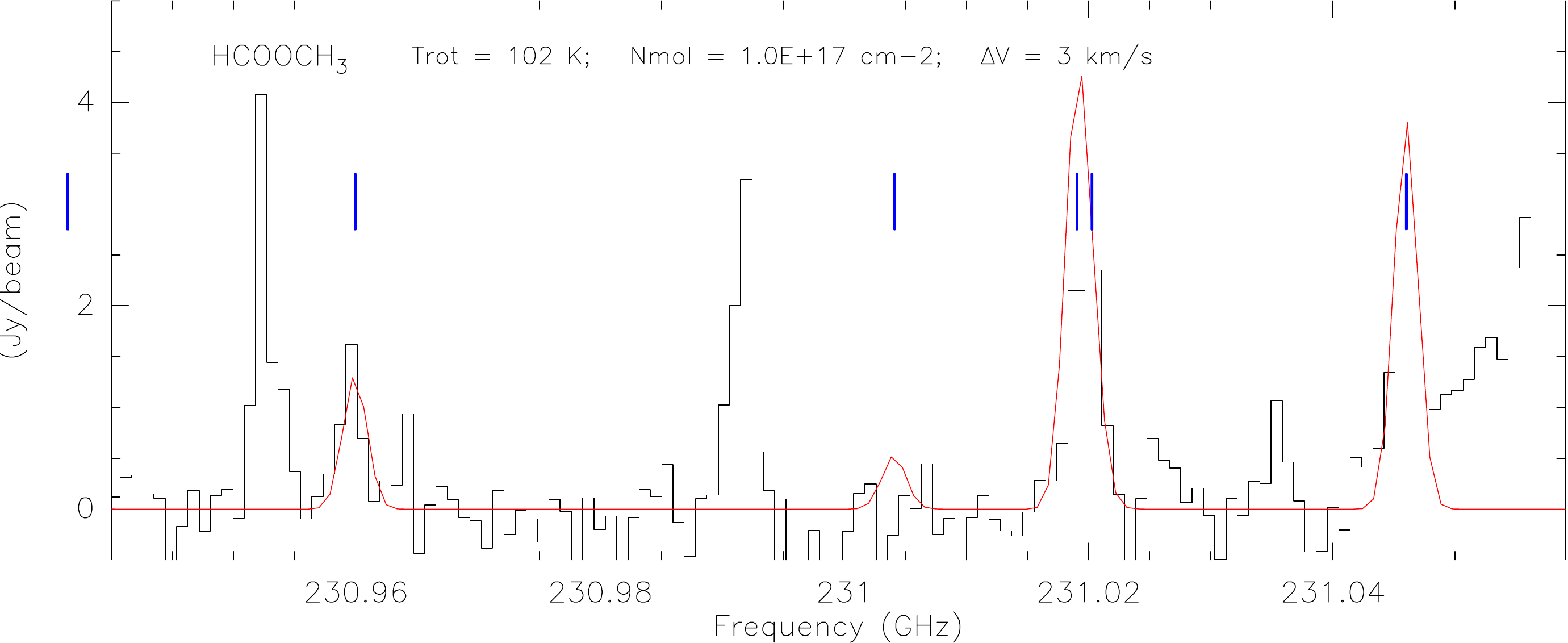}
&\includegraphics[width=9cm]{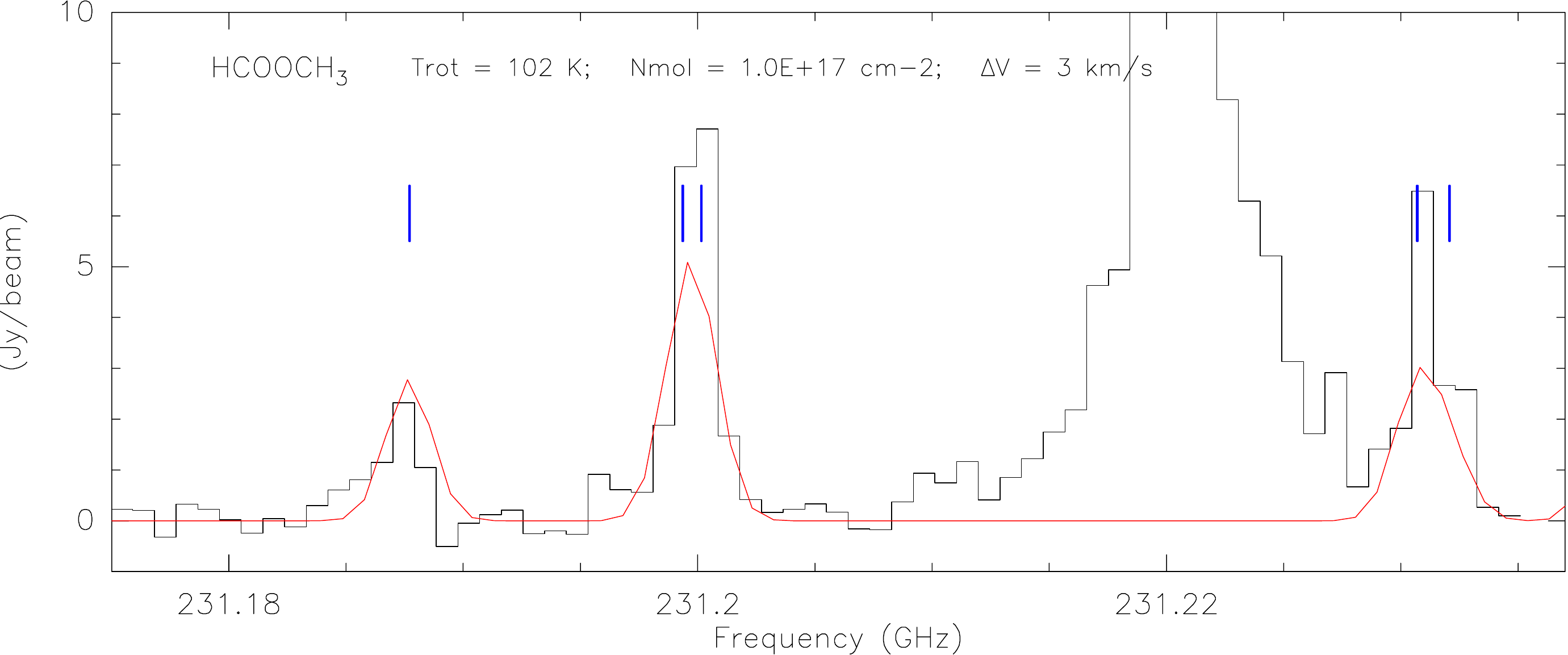}\\

\multicolumn{2}{l}{$\rm CH_3COCH_3$ @ SMA1   ($\rm T_{rot}$=126 K; $\rm N_{mol}=1.5\times10^{16}~cm^{-2}$; $\rm V_{lsr}=\rm 5.8~km\,s^{-1}$; $\rm \Delta V=3~km\,s^{-1}$)}\\
\includegraphics[width=9cm]{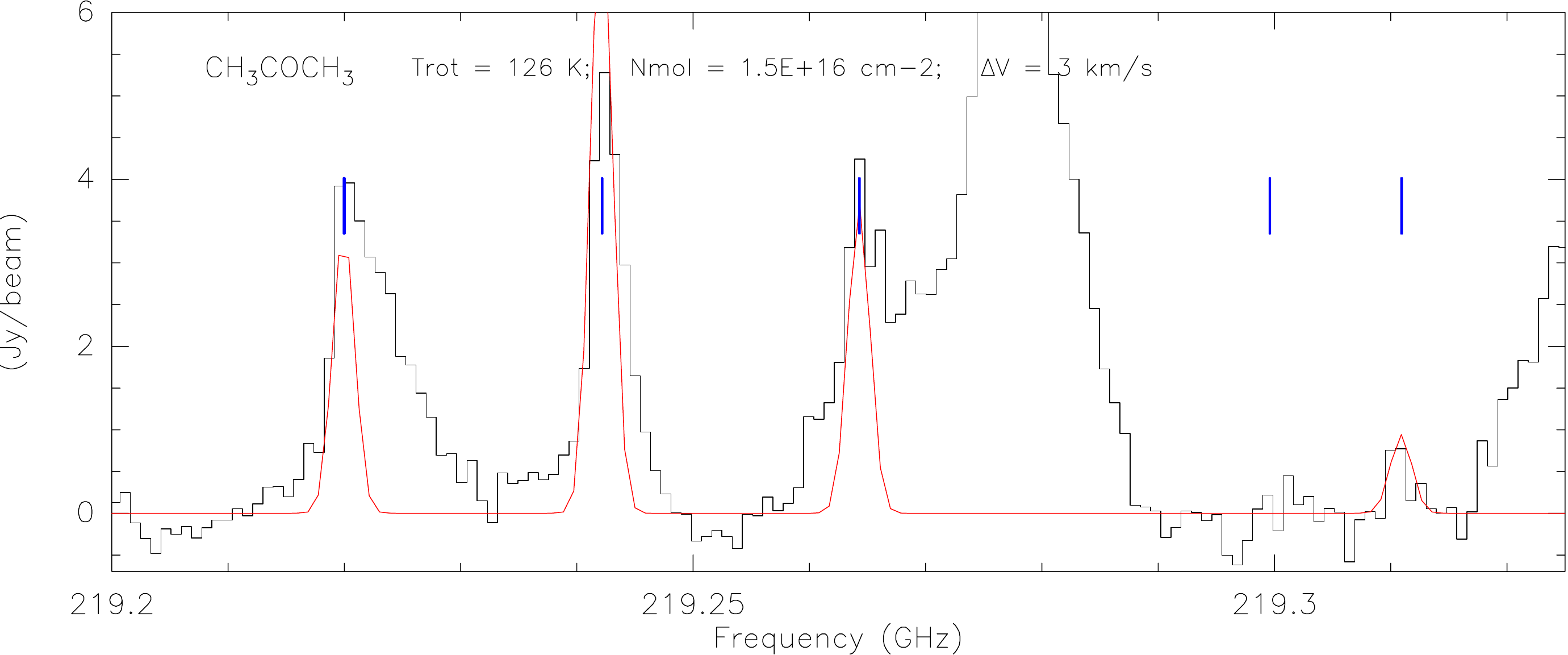}
&\includegraphics[width=9cm]{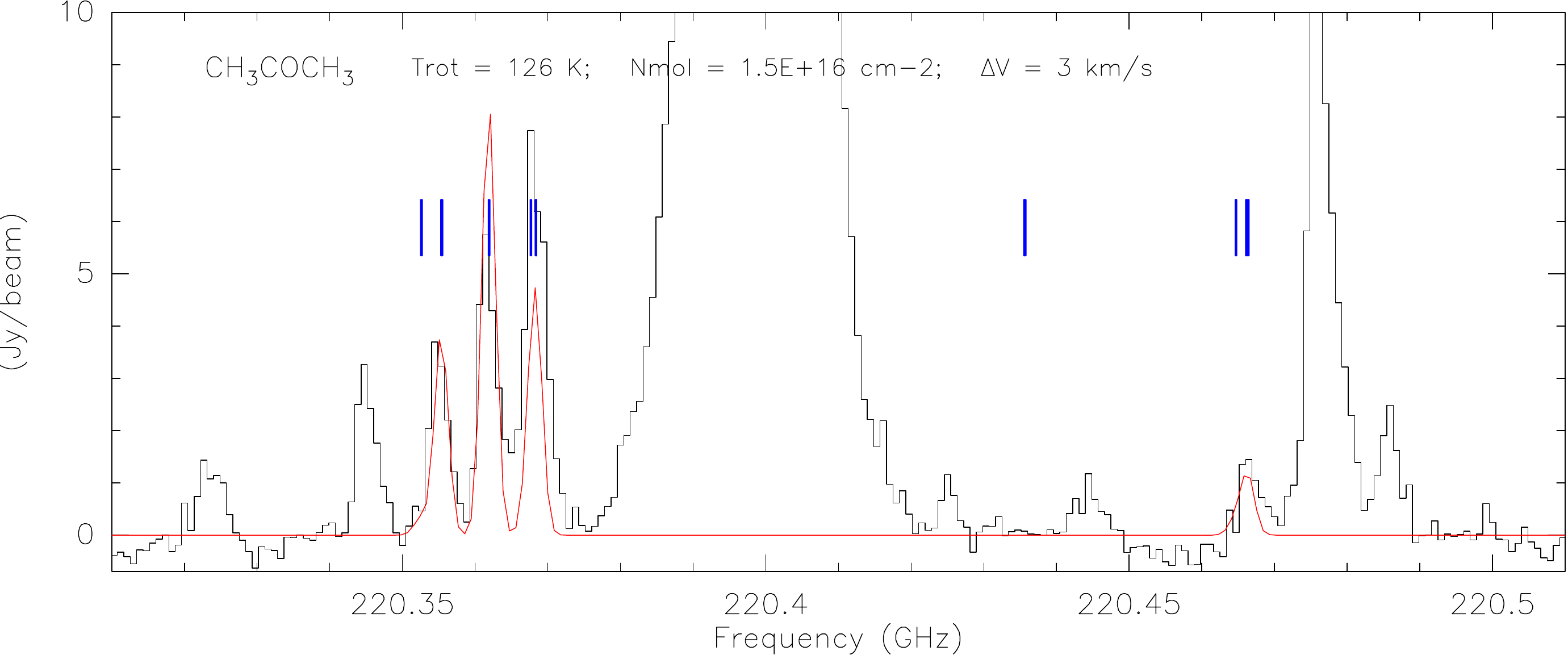}\\

\end{tabular}
\end{center}
\caption{(continued)}
\end{figure*}

\begin{figure*}[htb]
\ContinuedFloat
\begin{center}
\begin{tabular}{p{9cm} p{9cm}}
\includegraphics[width=9cm]{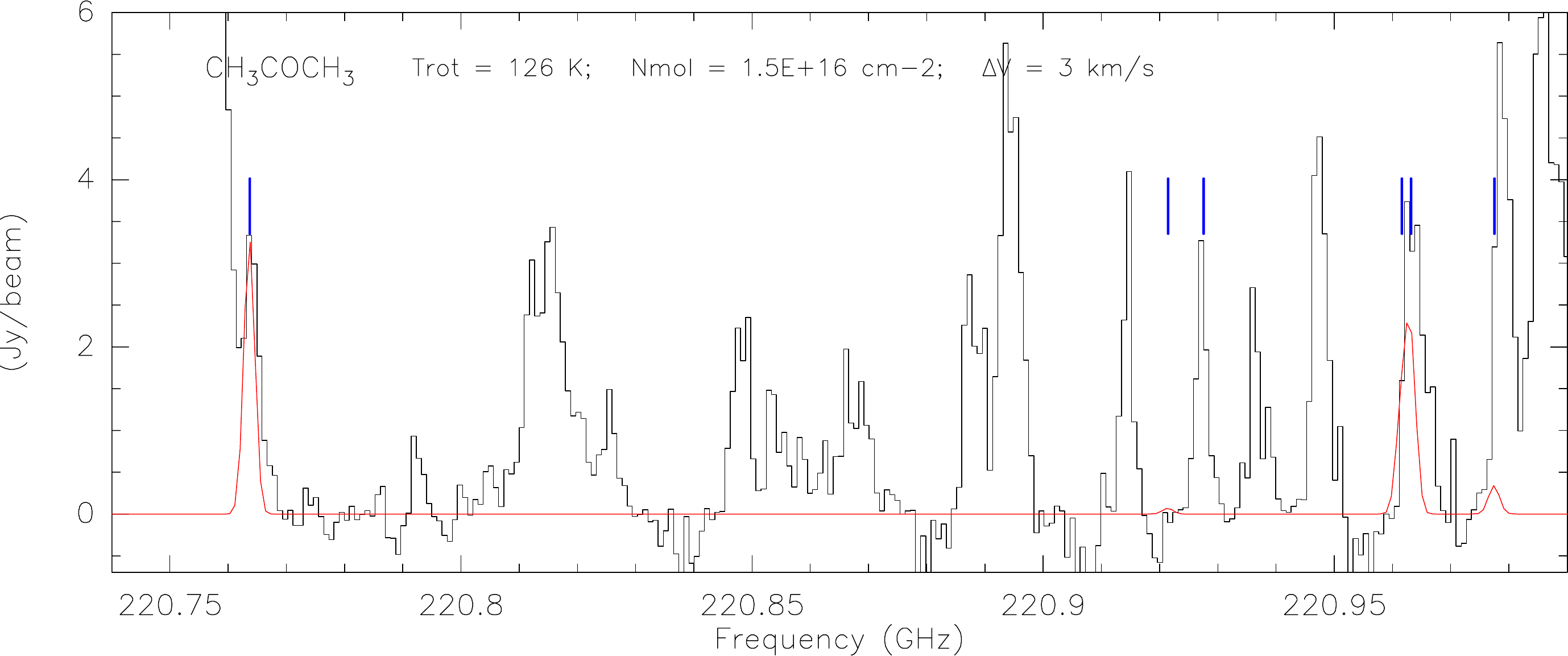}
&\includegraphics[width=9cm]{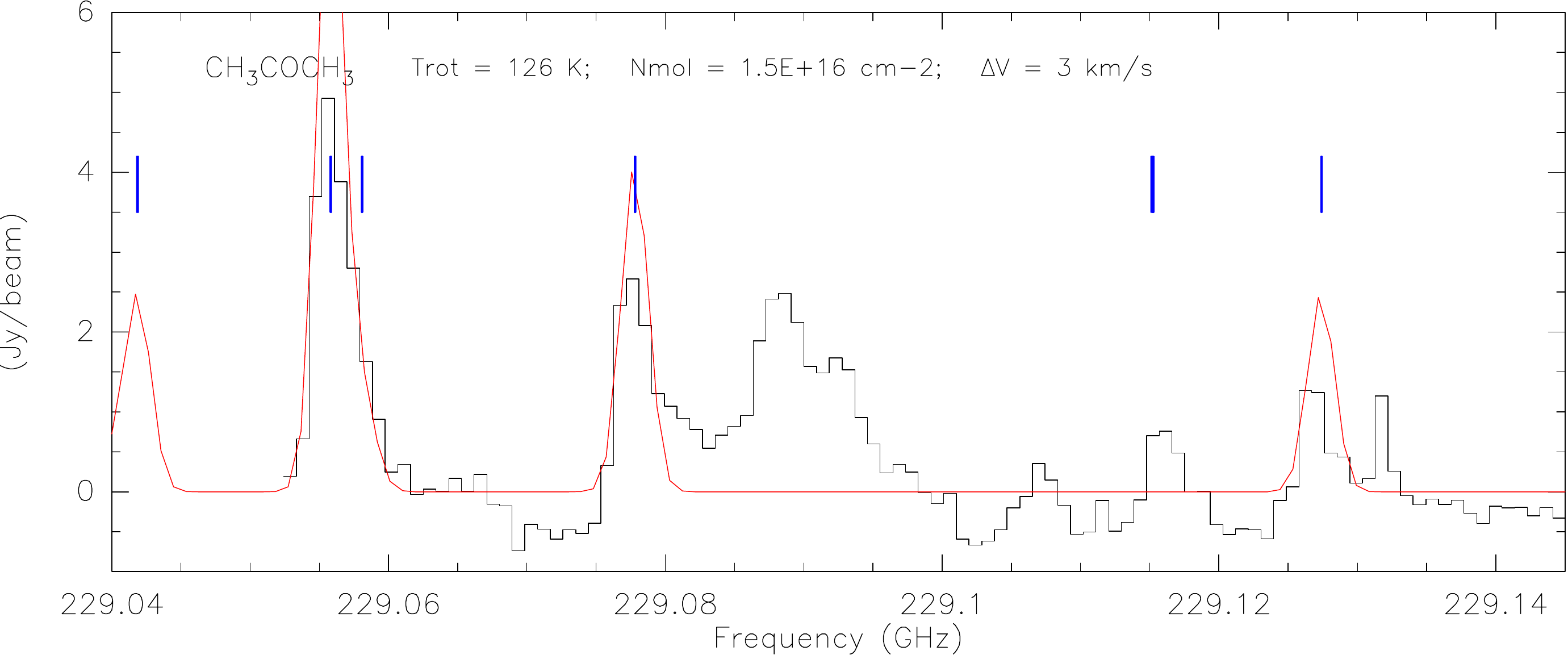}\\
\includegraphics[width=9cm]{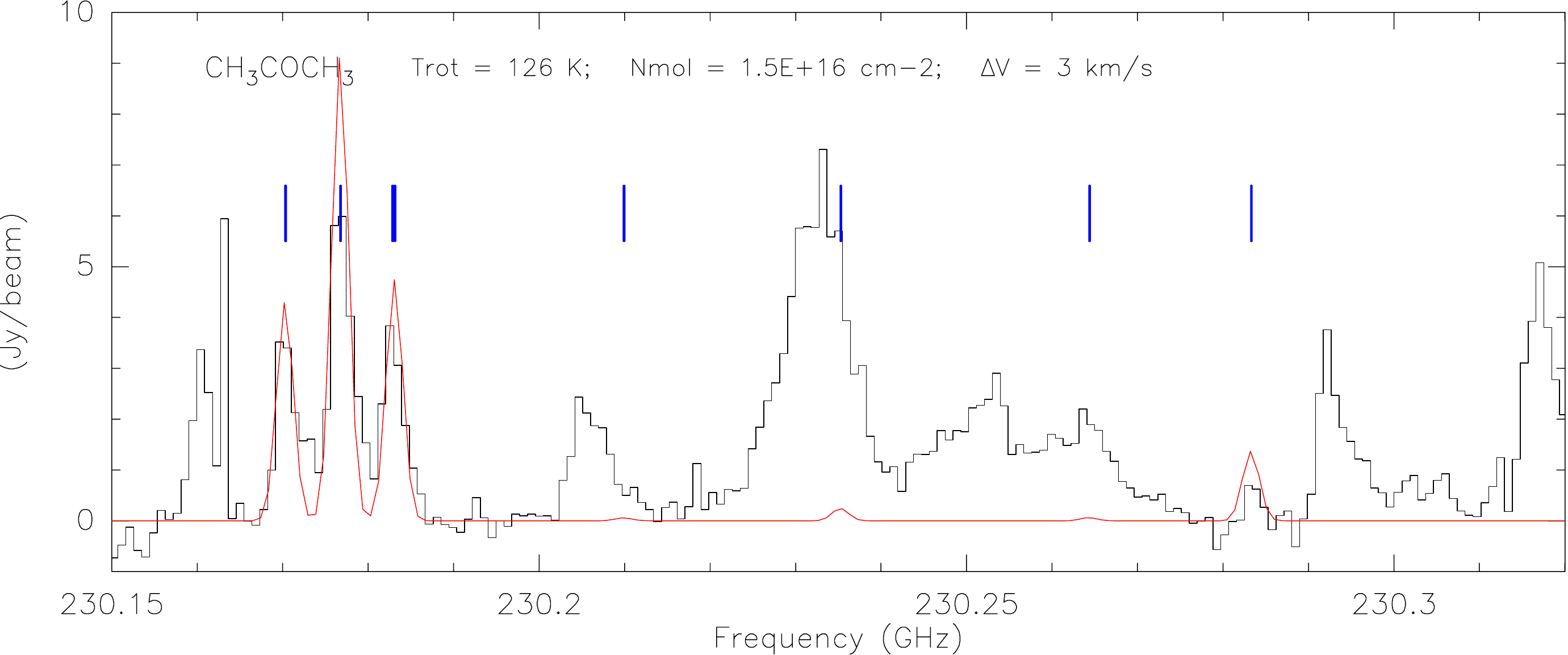}\\

\multicolumn{2}{l}{$\rm CH_3OCH_3$ @ mm3b ($\rm T_{rot}$=92 K; $\rm N_{mol}=2.0\times10^{17}~cm^{-2}$; $\rm V_{lsr}=\rm 5.6~km\,s^{-1}$; $\rm \Delta V=4~km\,s^{-1}$)}\\
\includegraphics[width=9cm]{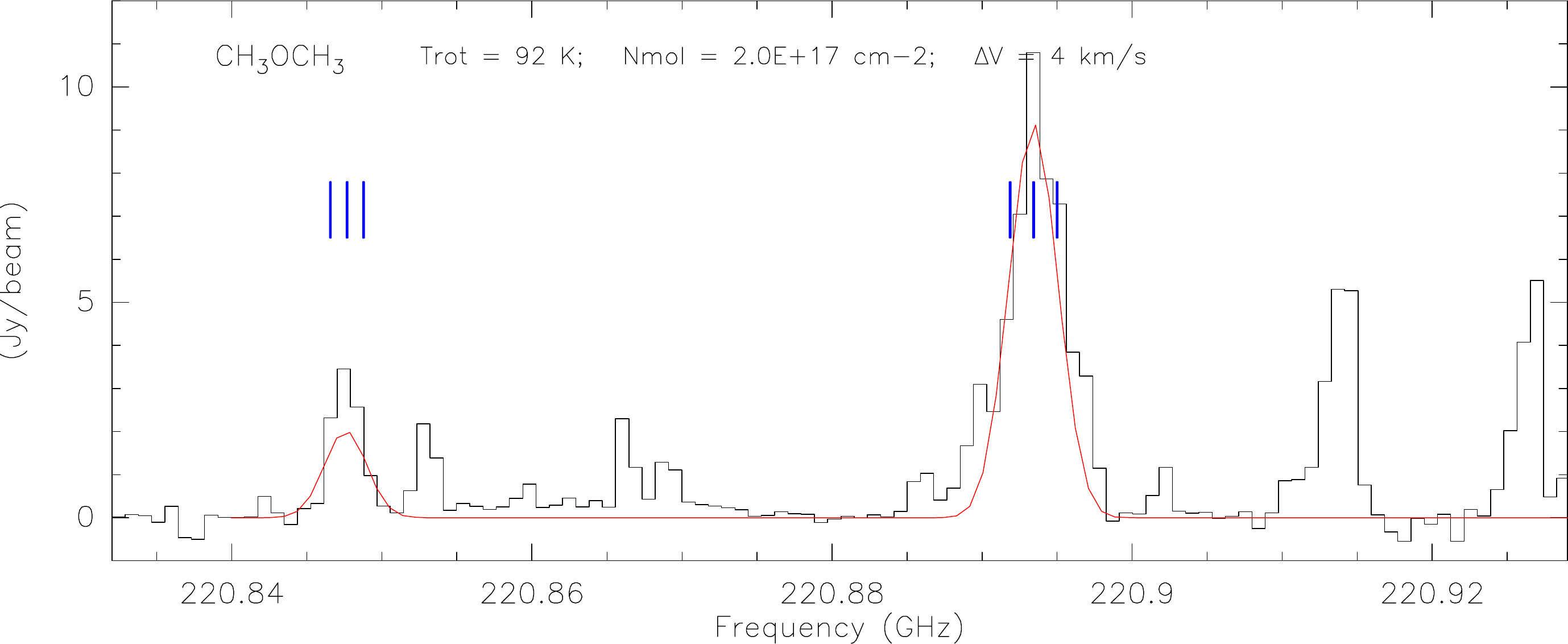}
&\includegraphics[width=9cm]{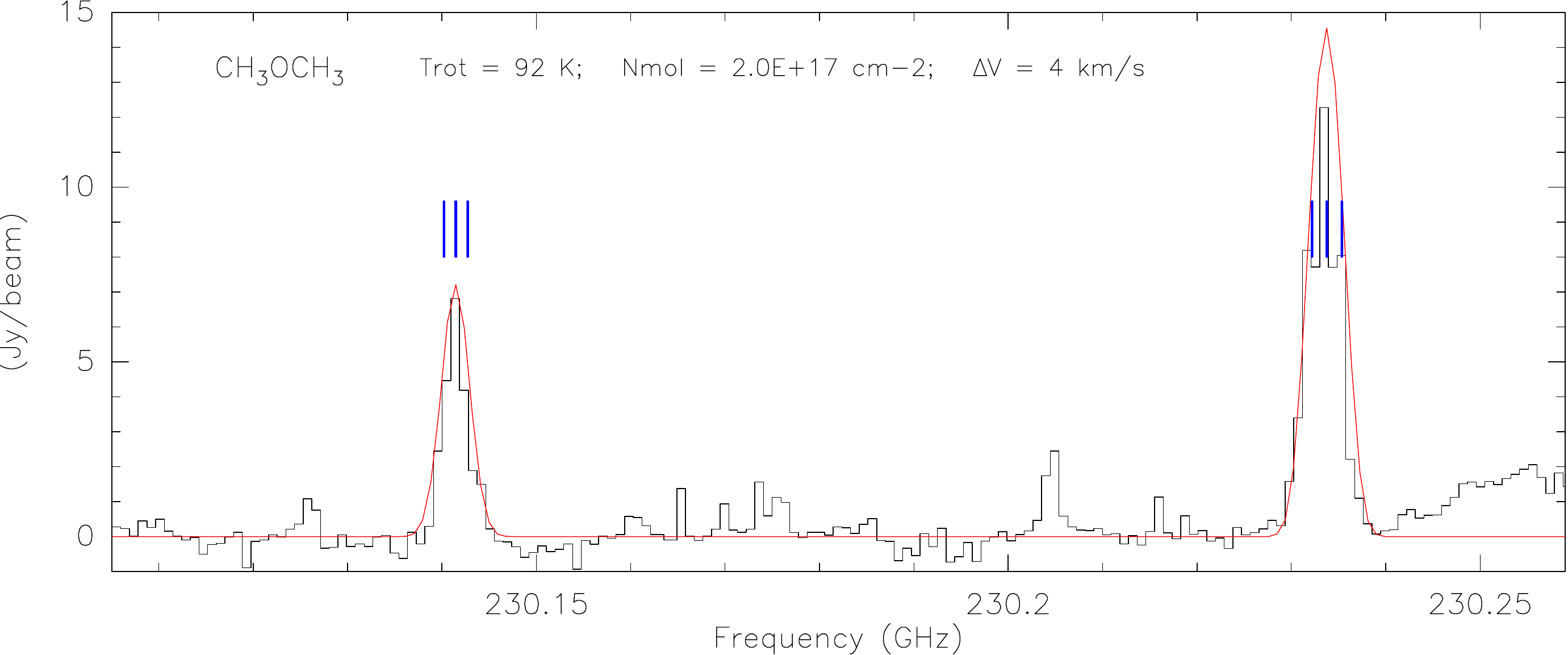}\\
\includegraphics[width=9cm]{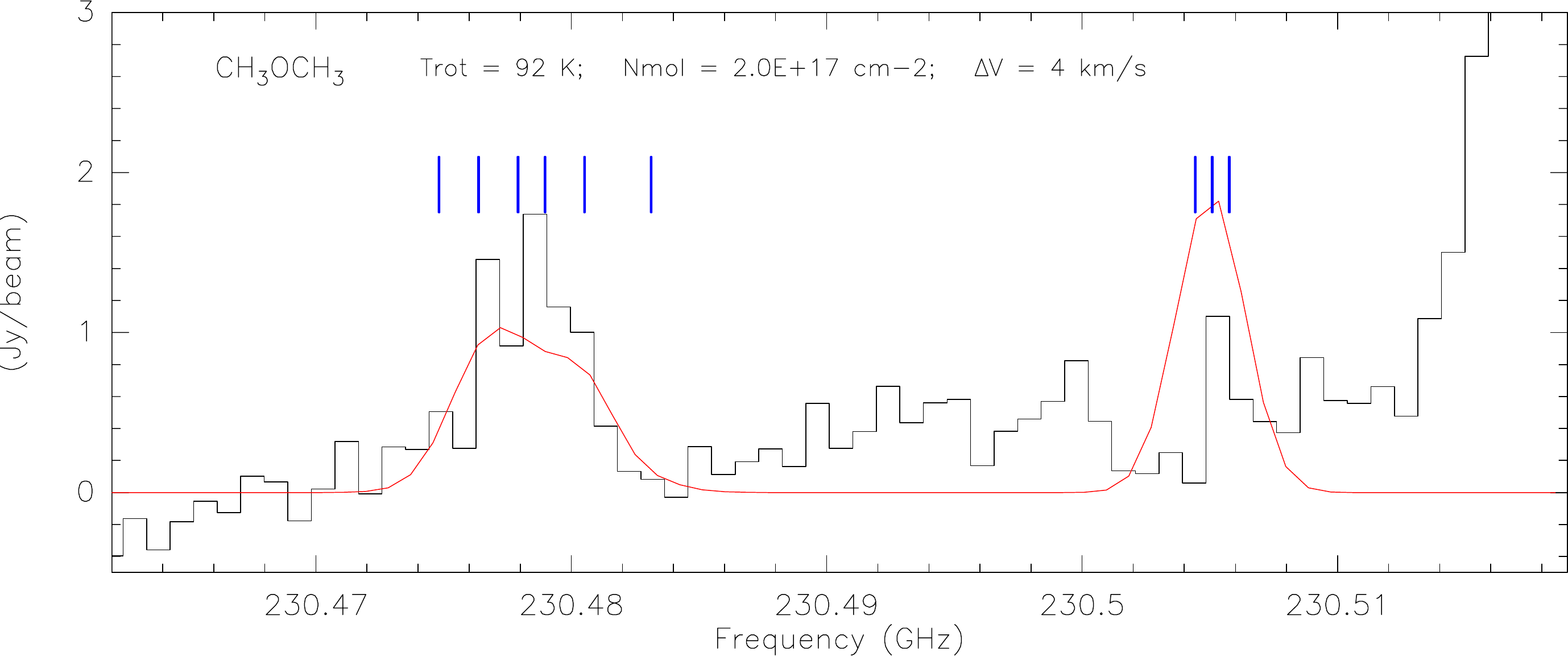}\\
\end{tabular}

\begin{tabular}{p{6cm} p{6cm} p{6cm}}
\multicolumn{3}{l}{$\rm CH_3CH_2OH$ @ near SMA1 \& mm3a ($\rm T_{rot}$=70 K; $\rm N_{mol}=1.5\times10^{16}~cm^{-2}$; $\rm V_{lsr}=\rm 8.4~km\,s^{-1}$; $\rm \Delta V=2.5~km\,s^{-1}$)}\\
\includegraphics[width=6cm]{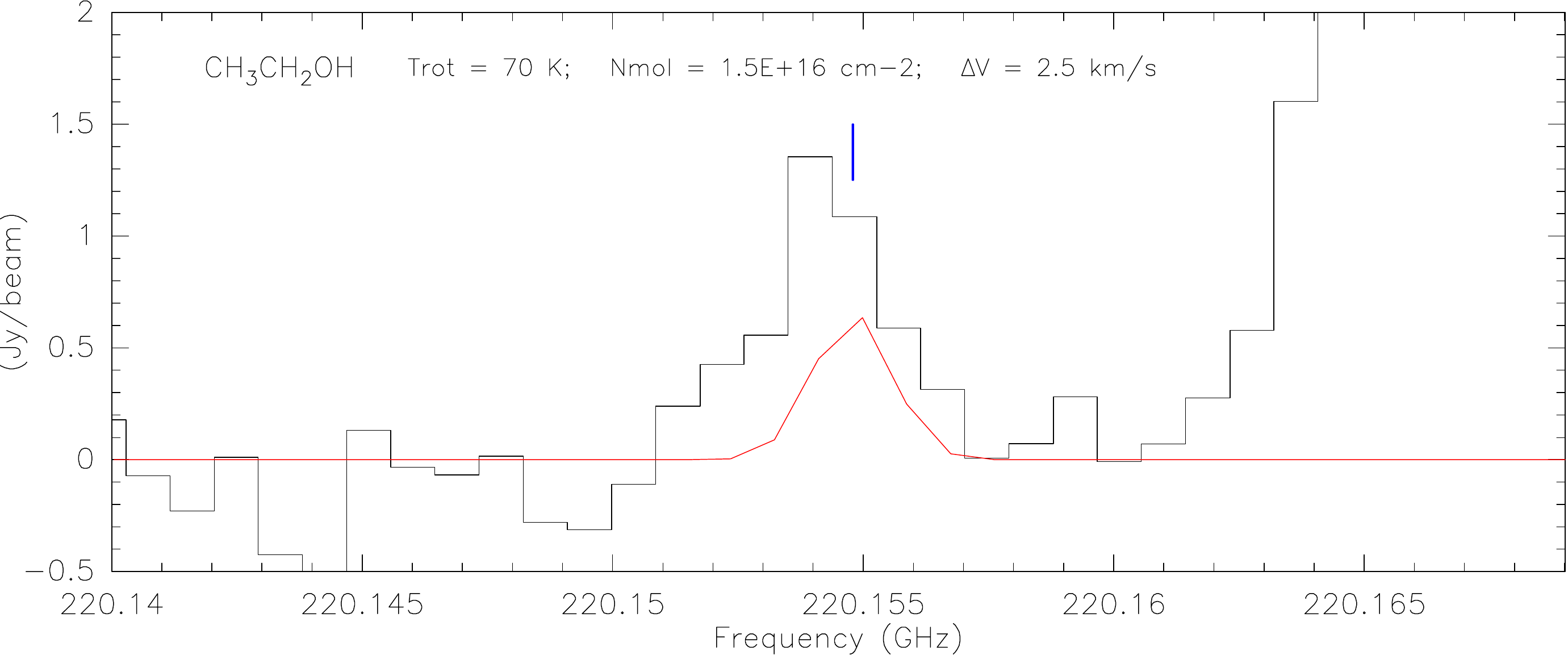}
&\includegraphics[width=6cm]{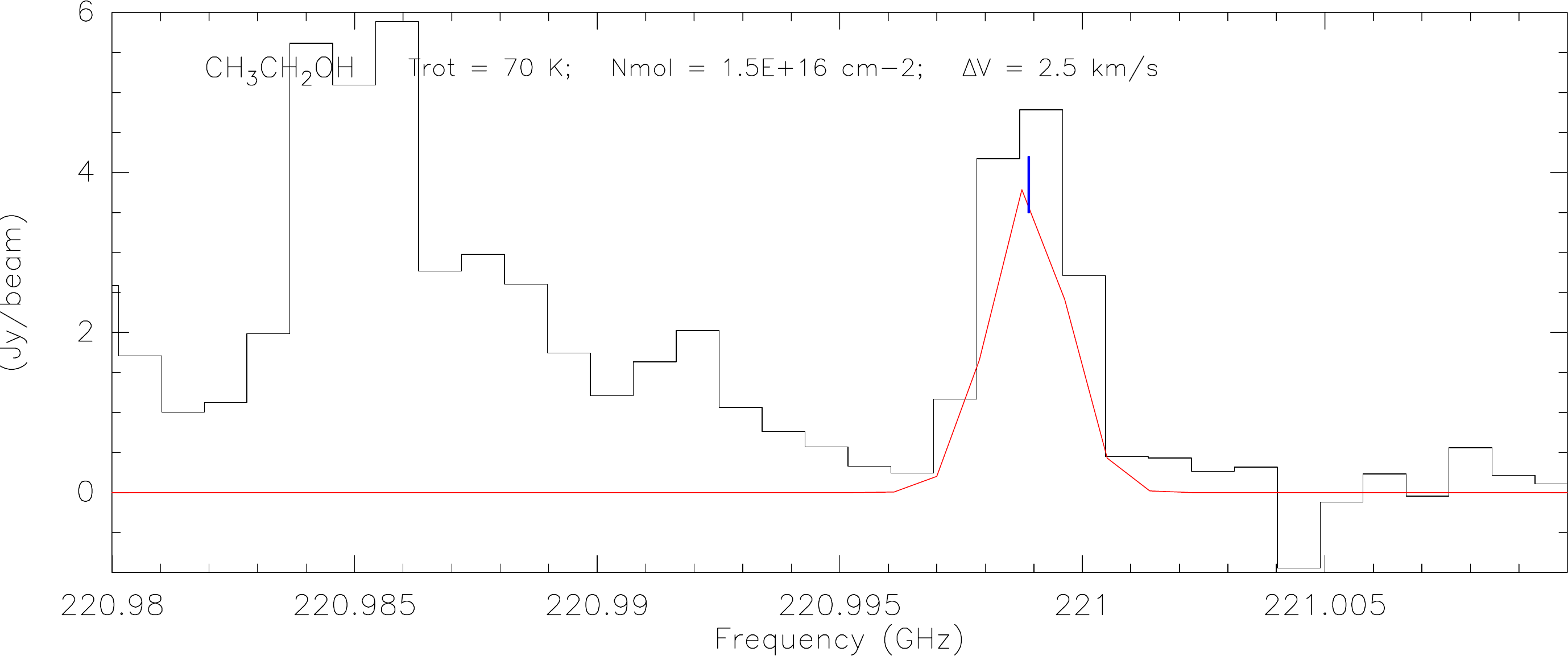}
&\includegraphics[width=6cm]{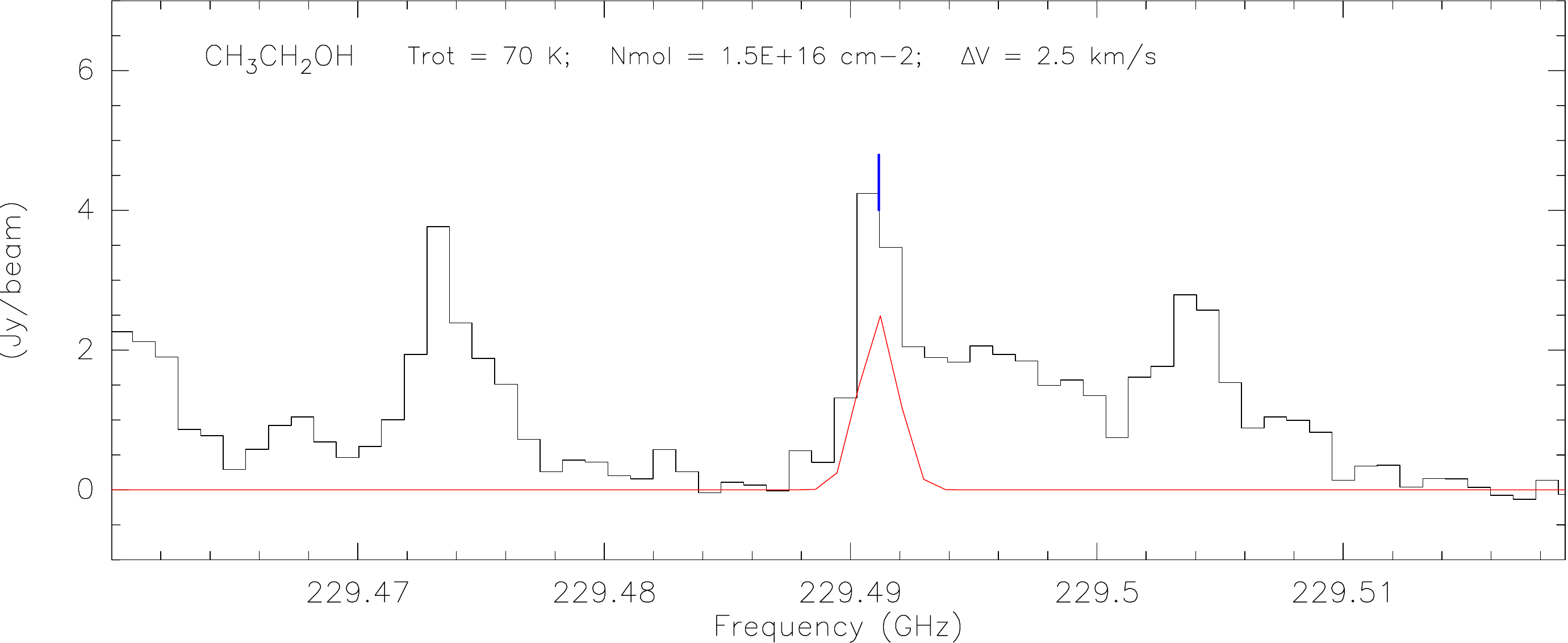}\\

\end{tabular}
\begin{tabular}{p{9cm} p{9cm}}
\includegraphics[width=9cm]{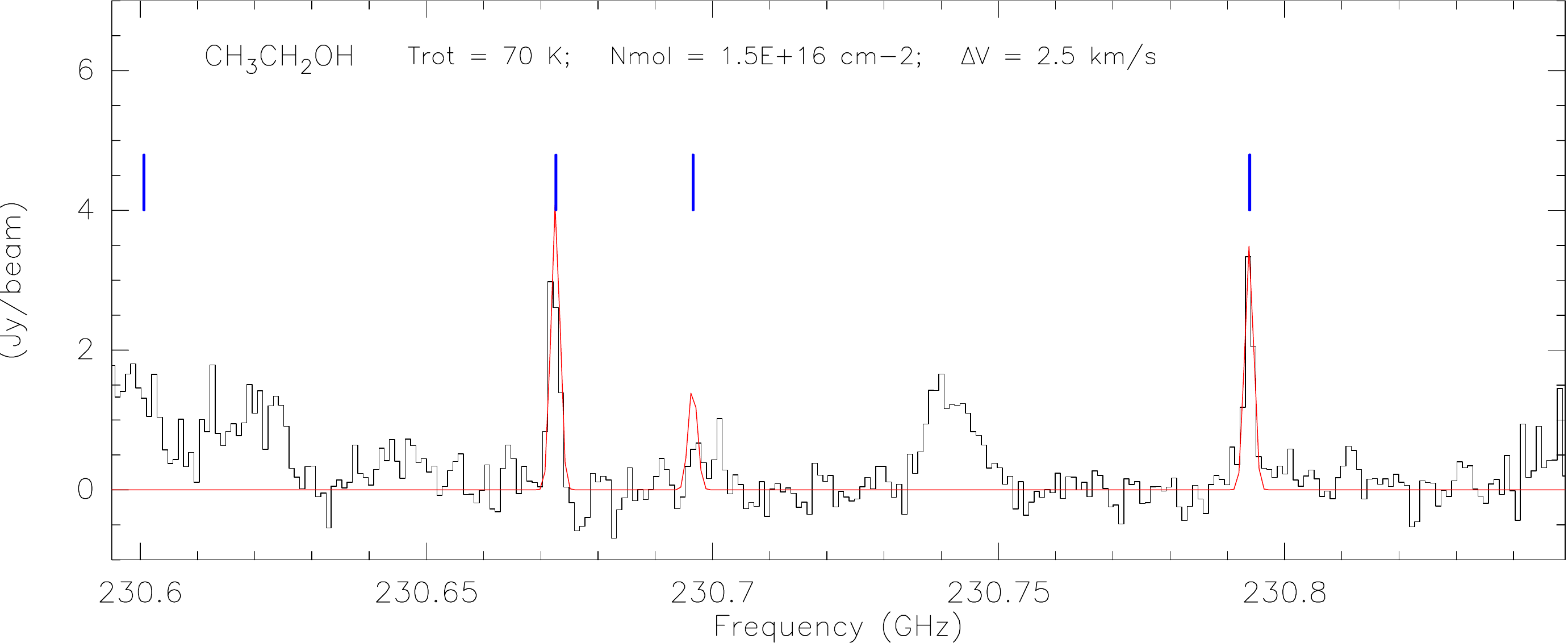}
&\includegraphics[width=9cm]{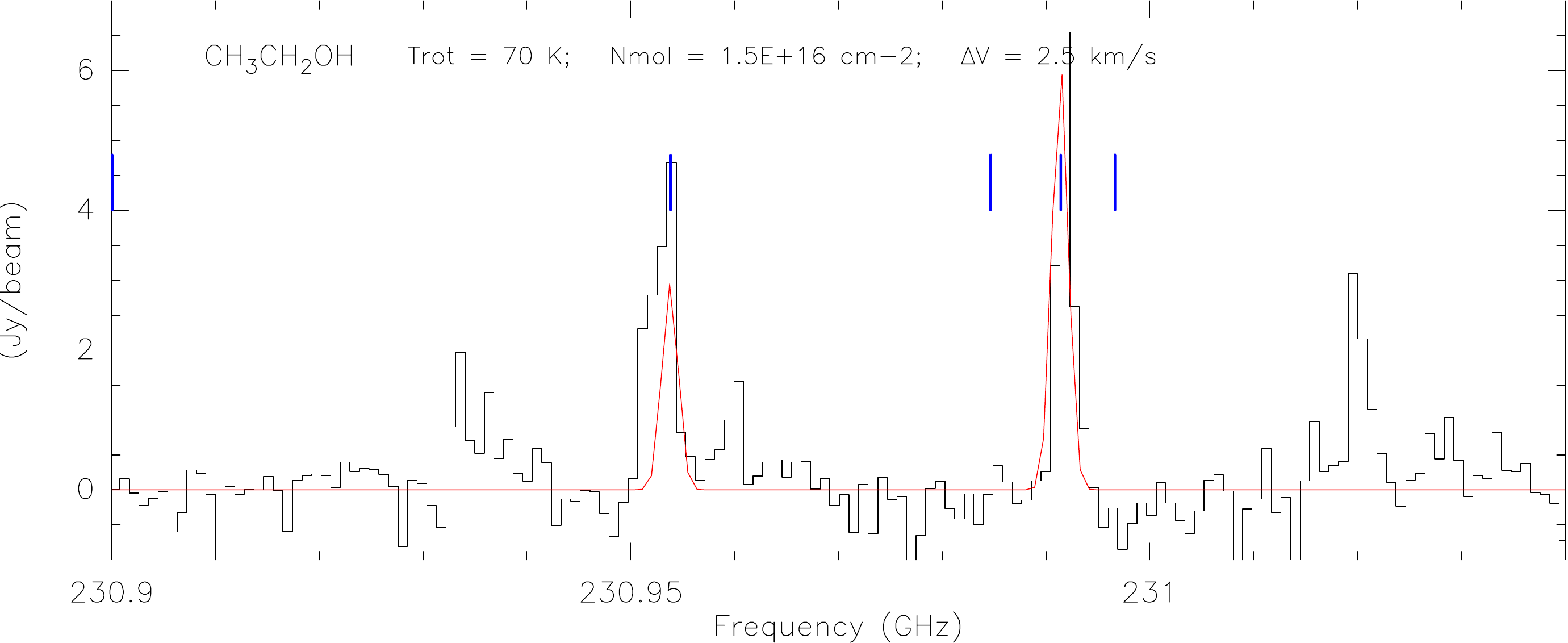}\\

\end{tabular}
\end{center}
\caption{(continued)}
\end{figure*}

\begin{figure*}[htb]
\ContinuedFloat
\begin{center}
\begin{tabular}{p{6cm} p{6cm} p{6cm}}
\multicolumn{3}{l}{$\rm CH_2CHCN$ @ North to hotcore ($\rm T_{rot}$=155 K; $\rm N_{mol}=5.0\times10^{15}~cm^{-2}$; $\rm V_{lsr}=\rm 4.9~km\,s^{-1}$; $\rm \Delta V=7~km\,s^{-1}$)}\\
\includegraphics[width=6cm]{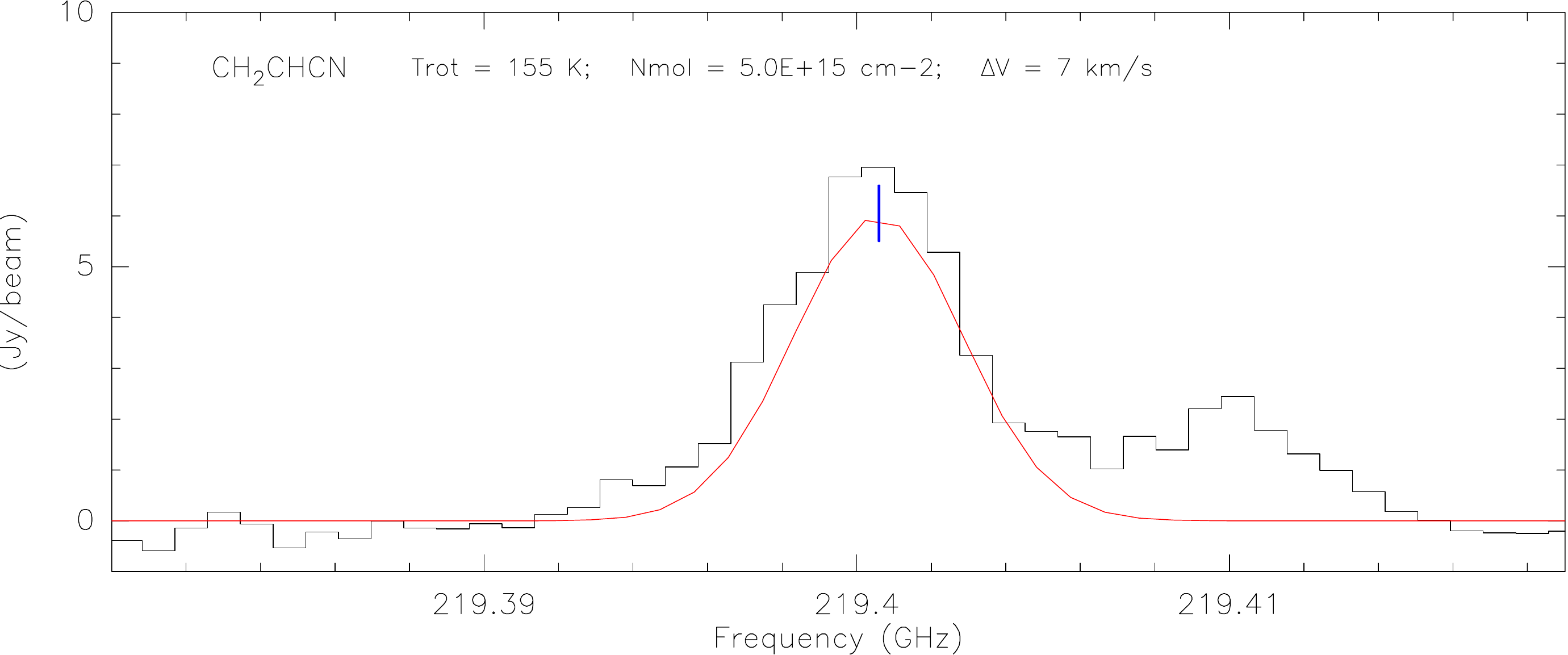}
&\includegraphics[width=6cm]{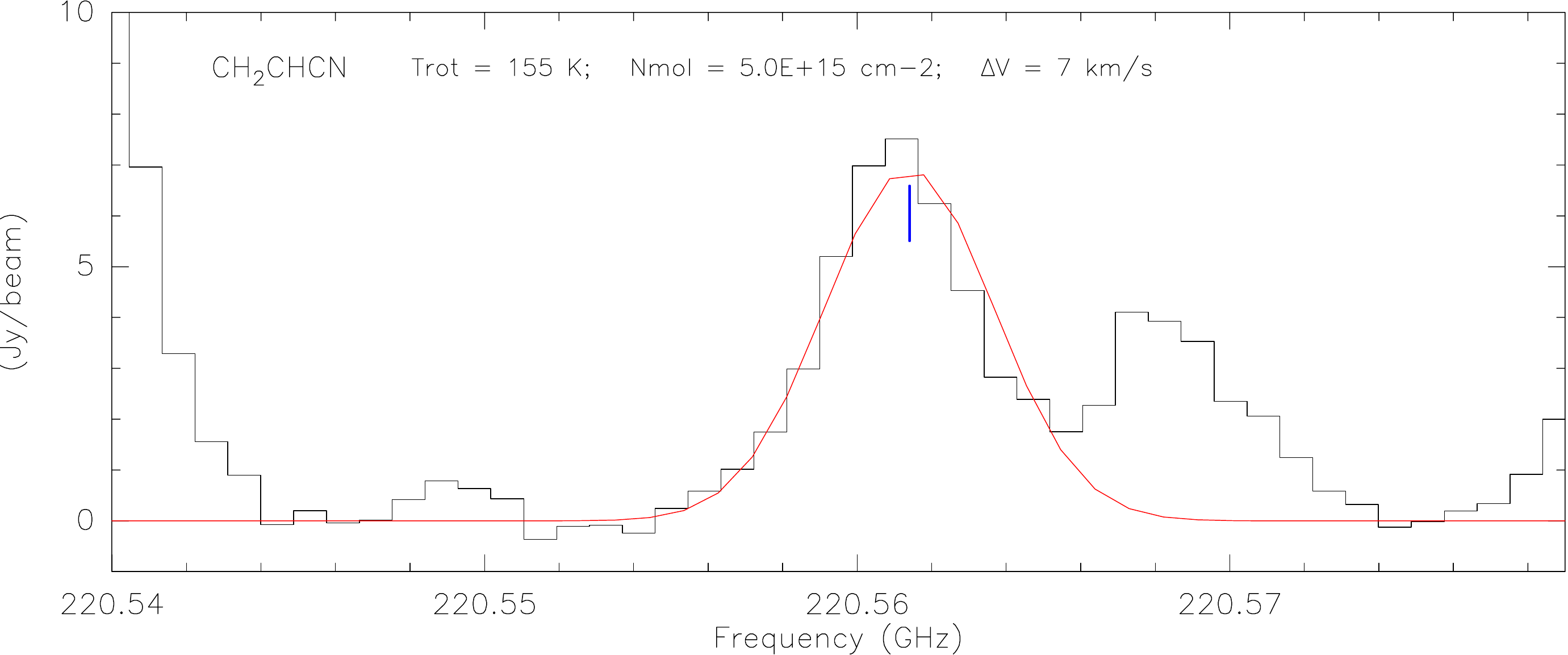}
&\includegraphics[width=6cm]{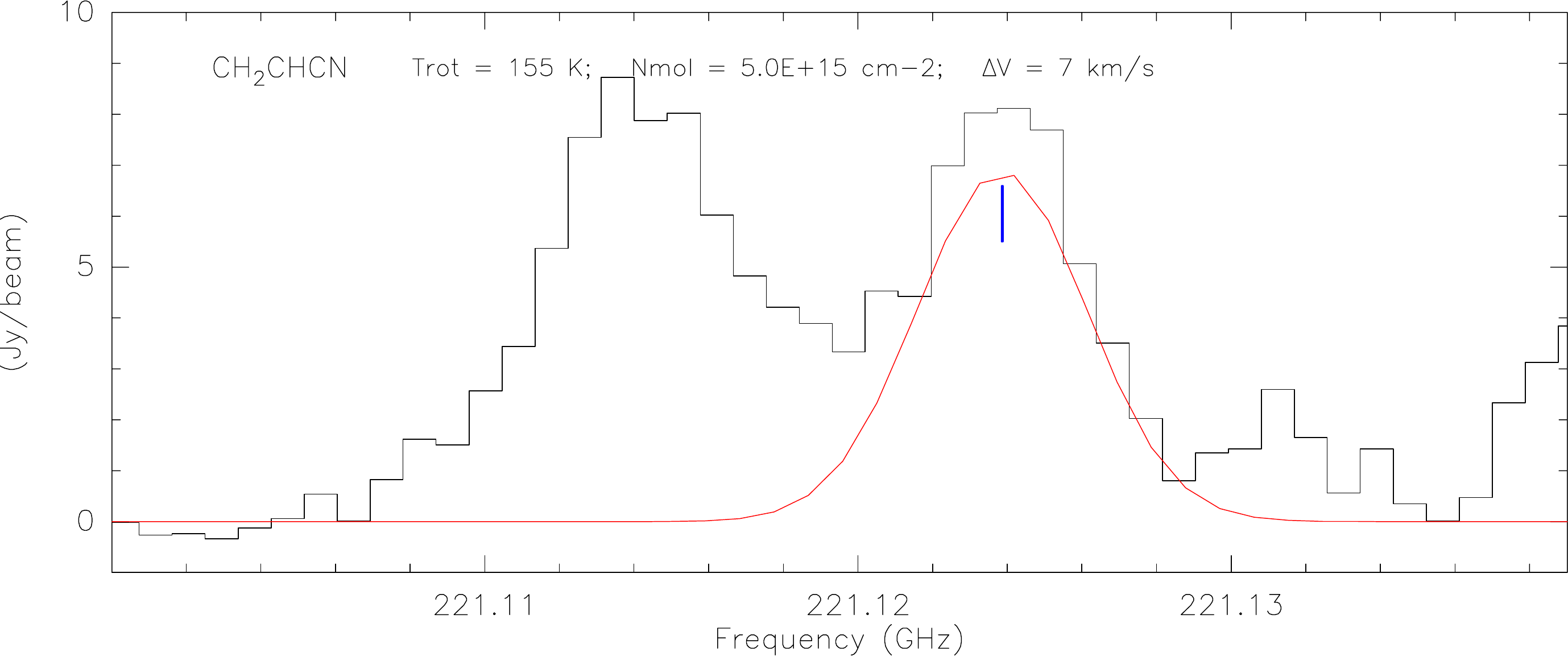}
\end{tabular}

\begin{tabular}{p{9cm} p{9cm} }
\includegraphics[width=9cm]{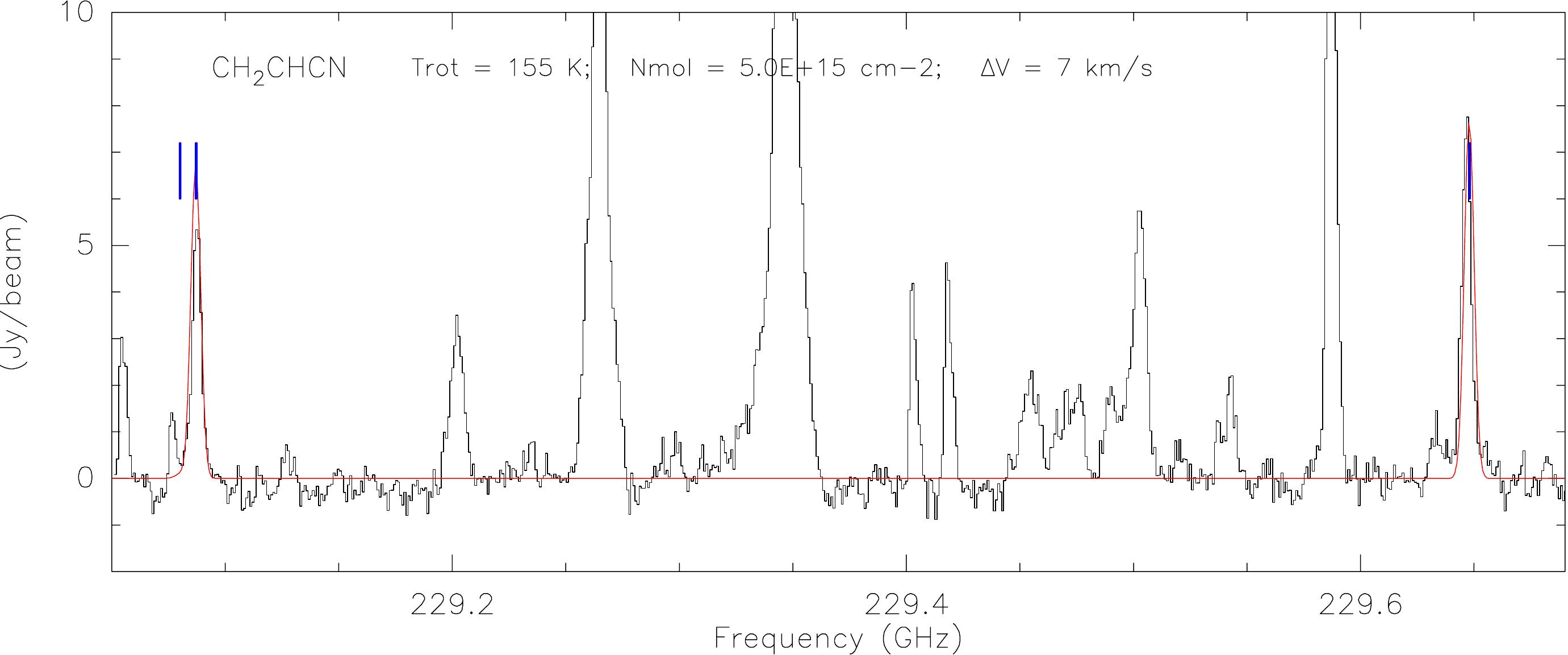}
&\includegraphics[width=9cm]{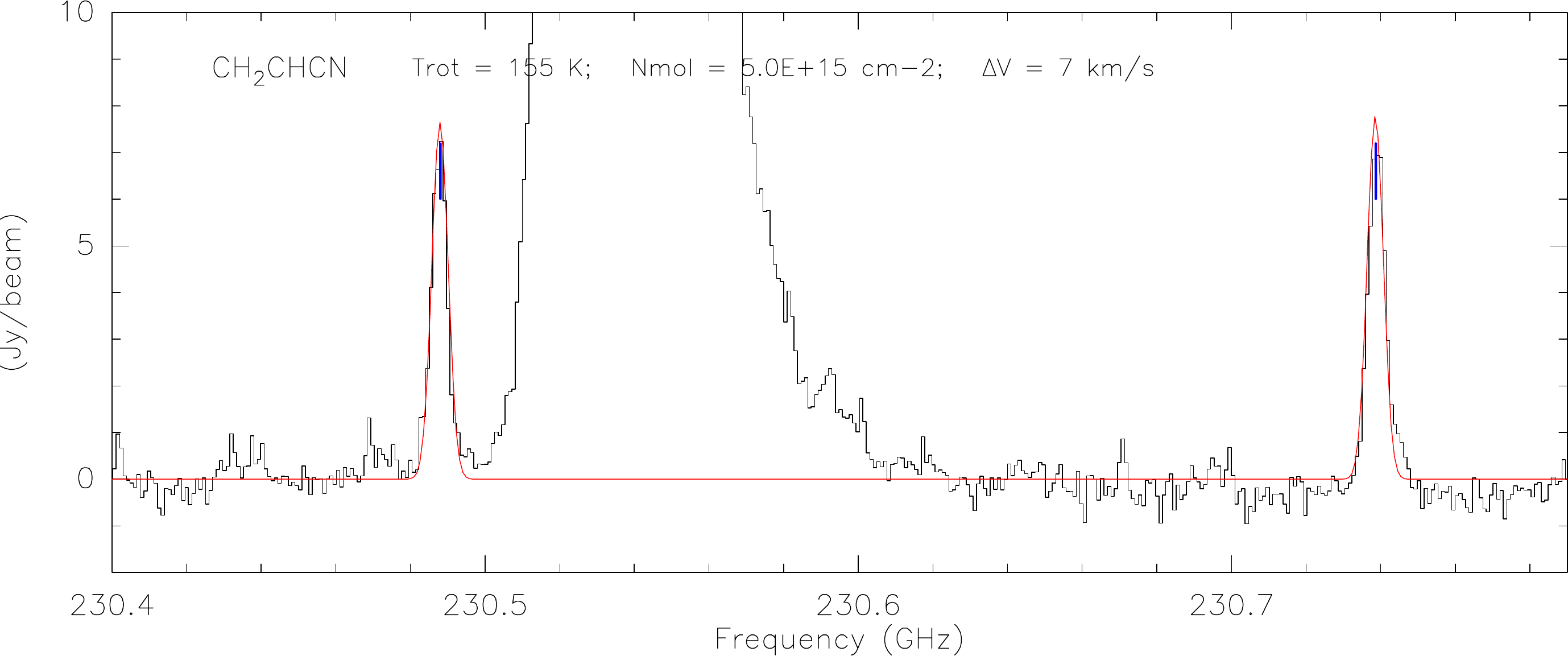}\\
\end{tabular}
%\end{center}
%\end{figure*}
\begin{tabular}{p{6cm} p{6cm} p{6cm}}
\multicolumn{3}{l}{$\rm CH_3CH_2CN$ @ hotcore ($\rm T_{rot}$=155 K; $\rm N_{mol}=1.5\times10^{16}~cm^{-2}$; $\rm V_{lsr}=\rm 5.1~km\,s^{-1}$; $\rm \Delta V=9~km\,s^{-1}$)}\\
\includegraphics[width=6cm]{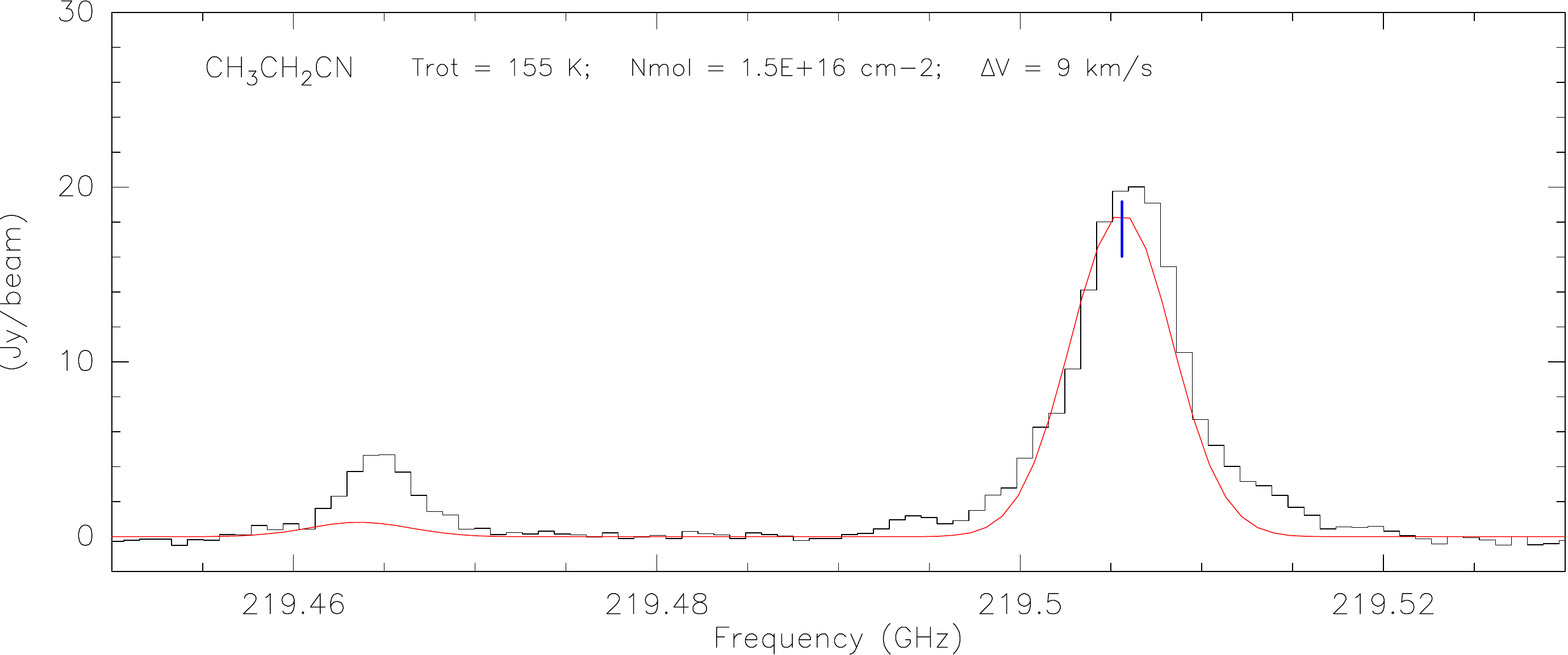}
&\includegraphics[width=6cm]{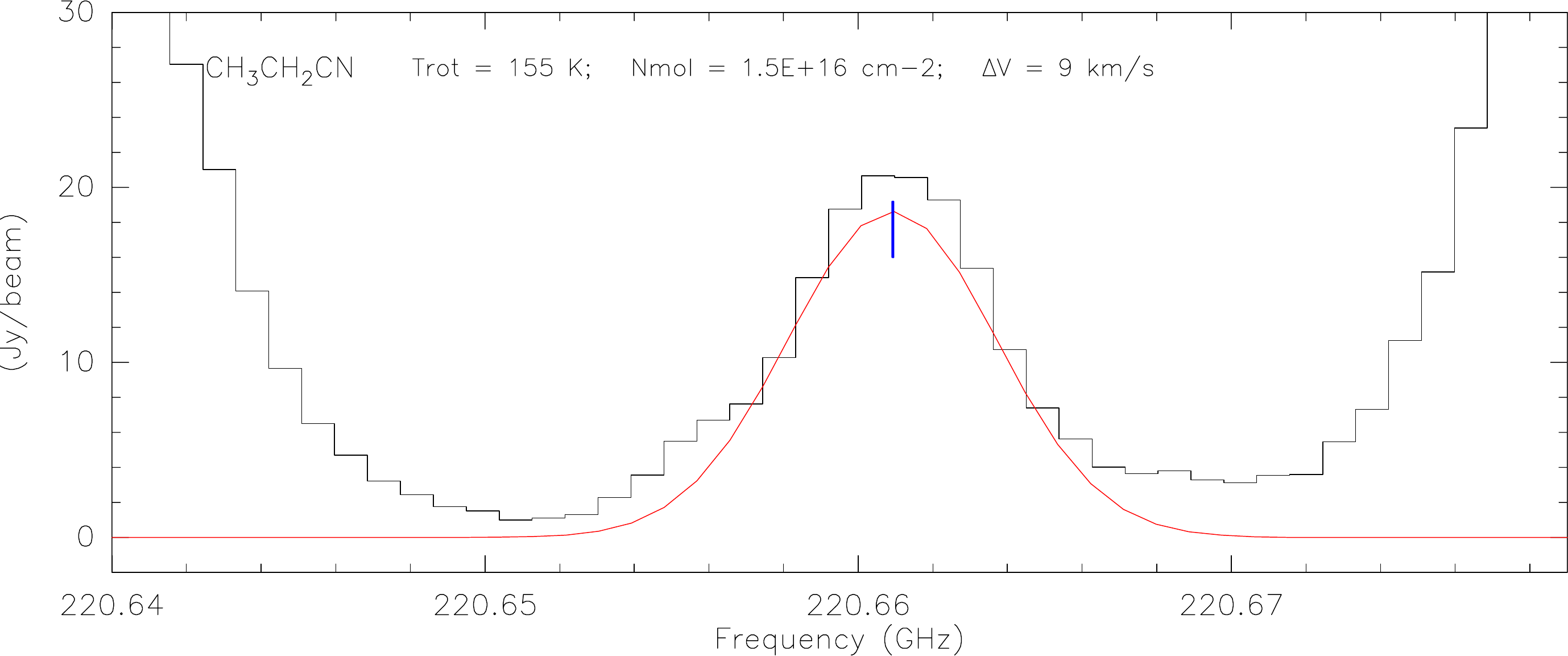}
&\includegraphics[width=6cm]{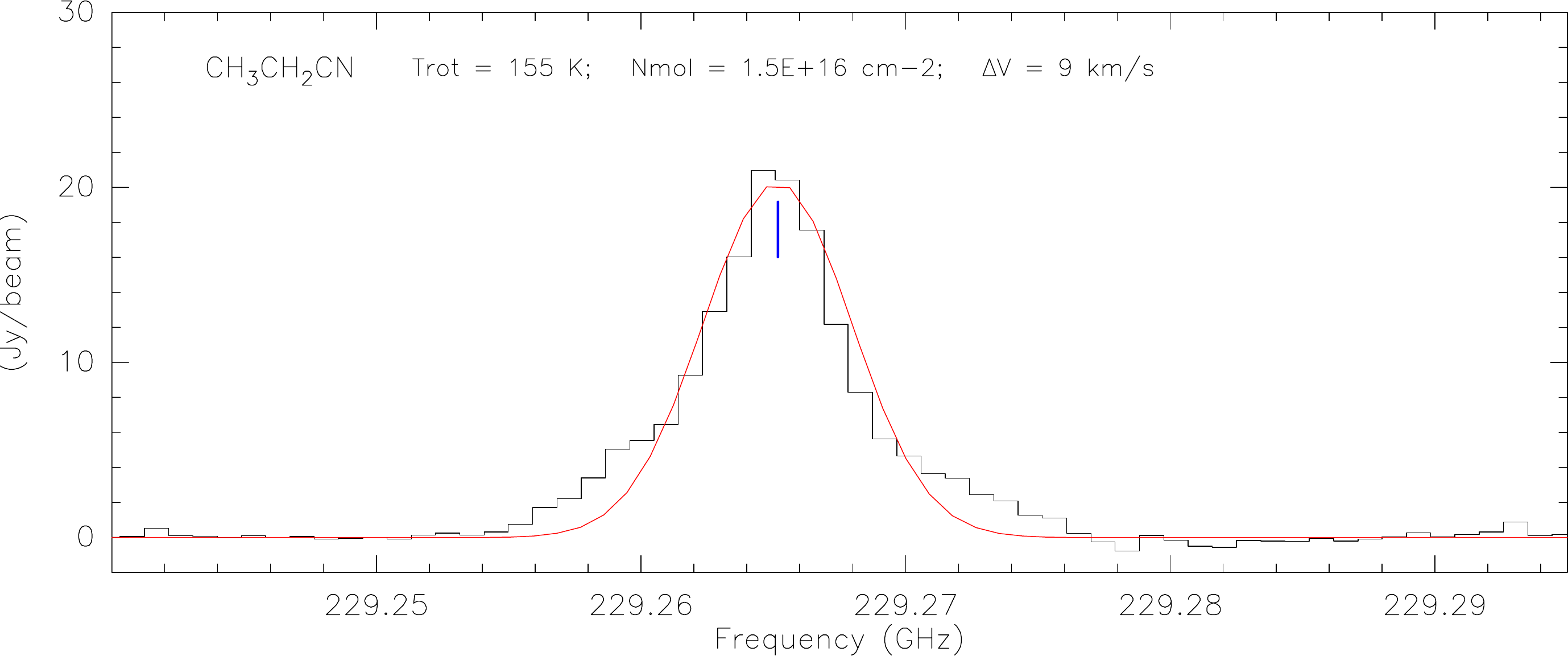}\\
\end{tabular}

%\begin{figure*}[htb]
%\begin{center}
\begin{tabular}{p{9cm} p{9cm}}
\multicolumn{2}{l}{$\rm CH_3CHO$ @ mm3b ($\rm T_{rot}$=102 K; $\rm N_{mol}=1.3\times10^{16}~cm^{-2}$; $\rm V_{lsr}=\rm 8.0~km\,s^{-1}$; $\rm \Delta V=5~km\,s^{-1}$)}\\
\includegraphics[width=9cm]{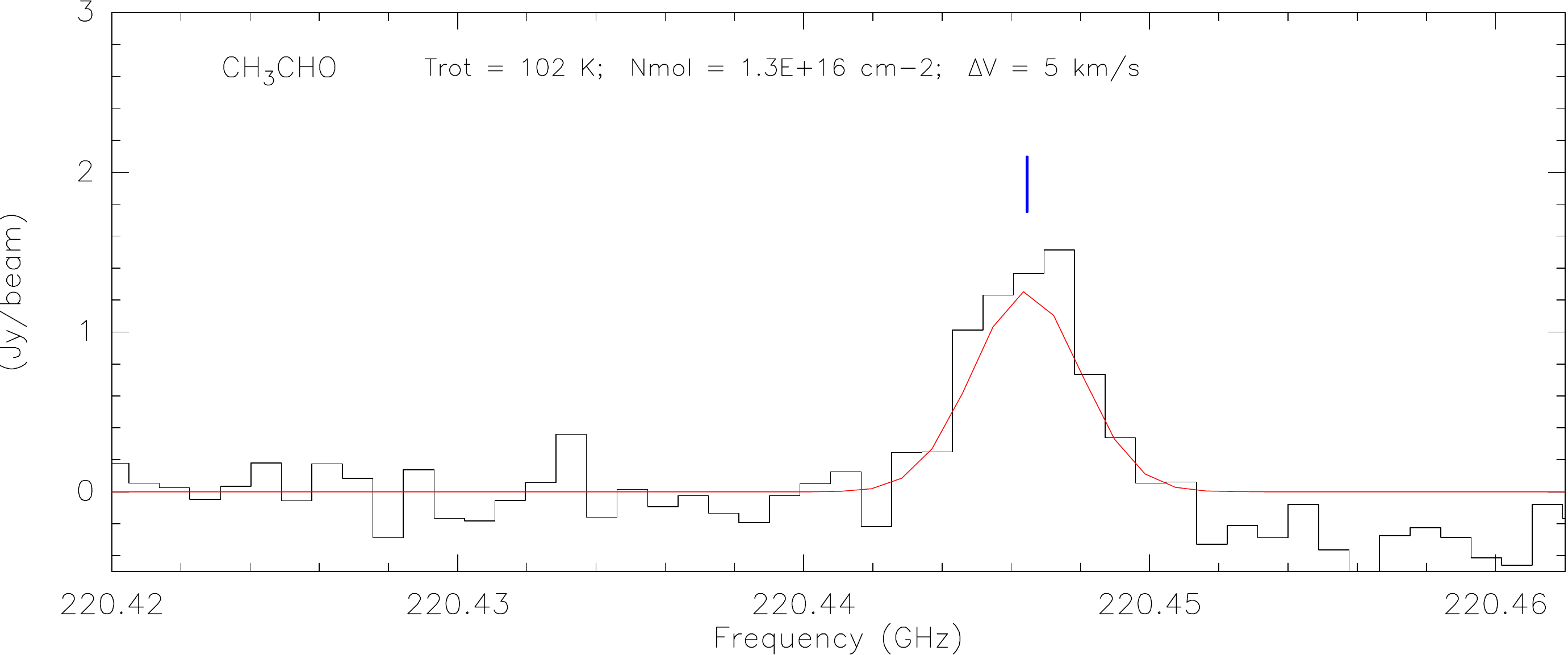}
&\includegraphics[width=9cm]{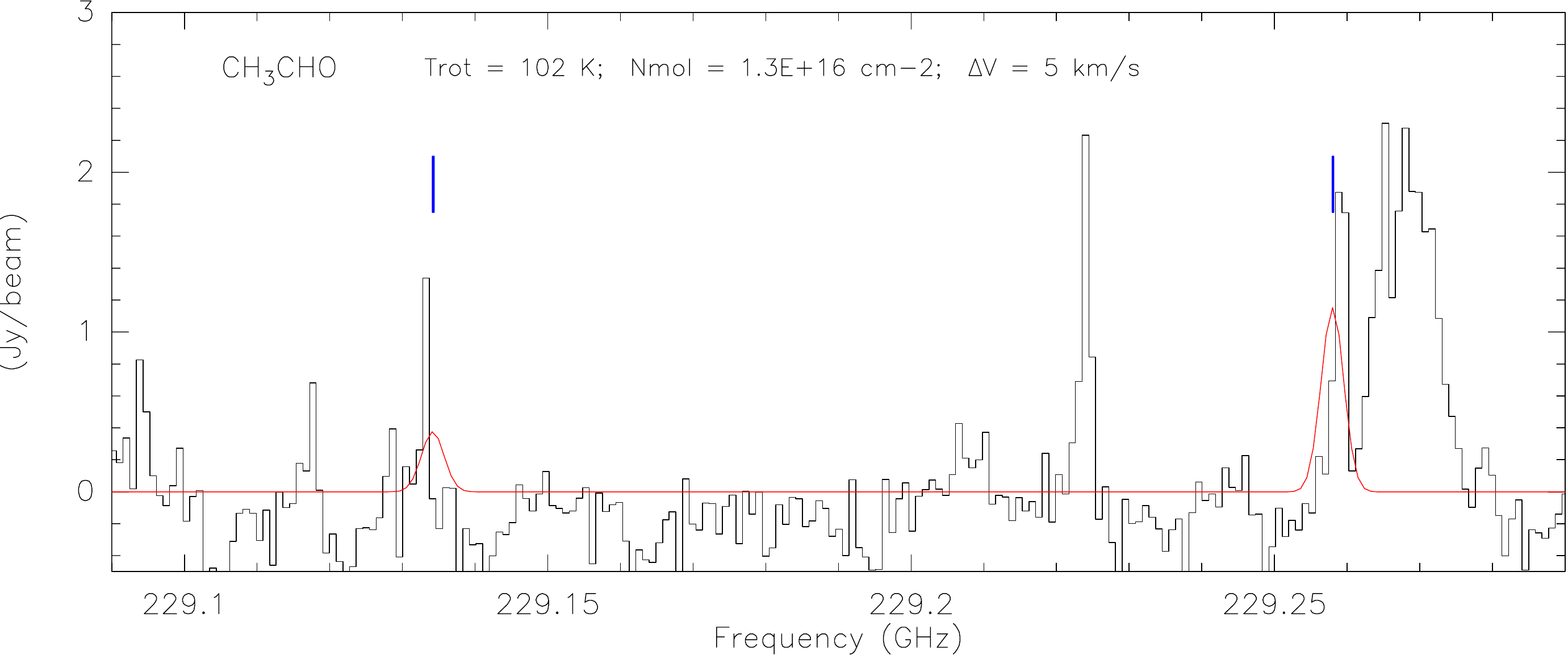}\\

\multicolumn{2}{l}{$\rm C_6H$ @ SMA1 ($\rm T_{rot}$=126 K; $\rm N_{mol}=1.0\times10^{15}~cm^{-2}$; $\rm V_{lsr}=\rm 4.0~km\,s^{-1}$; $\rm \Delta V=4~km\,s^{-1}$)}\\
\includegraphics[width=9cm]{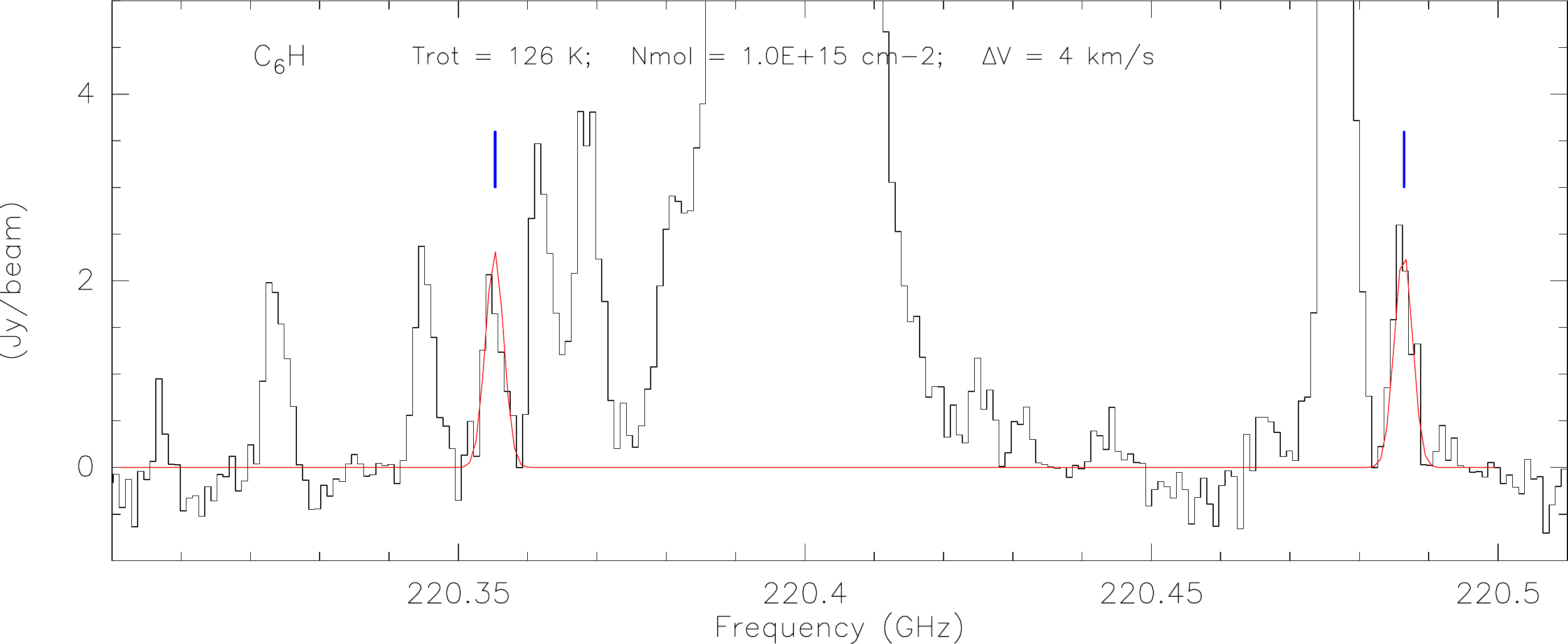}
&\includegraphics[width=9cm]{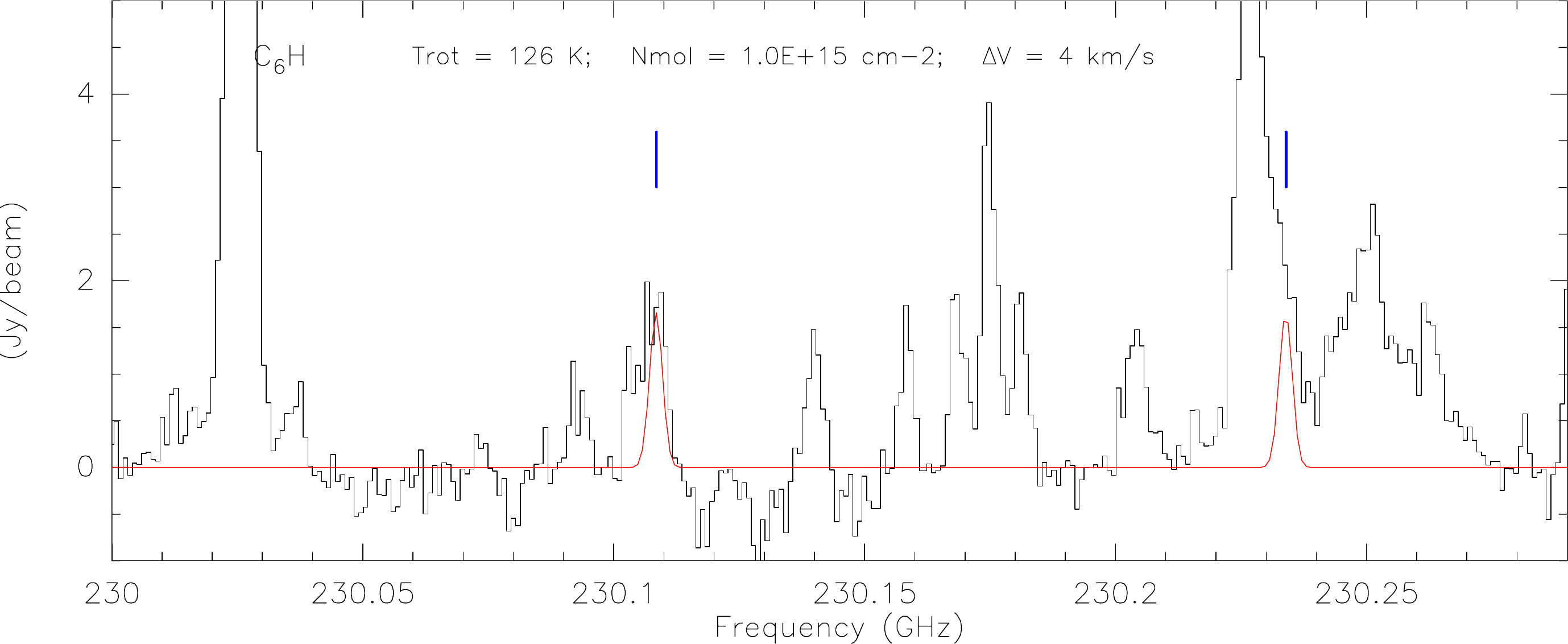}\\

\multicolumn{2}{l}{$\rm HC_7N$ @ North to hotcore  ($\rm T_{rot}$=155 K; $\rm N_{mol}=1.0\times10^{16}~cm^{-2}$; $\rm V_{lsr}=\rm 6.3~km\,s^{-1}$; $\rm \Delta V=5~km\,s^{-1}$)}\\
\includegraphics[width=9cm]{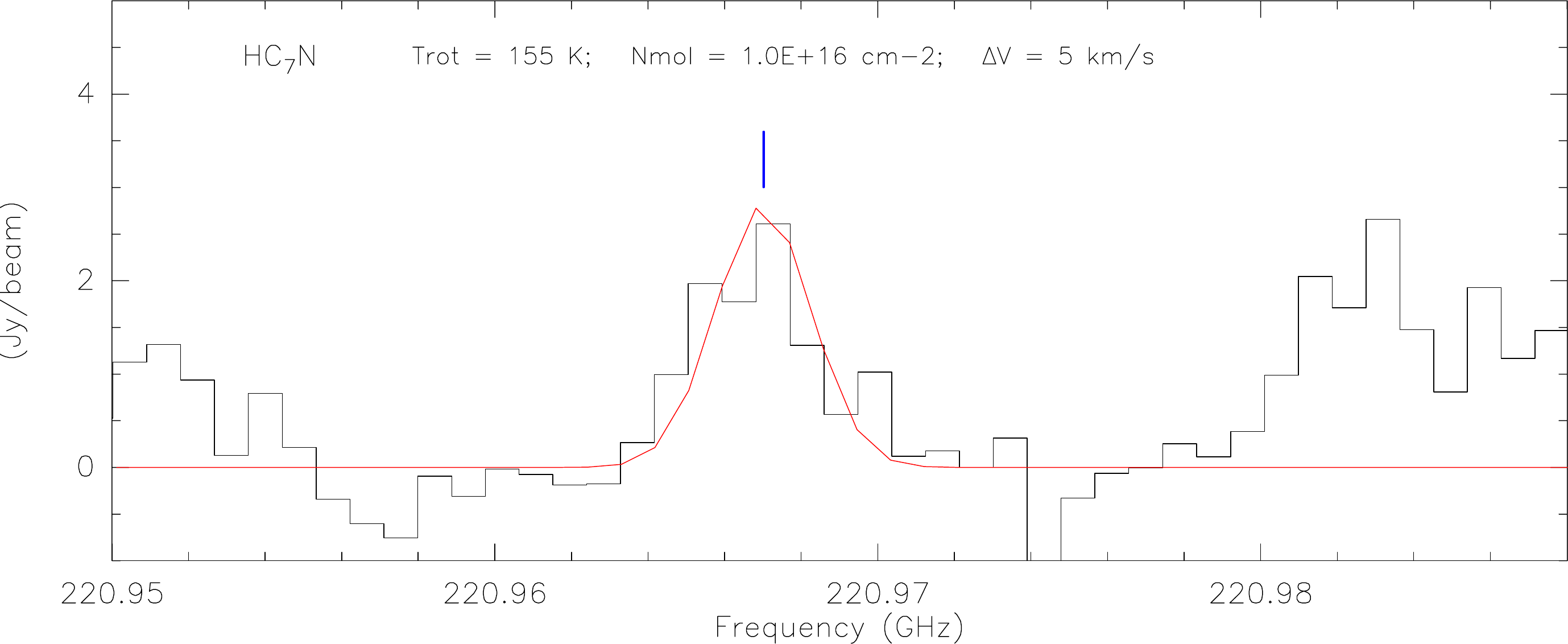}
&\includegraphics[width=9cm]{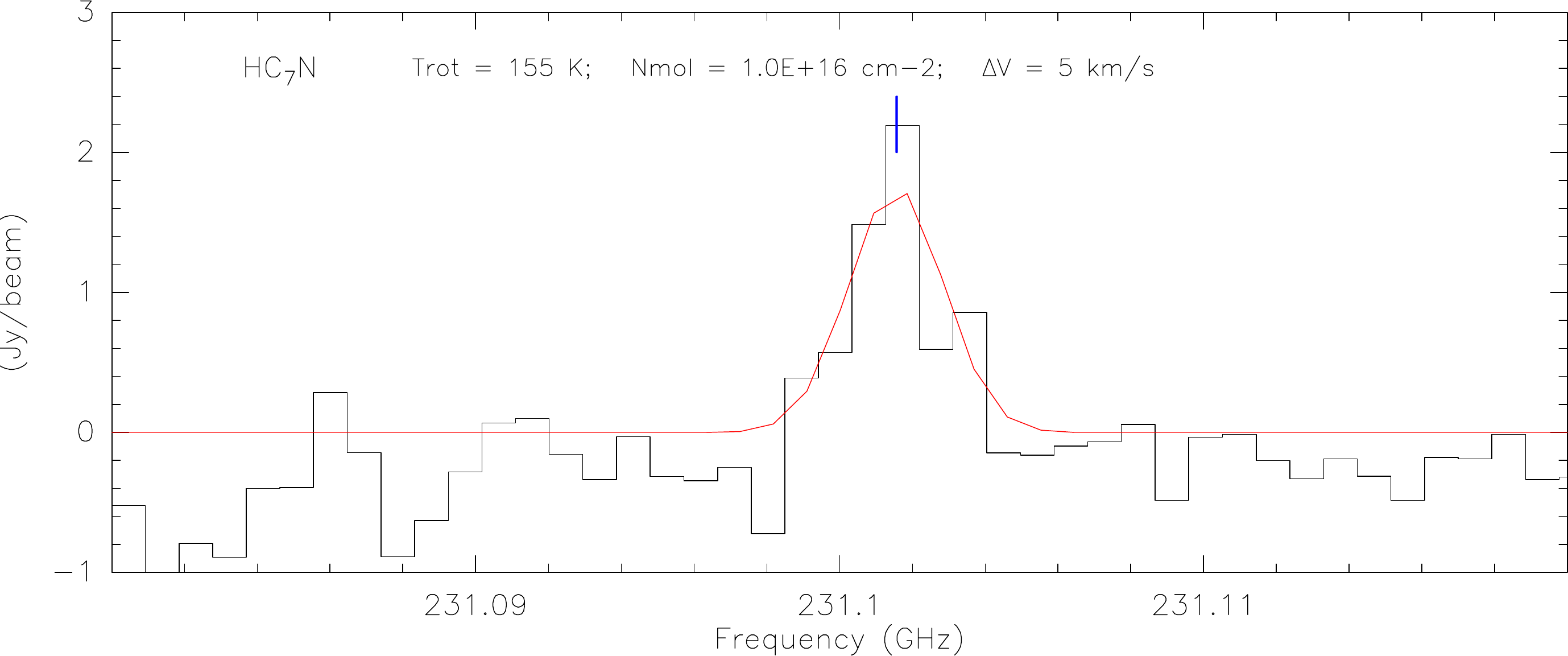}\\

\end{tabular}
\end{center}
\caption{(continued) }
\end{figure*}

%%%%%%%%%%%%%%%%%%%%%%%%%%
%%%%%%%%%%%%%%%%%%%%%%%%%%
\newpage

%\begin{longtable}{c}
%\begin{figure*} [htb]
%\small
%\begin{center}
\begin{figure*}[htb]
\begin{center}
\includegraphics[angle=270,scale=1.0] {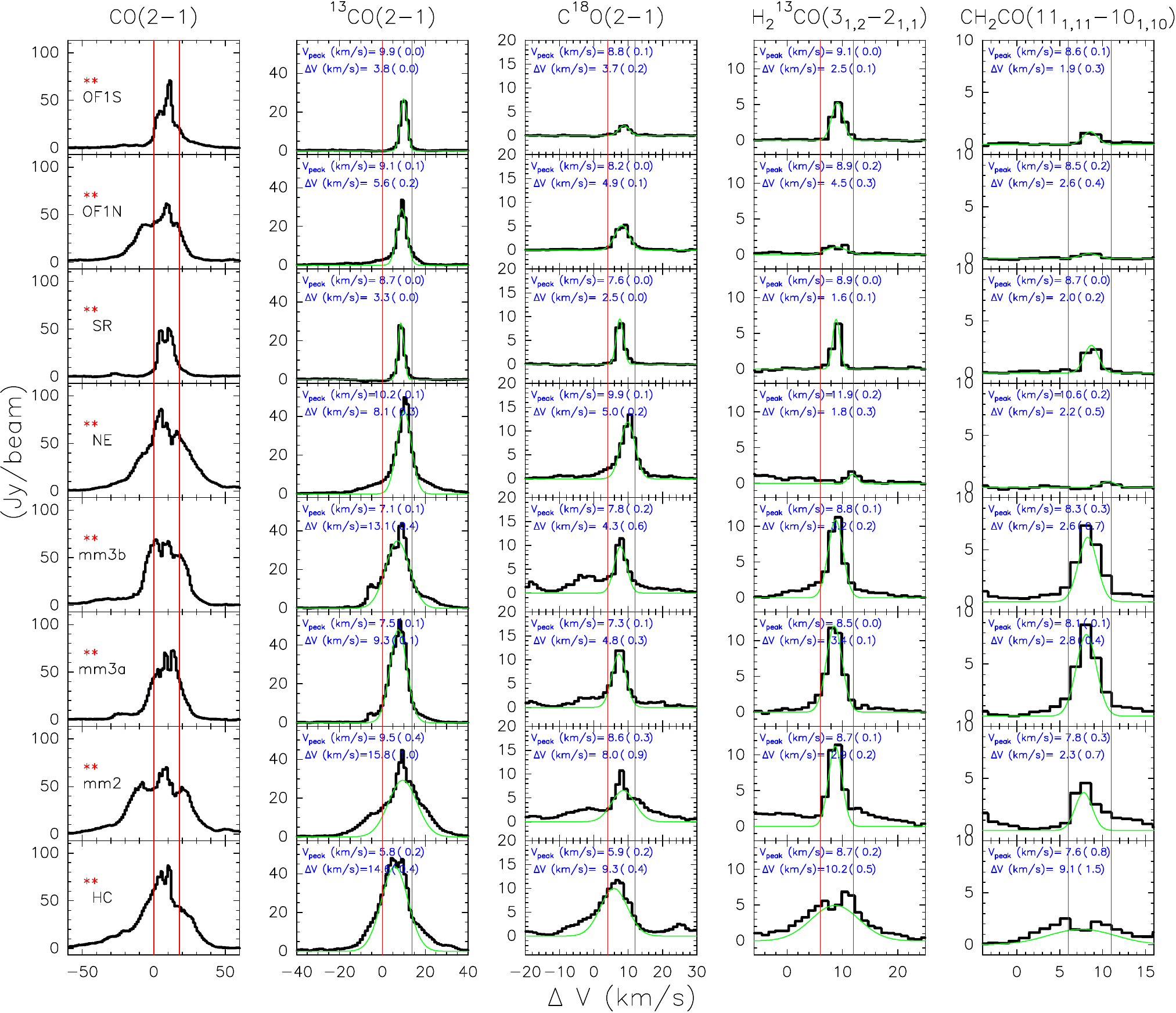}
\caption{Line profiles of identified species from each continuum peak. In measuring the main $\rm V_{peak}$ and $\rm \Delta V$ of each line at each substructure, lines in green and notes in blue (in the form of ``value(error)") are given by the Gaussian fittings using Gildas software. Multi-Gaussian fittings are used to $\rm CH_3CN$, $\rm CH_3^{13}CN$, $\rm HCOOCH_{3~(\nu=0)}$, and $\rm CH_3COCH_3$ due to blending, and to $\rm SO_2$, $\rm ^{34}SO_2$, and SO due to the second velocity component (the second velocity component of  HNCO measured from multi-Gaussian fittings are given in Figure~\ref{wing}). Single Gaussian fittings are used to the rest lines, while lines whose $\rm V_{peak}$ cannot be measured by Gaussian fittings owing to strong self-absorption are marked with ``**", and non-detections are not marked.  Two red vertical lines in each panel mark the velocity range over which the emission from each species is integrated to make the distribution maps shown in  Figures~\ref{into} and \ref{COMdis}.
}\label{velpro}
\end{center}
\end{figure*}

\begin{figure*}[htb]
\ContinuedFloat
\begin{center}

\includegraphics[angle=270,scale=1.0] {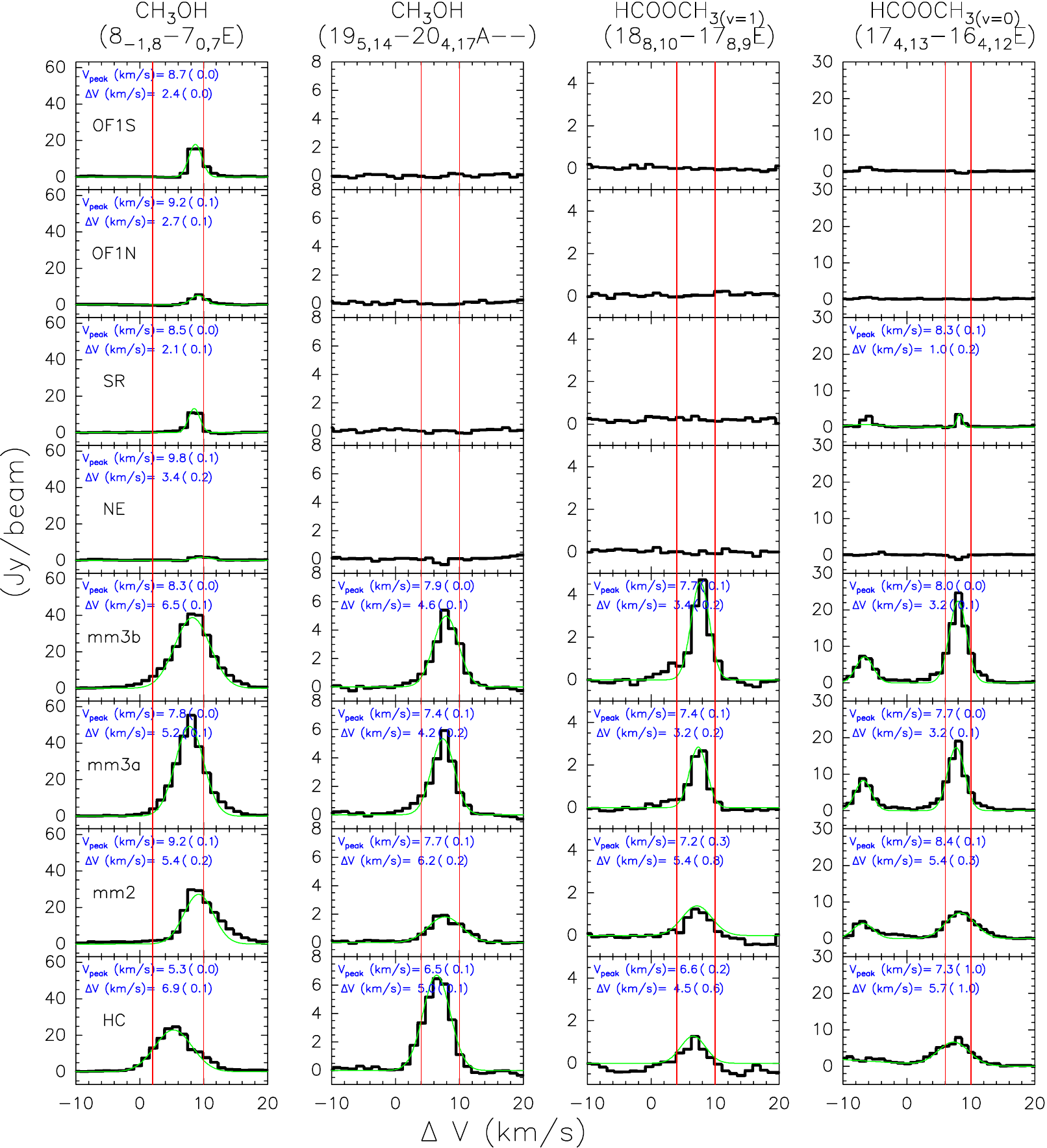}

\caption{(continued)}
\end{center}
\end{figure*} 

\begin{figure*}[htb]
\ContinuedFloat
\begin{center}

\includegraphics[angle=270,scale=1.0] {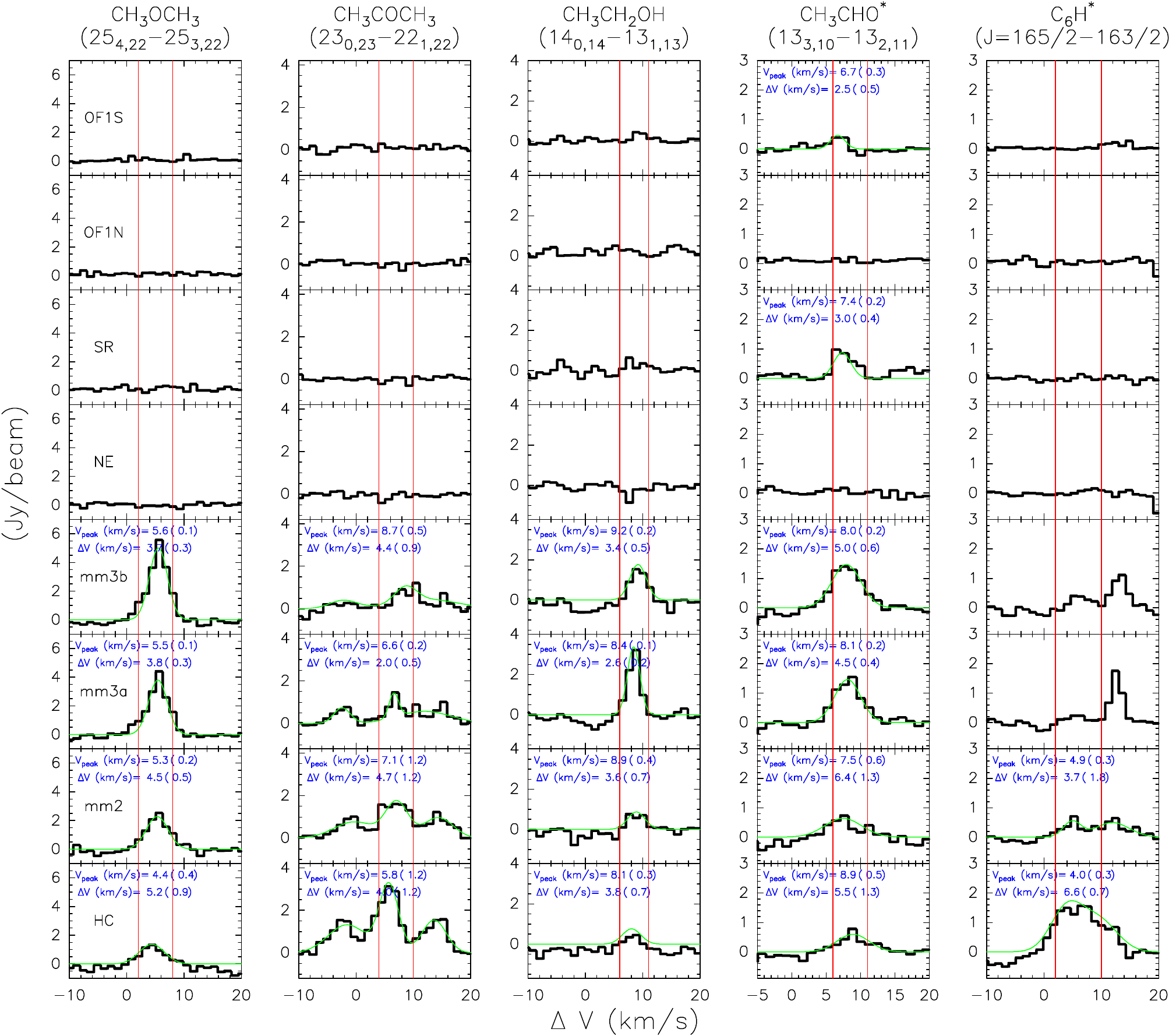}

\caption{(continued)}
\end{center}
\end{figure*}

\begin{figure*}[htb]
\ContinuedFloat
\begin{center}
\includegraphics[angle=270,scale=0.9] {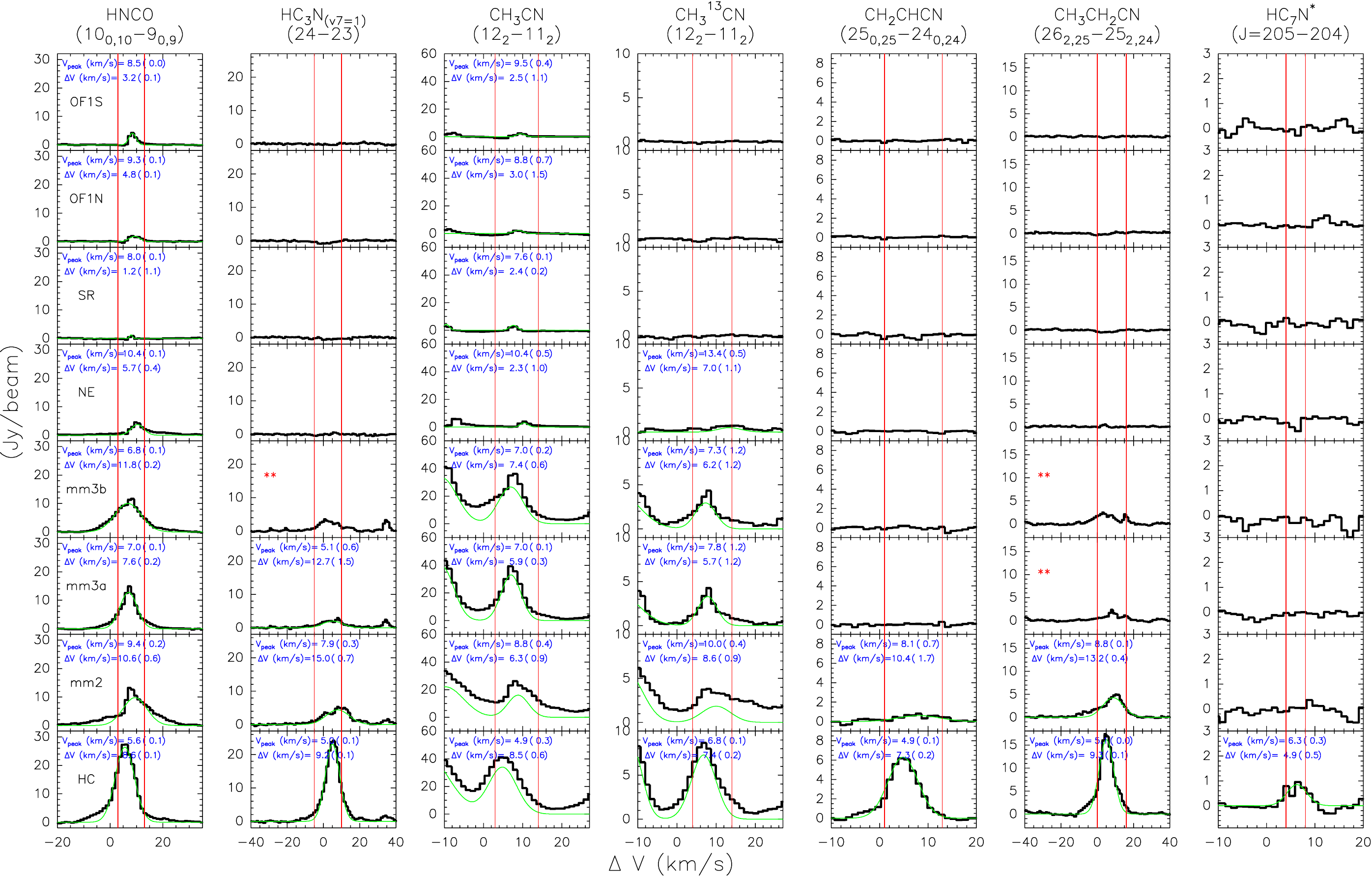}
\caption{(continued)}

\end{center}
\end{figure*} 

\begin{figure*}[htb]
\ContinuedFloat
\begin{center}
\includegraphics[angle=270,scale=1.0] {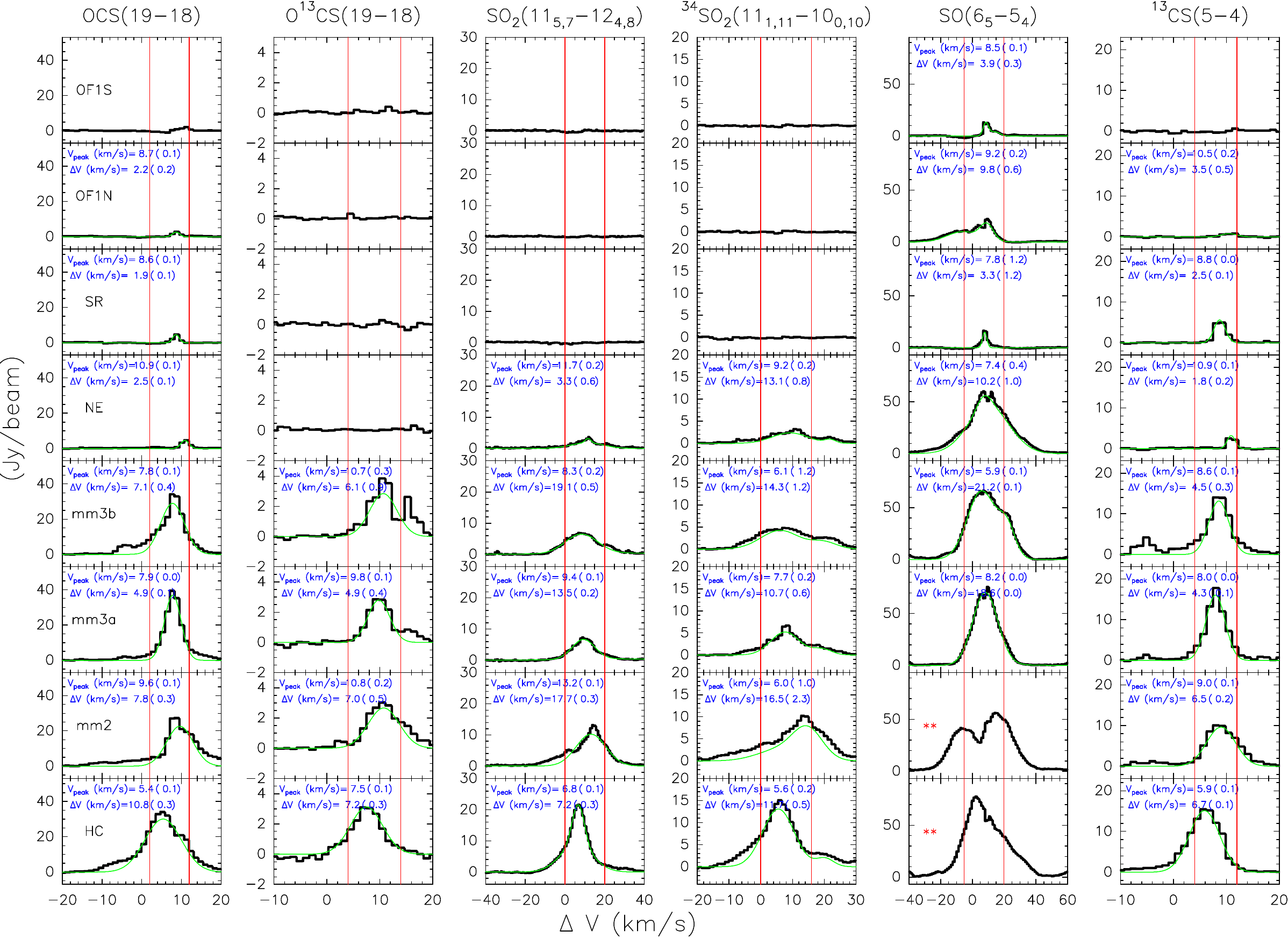}\\
\caption{(continued)}
\end{center}
\end{figure*}

%\end{longtable}

\newpage
\begin{figure*}
\includegraphics{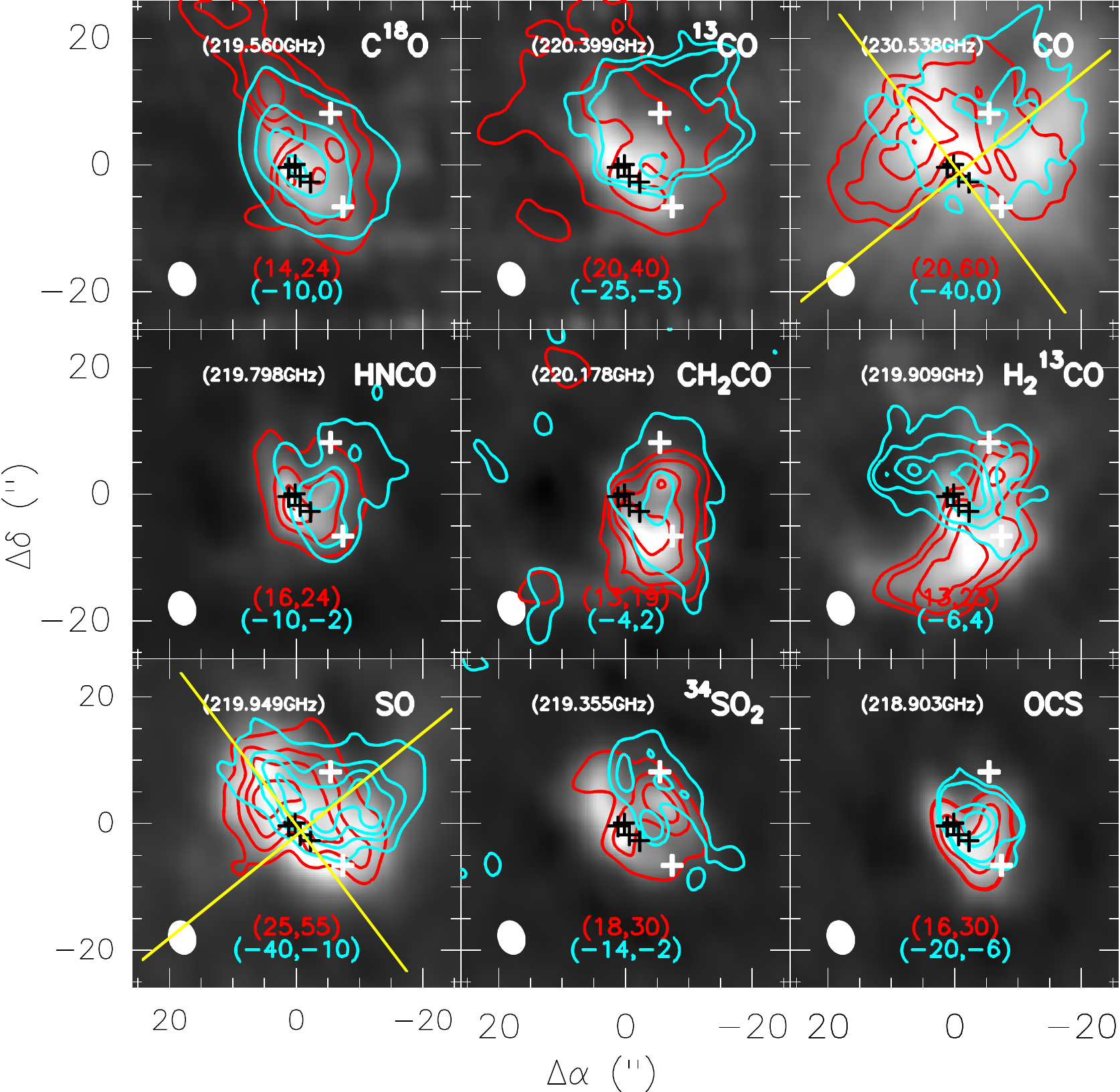}
\caption{ Line wing flux integration maps from the combined SMA-30\,m data at 1.3\,mm.  The grey maps show the intensity integrated ranging through the main emission of their linewidth in Figure~\ref{velpro}. Red and blue contours show the red-shifted and blue-shifted gas, with the intensity integrated as labelled. Yellow lines sketch the outflow directions. Black crosses denote the hotcore, SMA1, Source I, and Source N; white crosses denote the BN object and CR.}\label{shift}
\end{figure*} 

\newpage
\begin{figure*}
\begin{tabular}{ll}
\includegraphics[width=9cm, angle=0]{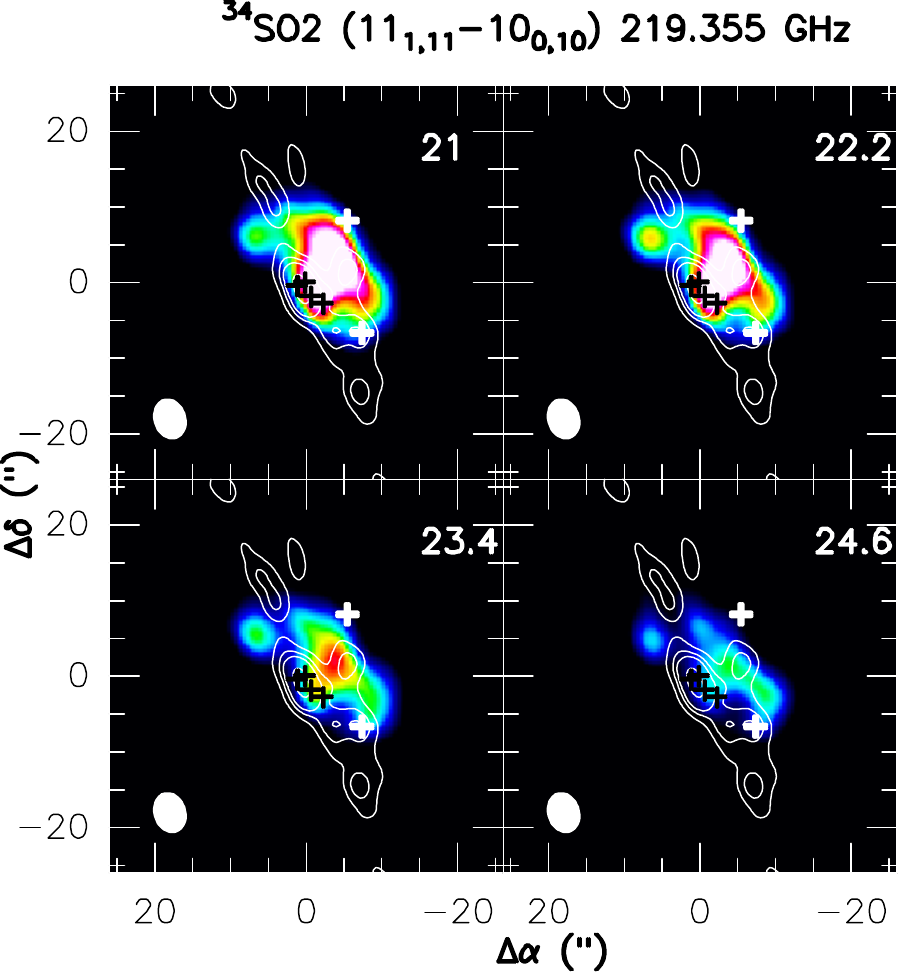}
&\includegraphics[width=9cm, angle=0]{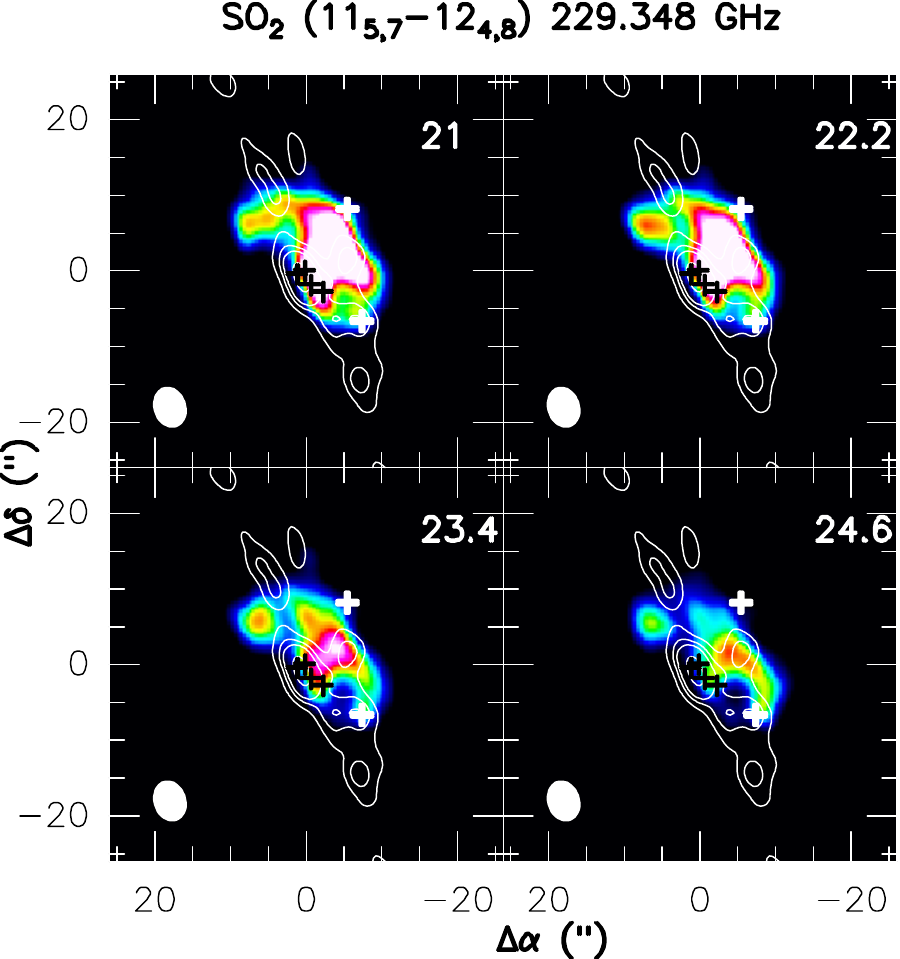}\\
\includegraphics[width=9cm, angle=0]{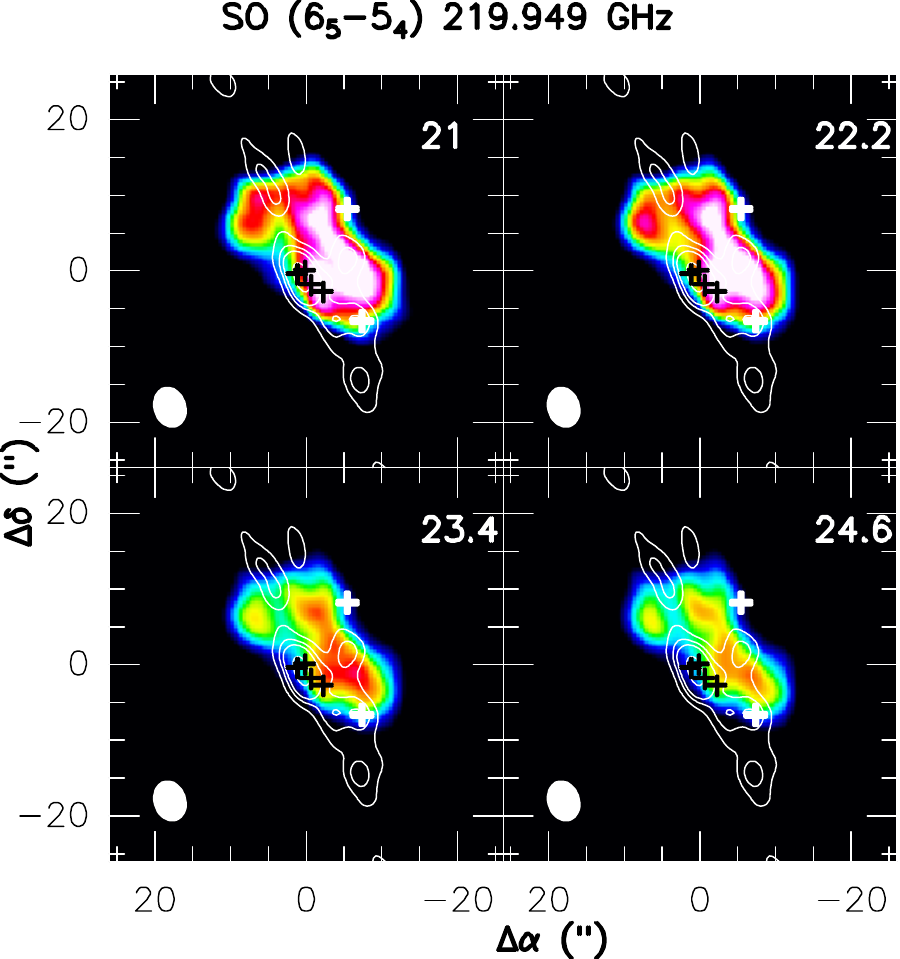}
\end{tabular}
\caption{ Channel maps of the second velocity component ($\rm V_{peak}\sim 22~km~s^{-1}$) of the strongest unblended $\rm ^{34}SO_2$ and  $\rm SO_2$ line, as well as of $\rm SO (6_5\rightarrow 5_4)$ .  
}\label{so2_cha}
\end{figure*}

\newpage
\begin{figure*}
\begin{tabular}{ll}
\includegraphics[width=9cm, angle=0]{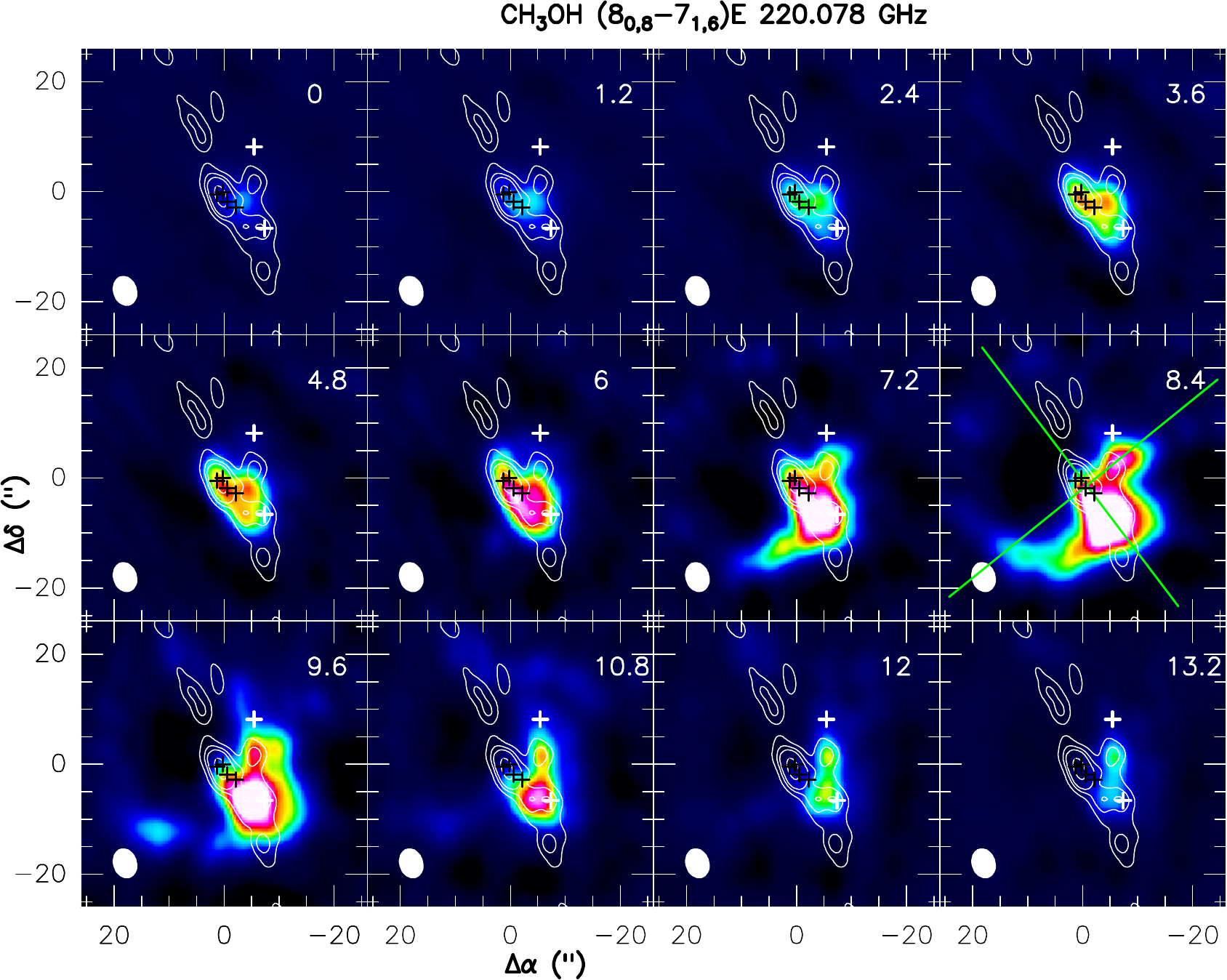}
&\includegraphics[width=9cm, angle=0]{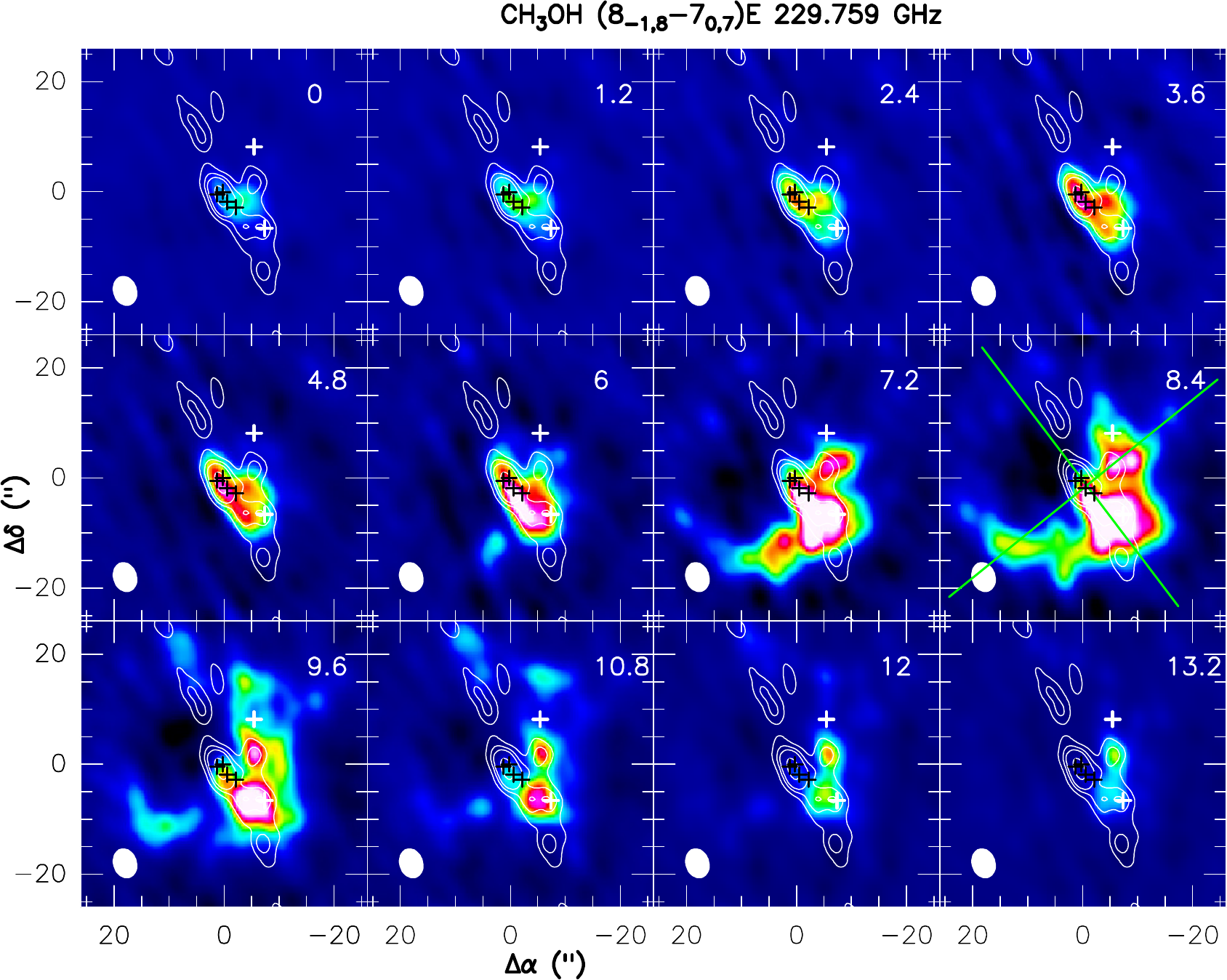}\\
\includegraphics[width=9cm, angle=0]{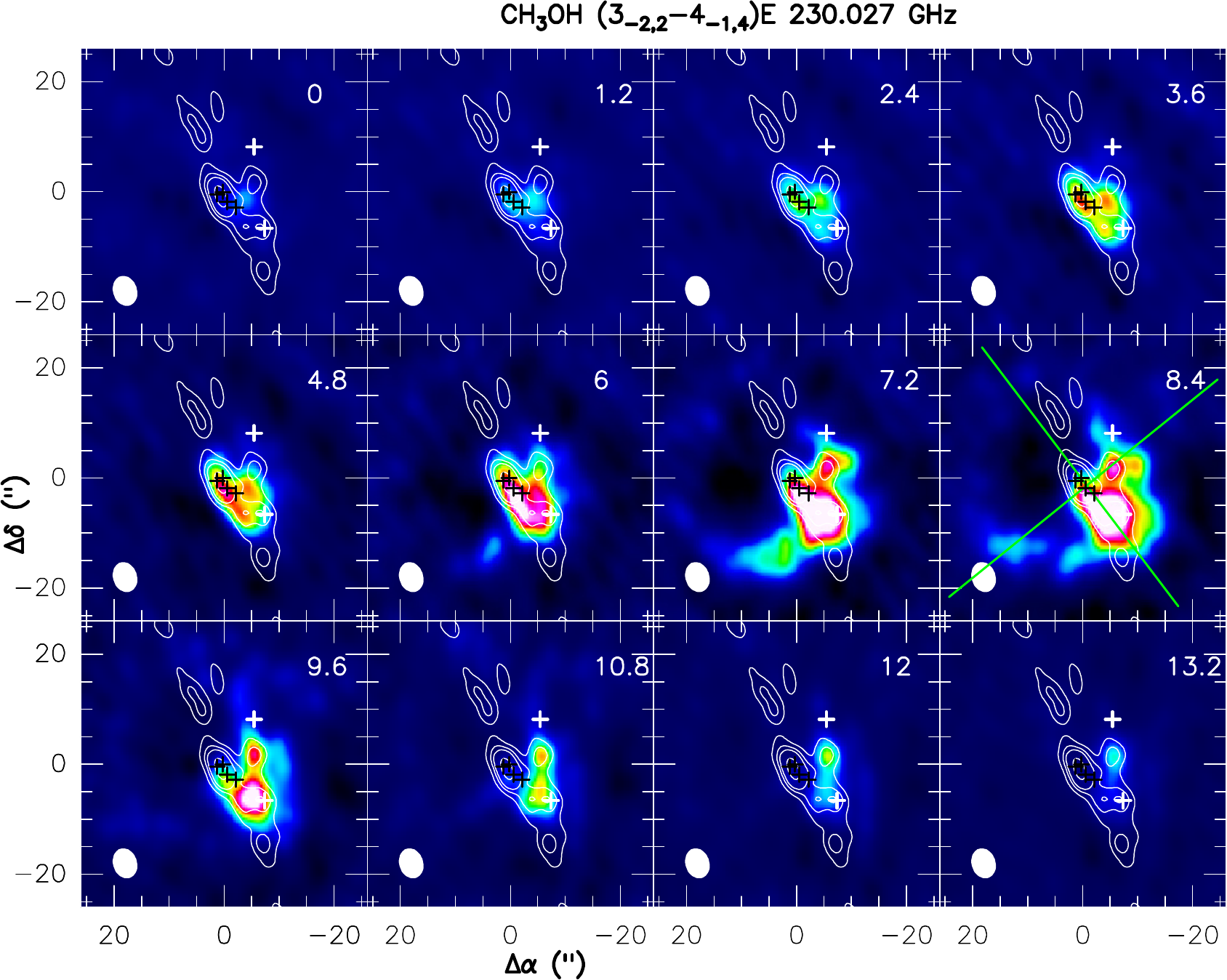}
&\includegraphics[width=9cm, angle=0]{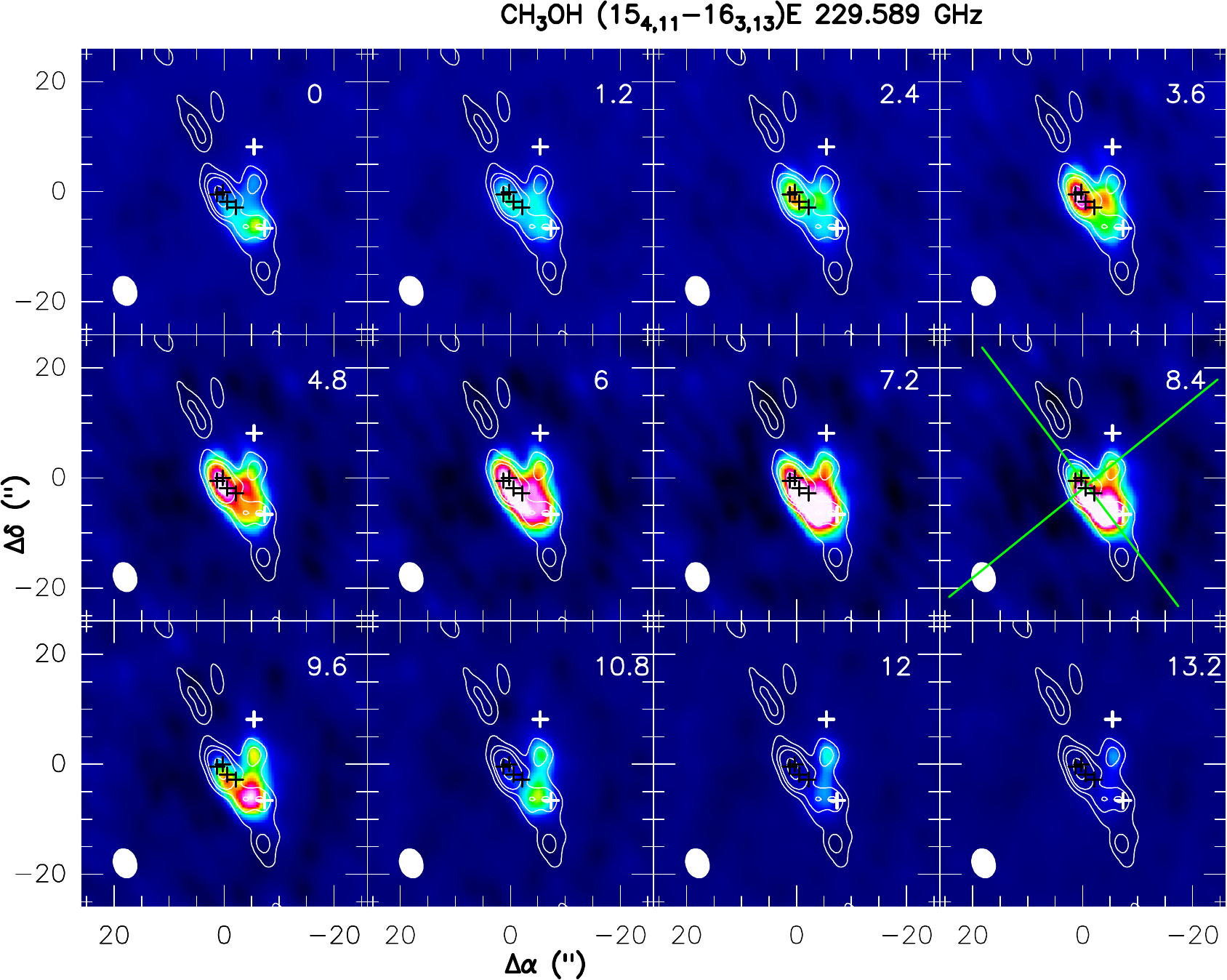}\\
\includegraphics[width=9cm, angle=0]{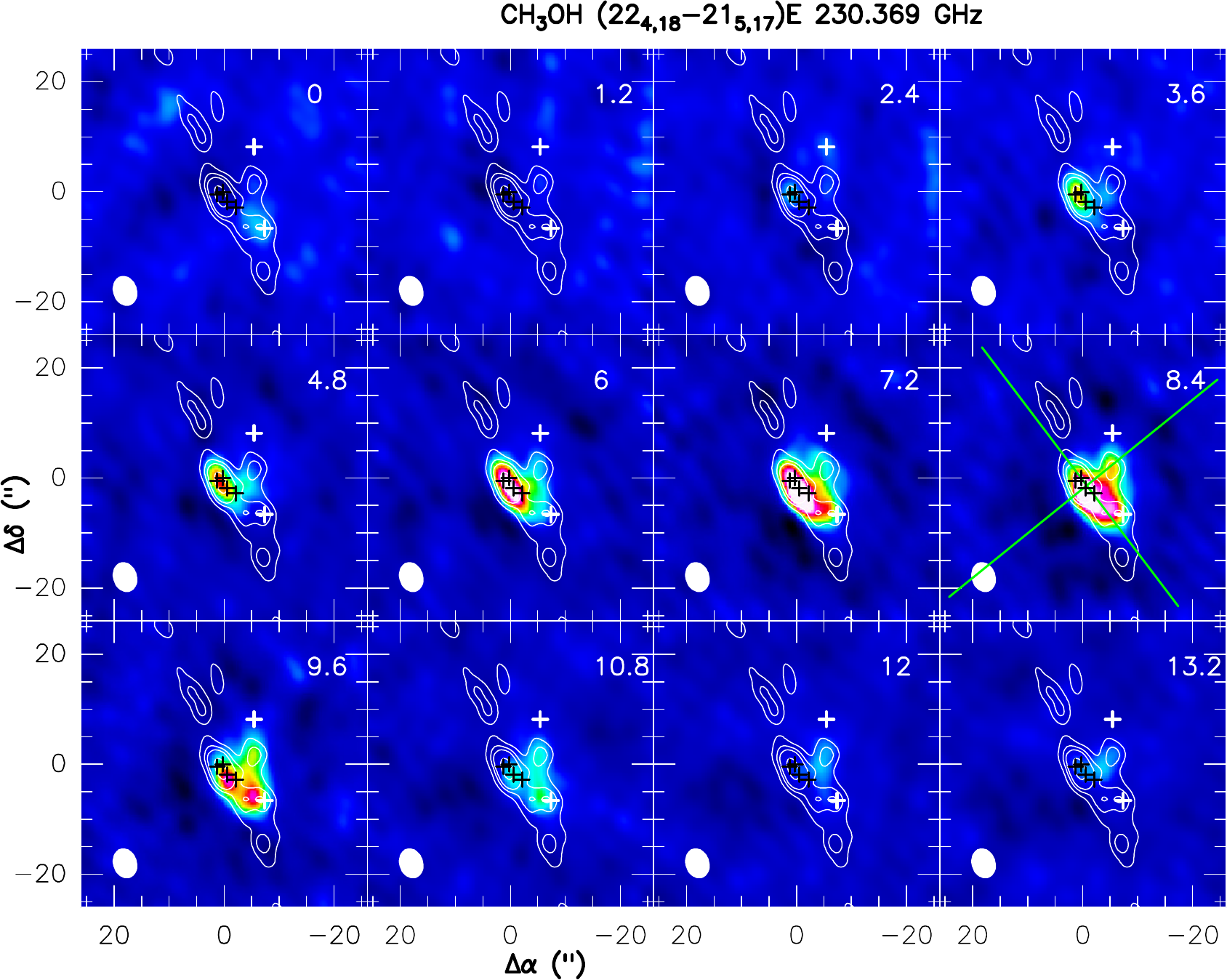}
&\includegraphics[width=9cm, angle=0]{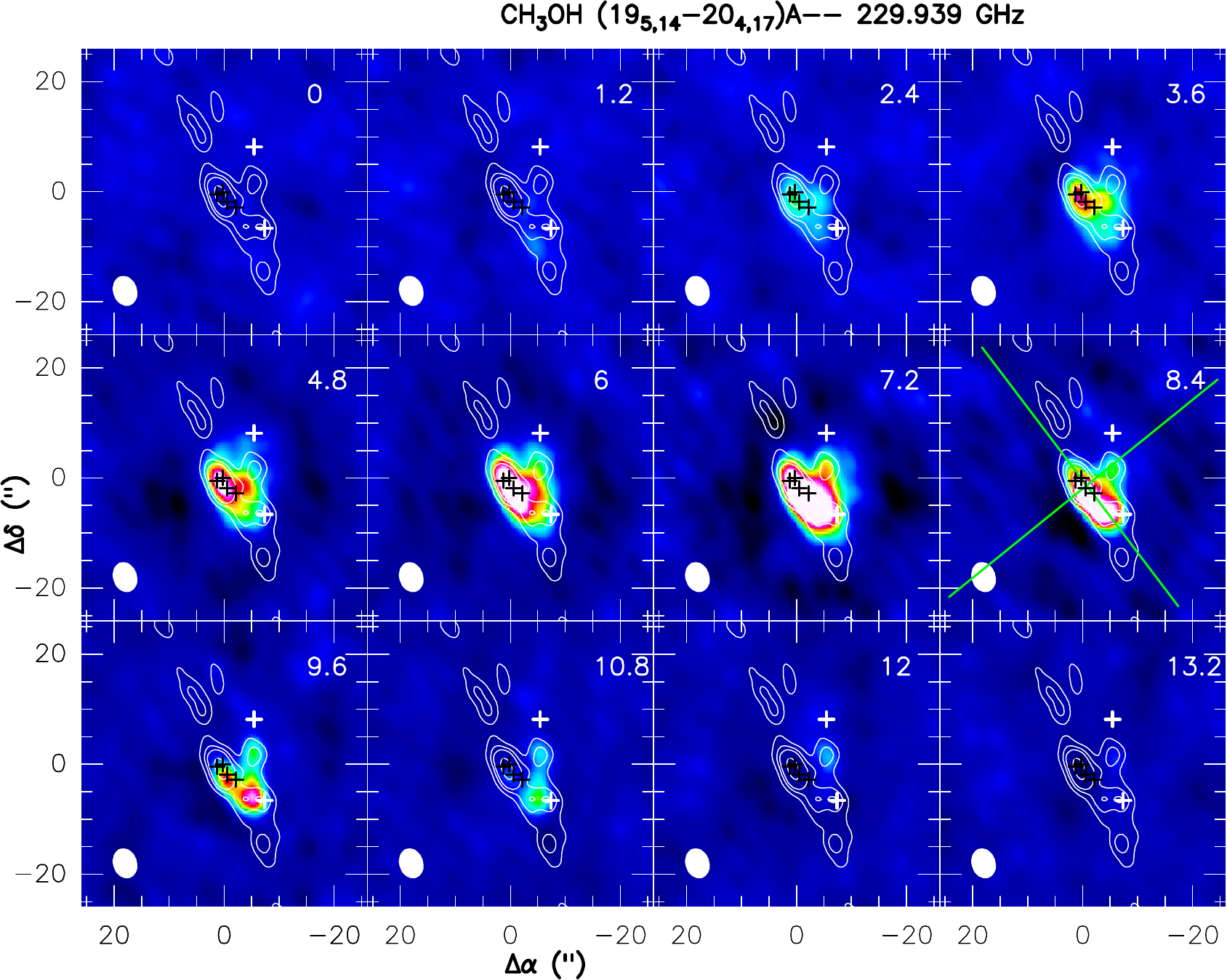}
\end{tabular}
\caption{ Channel maps of unblended $\rm CH_3OH$ lines show multiple velocity-dependent emission peaks:  $\rm 2.4-3.6~km\,s^{-1}$ towards SMA1, $\rm \sim 4.8-6~km\,s^{-1}$ towards HC, $\rm 7.2-9.6~km\,s^{-1}$ towards HC/mm3a/mm3b, $\rm 10.8~km\,s^{-1}$ towards mm3a/mm3b/mm2 and $\rm >12~km\,s^{-1}$ towards mm2. Green lines sketch the outflow directions. Black crosses denote the hotcore, SMA1, Source I, and Source N; white crosses denote the BN object and the CR.
}\label{ch3oh_cha}
\end{figure*}

%%%%%%%%%%%%%%%%%%%%%%%%%%
%%%%%%%%%%%%%%%%%%%%%%%%%%

\end{document}